\documentclass[
14.5pt, 
twoside,
english, 
doblespacing, 
parskip, 
headsepline, 
chapterinoneline, 
book]{MastersDoctoralThesis} 

\usepackage[utf8]{inputenc} 
\usepackage[T1]{fontenc} 

\usepackage{mathpazo} 
\usepackage[all]{xy}

\usepackage{amssymb,amsmath,amsfonts,amsthm}

\usepackage{caption}
\usepackage{subcaption}

\def\be{\begin{equation}}
\def\ee{\end{equation}}
\def\bea{\begin{eqnarray}}
\def\eea{\end{eqnarray}}

\theoremstyle{plain}
\newtheorem{theorem}{Theorem}[chapter]

\newtheorem{lemma}[theorem]{Lemma}
\newtheorem{proposition}[theorem]{Proposition}
\newtheorem{definition}[theorem]{Definition}
\theoremstyle{definition}

\newtheorem{example}[theorem]{Example}
\newcommand{\sect}[1]{\setcounter{equation}{0}\section{#1}}
\newcommand{\subsect}[1]{\subsection{#1}}

\renewcommand{\theequation}{\arabic{section}.\arabic{equation}}
\theoremstyle{remark}
\newtheorem*{remark}{Remark}

\def\dd{{\rm d}}
 \def\lh{{\mathcal{H}}_\omega}
  
  \def\qlhz{{\mathcal{H}}_{z,\omega} }
  \def\shc{\,{\rm sinhc}}
  \def\sinc{\,{\rm sinc}}
 
\def\1{\'{\i}}

\def\eee{{\rm e}}

 \def\xx{u}
 \def\yy{v}

\thesistitle{A geometric approach to Lie systems:  formalism of Poisson--Hopf algebra deformations} 
\supervisor{Dr. Otto Rutwig Campoamor-Stursberg, \qquad \qquad Dr. Francisco José Herranz Zorrilla} 
\examiner{} 
\degree{Doctor of Philosophy in Mathematics} 
\author{\textbf{Eduardo \textsc{Fern\'andez-Saiz}}} 
\addresses{} 

\subject{Mathematics} 
\university{Universidad Complutense de Madrid} 
\department{Department of Algebra, Geometry and Topology} 
\group{Research Group Name} 
\faculty{Faculty of Mathematical Science} 

\begin{document}

\frontmatter 

\pagestyle{plain} 


\begin{titlepage}
\begin{center}

\vspace*{.01\textheight}
{\scshape\LARGE \univname\par}\vspace{0.4cm} 
\includegraphics[height=7.0cm]{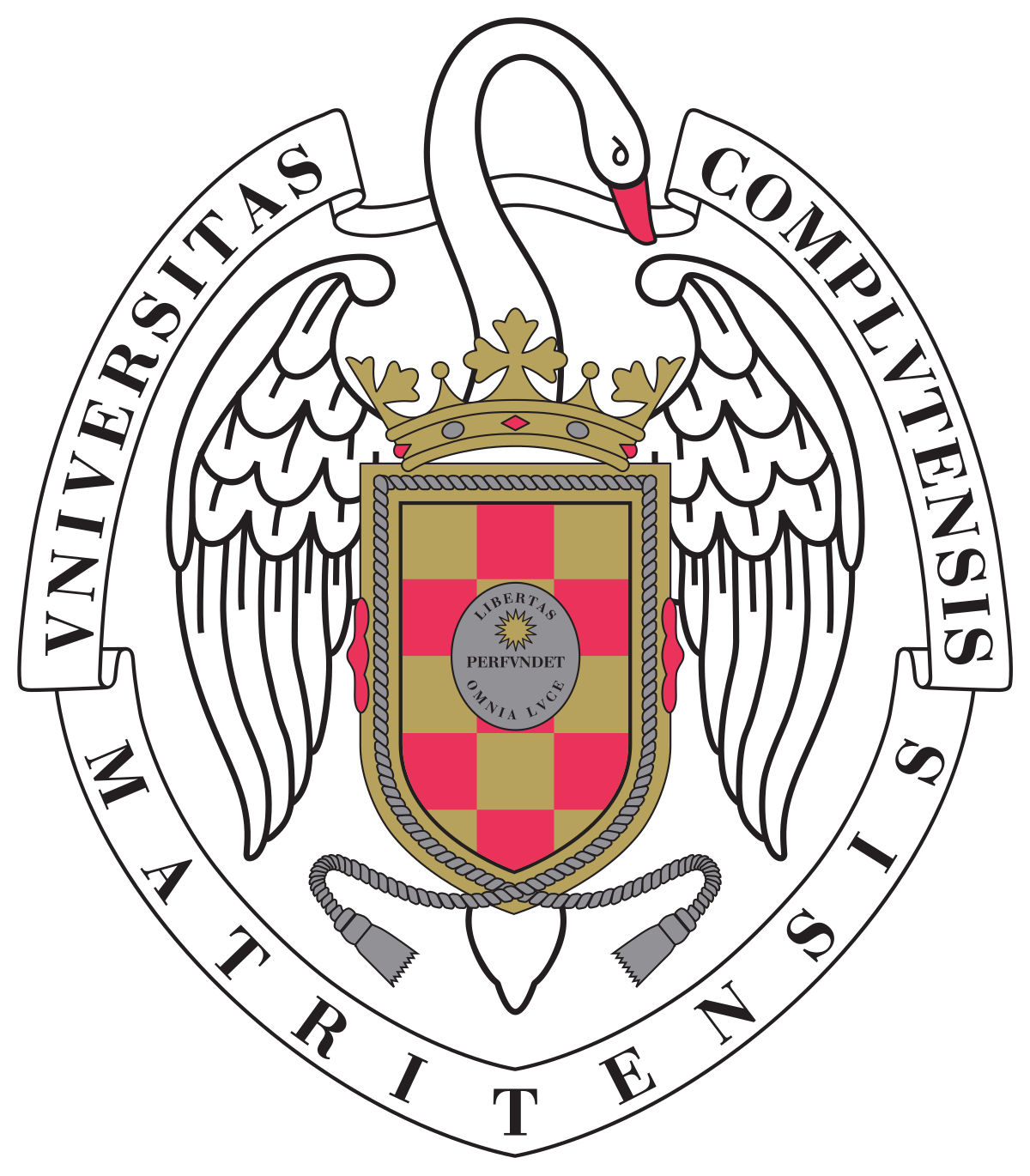}\\ 
\textsc{\Large Ph. D. Thesis}\\[0.2cm] 

\HRule \\[0.35cm] 
{\huge \bfseries \ttitle\par}\vspace{0.3cm} 
{\bfseries \LARGE (Un enfoque geométrico a los sistemas de Lie: formalismo de las deformaciones de álgebras de Poisson--Hopf)}
\HRule \\[0.40cm] 
 
\begin{minipage}[t]{0.30\textwidth}
\begin{flushleft} \large
\emph{Author:}\\
{\authorname} 
\end{flushleft}
\end{minipage}
\begin{minipage}[t]{0.4\textwidth}
\begin{flushright} \large
\emph{Supervisors:} \\
{\supname}\\
\end{flushright}
\end{minipage}\\[2cm]
 
\large \textit{A thesis submitted in fulfillment of the requirements\\ for the degree of \degreename}\\[0.40cm] 
 
Madrid, 2020


\vfill
\end{center}
\end{titlepage}


\begin{declaration}
\addchaptertocentry{\authorshipname} 
\vfill
\noindent D. \authorname, estudiante en el Programa de Doctorado Investigación Matemática,
de la Facultad de Ciencias Matemáticas de la Universidad Complutense de Madrid, como autor de la tesis presentada para la obtención del título de Doctor y titulada: A geometric approach to Lie systems: formalism of Poisson-–Hopf algebra deformations (Un enfoque geométrico a los sistemas de Lie: formalismo de las deformaciones de álgebras de Poisson--Hopf), y dirigida por: Dr. Otto Rutwig Campoamor Stursberg y Dr. Francisco José Herranz Zorrilla.
\vfill

DECLARO QUE:
\newline
La tesis es una obra original que no infringe los derechos de propiedad intelectual ni los derechos de propiedad industrial u otros, de acuerdo con el ordenamiento jurídico vigente, en particular, la Ley de Propiedad Intelectual (R.D. legislativo 1/1996, de 12 de abril, por el que se aprueba el texto refundido de la Ley de Propiedad Intelectual, modificado por la Ley 2/2019, de 1 de marzo, regularizando, aclarando y armonizando las disposiciones legales vigentes sobre la materia), en particular, las disposiciones referidas al derecho de cita.
\newline
\newline
Del mismo modo, asumo frente a la Universidad cualquier responsabilidad que
pudiera derivarse de la autoría o falta de originalidad del contenido de la tesis presentada de conformidad con el ordenamiento jurídico vigente.
\vfill
En Madrid, a 30 de septiembre de 2020
\vfill
Fdo.: Eduardo Fernández Saiz
\end{declaration}

\cleardoublepage


\vspace*{0.2\textheight}
\begin{flushright}
{\itshape By homely gifts and hindered Words\\
The human heart is told\\
Of Nothing -\\
``Nothing'' is the force\\
That renovates the World -}\bigbreak

\hfill Emily Dickinson
\end{flushright}

\newpage
\begin{center}
    {\bfseries \em \Large Abstract}\\ \vspace{0.5cm}
\end{center}
\addcontentsline{toc}{chapter}{Abstract/Resumen}
The notion of quantum algebras is merged with that of Lie systems in order to establish a new formalism called Poisson--Hopf algebra deformations of Lie systems. The procedure can be naturally applied to Lie systems endowed with a symplectic structure, the so-called Lie--Hamilton systems. This is quite a general  approach, as it can be applied to any quantum deformation and any underlying manifold.
One of its main features is that,  under quantum deformations, Lie systems  are  extended to generalized systems described by 
 involutive distributions.  As a consequence, a quantum deformed Lie system  no  longer has an underlying  Vessiot--Guldberg Lie algebra or a quantum algebra one, but keeps a Poisson--Hopf algebra structure that  enables us to obtain, in an explicit way, the $t$-independent constants of the motion from quantum deformed Casimir invariants, which are potentially useful in a further construction of the generalized notion of superposition rules.
We illustrate this     approach by considering  the non-standard quantum deformation of  $\mathfrak{sl}(2)$ applied  to well-known Lie systems, such as the  oscillator problem or Milne--Pinney  equation, as well as several  types of Riccati equations.  In this way, we obtain their new  generalized (deformed) counterparts that cover, in particular,  a new oscillator system  with a time-dependent frequency and a position-dependent mass.
Based on a recently developed procedure to construct Poisson--Hopf deformations of Lie--Hamilton systems ~\cite{Ballesteros5}, a novel unified approach 
to non\-equivalent deformations of Lie--Hamilton systems on the real plane with  a Vessiot--Guldberg Lie algebra isomorphic to $\mathfrak{sl}(2)$ is proposed. This, in particular, allows us to define a notion of Poisson--Hopf systems in dependence of a parameterized family of Poisson algebra representations \cite{BCFHL}. Such an approach is explicitly illustrated by applying it to the three non-diffeomorphic classes of  $\mathfrak{sl}(2)$ Lie--Hamilton systems.
Furthermore $t$-independent constants of motion are given as well.   Our methods can be employed to generate other Lie--Hamilton systems and their deformations for other Vessiot--Guldberg Lie algebras and their deformations. In addition, we study the deformed systems obtained from Lie--Hamilton systems associated to the oscillator algebra $\mathfrak{h}_4$, seen as a subalgebra of the 2-photon algebra $\mathfrak{h}_6$. As a particular application, we propose an epidemiological model of SISf type that uses the solvable Lie algebra $\mathfrak{b}_2$ as subalgebra of  $\mathfrak{sl}(2)$, by restriction of the corresponding quantum deformed systems. 
\vspace{1cm}


\newpage

\begin{center}

    {\bfseries \em \Large Resumen}\\ \vspace{0.5cm}
    
\end{center}

La noción de álgebras cuánticas se fusiona con la de sistemas de Lie para establecer un nuevo formalismo, las deformaciones del álgebra de Poisson--Hopf de los sistemas de Lie. El procedimiento puede aplicarse a sistemas de Lie dotados de una estructura simpléctica, los denominados sistemas de Lie--Hamilton. Este es un enfoque bastante general, ya que se puede aplicar a cualquier deformación cuántica y a cualquier variedad subyacente.
Una de sus principales características es que, bajo deformaciones cuánticas, los sistemas de Lie se extienden a distribuciones involutivas gene\-ra\-lizadas. Como consecuencia, un sistema de Lie deformado cuánticamente ya no tiene un álgebra de Vessiot--Guldberg Lie subyacente o un álgebra cuántica, sino que mantiene una estructura de álgebra de Poisson--Hopf que permite obtener, de manera explícita, las constantes del movimiento $t$-independientes a partir de los invariantes de Casimir deformados, que son potencialmente útiles en una construcción adicional de la noción generalizada de reglas de superposición.
Ilustramos este enfoque considerando la deformación cuántica no estándar de $\mathfrak{sl}(2)$ aplicada a sistemas de Lie conocidos, como el problema del oscilador o la ecuación de Milne--Pinney, así como varios tipos de ecuaciones de Riccati. De esta manera, se obtienen sus análogos generalizados (deformados) que dan lugar, en particular, a un nuevo sistema de tipo oscilatorio con una frecuencia dependiente del tiempo y una masa dependiente de la posición.
Basándonos en este procedimiento, desarrollado recientemente en \cite{Ballesteros5}, se presentan de modo unificado las deformaciones no equivalentes de los sistemas de Lie--Hamilton en el plano real con un álgebra de Lie de Vessiot--Guldberg isomorfa a $\mathfrak{sl}(2)$. Esto, en particular, nos permite definir una noción de sistemas de Poisson--Hopf en dependencia de una familia parametrizada de representaciones de álgebras de Poisson \cite{BCFHL}. Este enfoque se ilustra explícitamente aplicándolo a las tres clases no difeomórficas de los sistemas $\mathfrak{sl}(2)$ Lie--Hamilton. Se dan las constantes de movimiento independientes de $t$. Nuestros métodos se pueden emplear para generar otros sistemas de Lie--Hamilton y sus deformaciones para otras álgebras de Lie de Vessiot--Guldberg y sus deformaciones. An\'alogamente, se estudian los sistemas deformados a partir de los sistemas de Lie--Hamilton basados en el \'algebra del oscilador $\mathfrak{h}_4$, vista como sub\'algebra del \'algebra de Lie $\mathfrak{h}_6$, la llamada 2-photon algebra. Como aplicaci\'on adicional, se estudian modelos epidemiol\'ogicos del tipo SISf obtenidos como deformaci\'on cu\'antica de sistemas de Lie--Hamilton basados en el \'algebra resoluble en $\mathfrak{b}_2$, pero vista como sub\'algebra de $\mathfrak{sl}(2)$. 

\vspace{1cm}

\noindent \textbf{Palabras clave:} sistema de Lie, álgebra de Poisson--Hopf, coálgebra de Poisson, deformación cuántica.


\newpage

\section*{Breve extracto de la tesis}
\addcontentsline{toc}{chapter}{Breve extracto de la tesis}

\vspace{1cm}

\subsubsection{Introducci\'on y objetivos}

\vspace{0.5cm}

Desde su formulación original por Lie \cite{Lie}, los sistemas no autónomos de primer orden de ecuaciones diferenciales ordinarias que admiten una regla de superposición no lineal, los llamados sistemas de Lie, se han estudiado extensamente (ver \cite{C135, CGM00, Dissertations, PW} y sus referencias). El teorema de Lie \cite{Lie} establece que todo sistema de ecuaciones diferenciales de primer orden es un sistema de Lie si, y solo si, puede describirse como una curva en un álgebra de Lie de dimensión finita de campos vectoriales, la denominada  álgebra de Lie de Vessiot--Guldberg.
Aunque ser un sistema de Lie es más una excepción que una regla \cite{CGL10}, se ha demostrado que los sistemas de Lie son de gran interés dentro de las aplicaciones físicas y matemáticas (ver \cite{Dissertations} y sus referencias). Los sistemas de Lie que admiten un álgebra de Lie de Vessiot--Guldberg de campos vectoriales hamiltonianos, en relación con una estructura de Poisson, los sistemas de Lie--Hamilton, han encontrado incluso más aplicaciones que los sistemas de Lie estándar sin esta estructura geométrica asociada \cite{BBHLS, BCHLS13Ham, CLS13}. Los sistemas de Lie--Hamilton admiten un álgebra de Lie de dimensión finita adicional de funciones hamiltonianas, un álgebra de Lie--Hamilton, que permite la determinación algebraica de reglas de superposición y constantes del movimiento \cite{BHLS}.

La mayoría de los enfoques de los sistemas de Lie se basan  en la teoría de las álgebras de Lie y los grupos de Lie \cite{OV}. Sin embargo, el éxito de los grupos cuánticos \cite{Chari, Majid} y el formalismo de coálgebra dentro del análisis de sistemas superintegrables \cite{coalgebra2, coalgebra1}, y el hecho de que las álgebras cuánticas aparezcan como deformaciones de las álgebras de Lie sugirieron la posibilidad de ampliar la noción y técnicas de los sistemas de Lie--Hamilton más allá de la teoría de Lie. En esta memoria se propone un enfoque en esta dirección (capítulo \ref{Chapter3}, \cite{Ballesteros5}), donde se da un método para construir sistemas de Lie--Hamilton deformados cuánticamente (sistemas LH en resumen) mediante el formalismo de coálgebra y álgebras cuánticas.
La idea subyacente es utilizar la teoría de los grupos cuánticos para deformar sistemas de Lie y sus estructuras geométricas asociadas. Más exáctamente, la deformación transforma un sistema LH con su álgebra de Lie de Vessiot--Guldberg en un sistema hamiltoniano cuya dinámica está determinada por un conjunto (finito) de generadores de una distribución generalizada de Stefan--Sussmann, sección \ref{SecDefLHsystem}. Mientras tanto, el álgebra inicial de Lie--Hamilton (álgebra LH en resumen) se identifica con un álgebra de Poisson--Hopf. Las estructuras deformadas permiten la construcción explícita de constantes de movimiento $t$-independientes mediante técnicas de álgebras cuánticas para el sistema deformado.

Este trabajo tiene como objetivo ilustrar el enfoque introducido en \cite{Ballesteros5, BCFHL} para construir deformaciones de los sistemas LH. Esto abarca, fundamentalmente, los siguientes objetivos: desarrollar el formalismo basado en deformaciones de estructuras de Poisson--Hopf, ya que estas estructuras permiten una sistematización adicional que abarca los sistemas LH no equivalentes entre sí, correspondientes a las álgebras isomórficas de LH, y ofrecer un procedimiento general para la obtención de nuevos sistemas LH.

Específicamente, se aporta un algoritmo para la consecución de deformaciones sistemas LH, lo cual pone de manifiesto la importancia del formalismo previamente desarrollado. Además, mostramos que las deformaciones de Poisson--Hopf de los sistemas LH basadas en un álgebra LH isomorfas a $\mathfrak{sl}(2)$ (teorema \ref{MT}), se pueden describir genéricamente, facilitando así las funciones hamiltonianas deformadas y los campos vectoriales hamiltonianos deformados asociados; una vez estudiado el sistema no deformado.

Igualmente, se proporciona un nuevo método para construir sistemas LH con un álgebra LH isomorfa a un álgebra $\mathfrak{g}$ de Lie fija, sección \ref{SecDefsl2}. Nuestro enfoque se basa en el uso de la foliación simpléctica en $\mathfrak{g}$ inducida por el corchete de Kirillov--Konstant--Souriau, teorema \ref{MT2}. Como caso particular, se muestra explícitamente cómo nuestro procedimiento explica la existencia de tres tipos de sistemas LH en el plano relacionado con un álgebra LH isomorfa a $\mathfrak{sl}(2)$. Esto se debe al hecho de que cada uno de los tres tipos diferentes corresponde a uno de los tres tipos de hojas simplécticas en $\mathfrak{sl}(2)$. Análogamente, se puede generar el único tipo de sistemas LH en el plano admitiendo un álgebra de Lie de Vessiot--Guldberg isomorfa a $\mathfrak{so}(3)$.

Nuestra sistematización nos permite dar directamente el sistema deformado de Poisson--Hopf a partir de la clasificación de sistemas LH \cite{BBHLS, BHLS}, sugiriendo además una noción de sistemas Poisson--Hopf Lie basados en una familia $z$-parametrizada de morfismos del álgebra de Poisson (definición \ref{defPHsistem} en la sección \ref{PHDefLHSstem}).

\vspace{0.5cm}

\subsubsection{Resultados m\'as relevantes y estructura}

\vspace{0.5cm}

Presentamos un  procedimiento genérico que nos permite introducir la noción de deformación cuántica de
sistemas LH, y basados en la noción de distribuciones involutivas en el sentido de Stefan--Sussman.
La existencia de principios de superposición no lineal para sistemas no autónomos de primer orden de ecuaciones diferenciales ordinarias constituye una propiedad estructural que surge naturalmente del enfoque desde la teoría de grupos para las ecuaciones diferenciales iniciado por Lie, en el contexto del desarrollo del programa geométrico basado en grupos de transformación, así como de la clasificación analítica de ecuaciones diferenciales desarrollada por Painlevé y Gambier, entre otros, dando lugar a lo que hoy se conoce como la teoría de los sistemas de Lie \cite{Lie, VES}.

Recordamos que, más allá de los sistemas superintegrables \cite{coalgebra2, coalgebra1}, las coálgebras se han aplicado recientemente a deformaciones bi-hamiltonianas integrables de sistemas de Lie--Poisson \cite{Ballesteros1} y a deformaciones integrables de los sistemas de Rössler y Lorenz \cite{Ballesteros2}.
Este trabajo propone una gene\-ra\-lización de los sistemas de Lie en esta línea. La idea es considerar sistemas de Lie deformados cuánticamente que poseen una estructura de álgebra de Poisson--Hopf que reemplaza el álgebra de Lie de Vessiot--Guldberg del sistema inicial, lo que nos permite una construcción de constantes del movimiento $t$-independientes expresadas en términos de Casimires invariantes deformados.

La estructura de la tesis es la siguiente. Los capítulos \ref{Chapter1} y \ref{Chapter2} están dedicados a la introducción de los principales aspectos de los sistemas LH y las álgebras de Poisson--Hopf, as\'{\i} como una s\'{\i}ntesis de sus propiedades fundamentales. 
El formalismo general para construir deformaciones del tipo Poisson--Hopf de sistemas LH \cite{Ballesteros5} se analiza con detalle en el capítulo \ref{Chapter3}. En \'el se presenta un procedimiento algor\'{\i}tmico esquematizado para determinar tanto las deformaciones como las constantes del movimiento. 

En el capítulo \ref{Chapter4} se presentan la deformaciones (no estándar)  de sistemas basados en el \'algebra de Lie simple $\mathfrak{sl}(2)$. Un enfoque unificado de las deformaciones de los sistemas Poisson--Hopf Lie con un álgebra LH isomorfa a $\mathfrak{g}$ se estudia en la sección \ref{SeccMEthodConst}. Dicho procedimiento se ilustra explícitamente, aplicándolo a las tres clases no difeomorfas de sistemas LH $\mathfrak{sl}(2)$ en el plano, obteniendo así su correspondiente deformación. En el cap\'{\i}tulo  \ref{Chapter5} se estudian deformaciones de ecuaciones diferenciales relevantes. En primer lugar, se consideran las deformaciones de la ecuaci\'on de Milne--Pinney, de las cuales se derivan nuevos sistemas de tipo oscilatorio con una masa dependiente de la posici\'on. Como segundo tipo relevante, se estudian las ecuaciones de Riccati, espec\'{\i}ficamente las ecuaciones compleja y acoplada. En el cap\'{\i}tulo \ref{Chapter6} se considera el problema de las deformaciones cu\'anticas de sistemas LH basados en el \'algebra del oscilador  $\mathfrak{h}_4$, con \'enfasis especial en su estructura como sub\'algebra de la llamada 2-photon \'algebra  $\mathfrak{h}_6$. Esto nos permite deducir deformaciones del oscilador arm\'onico amortiguado  (sección \ref{h4 damped oscilator}). El \'algebra resoluble af\'{\i}n  $\mathfrak{b}_2$,  vista como sub\'algebra de  $\mathfrak{sl}(2)$, se aplica en el cap\'{\i}tulo  \ref{Chapter8} para proponer un nuevo modelo epidemiol\'ogico alternativo. Las t\'ecnicas desarrolladas con anterioridad se utilizan para construir y analizar un modelo SISf deformado, del que se obtienen las constantes del movimiento. Finalmente, en las conclusiones resumimos los resultados obtenidos y comentamos v\'{\i}as futuras de trabajo o actualmente en proceso de ejecuci\'on.

\vspace{0.5cm}

\subsubsection{Conclusiones}

\vspace{0.5cm}

En este trabajo hemos propuesto una noci\'on de deformaci\'on de sistemas LH basada en las \'algebras de Poisson--Hopf. Este enfoque difiere radicalmente de otros m\'etodos empleados en la teor\'{\i}a de sistemas de Lie ~\cite{BCHLS13Ham, CGL10,Dissertations,CLS13, PW}, dado que las deformaciones no corresponden formalmente a sistemas de Lie, sino a una noci\'on extendida que precisa de una estructura de Hopf, de modo que el sistema sin deformar se obtiene por un paso al l\'{\i}mite en el cual el par\'ametro de deformaci\'on desaparece. La introducci\'on de una estructura de Poisson--Hopf permite una generalizaci\'on de estos sistemas, en el sentido de que el \'algebra finito-dimensional de Vessiot--Guldberg  se reemplaza por una distribuci\'on involutiva en el sentido de  Stefan--Sussman (cap\'{\i}tulo 3). 

En el cap\'{\i}tulo 4 se estudia el an\'alogo de las deformaciones cu\'anticas de  $\mathfrak{sl}(2)$, estableciendo expl\'{\i}citamente las constantes del movimiento para los sistemas deformados cu\'anticamente.  Los tres sistemas de Lie planos no equivalentes basados en el \'algebra de Lie $\mathfrak{sl}(2)$ se describen de modo unificado, lo que proporciona una bella interpretaci\'on geom\'etrica de estos sistemas y sus corres\-pondientes deformaciones cu\'anticas. El cap\'{\i}tulo 5 est\'a dedicado al an\'alisis de sistemas espec\'{\i}ficos de ecuaciones diferenciales y su contrapartida deformada. Consideramos en primer lugar la ecuaci\'on de Milne--Pinney, cuyas deformaciones nos proporcionan nuevos sistemas de tipo oscilatorio con la particularidad de que la masa de la part\'{\i}cula es dependiente de la posici\'on, y donde se obtienen expl\'{\i}citamente las constantes del movimiento. Merece la pena observar que la deformaci\'on est\'andar o de Drinfel'd--Jimbo de $\mathfrak{sl}(2)$ no lleva a un oscilador del tipo mencionado, ya que, en este caso, la deformaci\'on viene descrita por $\!\shc(z q p)$ en lugar de  $\!\shc(z q^2)$. Esto se deduce de la correspondiente realizaci\'on simpl\'ectica dada en ~\cite{BCFHL}. Este hecho justifica que hayamos escogido la deformaci\'on no-est\'andar de $\mathfrak{sl}(2)$, para obtener aplicaciones f\'{\i}sicas.  A pesar de ello, la deformaci\'on de Drinfel'd--Jimbo podr\'{\i}a proporcionar informaci\'on suplementaria para la ecuaci\'on  de Milne--Pinney, dando lugar a sistemas no equivalentes a los estudiados en la memoria. En cualquier caso, el m\'etodo nos sugiere un enfoque alternativo de los sistemas con masa no constante, para los cuales los m\'etodos cl\'asicos son de limitado alcance. Un segundo tipo que se ha estudiado corresponde a las ecuaciones compleja y acoplada de Riccati, que se han analizado exhaustivamente en la lite\-ratura. Para ellos se obtienen la versi\'on deformada y las constantes del movimiento. Los resultados principales de los cap\'{\i}tulos 
 3-5 han sido publicados en \cite{Ballesteros5} y \cite{BCFHL}.   En el cap\'{\i}tulo 6 nos centramos en sistemas de tipo oscilatorio obtenidos a partir de los sistemas LH deformados basados en el \'algebra de Lie $\mathfrak{h}_4$, vista como sub\'algebra de la llamada 2-photon algebra $\mathfrak{h}_6$. En particular, estas deformaciones se obtienen como restricci\'on de los correspondientes sistemas deformados para $\frak{h}_6$. Un ejemplo ilustrativo de este tipo de sistemas viene dado por el oscilador amortiguado deformado. Resta determinar un principio de superposici\'on para tales sistemas, un problema actualmente en ejecuci\'on. En el cap\'{\i}tulo 7 se usa el \'algebra resoluble af\'{\i}n $\mathfrak{b}_2$, vista como sub\'algebra de $\mathfrak{sl}(2)$, para obtener sistemas deformados aplicables en el contexto de los modelos epidemiol\'ogicos. Esta idea constituye una novedad, dado que los m\'etodos empleados habitualmente en este contexto son de naturaleza estoc\'astica. Los resultados de este cap\'{\i}tulo has sido enviados recientemente para su publicaci\'on. 
 
 \medskip

Existe una pl\'etora de posibilidades y aplicaciones que emergen del formalismo de deformaciones de Poisson--Hopf. Aunque los resultados han sido principalmente considerados en el plano, para el cual existe una clasificaci\'on expl\'{\i}cita de sistemas LH  \cite{BBHLS,BCHLS13Ham}, el m\'etodo es v\'alido para variedades arbitrarias y \'algebras de Vessiot--Guldberg de dimensiones mayores.  Un estudio sistem\'atico de estos sistemas sin duda dar\'a lugar a nuevas propiedades de los sistemas deformados que merecen ser analizadas con detalle. En particular, las propiedades din\'amicas de sistemas espec\'{\i}ficos de ecuaciones diferenciales pueden estudiarse mediante estas t\'ecnicas, donde se espera que nuevas propiedades sean descubiertas. 

\medskip
En relaci\'on con la actual pandemia COVID-19, podemos preguntarnos si existe una descripci\'on en t\'erminos de los modelos SISf. Este modelo es una primera aproximaci\'on para procesos de infecci\'on primarios, en los cuales se consideran dos tipos de estados en la poblaci\'on: los infectados y los susceptibles de infecci\'on. El modelo no contempla la posibilidad de adquirir inmunidad. Parece que la COVID-19 est\'a sujeta a ciertos tipos de inmunidad, aunque s\'olo hasta un treinta por ciento de la poblaci\'on. En este sentido, un modelo SIR que considera individuos inmunes no es un mode\-lo apropiado para esta situaci\'on. Ser\'{\i}a interesante tener un modelo que contemple individuos inmunes y no inmunes simult\'aneamente. Actualmente estamos buscando un modelo hamiltoniano estoc\'astico que contemple estas variables. 

\section*{Trabajos futuros}

Una de las cuestiones principales a resolver es si el enfoque de Poisson--Hopf proporciona un m\'etodo efectivo para deducir un an\'alogo deformado de los principios de superposici\'on para sistemas LH deformados. Asimismo, ser\'{\i}a interesante saber si una tal descripci\'on puede aplicarse simult\'aneamente a sistemas no equivalentes, como una extrapolaci\'on de la noci\'on de contracci\'on a los sistemas de Lie. Otros aspecto relevante es la posibilidad de obtener una descripci\'on unificada de estos sistemas a partir de ciertos sistemas "can\'onicos" fijos, lo que implicar\'{\i}a una primera sistematizaci\'on de los sistemas LH desde una perspectiva m\'as amplia que la de las \'algebras de Lie finito-dimensionales. Algunas v\'{\i}as de trabajo futuro pueden resumirse como sigue: 

\begin{itemize}

\item En la clasificación de los sistemas LH en el plano juega un papel central la llamada 2-photon álgebra $\mathfrak{h}_6$, ya que es el álgebra de máxima dimensión que puede aparecer con las propiedades de un álgebra LH. El estudio de sus deformaciones cuánticas es, por tanto, una cuestión fundamental para completar el análisis de las deformaciones de los sistemas LH en el plano. Cabe señalar que existen esencialmente dos posibilidades diferentes para estas deformaciones, dependiendo de la estructura de dos subálgebras prominentes, el álgebra $\mathfrak{h}_4$ y $\mathfrak{sl}(2)$, que dan lugar a sistemas y deformaciones con diferentes propiedades. El primer caso, basado en el \'algebra del oscilador $\mathfrak{h}_4$, ha sido parcialmente considerado en el cap\'{\i}tulo 6. Sin embargo, resta a\'un obtener una regla de superposici\'on efectiva, cuya implementaci\'on estamos analizando actualmente. El an\'alisis debe completarse identificando clases particulares de sistemas de ecuaciones diferenciales que puedan deformarse mediante este procedimiento, y que puedan interpretarse como perturbaciones del sistema inicial. El segundo caso, basado en la extensi\'on  de los resultados obtenidos para $\mathfrak{sl}(2)$ al \'algebra $\mathfrak{h}_6$, es estructuralmente muy distinto debido a la naturaleza de la deformaci\'on cu\'antica. Esperamos que nuevos sistemas con propiedades de inter\'es surgan de este an\'alisis.  Desde el punto de vista de las aplicaciones, estos sistemas tienen muchas propiedades interesantes, tales como nuevos sistemas del tipo Lotka--Volterra o sistemas de tipo oscilatorio con masas y frecuencias dependientes de la posici\'on y el tiempo, pero cuya din\'amica pueda caracterizarse mediante la existencia de un procedimiento para la determinaci\'on exacta de las constantes del movimiento y las reglas de superposici\'on. En an\'alisis completo de los sistemas LH deformados basados en el \'algebra $\mathfrak{h}_6$ est\'a actualmente en proceso, para ser enviado pr\'oximamente para su publicaci\'on.

\item Por otro lado, cabe observar que, actualmente, no existe una clasificación de los sistemas LH para las dimensiones $n\geq 3$. Un problema de interés que surge en este contexto es analizar la posibilidad de generar nuevos sistemas LH, tanto clásicos como deformados, mediante la extensión de los sistemas en el plano, en combinación con las proyecciones de las realizaciones de campos vectoriales. En este contexto, se sabe que las proyecciones de realizaciones del álgebra de Lie asociadas con una representación lineal dan lugar a realizaciones no lineales. Analizando la cuestión desde la perspectiva de las álgebras funcionales (hamiltonianas), es concebible la existencia de estructuras simplécticas compatibles que dan lugar a sistemas LH en dimensiones superiores, así como una dependencia de dichas formas simplécticas. Criterios de este tipo pueden combinarse con deformaciones cuánticas de álgebras de Lie conocidas, para obtener nuevas aplicaciones de estas en el contexto de ecuaciones diferenciales.

\item Como complemento al modelo SISf basado en $\mathfrak{b}_2$ como sub\'algebra de $\mathfrak{sl}(2)$, es razonable desarrollar el modelo correspondiente a la misma \'algebra, pero vista como una sub\'algebra de $\mathfrak{h}_4$. Nuevamente, el car\'acter esencialmente distinto de la deformaci\'on nos lleva a sistemas con propiedades muy diferentes, lo que sugiere comparar ambos modelos con detalle, analizando las soluciones num\'ericas deducidas de ambos enfoques. Un primer paso en esta direcci\'on est\'a actualmente en proceso. 

\item Desear\'{\i}amos asimismo extender nuestro estudio a modelos epidemiol\'ogicos m\'as complicados, aunque a primera vista no hayamos localizado nuevos sistemas de Lie, al menos en su forma usual. Sospechamos que la descripci\'on hamiltoniana de los modelos compartimentales pueden asociarse a sistemas de Lie, como muestra el ejemplo desarrollado. En particular, debe analizarse con m\'as detalle como las soluciones del modelo deformado \eqref{dssis} permiten recuperar las soluciones del modelo sin deformar, cuando el par\'ametro de deformaci\'on tiende a cero. Se precisa un an\'alisis m\'as detallado para determinar si tales modelos integrables mode\-lizan procesos distintos a los infecciosos. En particular, nos interesar\'{\i}a saber si es posible modelizar la din\'amica subat\'omica mediante hamiltonianos deformados del tipo \eqref{defham}. Existe una teoría estocástica de los sistemas de Lie desarrollada en \cite{Ortega} que podría ser otro punto de partida para tratar estos sistemas. En el presente trabajo tuvimos la suerte de encontrar una teoría con fluctuaciones que coincidían con una expansión estocástica, pero esto es más una excepción que una regla. De hecho, parece que la forma más factible de proponer modelos estocásticos es utilizar la teoría estocástica de Lie en lugar de esperar un destello de suerte con las fluctuaciones. Como hemos dicho, encontrar soluciones particulares no es en absoluto trivial. La búsqueda analítica es una tarea muy atroz. Creemos que para ajustar soluciones particulares en el principio de superposición, es posible que sea necesario calcular estas soluciones particulares numéricamente. En \cite{piet} se pueden idear algunos métodos numéricos específicos para soluciones particulares de sistemas de Lie.

\item Finalmente, a partir de la ecuación de Chebyshev, se ha demostrado que el punto de simetrías de Noether de esta ecuación se puede expresar para $n$ arbitrario en términos de los polinomios de Chebyshev $T_ {n}(x), U_ {n}(x)$ de primer y segundo tipo, respectivamente.
Además, se ha observado que la realización genérica del álgebra de simetría de puntos de Lie $\mathfrak{sl}(3, \mathbb{R})$ puede ampliarse a ecuaciones diferenciales ordinarias de segundo orden lineales más generales y las soluciones  se pueden expresar en términos de funciones trigonométricas o hiperbólicas. En particular, los conmutadores de las simetrías de puntos genéricos muestran que varias de las relaciones algebraicas de las soluciones generales surgen realmente como consecuencia de la simetría. Las mismas conclusiones son válidas para la estructura de la subálgebra de cinco dimensiones de las simetrías de Noether. Se ha demostrado que la realización de los genera\-dores de simetría sigue siendo válida para ecuaciones diferenciales de tipo hipergeométrico, lo que nos permite obtener realizaciones de $\mathfrak{sl}(3, \mathbb{R})$ en términos de funciones hipergeométricas en general y varios polinomios ortogonales en particular, como los polinomios de Chebyshev o Jacobi. Otro hecho notable que surge de este análisis es que los términos forzados son siempre independientes de las "velocidades" $y_ {1}^{\prime}, y_ {2}^{\prime}$. Esto es nuevamente una consecuencia de la realización genérica elegida, y la cuestión de si otras realizaciones genéricas en términos de la solución general de la EDO (o sistema) permiten determinar términos forzosos que dependen explícitamente de las derivadas, e incluso conducen a ecuaciones diferenciales autónomas (sistemas), surge de manera natural. En este contexto, sería deseable obtener una realización de $\mathfrak{sl}(3, \mathbb{R})$ que no solo permita describir genéricamente las simetrías de punto y Noether de los polinomios de Jacobi, sino que también se aplique a las ecuaciones diferenciales asociadas a la familias restantes de polinomios ortogonales, específicamente los polinomios de Laguerre y Hermite. Esto permitiría construir más ecuaciones y sistemas no lineales que posean una subálgebra de simetrías de Noether, los generadores de las cuales se den en términos de estos polinomios ortogonales.  
\end{itemize}


\setcounter{tocdepth}{1}

\chapter*{Agradecimientos}

\addchaptertocentry{Agradecimientos} 
\begin{flushright}
{\itshape “Entre los pecados mayores que los hombres cometen, aunque algunos dicen que es la soberbia, yo digo que es el ser desagradecido, ateniéndome a lo que suele decirse: que de los desagradecidos está lleno el infierno. Este pecado, en cuanto me ha sido posible, he procurado yo huir desde el instante que tuve uso de razón, y si no puedo pagar las buenas obras que me hacen con otras obras, pongo en su lugar los deseos de hacerlas, y cuando estos no bastan, las publico, porque quien dice y publica las buenas obras que recibe, también las recompensara con otras, si pudiera; porque por la mayor parte los que reciben son inferiores a los que dan, y así es Dios sobre todos, porque es dador sobre todos, y no pueden corresponder las dádivas del hombre a las de Dios con igualdad, por infinita distancia, y esta estrecheza y cortedad en cierto modo la suple el agradecimiento.”}\bigbreak

\hfill M. de Cervantes, {\em El ingenioso hidalgo don Quijote de la Mancha}
\end{flushright}

\vspace{1cm}

Esta tesis ha sido realizada gracias al contrato predoctoral (CT45/15-CT46/15) otrogado por la Universidad Complutense de Madrid, a la Universidad de Burgos por su financiación durante el primer año y a los diversos proyectos de investigación: MTM2016-79422-P (AEI/FEDER, EU) y MTM2013-48320- P.
\vspace{1cm}

Ha sido un duro y satisfactorio período de aprendizaje, por lo que me gustaría agradecer a todas las personas que enriquecieron de una forma u otra esta etapa de mi vida.

Antes de nada, deseo agradecer a mis dos directores de tesis por su inmensa generosidad e inagotable paciencia, no habría llegado tan lejos. He aprendido más de lo que imaginaba junto a vosotros tanto en el plano personal como en el profesional. Por la confianza y dedicación que me habéis brindado cada día.

Quiero agradecer a la Universidad de Burgos, en especial a Ángel Ballesteros por su paciencia, sus siempre acertadas palabras y su buen hacer, pues me otorgaron una gran confianza para continuar mi camino; as\'{\i} como a todo el departamento de Física por hacerme sentir en casa, y como olvidarme del "Patillas" y del variopinto elenco de personas que lo frecuentan. A la Universidad de Varsovia por su gran acogida durante mi estancia de la mano de Javier de Lucas, qui\'en tan amablemente compartió conocimientos y numerosas discusiones que sin duda han ayudado a mi formación matemática.
No tengo más que palabras buenas, llenas de cariño y grandes recuerdos, para describir a las personas que componen la Red de Geometría, Mecánica y Control y los momentos que me han regalado, solo mencionaré en representación de todos los componentes de este grupo a Cristina por su amistad y apoyo en todo momento, siento mucho no dar más nombres en pero mi mala memoria no har\'{i}a justicia a los omitidos, ha sido una suerte haberos conocido. 
A la Universidad Complutense de Madrid por mi formación, y por haber sido mi hogar durante tantos años, sin olvidar a las grandes personas que componen el servicio de limpieza y cafetería por apoyarnos y hacernos la vida más fácil. A José M. Ancochea quien me ofreció la oportunidad de disfrutar la experiencia de compartir asignatura. En especial a las personas que empezaron siendo compañeros y compañeras de clase y continuaron siendo grandes amistades como David, Patri, Javi, Judit, Irene y Amanda; así como los que dejaron su huella más tarde, gracias Jorge, Luismi, Rober, Luis, Paco, Pedro, Juan... Y tantas otras personas que de diferentes maneras me ayudaron a recorrer esta senda.

Mis amigos de toda la vida, como agradecer a Rober, Alex, Fran, Carlos, Bea, Cris y compañia, todo lo que me habéis enseñado sobre la vida y lo que hemos compartido juntos... no tiene precio el apoyo que me ofrecéis cada día, recordandome siempre que es mejor caer que no intentar volar. Esas conversaciones de madrugada, viviendo nuestros sueños aprovechando aquellas noches de insomnio. Al dojo de Aikido Miguel Hernández por aportarme la serenidad y el equilibrio mental que necesitaba. Gracias, Alba, por mostrarme que incluso siguiendo los senderos oscuros y salvajes entre las rocas florece el brillo de un nuevo día. 

A mi familia, que merecen una mención especial pues sois un apoyo y un ejemplo a seguir, gracias a mi madre, a mi hermano y a Yuki por vuestra paciencia, compañia y comprensión siempre seréis los pilares más importantes en el laberinto de la vida.

Por último, por haber sido el precursor de esta aventura en la que me embarqué al empezar el grado, gracias a Paco Conejero, desde que te conocí supe que debía seguir este camino. Te encargaste de recordarmelo y animarme en cada taza de café.
\vspace{1cm}

Muchas gracias a todas las personas que se cruzaron en mi vida, espero no ofender a las que no aparecen de manera explícita pues no es por falta de cariño o agradecimiento. Siempre podéis llamarme y os invitaré a una cerveza como compensación.

\begin{flushright}
Eternamente agradecido,\bigbreak

\hfill Edu.
\end{flushright}


\tableofcontents 

\listoffigures 

\listoftables 

\dedicatory{A mi madre y a mi hermano\ldots} 


\mainmatter 

\pagestyle{thesis} 


\chapter*{Introduction}
\addcontentsline{toc}{chapter}{Introduction}

A {\it Lie system} is a nonautonomous
system of first-order ordinary differential equations whose general solution can be written as a function (a so-called {\it superposition rule}) depending on a certain number of particular solutions and some significant constants \cite{DAV,LS,VES}. Superposition rules constitute a structural property that emerges naturally from the group-theoretical approach to differential equations initiated by Sophus Lie, Vessiot, and Guldberg, within the context of the development of the geometric program based on transformation groups, as well as from the analytic classification of differential equations developed by Painlev\'e and  Gambier, among others. Indeed, Lie proved that every Lie system can be described by a finite-dimensional Lie algebra of vector fields, a {\it Vessiot--Guldberg Lie algebra} \cite{LS}, and Vessiot used Lie groups to derive superposition rules \cite{VES}. 

In the frame of physical problems, it was not until the 80's when the power of superposition rules and Lie systems was fully recognized \cite{PW}, motivating a systematic analysis of their applications in classical dynamics and their potential generalization to quantum systems (see \cite{CGM00,CGM07,Dissertations,PW} and references therein). 

Although Lie systems, as well as their refinements and generalizations, represent a valuable auxiliary tool in the integrability study of physical systems, it seems surprising that the methods  employed have always remained within the limitations of Lie group and distribution theory, without considering  other  frameworks that have turned out to be a very successful approach to integrability, such as quantum groups and Poisson--Hopf algebras ~\cite{Abe,coalgebra2,coalgebra1,Chari,Majid}. We recall that, beyond superintegrable systems ~\cite{coalgebra2,coalgebra1}, Poisson coalgebras have been recently applied to integrable bi-Hamiltonian deformations of Lie--Poisson systems ~\cite{Ballesteros1} and to integrable deformations of R\"ossler and Lorenz systems~\cite{Ballesteros2}.

This work presents a novel generic procedure for the Poisson--Hopf algebra deformations of  {\it Lie--Hamilton (LH) systems}, namely Lie systems endowed with a Vessiot--Guldberg Lie algebra of Hamiltonian vector fields with respecto to a Poisson structure \cite{CLS13}. LH systems possess also a finite-dimensional Lie algebra of functions, a so-called {\it  LH algebra}, governing their dynamics \cite{CLS13}. The proposed approach is based on the Poisson coalgebra formalism extensively used in the context of superintegrable systems, together with the notion of involutive distributions in the sense of Stefan--Sussman (see \cite{Pa57,Va94,WA} for details). The main point will be to consider a  Poisson--Hopf algebra structure that replaces the LH algebra of the non-deformed LH system, thus  allowing us to obtain an explicit construction of $t$-independent constants of the motion, that will be expressed in terms of the deformed Casimir invariants. Moreover, the deformation will generally transform the Vessiot--Guldberg Lie algebra of the LH system into a set of vector fields generating an integrable distribution in the sense of Stefan--Sussman. Consequently, the deformed LH systems are not, in general, Lie systems anymore.

The novel approach is presented in the chapter \ref{Chapter3}, where the basic properties of LH systems and Poisson--Hopf algebras are reviewed (for details on the general theory of Lie and LH systems, chapters \ref{Chapter1} and \ref{Chapter2}; the reader is referred to ~\cite{BBHLS,BCHLS13Ham, BHLS,C132,C135,CGL10,CGL11,CGM00,CGM07,Dissertations,CLS13,HLT, Ibragimov16,Ibragimov17,LS,PW}).
To illustrate this construction, we consider the Poisson--Hopf algebra analogue of the so-called non-standard quantum deformation of  $\mathfrak{sl}(2)$~\cite{non, Ohn, Shariati} together with its deformed Casimir invariant, chapter \ref{Chapter4}. 
 
Afterwards, relevant examples of deformed LH systems that can be extracted from this deformation are given. Firstly, the non-standard deformation of the Milney--Pinney equation is presented in part \ref{Part3}, where this deformation is shown to give rise to a new oscillator system with a position-dependent mass and a time-dependent frequency (chapter \ref{Chapter5}), whose (time-independent) constants of the motion are also explicitly deduced.
In secction \ref{RiccatiSeccion} several deformed (complex and coupled) Riccati equations are obtained as a straightforward application of the formalism here presented. We would like to stress that, albeit these applications are carried out on the plane, thus provinid us a deeper insight in the proposed formalism, the method here presented is by no means constrained dimensionally, and its range of applicability goes far beyond the particular cases considered here. 


Since its original formulation by Lie \cite{LS}, Lie systems, have been studied extensively (see \cite{C135,CGM00,Dissertations,Ol,Pa57,VES,PW} and references therein). The Lie theorem \cite{CGM07,LS} states that every system of first-order differential equations is a Lie system if and only if it can be described as a curve in a finite-dimensional Lie algebra of vector fields,  referred to as {\it Vessiot--Guldberg Lie algebra}.

Although being a Lie system is rather an exception than a rule \cite{CGL11}, Lie systems have   been shown to be of great interest within physical and mathematical applications (see \cite{Dissertations} and references therein). Surprisingly, Lie systems admitting a Vessiot--Guldberg Lie algebra of Hamiltonian vector fields relative to a Poisson structure, the {\it Lie--Hamilton systems}, have found even more applications than standard Lie systems with no associated geometric structure \cite{BBHLS,BCHLS13Ham,BHLS, CLS13}. Lie--Hamilton systems admit an additional finite-dimensional Lie algebra of Hamiltonian functions, a {\it Lie--Hamilton algebra}, that allows us to deduce an algebraic determination of superposition rules and constants of the motion of the system~\cite{BHLS}.

Apart from the theory of quasi-Lie systems \cite{CGL11} and superposition rules for nonlinear operators, most approaches to Lie systems rely strongly on the theory of Lie algebras and Lie groups. However,  the success of quantum groups \cite{Chari,Majid} and the coalgebra formalism within the analysis of superintegrable systems \cite{coalgebra2,coalgebra1}, and the fact that quantum algebras appear as deformations of Lie algebras, suggested the possibility of extending the notion and techniques of Lie--Hamilton systems beyond the range of application of the Lie theory. An approach in this direction was recently proposed in   
\cite{BCFHL}, where a method to construct quantum deformed Lie--Hamilton systems  (LH systems in short) by means of the coalgebra formalism and quantum algebras was given. 

The underlying idea is to use the theory of quantum groups to deform Lie systems and their associated structures. More exactly, the deformation transforms a LH system with its Vessiot--Guldberg Lie algebra into a Hamiltonian system whose dynamics is determined by a set of generators of a Steffan--Sussmann distribution. Meanwhile, the initial Lie--Hamilton algebra (LH algebra in short)  is mapped into a Poisson--Hopf algebra. The deformed structures allow for the explicit   construction of $t$-independent constants of the motion through quantum algebra techniques for the deformed system. 

We illustrate how the approach introduced in \cite{BCFHL} to construct deformations of LH systems via Poisson--Hopf structures allows a further systematization that encompasses the nonequivalent LH systems corresponding to isomorphic LH algebras. 
Specifically, we show that Poisson--Hopf deformations of LH systems based on a LH algebra isomorphic to $\mathfrak{sl}(2)$ can be described 
generically, hence providing the deformed Hamiltonian functions and the corresponding deformed Hamiltonian vector fields, once the corresponding 
counterpart  of the non-deformed system is known. This  provides a new method to construct LH systems with a LH algebra
isomorphic to a fixed Lie algebra $\mathfrak{g}$. The approach is based on the symplectic foliation in $\mathfrak{g}^*$ induced by the Kirillov--Konstant--Souriou bracket on $\mathfrak{g}$. As a particular case, it is explicitly shown how our procedure explains the existence of three types of LH systems on the plane related to a LH  algebra isomorphic to $\mathfrak{sl}(2)$. This is due to the fact that each one of the three different types corresponds to one of the three types of symplectic leaves in $\mathfrak{sl}^*(2)$. In analogy, one can generate the only LH system on the plane admitting a Vessiot--Guldberg Lie algebra isomorphic to $\mathfrak{so}(3)$. This systematization enables us to give directly the Poisson--Hopf deformed system from the classification of LH systems~\cite{BBHLS,BHLS},
further suggesting a notion of Poisson--Hopf Lie systems based on a $z$-parameterized family of Poisson algebra morphisms. Our methods seem to be extensible to study also  LH systems and their deformations on other more general manifolds.

The structure of the thesis goes as follows. Chapters \ref{Chapter1} and \ref{Chapter2} are devoted to review the main aspects of LH systems and Poisson--Hopf algebras. The general procedure to construct Poisson--Hopf algebra deformations of LH systems~\cite{BCFHL}, and other properties of the underlying formalism are given in Chapter \ref{Chapter3}. In \ref{Chapter4},  the (non-standard) Poisson--Hopf algebra deformation of LH systems based on the simple Lie algebra  $\mathfrak{sl}(2)$ are analyzed in detail, while  
the unified approach to   deformations of Poisson--Hopf Lie systems with a LH algebra isomorphic to a fixed Lie algebra $\mathfrak{g}$ is treated in Section \ref{SeccMEthodConst}. The procedure is explicitly illustrated, considering the three non-diffeomorphic classes of $\mathfrak{sl}(2)$-LH systems on the plane, from which the corresponding 
deformations are described in detail. Two general types of differential equations are considered in the context of their quantum deformations. First of all, the deformed Milne--Pinney equation is analyzed in detail, from which new oscillator systems with a position-dependent mass are derived. As a second relevant case, deformations of the Riccati equations are studied, specifically the deformed complex and deformed coupled Riccati equations.  In Chapter \ref{Chapter6} the problem of quantum deformations of LH systems based on the oscillator algebra $\mathfrak{h}_4$ is considered, with special emphasis of its structure as a subalgebra of the so-called 2-photon algebra $\mathfrak{h}_6$. This in particular leads to quantum deformations of the damped harmonic oscillator. In Chapter \ref{Chapter8} we use the solvable Lie subalgebra $\mathfrak{b}_2$  of  $\mathfrak{sl}(2)$  to propose a new and alternative epidemiological model. The techniques discussed in previous chapters are applied to analyze a defomed SISf model, which in particular is obtained by restriction of the corresponding $\mathfrak{sl}(2)$-deformed system, and for which the constants of the motion are explicitly constructed. 
Finally, in the Conclusions we summarize the results and outline some future work to be accomplished or already in progress.

\part{Formalism of Poisson--Hopf Algebra Deformations}
\chapter{Lie Systems and Poisson--Hopf Algebras} 

\label{Chapter1} 


\newcommand{\keyword}[1]{\textbf{#1}}
\newcommand{\tabhead}[1]{\textbf{#1}}
\newcommand{\code}[1]{\texttt{#1}}
\newcommand{\file}[1]{\texttt{\bfseries#1}}
\newcommand{\option}[1]{\texttt{\itshape#1}}
\renewcommand{\theequation}{1.\arabic{equation}}

\section{Lie systems}

\bigskip

Lie systems, besides their undeniable interest within Geometry, play a relevant role in many applications in Biology, Cosmology, Control Theory, Quantum Mechanics, among other disciplines. A specially interesting case is when a systems of first-order ordinary differential equations, which is the prototype of Lie system, can be endowed with a compatible symplectic structure, leading to the notion of Lie--Hamilton systems, of special interest within the frame of Classical Mechanics \cite{DAV,GOL}.  Lie--Hamilton systems and their fundamental properties are conveniently described in terms of a $t$-dependent vector field that describes the dynamics. An illustrative example for this type of systems is given by the second-order Riccati equation.


\bigskip
\bigskip

\subsection{Lie systems and superposition rules}

General properties of Lie systems, as well as additional geometrical applications, can be e.g. found in \cite{Dissertations,WA}.  

\bigskip

\begin{definition}
A superposition rule for a system ${\bf X}$ defined on an $n$-dimensional manifold $M$ is a map $$\Phi: M^k \times M \longrightarrow M$$ such that ${x}(t):=\Phi(x_{(1)} (t), \dots, x_{(k)}(t); \lambda)$ is a general solution of the system ${\bf X}$, where $x_{i}(t)$ are particular solutions and $\lambda$ is a point of the manifold $M$, corresponding to the initial condition of the Cauchy problem.
\end{definition}

\bigskip

Let $M$ be an $n$-dimensional manifold and let $\pi_i$ be the projections  $\pi_1: TM \longrightarrow M$  with $\pi_1(x,v):=x$ and $\pi_2: \mathbb{R} \times M \longrightarrow M$ with $\pi_2(t,x):=x$. A smooth map ${\bf X}:\Omega \subseteq \mathbb{R}\times M\rightarrow TM$, where $\Omega$ is an open subset of $\mathbb{R}\times M$, is called a \textit{$t$-dependent vector field} if the diagram \[\xymatrix{
\Omega  \ar[r]^-{\bf X}\ar[rd]^-{\pi_{2}}  
&   TM \ar[d]^{\pi_{1}} \\
 & M }
\]
 is commutative, i.e. $\pi_{1}\circ {\bf X} =\pi_{2}$ in $\Omega$. We observe that defining $\Omega_{t}:=\{x\in M / (t,x) \in \Omega \}$ for each $t\in \mathbb{R}$, $\Omega_{t}$ is not empty and recovers the usual notion of vector field,  denoted by ${\bf X}_{t}$. The notion of $t$-dependent vector field is very useful in the geometric theory of Lie systems and will play a relevant role in the formalism that will be developed.
 
\bigskip
 
\begin{remark}
It follows that if ${\bf X}$ is a  $t$-dependent vector field, then it is equivalent to a linear morphism ${\bf X}: \mathcal{C}^{\infty}(M)\rightarrow\mathcal{C}^{\infty}(\mathbb{R}\times M)$ defined as ${\bf X}(f)(t,x):=({\bf X}_{t}f)(x)$, for all $(t,x)\in \mathbb{R\times M}$, such that it satisfies the Leibniz rule in $\mathcal{C}^{\infty}(M)$ at each point of $\mathbb{R}\times M$.
\end{remark}

\bigskip

Let $\widetilde{\bf X}$ be a vector field over $\mathbb{R}\times M$ such that $\iota_{\widetilde{\bf X}}dt=1$\footnote{Let $M$ be a smooth manifold, ${\bf X}\in \mathfrak{X}(M)$ and $\omega \in \Omega^{p}(M)$, then $\iota_{{\bf X}}\omega$ is the contraction of a differential form $\omega$ with respect to the vector field ${\bf X}$.} and $(\widetilde{{\bf X}}\, \pi_{2}^* f)(t,x)={\bf X}(f)(t,x)$. If the $t$-dependent vector field ${\bf X}$ is given by \begin{equation} \label{vectofield}
   {\bf X}=\sum_{j=1}^{n}{\bf X}_{j}(t,x)\frac{\partial}{\partial x_{j}},
\end{equation} 
in local coordinates, then $\widetilde{{\bf X}}$ has the expression $$\widetilde{{\bf X}}=\frac{\partial}{\partial t}+\sum_{j=1}^{n}{\bf X}_{j}(t,x)\frac{\partial}{\partial x_{j}},$$ called the \textit{autonomization of ${\bf X}$}. An integral curve of the $t$-dependent vector field ${\bf X}$ is a map $\gamma:\mathbb{R}\rightarrow \mathbb{R}\times M$ such that $\pi_2\circ \gamma$, where $\gamma$ corresponds to an integral curve of $\widetilde{{\bf X}}$. It follows in particular that for any $t$ with  $\gamma(t)=(t,x(t))$ the components of $x(t)$ satisfy the following first order system: 
\begin{equation}
\frac{dx_{j}}{dt}={\bf X}_{j}(t,x),\quad i=1,\dots n.\label{eq:system}
\end{equation}
This is the so-called associated system of ${\bf X}$. It is straightforward to verify that an arbitrary system of first-order ordinary differential equations of the type \eqref{eq:system} determines a $t$-dependent vector field \eqref{vectofield}, the integral curves of which satisfy the system. This establishes  a one-to-one correspondence between systems of type \eqref{eq:system} and $t$-dependent vector fields.

\bigskip
If $\mathcal{A}$ is a family of vector fields on an $n$-dimensional manifold $M$, then $Lie(\mathcal{A})$ denotes the Lie algebra spanned by the vector fields and their successive commutators $$\mathcal{A}, [\mathcal{A},\mathcal{A}],[\mathcal{A}[\mathcal{A},\mathcal{A}]],[\mathcal{A},[\mathcal{A},[\mathcal{A},\mathcal{A}]]],\dots$$ where $[\mathcal{A},\mathcal{B}]$ 
is a shorthand notation for the brackets 
$\{[{\bf X},{\bf Y}]/{\bf X}\in \mathcal{A} \ \ and\ \ {\bf Y}\in \mathcal{B}\}$. It follows from the construction that the Lie algebra $Lie(\mathcal{A})$ is the smallest Lie algebra of vector fields containing the set $\mathcal{A}$. The main result concerning the classical theory of Lie-systems is given by the Lie--Scheffers theorem (for more details see \cite{LS}).

\bigskip

\begin{theorem}[Lie--Scheffers Theorem] \label{LSTheorem} A system of first-order ordinary differential equations \eqref{eq:system} admits a superposition rule if and only if there exist smooth functions $\beta_{j}(t)$ such that associated t-dependent vector field has the form 
\begin{equation}
    {\bf X}(t,x)=\sum_{j=1}^{\ell}\beta_{j}{\bf X}_{j}(x),
\end{equation} 
and such that the vector fields ${\bf X}_{j}$ (in a manifold $M$) span an $\ell $-dimensional real Lie algebra $V^{X}:=Lie(\{{\bf X}_{j}/j=1,\dots,\ell\})$.
\end{theorem}

\bigskip

The Lie algebra of vector fields $V^{X}$ is usually called a \textit{Vessiot--Guldberg Lie algebra}, where in addition the following numerical constraint must be satisfied:
\begin{equation}
\label{LieCondition}
    dim(V^{X})\leq m\cdot n=\dim(M).
\end{equation}   
The scalar $m$ corresponds to the number of particular solutions of the system that are required for establishing a superposition rule. The relation \eqref{LieCondition} is also known as the  \textit{Lie condition}. It should be emphasized that a given system may admit different superposition rules, as a Vessiot--Guldberg algebra is not an invariant of the system, as happens e.g. with the Lie algebra of Lie-point symmetries. Thus, a specific system may admit nonisomorphic  Vessiot--Guldberg algebras, and hence different superposition rules. 

\bigskip

\begin{definition}
A first-order system of differential equations that admits a superposition rule is a Lie system. 
\end{definition}

\bigskip
\bigskip

\begin{example}
Let $\bf X$ be a $t$-dependent vector field over $\mathbb{R}^2$ that expressed in local coordinates is given by 
 \begin{equation}
    {\bf X}=-\frac{1}{t}X_{1}+X_{2}+\eta^{2}X_{3},
\end{equation} 
where 
\begin{equation}
    {\bf X}_{1}=y\frac{\partial}{\partial y}, \quad {\bf X}_{2}=y\frac{\partial}{\partial x},\quad {\bf X}_{3}=x\frac{\partial}{\partial y}, \quad {\bf X}_{4}=x\frac{\partial}{\partial x}+y\frac{\partial}{\partial y}.
\end{equation}
These  vector fields span a Vessiot--Guldberg Lie algebra $V^{X}$ isomorphic to $gl(2)$, with domain $\mathbb{R}^{2}_{x\neq 0}$ (for more details see Appendix \ref{AppendixB}) and the associated system takes the form \begin{equation}
    \frac{dx}{dt}=y, \frac{dy}{dt}=-\frac{2}{t}y+\eta^{2}x.
\end{equation}
Hence, this system is a Lie system.

\end{example}

\bigskip
\bigskip


\subsection{Lie--Hamilton systems}

\bigskip

\begin{definition}
 A Lie system ${\bf X}$ is, furthermore, a Lie--Hamilton system ~\cite{BBHLS,BCHLS13Ham,BHLS,Dissertations,CLS13, HLT} if it admits a Vessiot--Guldberg Lie algebra $V$ of Hamiltonian vector fields relative to a Poisson structure. This amounts to the existence, around each generic point of $M$, of a symplectic form, $\omega$, such that:
\be
\mathcal{L}_{ {\bf X}_i}\omega=0
\label{der}
\ee
for a basis ${  {\bf X}}_1,\ldots,{  {\bf X}}_\ell$ of $V$ (cf. \cite{BBHLS}). Then each vector field $ {\bf X}_i$ admits a Hamiltonian function $h_i$ given by the rule:
\be 
\iota_{{ {\bf X}}_i}\omega={\rm d}h_i,
\label{contract3}
\ee
where $\iota_{ {\bf X}_i}\omega$ stands for the contraction of the vector field ${\bf X}_i$ with the symplectic form $\omega$.
\end{definition}

\bigskip

Since $\omega$ is non-degenerate, every function $h$ induces a unique associated Hamiltonian vector field ${ {\bf X}}_h$. This fact gives rise to a Poisson bracket on $\mathcal{C}^\infty(M)$ given by
 \begin{equation}\label{LB}
 \{\cdot,\cdot\}_\omega\ :\ \mathcal{C}^\infty\left(M\right)\times \mathcal{C}^\infty\left(M\right)\ni (f_1,f_2)\mapsto X_{f_2} f_1\in \mathcal{C}^\infty\left(M\right),
 \end{equation}
turning $(\mathcal{C}^\infty(M),\{\cdot,\cdot\}_\omega)$ into a Lie algebra. The space ${\rm Ham}(\omega)$ of  Hamiltonian vector fields on $M$ relative to $\omega$ is also a Lie algebra relative to the commutator of vector fields. Moreover, we have the following exact sequence of Lie algebras morphisms (see \cite{Va94})
\begin{equation}\label{seq}
0\hookrightarrow \mathbb{R}\hookrightarrow (\mathcal{C}^\infty(M),\{\cdot,\cdot\}_\omega)\stackrel{\phi}{\longrightarrow} ({\rm Ham}(\omega),[\cdot,\cdot])\stackrel{\pi}{\longrightarrow} 0,
\end{equation} 
where $\pi$ is the projection onto $0$ and $\phi$ maps each $f\in C^\infty(M)$ into the Hamiltonian vector field ${ {\bf X}}_{-f}$.
In view of the sequence (\ref{seq}), the Hamiltonian functions $ h_1,\ldots,h_\ell$ and their successive Lie brackets with respect to (\ref{LB}) span a finite-dimensional Lie algebra of functions contained in $\phi^{-1}(V)$. This Lie algebra is called a
   {\em  Lie--Hamilton (LH) algebra}  $\lh$  of  ${\bf X}$ (see \cite{CLS13,HLT} and references therein). 
   
\bigskip
   
Let ${\bf X}$ be a system on an $n$-dimensional manifold $M$. A function  $f\in \mathcal{C}^{\infty}(TM)$ is called a constant of the motion\footnote{The space of constant of the motion for a system form a $\mathbb{K}-$algebra.} of the system ${\bf X}$ if it is a first integral of the vector field $\widetilde{{\bf X}}$, i.e., $$\widetilde{{\bf X}}f=0.$$

\bigskip
\bigskip

\subsubsection{Damped  harmonic oscillator}\label{dampedexample}

Consider a $t$-dependent one-dimensional damped harmonic oscillator of the form
\begin{equation}\label{DampedSys}
\begin{aligned}
\frac{dx}{dt}&=a(t)x+b(t)p+f(t),\\
\frac{dp}{dt}&=-a(t)p-c(t)x-d(t),
\end{aligned}\qquad (x,p)\in T^*_x\mathbb{R},
\end{equation}
for arbitrary $t$-dependent functions $a(t),b(t),c(t),d(t),f(t)$. The system (\ref{DampedSys}) is associated with the $t$-dependent vector field
$$\label{DynDamped}
{\bf X}_{t}=f(t){\bf X}_{1}-d(t){\bf X}_{2}+a(t) {\bf X}_{3}+b(t){\bf X}_{4}-c(t){\bf X}_{5},
$$ where $$
\begin{gathered}
{\bf X}_{1}=\frac{\partial}{\partial x},\,\quad
{\bf X}_{2}=\frac{\partial}{\partial p},\,\quad
{\bf X}_{3}=x\frac{\partial}{\partial x}-p\frac{\partial}{\partial p},\,\quad
{\bf X}_{4}=p\frac{\partial}{\partial x},\,\quad
{\bf X}_{5}=x \frac{\partial}{\partial p},\,
\end{gathered} 
$$are such that $ \left\langle {\bf X}_{1}, {\bf X}_{2} \right\rangle \simeq \mathbb{R}^{2} $ and $ \left\langle {\bf X}_{3}, {\bf X}_{4}, {\bf X}_{5} \right\rangle \simeq \mathfrak{sl}(2) $. Moreover $\langle {\bf X}_1,\ldots,{\bf X}_5\rangle\simeq \mathfrak{sl}(2,\mathbb{R})\ltimes \mathbb{R}^2$ and ${\bf X}$ becomes a Lie system related to a Vessiot--Guldberg Lie algebra $V_{do}$ isomorphic to $\mathfrak{sl}(2,\mathbb{R})\ltimes \mathbb{R}^2$.

\medskip
It can be easily shown that  $V_{do}$ consists of Hamiltonian vector fields with respect to the symplectic form $\omega=\mathsf dx\wedge \mathsf dp$ on $T^*\mathbb{R}$. In fact, the Hamiltonian functions associated to the vector fields ${\bf X}_1,\ldots,{\bf X}_5$ are given by 
$$
\begin{gathered}
h_{1}= p,\,\qquad
h_{2}= -x,\,\qquad
h_{3}= xp,\,\qquad
h_{4}= \frac{1}{2}p^{2},\,\qquad
h_{5}=- \frac{1}{2}x^{2},
\end{gathered}
$$
respectively. The functions $h_1,\ldots, h_5$ along with $h_0$ span a Lie algebra $\mathcal{H}_{\omega}\simeq\mathfrak{h}_{6}$ with respect to the standard Poisson bracket on $T^*\mathbb{R}$.

The constants of the motion for the damped harmonic oscillator equations can be obtained by applying the coalgebra formalism introduced in \cite{BCHLS13Ham}. Using the Casimir invariants of the underlying Lie algebra, it follows that these constants of the motion of the Lie system (\ref{DampedSys}) are  given by~\cite{BHLS}  
\begin{equation*}
\begin{gathered}
F^{(1)}=0, \qquad F^{(2)}=0,\\
F^{(3)}=\left((x_{(2)}-x_{(3)})p_{(1)}+(x_{(3)}-x_{(1)})p_{(2)}+(x_{(1)}-x_{(2)})p_{(3)}\right)^{2}.
\end{gathered}
\end{equation*}
By permutation of the indices corresponding to the variables of the non-trivial invariant $F^{(3)}$, we can find additional constants of the motion:
\begin{equation*}
\begin{aligned}
F^{(3)}_{14}=\left((x_{(2)}-x_{(3)})p_{(4)}+(x_{(3)}-x_{(4)})p_{(2)}+(x_{(4)}-x_{(2)})p_{(3)}\right)^{2}, \\
F^{(3)}_{24}=\left((x_{(3)}-x_{(4)})p_{(1)}+(x_{(4)}-x_{(1)})p_{(3)}+(x_{(1)}-x_{(3)})p_{(4)}\right)^{2}.
\end{aligned}
\end{equation*}
In order to derive a superposition rule, we just need to obtain the value of $p_{(1)}$ from the equation $k_1=F^{(3)}$, where $k_1$ is a real constant; and then plug this value into the equation $k_2=F^{(3)}_{24}$ to obtain~\cite{BHLS} $$
\begin{gathered}
x_{(1)}=x_{(3)}+\frac{(x_{(4)}-x_{(3)})\sqrt[]{k_1}+(x_{(2)}-x_{(3)})\sqrt[]{k_2}}{\sqrt[]{F^{(3)}_{14}}},\\
p_{(1)}=\frac{\sqrt[]{k_1}}{x_{(2)}-x_{(3)}}+p_{(3)}+\frac{\sqrt[]{k_2}(p_{(2)}-p_{(3)})}{\sqrt[]{F^{(3)}_{14}}}+\frac{\sqrt[]{k_1}(p_{(2)}-p_{(3)})(x_{(4)}-x_{(3)})}{\sqrt[]{F^{(3)}_{14}}(x_{(2)}-x_{(3)})}.
\end{gathered}
$$

\bigskip
\bigskip

Clearly, the superposition rule obtained above is merely one among many possible superposition rules, that could be derived considering other choices of the constants of the motion.  

\bigskip
\bigskip


\subsubsection{A second-order Riccati equation in Hamiltonian form}

\bigskip

Another relevant application of Lie--Hamilton systems is given by the class of Riccati equations, that have been analyzed in detail in \cite{CLS13}. 
The most general class of second-order Riccati equations is given by the family of second-order 
differential equations of the form
\begin{equation}\label{NLe}
\frac{d^2x}{dt^2}+(f_0(t)+f_1(t)x)\frac{dx}{dt}+c_0(t)+c_1(t)x+c_2(t)x^2+c_3(t)x^3=0,
\end{equation}
with
$$
f_1(t)=3\sqrt{c_3(t)},\qquad f_0(t)=\frac{c_2(t)}{\sqrt{c_3(t)}}-\frac{1}{2c_3(t)}\frac{dc_3}{dt}(t), \qquad c_3(t)\neq 0.
$$

\bigskip

These equations arise by reducing third-order linear differential equations through a dilation symmetry and a time-reparametrization. Their interest is due to their use in the study of several physical and mathematical problems
 \cite{CLS13,CRS05}.

It was recently discovered~\cite{CRS05a} that every second-order Riccati equation (\ref{NLe}) admits a $t$-dependent non-natural regular Lagrangian of the form
$$
L(t,x,v)=\frac{1}{v+U(t,x)},
$$
with $U(t,x)=a_0(t)+a_1(t)x+a_2(t)x^2$ and $a_0(t),a_1(t),a_2(t)$ being certain functions related to the
 $t$-dependent coefficients of (\ref{NLe}), see \cite{CRS05}. Therefore,
\begin{equation}\label{Legtrricsec}
p=\frac{\partial L}{\partial v}=\frac{-1}{(v+U(t,x))^2},
\end{equation}
and the image of the Legendre transform $\mathbb{F}L:(t,x,v)\in \mathcal{W}\subset \mathbb{R}\times {\rm T}\mathbb{R}\mapsto (t,x,p)\in \mathbb{R}\times {\rm T}^*\mathbb{R}$, where $\mathcal{W}=\{(t,x,v)\in \mathbb{R}\times {\rm T}\mathbb{R}\mid v+U(t,x)\neq 0\}$, is the open submanifold $\mathbb{R}\times\mathcal{O}$ where $\mathcal{O}=\{(x,p)\in {\rm T}^*\mathbb{R}\mid p< 0\}$. The Legendre transform is not injective, as $(t,x,p)=\mathbb{F}L(t,x,v)$ for $v={\pm 1}/{\sqrt{-p}}-U(t,x)$. Nevertheless, it can become so by restricting it to the open set $\mathcal{W}_+=\{(t,x,v)\in \mathbb{R}\times {\rm T}\mathbb{R}\mid v+U(t,x)>0\}$. In such a case, $v=1/{\sqrt{-p}}-U(t,x)$ and we can define over $\mathbb{R}\times\mathcal{O}$ the $t$-dependent Hamiltonian
\begin{equation*}
 h(t,x,p)=p\left(\frac 1{\sqrt{-p}}-U(t,x)\right)-\sqrt{-p}=-2\sqrt{-p}- p\, U(t,x).
\end{equation*}
Its Hamilton equations read
\begin{equation}
\left\{
\begin{aligned}\label{Hamil}
\frac{dx}{dt}&=\frac{\partial h}{\partial p}=\frac{1}{\sqrt{-p}}-U(t,x)=\frac{1}{\sqrt{-p}}-a_0(t)-a_1(t)x-a_2(t)x^2,\\
\frac{dp}{dt}&=-\frac{\partial h}{\partial x}= p\frac{\partial U}{\partial x}(t,x)= p(a_1(t)+2a_2(t)x).
\end{aligned}\right.
\end{equation}

\bigskip

Since the general solution $x(t)$ of every second-order Riccati equation (\ref{Legtrricsec}) can be recovered from the general solution $(x(t),p(t))$ of its corresponding system (\ref{Hamil}), the analysis of the latter provides information about general solutions of second-order Riccati equations.

\bigskip

The relevant point is to observe that the system (\ref{Hamil}) is actually a Lie system as shown in~\cite{CLS12a}. Indeed, consider the vector fields over $\mathcal{O}$ of the form
\begin{equation}\label{VF}
\begin{gathered}
{\bf X}_1=\frac{1}{\sqrt{-p}}\frac{\partial}{\partial x},\quad\quad
{\bf X}_2=\frac{\partial}{\partial x},\quad\quad
{\bf X}_3=x\frac{\partial}{\partial x}-p\frac{\partial}{\partial p},\quad\quad
{\bf X}_4=x^2\frac{\partial}{\partial x}-2xp\frac{\partial}{\partial p},\\
{\bf X}_5=\frac{x}{\sqrt{-p}}\frac{\partial}{\partial x}+2\sqrt{-p}\frac{\partial}{\partial p}.
\end{gathered}
\end{equation}
Their non-vanishing commutation relations read
\begin{equation}\label{ComRel}
\begin{gathered}
\left[{\bf X}_1,{\bf X}_3\right]=\frac 12{\bf X}_1,\qquad [{\bf X}_1,{\bf X}_4]={\bf X}_5,\qquad
[{\bf X}_2,{\bf X}_3]={\bf X}_2,\qquad [{\bf X}_2,{\bf X}_4]=2{\bf X}_3,\\\left[{\bf X}_2,{\bf X}_5\right]={\bf X}_1,
\qquad \left[{\bf X}_3,{\bf X}_4\right]={\bf X}_4,\qquad [{\bf X}_3,{\bf X}_5]=\frac 12 {\bf X}_5,\\
\end{gathered}
\end{equation}
and therefore span a five-dimensional Lie algebra $V$ of vector fields. The $t$-dependent 
vector field ${\bf X}_t$ associated to the system (\ref{Hamil}) is given by~\cite{CLS12a}
\begin{equation}
{\bf X}_t={\bf X}_1-a_0(t){\bf X}_2-a_1(t){\bf X}_3-a_2(t){\bf X}_4.\label{F2}
\end{equation}

In view of expressions (\ref{ComRel}) and (\ref{F2}), the system (\ref{Hamil}) is a Lie system. Note also that a similar result would have been obtained by restricting the Legendre transform over the open set $\mathcal{W}_-=\{(t,x,v)\in\mathbb{R}\times{\rm T}\mathbb{R}\mid v+U(t,x)<0\}$.

\bigskip

Next, the vector fields (\ref{VF}) 
  are additionally Hamiltonian vector fields
relative to the Poisson bivector
$\Lambda=\partial/\partial x\wedge\partial/\partial p$ on $\mathcal{O}$. Indeed, they admit the 
Hamiltonian functions 
\begin{eqnarray}\label{equFun}
h_1=-2\sqrt{-p}, \quad  h_2=p,\quad h_3=xp,\quad h_4=x^2p,\quad  h_5=-2x\sqrt{-p},
\end{eqnarray}
which span along with $h_6=1$ a six-dimensional real Lie algebra of functions with respect to the Poisson bracket induced
by $\Lambda$ (see \cite{CLS12a} for details).
 
 \bigskip

The above action enables us to write the general solution $\xi(t)$ of system (\ref{Hamil}) in the form $\xi(t)=\Phi(g(t),\xi_0)$, 
where $\xi_0\in \mathcal{O}$ and $g(t)$ is the solution of the equation
\begin{equation}\label{HamGrup}
\frac{dg}{dt}=-\left({\bf X}^R_1(g)-a_0(t){\bf X}^R_2(g)-a_1(t){\bf X}^R_3(g)-a_2(t){\bf X}^R_4(g)\right),\qquad g(0)=e,
\end{equation}
on $G$, with the ${\bf X}^R_\alpha$ being a family of right-invariant vector fields over $G$ whose vectors 
${\bf X}^R_\alpha(e)\in T_eG$ close on the same commutation relations as the ${\bf X}_\alpha$ (cf. \cite{Dissertations}). 

Let us now apply to Lie systems (\ref{HamGrup}) the reduction theory for Lie systems. Since $T_eG\simeq \mathbb{R}^2\oplus_s
 \mathfrak{sl}(2,\mathbb{R})$, a particular solution of a Lie system of the form (\ref{HamGrup}) but over $SL(2,\mathbb{R})$, which amounts to integrating (first-order) Riccati equations (cf. \cite{Dissertations}), provides us with a transformation which maps system (\ref{HamGrup}) into an easily integrable Lie system over $\mathbb{R}^2$. In short, the explicit determination of the general solution of a second-order Riccati equation reduces to solving Riccati equations. 
 
\bigskip

In order to determine a superposition rule for the system  (\ref{Hamil}), it suffices to consider 
 two common functionally independent first-integrals for the 
diagonal prolongations $\widetilde{{\bf X}}_{1},\widetilde{{\bf X}}_{2},\widetilde{{\bf X}}_{3},\widetilde{{\bf X}}_{4},\widetilde{{\bf X}}_{5}$ 
to a certain ${\rm T}^*\mathbb{R}^{(m+1)}$, provided that these prolongations to ${\rm T}^*\mathbb{R}^{m}$ are linearly independent at a generic point. In the present case, 
  it can be easily verified that $m=4$. The resulting first-integrals  (see \cite{BHLS,CLS13,CL11Sec}) are explicitly given by 
\begin{equation*}
\begin{aligned}
F_0=(x_{(2)}-x_{(3)})\sqrt{p_{(2)}p_{(3)}}+(x_{(3)}-x_{(1)})\sqrt{p_{(3)}p_{(1)}}+(x_{(1)}-x_{(2)})\sqrt{p_{(1)}p_{(2)}},\\
F_1=(x_{(1)}-x_{(2)})\sqrt{p_{(1)}p_{(2)}}+(x_{(2)}-x_{(0)})\sqrt{p_{(2)}p_{(0)}}+(x_{(0)}-x_{(1)})\sqrt{p_{(0)}p_{(1)}},\\
F_2=(x_{(1)}-x_{(3)})\sqrt{p_{(1)}p_{(3)}}+(x_{(3)}-x_{(0)})\sqrt{p_{(3)}p_{(0)}}+(x_{(0)}-x_{(1)})\sqrt{p_{(0)}p_{(1)}}.
\end{aligned}
\end{equation*}
Note that given a family of solutions $(x_{(i)}(t),p_{(i)}(t))$, with $i=0,\ldots,3$, of (\ref{Hamil}), then $d\bar F_j/dt=\widetilde {\bf X}_tF_j=0$ for $j=0,1,2$ and $\bar F_j=F_j(x_{(0)}(t),p_{(0)}(t),\ldots,x_{(3)}(t),p_{(3)}(t))$.

In order to derive a superposition rule, it remains to obtain the value of $p_{(0)}$ from the equation $k_1=F_1$, where $k_1$ is a real constant. Proceeding along these lines and from the results given in~\cite{BHLS}, we get
$$
\sqrt{-p_{(0)}}=\frac{k_1+(x_{(2)}-x_{(1)})\sqrt{p_{(1)}p_{(2)}}}{(x_{(2)}-x_{(0)})\sqrt{-p_{(2)}}+(x_{(0)}-x_{(1)})\sqrt{-p_{(1)}}},
$$
and then plug this value into the equation $k_2=F_2$ to have
\begin{equation*}
\begin{aligned}
x_{(0)}&=\frac{k_1\Gamma(x_{(1)},p_{(1)},x_{(3)},p_{(3)})+k_2\Gamma(x_{(2)},p_{(2)},x_{(1)},p_{(1)})-F_0 x_{(1)} \sqrt{-p_{(1)}}}
{k_1(\sqrt{-p_{(1)}}-\sqrt{-p_{(3)}})+k_2(\sqrt{-p_{(2)}}-\sqrt{-p_{(1)}})-\sqrt{-p_{(1)}}F_0},\\
p_{(0)}&=-\left[{k_1/F_0(\sqrt{-p_{(3)}}-\sqrt{-p_{(1)}})+k_2/F_0(\sqrt{-p_{(1)}}-\sqrt{-p_{(2)}})+\sqrt{-p_{(1)}}}\right]^2,\\
\end{aligned}
\end{equation*}
where $\Gamma(x_{(i)},p_{(i)},x_{(j)},p_{(j)})=\sqrt{-p_{(i)}}x_{(i)}-\sqrt{-p_{(j)}}x_{(j)}$. The above expressions give us a superposition rule $\Phi:(x_{(1)},p_{(1)},x_{(2)},p_{(2)},x_{(3)},p_{(3)};k_1,k_2)\in{\rm T}^*\mathbb{R}^3\times\mathbb{R}^2\mapsto (x_{(0)},p_{(0)})\in {\rm T}^*\mathbb{R}$ for the system (\ref{Hamil}). Finally, since every $x_{(i)}(t)$ is a particular solution for (\ref{NLe}), the map $\Upsilon=\tau\circ \Phi$ furnishes the general solution of second-order Riccati equations in terms of three generic particular solutions
 $x_{(1)}(t),x_{(2)}(t),x_{(3)}(t)$ of (\ref{NLe}), the corresponding $p_{(1)}(t),p_{(2)}(t),p_{(3)}(t)$ and two real constants
  $k_1,k_2$.


\bigskip
\bigskip

\section{Poisson--Hopf algebras} \label{PHALGRAsection}

\bigskip

The Hopf structure \cite{Abe,MilnorMoore, Swee} is introduced through relevant examples, such as the universal enveloping algebra of a Lie algebra. The Hopf algebra structure originally appeared in the context of group cohomology, from which its use has been expanded to constitute nowadays an essential tool for the algebraic analysis. The next important definition will be the notion of Poisson structure and its compatibility with the Hopf algebras. In this section all vector spaces are defined over $\mathbb{C}$, where $V\otimes W$ denotes the tensor product of the two vector spaces $V$ and $W$.

\bigskip
\bigskip

\subsection{Hopf algebras}

\bigskip

\begin{definition}
 An algebra is a pair $(A,m)$, where $A$ is a linear space and $m:A\otimes A\rightarrow A$ is a bilinear map. Furthermore, $(A,m)$ is an algebra with unit if there is an element in $A$ such that $m(1,a)=m(a,1)$ for all $a\in A$. The algebra is associative  if for arbitrary  $x,y,z\in A$ the identity $m(m(x,y),z)=m(x,m(y,z))$ holds.
\end{definition}

\bigskip

If the map $m$ is the inner product on the tensor product, it is possible define the unit map $\eta: \mathbb{C}\rightarrow A$ such that $\eta (1)=1_{A}$. Then, the properties of algebra are

\begin{eqnarray*}
     m \circ (\eta \circ id_{A})=id_{A}=m\circ (id_{A}\circ \eta)\\
     m\circ (m \otimes id_{A})=id_{A}=m\circ (id_{A}\otimes m).
\end{eqnarray*}

\bigskip

\begin{definition}
 
 A coassociative coalgebra with counit is a linear space $A$ endowed with two linear maps: the coproduct   $\Delta: A\rightarrow A\otimes A$ and the counit $\epsilon : A \rightarrow \mathbb{C}$ such that \begin{eqnarray}
 (id\otimes \Delta)\circ \Delta=(\Delta \otimes id)\circ \Delta,\\
 (id\otimes \epsilon)\circ \Delta=id=(\epsilon \otimes id)\circ \Delta,
 \end{eqnarray}
 moreover, the coalgebra is cocommutative if the following diagram is commutative
 
 \begin{equation*}
     \xymatrix{{A} \ar[r]^{\Delta}  \ar[d]_{\Delta} & {A\otimes A}\ar[d]_{id\otimes \Delta} \\
{A \otimes A} \ar[r]_{\Delta\otimes id} &
{A \otimes A \otimes A.}}
 \end{equation*}
 
\end{definition}

\bigskip

Coalgebras can be described essentially by a dualization process. More specifically, from the point of view of the commutative diagrams, they are obtained reversing the direction of the corresponding maps. 

In view of this definitions, a \textit{bialgebra} $(A,m,\Delta)$ is a linear space where $(A,m)$ is an algebra and $(A,\Delta)$ is a coalgebra, such that it is verified \begin{eqnarray}
    \Delta (xy)&=&\Delta (x)\Delta (y),\nonumber\\
    \Delta (1)&=&1 \otimes 1,\\
    \epsilon (xy)&=&\epsilon (x) \epsilon (y)\nonumber
\end{eqnarray}
for all $x,y\in A$, i.e. the coproduct and the counit are algebra homomorphisms. 

\bigskip

The existence of a compatible coproduct with the product in an algebra allows the representation of $A$ over vector the space $V\otimes V'$, once the representations of $A$ over $V$ and $V'$ are known. The representation theory can be framed in this context, the map $D:A\otimes V\rightarrow V$ is a representation of $A$ on $V$ if it satisfies \begin{eqnarray*}
D\circ(id_{A}\otimes D)&=&D\circ(m\otimes id_{V}),\\
D\circ (\eta \otimes id_{V})&=&id_{V},
\end{eqnarray*}
i.e. $D$ is a representation consistent with the inner product and unit.

\bigskip

\begin{definition}
 A bialgebra $(A,m,\Delta)$ is a Hopf algebra if there exist a linear map $\gamma: A\rightarrow A$, so-called antipode, such that \begin{equation}\label{Hopf condition}
     m\circ (id_{A}\otimes \gamma)\circ \Delta=\eta \circ \epsilon=m\circ(\gamma \otimes id_{A})\circ \Delta.
 \end{equation}
\end{definition}

\bigskip

Let $(A,m,\Delta)$ be a bialgebra, if the antipode map exists, then it is unique. Hence, the Hopf algebra structure compatible with the bialgebra is unique. Furthermore, the antipode can be defined as the linear map such that the following diagram is commutative 

\begin{equation} \label{diagramHopf}  \medskip
   \centerline{ {\xymatrix@C=0.4em{&A\otimes A\ar[rr]^{{\rm id_{A}}\,\otimes\,  \gamma}&&A\otimes A\ar[rd]^m&&\\
   A\ar[rd]^\Delta\ar[rr]^{\epsilon}\ar[ur]^\Delta&&\mathbb{K}\ar[rr]^{\eta}&&A\\
   &A\otimes A\ar[rr]^{\gamma\,\otimes\, {\rm id_{A}}}&&A\otimes A\ar[ru]^m&}	}}
 \end{equation}
 where $\mathbb{K}$ is $\mathbb{R}$ or $\mathbb{C}$. The following proposition puts together the most relevant properties of Hopf algebras.
 
 \bigskip
 
 \begin{proposition}
 Let $(A,m,\Delta)$ be a Hopf algebra. Then for all $x,y\in A$ it is verified
 \begin{eqnarray*}
 \gamma(1)&=&1,\\
 \epsilon(\gamma(x))&=&\epsilon(x),\\
 \gamma(xy)&=&\gamma(y)\gamma(x),\\
 (\Delta \circ \gamma)(x)&=&((\gamma \otimes \gamma)\circ \sigma \circ\Delta)(x),
 \end{eqnarray*}
 where $\sigma(x\otimes y)=y\otimes x$  is a permutation. Hence, $\gamma$ is an antihomomorphism and anticohomomorphism. Moreover, the antipode map can be non-invertible.
 \end{proposition}
 
 \bigskip
 \bigskip

\begin{example}[Lie algebra]\cite{Jacobson}
Let $\mathfrak{g}$ be a Lie algebra, then the \textit{universal enveloping algebra} of $\mathfrak{g}$,\footnote{$\mathcal{U}(\mathfrak{g})$ is an algebra together with a morphism of Lie algebras $i:\mathfrak{g}\rightarrow \mathcal{U}(\mathfrak{g})$ such that $F:\mathfrak{g}\rightarrow \mathfrak{g}'$ is a morphism of Lie algebras. Then, there is a single morphism $\widetilde{F}: \mathcal{U}(\mathfrak{g})\rightarrow \mathfrak{g}'$ that verifies $\widetilde{F}\circ i=F$.} denoted $\mathcal{U}(\mathfrak{g})$, is a general associative algebra that is obtained  through the following quotient $$\mathcal{U}(\mathfrak{g})=\mathcal{T}(\mathfrak{g})/\mathcal{I},$$ where $\mathcal{I}$ is an ideal spanned by $\{XY-YX-[X,Y]/X,Y\in \mathfrak{g}\}$ and the tensor algebra of $\mathfrak{g}$, $\mathcal{T}(\mathfrak{g})$ denotes the graded algebra $\bigoplus_{k}\mathcal{T}^{k}(\mathfrak{g})$. Given a generic element $X\in\mathfrak{g}$ such that \begin{eqnarray*}
&&\Delta(X)=1\otimes X+X \otimes 1, \quad \Delta (1)=1\otimes 1, \\
&&\epsilon (X)=0, \quad \epsilon (1)=1, \\
&&\gamma(X)=-X,
\end{eqnarray*} 
then we can extend the mappings to  $\mathcal{U}(\mathfrak{g})$. This endowes  the universal enveloping algebra with a Hopf algebra structure\footnote{The elements $X\in\mathfrak{g}$ are called primitive elements of the Hopf algebra if they can be written as $X \otimes 1 + 1\otimes X$.}.\end{example}

\bigskip
\bigskip

\begin{remark}[Friedrichs' theorem]
Only the generators of a Lie algebra can be primitive elements of the universal enveloping algebra. For more details see \cite{Postnikov}.
\end{remark}

\bigskip

\begin{theorem}[Poincar\'e--Birkhoff--Witt]\cite{Knapp}
Let $\mathfrak{g}$ be a Lie algebra and $\{X_{1},\dots,X_{n}\}$ a basis of $\mathfrak{g}$, then the set $\{X_{1}^{k_{1}}\cdots X_{n}^{k_{n}}/k_{j}\in \mathbb{N}\}$ is a basis of the universal enveloping algebra $\mathcal{U}(\mathfrak{g})$. 
\end{theorem}

\bigskip
\bigskip


\subsection{Poisson algebras}

\bigskip

\begin{definition}
 A Poisson algebra is a vector space $A$ over field $\mathbb{K}$ endowed with two bilinear maps: $m: A\otimes A \rightarrow A,$ commutative, and the Poisson bracket\footnote{The Poisson bracket induces a derivation in the product $m$.} $\{\cdot,\cdot\}: A\otimes A \rightarrow A,$ where $(A,m)$ is an associative algebra and $(A,\{\cdot,\cdot\})$  is a Lie algebra. Then, the Poisson bracket obeys the Jacobi identity and it is antisymmetric.
\end{definition}

\bigskip

 Let $M$ be an $n$-dimensional manifold and consider its ring of smooth functions $\mathcal{C}^{\infty}(M)$. According to the last definition, a bivector $\Lambda \in \mathfrak{X}^2(M)$ induces a Poisson bracket on $\mathcal{C}^{\infty}(M)$ and it is an anti-symmetric biderivation. If $(x_1,\dots,x_n)$ are a local coordinates on $M$, a bivector $\Lambda$ turns out to be $$\Lambda=\lambda^{ij}(x)\frac{\partial}{\partial x_i}\wedge\frac{\partial}{\partial x_j},\ \ i<j,\ \ \lambda^{ij}\in \mathcal{C}^{\infty}(M),$$ hence the bracket takes the form \begin{equation}\label{poisson bracket def}
     \{f,g\}:=\Lambda(f,g)=\lambda^{ij}(x)\frac{\partial f}{\partial x_i}\frac{\partial g}{\partial x_j},
 \end{equation} for all $f,g \in \mathcal{C}^{\infty}(M)$. Note also that, in a general context, the last condition is the vanishing of the Schouten–-Nijenhuis bracket $[[\Lambda,\Lambda]]=0$ \cite{Schouten}. Moreover, the tensor $\lambda^{ij}$ can be degenerate, then this Poisson structure exists in odd-dimensional manifolds.

In this theoretical environment, a function $C\in \mathcal{C}^{\infty}(M)$ is a \textit{Casimir element} of the Poisson algebra $(\mathcal{C}^{\infty}(M),\Lambda)$ if it satisfies $$\{C,f\}=0$$ for all $f\in\mathcal{C}^{\infty}(M)$, i.e. the Casimir element belongs to the center of the Poisson algebra.

\bigskip

\begin{remark}[Racah 1951]
It was first shown by Racah that for semisimple Lie algebras $\frak{s}$, the number of functionally independent Casimir invariants (operators) coincide with the the rank of $\frak{s}$, 
i.e, the dimension of a Cartan subalgebra. 
\end{remark}

\bigskip
\bigskip

\begin{example} \cite{BCFHL}
Consider the Lie algebra $\mathfrak{sl}_2$ of traceless real $2\times 2$ matrices. This algebra admits the basis $\{e_1,e_2,e_3\}$ with commutation relations 
\begin{equation}\nonumber
[e_1,e_2]=e_1,\qquad [e_1,e_3]=2e_2,\qquad [e_2,e_3]=e_3.
\end{equation}
The elements of $\mathfrak{sl}_2$ can be considered as linear functions $v_1,v_2,v_3$ on $\mathfrak{sl}^*_2$, respectively. In this case, the corresponding Poisson bracket is given by 
$$
\{v_1,v_2\}=v_1,\qquad \{v_1,v_3\}=2v_2,\qquad \{v_2,v_3\}=v_3.
$$
This amounts to the Poisson bivector
$$
\Lambda=v_1\frac{\partial}{\partial v_1}\wedge \frac{\partial}{\partial v_2}+2v_2\frac{\partial}{\partial v_1}\wedge \frac{\partial}{\partial v_3}+v_3\frac{\partial}{\partial v_2}\wedge \frac{\partial}{\partial v_3}.
$$
The Poisson structure on $\mathfrak{g}^*$ admits a Casimir given by the function
$$
C=v_1v_3-v_2^2.
$$
The surfaces $S_k$, where $C$ takes a constant value $k$, are one-sided hyperboloids for $k>0$, two-sided hyperboloids when $k>0$, and cones for $k=0$. The Poisson bivector on the neighbourhood of a generic point of such surfaces reads
$$
\Lambda_{\mathfrak{g}}=v_1\frac{\partial}{\partial v_1}\wedge \frac{\partial}{\partial v_2},
$$
in the coordinate system $\{v_1,v_2,C\}$. As the canonical Poisson bivector on $\mathbb{R}^2\simeq T^*\mathbb{R}$ is given by 
$$
\Lambda_{\mathbb{R}^2}=\frac{\partial}{\partial x}\wedge\frac{\partial}{\partial y},
$$
it turns out that $\Lambda$ and $\Lambda_{\mathfrak{g}}$ locally describe the same Poisson bivector, but in different coordinates. Hence, there exits a Poisson algebra morphism $\phi:C^\infty(S_k)\rightarrow C^\infty(\mathbb{R}^2)$ given by
$$
\phi(v_1)=x,\qquad \phi(v_2/v_1)=y,\qquad \phi(C_z)=k.
$$
Therefore, 
$
\phi(v_3)=k/x+y^2x,
$  where $k$ is an arbitrary constant.
\end{example}

\bigskip
\bigskip


 \begin{example}   Consider the Lie--Hamilton algebra $\overline{\mathfrak{iso}(2)}=\langle e_0,e_1,e_2,e_3\rangle$, where $\{e_0,e_1,e_2,e_3\}$ is a basis with commutation relations
$$
[e_1,e_2]=e_0,\qquad [e_1,e_3]=e_2,\qquad [e_2,e_3]=-e_1
$$
and $e_4$ spans the center of $\overline{\mathfrak{iso}(2)}$. Considering the elements of the above basis as linear functions $\{v_0,v_1,v_2,v_3\}$ on the dual to $\mathfrak{iso}(2)$,  the corresponding Poisson bivector reads
$$
\Lambda=v_0\frac{\partial}{\partial v_1}\wedge\frac{\partial }{\partial v_2}+v_2\frac{\partial}{\partial v_1}\wedge \frac{\partial}{\partial v_3}-v_1\frac{\partial}{\partial v_2}\wedge\frac{\partial}{\partial v_3}.
$$
This Poisson algebra admits two Casimirs
$$
C_1:=v_0,\qquad C_2:=2v_0v_3-v_1^2-v_2^2
$$
which allow us to restrict the Poisson bracket to the surfaces $S_{\kappa_1,\kappa_2}$ where the Casimirs take constant values $\kappa_1,\kappa_2$. These are symplectic manifolds where the Poisson bivector $\Lambda$ can be mapped into a canonical form. In fact, the use of the coordinates $v_1,v_2,C_1,C_2$ leads to the Poisson bivector in the form
$$
\Lambda=\frac{\partial}{\partial (v_1/v_0)}\wedge \frac{\partial}{\partial v_2}.
$$
Hence, this leads to a representation of the Poisson bracket in terms of functions
$$
\phi(v_1/v_0)=x,\qquad \phi(v_2)=y,\qquad \phi(C_1)=k_1,\qquad \phi(C_2)=k_2.
$$
In consequence, a possible morphism of Poisson algebras is given by
$$
\phi(v_0)=k_1,\qquad \phi(v_1)=k_1 x,\qquad \phi(v_2)=y,\qquad \phi(v_3)=\frac{k_2+k_1^2x^2+y^2}{2k_1}.
$$
\end{example}

\bigskip
\bigskip

\begin{example} Consider now the Lie algebra $\mathfrak{so}(3)=\langle e_1,e_2,e_3\rangle$, where the basis $\{e_1,e_2,e_3\}$ satisfies the commutation relations
$$
[e_1,e_2]=e_3,\qquad [e_2,e_3]=e_1,\qquad [e_3,e_1]=e_2.
$$
Following the same methods sketched in the previous example, let us consider the Poisson algebra $(C^\infty(\mathfrak{so}(3)^*),\{\cdot,\cdot\})$. In this case, one obtains that the Poisson bivector associated with this Poisson manifold reads
$$
\Lambda=v_3\frac{\partial}{\partial v_1}\wedge\frac{\partial}{\partial v_2}+v_1\frac{\partial}{\partial v_2}\wedge\frac{\partial}{\partial v_3}+v_2\frac{\partial}{\partial v_3}\wedge\frac{\partial}{\partial v_1}
$$
and
$$
C_1=v_1^2+v_2^2+v_3^2
$$
becomes a Casimir function for this Poisson algebra. Therefore, the symplectic foliation for the Lie algebra is given by surfaces parametrized by the value of $C_1$. Moreover, the Poisson bivector becomes a symplectic structure on each leaf admitting a canonical form. This canonical form can be obtained by writing $\Lambda$ in the coordinate system $v_1,v_2,C$, then
$$
\Lambda=\sqrt{k-v_1^2-v_2^2}\frac{\partial}{\partial v_1}\wedge \frac{\partial}{\partial v_2}.
$$
To write the above in a canonical form, it is enough to introduce new variables $v_1=r\cos\varphi$ $v_2=r\sin\varphi$. This is natural as the symplectic leaves are spheres. 
Consequently,
$$
\Lambda=\frac{\sqrt{k-r^2}}{r}\frac{\partial}{\partial r}\wedge\frac{\partial}{\partial \varphi}=\frac{\partial}{\partial (-\sqrt{k-r^2})}\wedge\frac{\partial}{\partial \varphi}.
$$
Hence, it is enough to define the Poisson morphism as
$$
\phi(-\sqrt{k-r^2})=x,\qquad \phi(\varphi)=y.
$$
Undoing the previous changes of coordinates, we obtain
$$
h_1=-\sqrt{k-x^2}\cos(y),\qquad 
h_2=\sqrt{k^2-x^2}\sin(y),\qquad
h_3=x,
$$ where $h_i:=\phi(v_i)$ for $i=1,2,3$.
\end{example}

\bigskip
\bigskip

\begin{remark}[Schur's lemma]\cite{Shur}
Let $\mathfrak{g}$ be a Lie algebra  and $\phi : \mathfrak{g}\rightarrow End(V)$ a representation of $\mathfrak{g}$. If this representation is irreducible, then $\phi$ can be extended to a representation of the universal enveloping algebra $\mathcal{U}(\mathfrak{g)}$. In view of this result, any $x$ in the center of $\mathcal{U}(\mathfrak{g})$ has as its image $\phi(x)$ a multiple of the identity.
\end{remark}




\bigskip
\bigskip

 \subsection{Poisson--Hopf algebras}
 
 \bigskip
 
  Consider the Poisson algebras $(A,\{\cdot,\cdot \}_A)$ and $(B,\{\cdot,\cdot \}_B)$. A linear map $f:A\rightarrow B$ is a \textit{homomorphism of Poisson algebras} if  
  \begin{eqnarray}
 && f(xy)=f(x)f(y),\nonumber\\
 &&  f(\{x,y\})=\{f(x),f(y)\},\nonumber
  \end{eqnarray}
  for all $x,y \in A$. In this context, we define the Poisson structure on the tensor product $A\otimes B$ as $$\{x_1 \otimes y_1,x_2 \otimes y_2\}_{A\otimes B}:=\{x_!,x_2\}_A \otimes y_1 y_2+x_1 x_2 \otimes \{y_1,y_2\}_B.$$ Moreover, this Poisson bracket is the only Poisson structure such that, when it is projected onto either $A$ or $B$, the original Poisson brackets are recovered \cite{tjin}.
  
  \bigskip
  \bigskip
  
\begin{example}[Symmetric coalgebra]
The \textit{symmetric algebra} $S(\mathfrak{g})$ of a (finite dimensional) Lie algebra 
$\mathfrak{g}$ is the smallest commutative algebra containing $\mathfrak{g}$.
The second tensor power $\mathfrak{g}\otimes \mathfrak{g}$ of the Lie algebra is the space of real valued bilinear maps on the dual space. Recursively, the $k^{th}$ tensor power $\mathfrak{g}^{\otimes k}$ is the space of real valued k-linear maps. Taking the direct sum of the tensor powers of all orders, we arrive at the tensor algebra $\mathcal{T}(\mathfrak{g})$ of $\mathfrak{g}$. Here, the multiplication is 
\begin{equation}
\mathcal{T}(\mathfrak{g})\times \mathcal{T}(\mathfrak{g}) \longrightarrow \mathcal{T}(\mathfrak{g}), \qquad (v,u)\mapsto v\otimes u.
\end{equation}
We consider a basis $\{x_1,\dots, x_r\}$ of the Lie algebra 
$\mathfrak{g}$.
The space generated by the elements 
\begin{equation}
x_i\otimes x_j-x_j \otimes x_i
\end{equation}
 is an ideal, denoted by $\mathcal{R}$. The quotient space $\mathcal{T}(\mathfrak{g})/\mathcal{R}$ is called a symmetric algebra and denoted by $S(\mathfrak{g})$. 
The elements of $S(\mathfrak{g})$ can be regarded as polynomial functions on $\mathfrak{g}^*$. Therefore, this space can be endowed  with an appropriate Poisson bracket that makes $S(\mathfrak{g})$ into a Poisson algebra. It can be shown that a coalgebra structure can always be defined on $S(\mathfrak{g})$  introducing the comultiplication 
\begin{equation}\label{Con}
{\Delta} : S(\mathfrak{g})\rightarrow
S(\mathfrak{g}) \otimes S(\mathfrak{g}), \qquad 
{\Delta}(x)=x\otimes 1+1\otimes x,\quad \forall  x\in\mathfrak {g}\subset S(\mathfrak{g}),
\end{equation}
which is a Poisson algebra homomorphism. This makes $S(\mathfrak{g})$ into a Poisson-Hopf algebra. Furthermore, in the light of the coassociatity condition 
\begin{equation}
\Delta^{(3)}:=(\Delta \otimes {\rm Id}) \circ \Delta=({\rm Id} \otimes \Delta) \circ \Delta,
\end{equation}
we can define the third-order coproduct  \begin{equation}\label{3co}
\Delta^{(3)}: S(\mathfrak{g})\rightarrow
S(\mathfrak{g}) \otimes S(\mathfrak{g})\otimes S(\mathfrak{g}), \qquad {\Delta}^{(3)}(x)=x\otimes 1\otimes 1 +1\otimes x\otimes 1+1\otimes 1\otimes x
\end{equation}
for all $x\in\mathfrak {g}$, where $\mathfrak g$ is understood as a subset of $S(\mathfrak{g})$. The {$m$th-order coproduct} map 
can be defined, recursively, as  
\begin{equation}\label{copr}
\Delta ^{(m)}:  S(\mathfrak{g})\rightarrow   S^{(m)}(\mathfrak{g}), \qquad {\Delta}^{(m)}:= ({\stackrel{(m-2)-{\rm times}}{\overbrace{{\rm
Id}\otimes\ldots\otimes{\rm Id}}}}\otimes {\Delta^{(2)}})\circ \Delta^{(m-1)},\qquad m\ge 3,
\end{equation}
which, clearly, is also a Poisson algebra homomorphism.
\end{example}  

\bigskip
\bigskip

 \begin{definition}
Let $(A,m,\Delta)$ be a Hopf algebra and $(A, \{\cdot,\cdot \})$ a Poisson structure for $A$, then the triple  $(A,m,\Delta, \{\cdot,\cdot \})$ is a Poisson--Hopf algebra if the coproduct $\Delta$ is a homomorphism of Poisson algebras, i.e. 
\begin{equation*}
    \Delta(\{x,y\}_A)=\{\Delta(x),\Delta(y)\}_{A\otimes A},
\end{equation*} for all $x,y\in A$. If the antipode map does not exist, the triple is a Poisson bialgebra  \cite{Chari,Dri,  tjin}.
 \end{definition}
 
 \bigskip


\chapter{Poisson Manifolds and Quantum Groups} 

\label{Chapter2} 

\renewcommand{\theequation}{2.\arabic{equation}}
\section{Symplectic Geometry: Poisson Manifolds}

\bigskip

\subsection{Poisson bivectors and symplectic forms}

\bigskip

If  $M$ is a smooth manifold such that its ring of functions is a Poisson algebra $(\mathcal{C}^{\infty}(M), \{\cdot,\cdot\})$, then $M$ is called a \textit{Poisson manifold} \cite{Va94}. Clearly, if $\{\cdot,\cdot\}$ is a Poisson bracket, then $\{f,\cdot\}$ defines a vector field ${\bf X}_f \in \mathfrak{X}(M)$ for all $f\in \mathcal{C}^{\infty}(M)$. The vector field  ${\bf X}_f$ is well-defined and satisfies the commutator  $$\{f,g\}={\bf X}_f g=-{\bf X}_g f=dg({\bf X}_f)=-df({\bf X}_g).$$ Such a  vector field ${\bf X}_f$ is called a \textit{Hamiltonian vector field}. In these conditions, it can be easily verified that a Poisson bivector is a skew-bilinear form on $T^*(M)$.

\bigskip

In particular, if ${\bf X}_f,{\bf X}_g\in \mathfrak{X}(M)$ are Hamiltonian vector fields, they satisfy the identity  $$[{\bf X}_f,{\bf X}_g](h)=\{f,\{g,h\}\}-\{g,\{f,h\}\}=\{\{f,g\},h\}={\bf X}_{\{f,g\}}(h).$$ In order words, for all functions $f,g\in\mathcal{C}^\infty(M)$, the Hamiltonian vector fields satisfy the commutator condition \begin{equation}
[{\bf X}_f,{\bf X}_g]={\bf X}_{\{f,g\}}.
\end{equation} 

\bigskip

\begin{lemma}
If $(M,\{\cdot,\cdot\})$ is a Poisson manifold, then \begin{equation}\label{Lie derivate of bivector}
\mathcal{L}_{{\bf X}_f}\Lambda=0,
\end{equation} 
for all $f\in\mathcal{C}^\infty(M)$.
\end{lemma}

Whenever $(M,\omega)$ is a symplectic manifold,  the preceding equation \eqref{Lie derivate of bivector} turns out to adopt the particular $$\mathcal{L}_{{\bf X}_f}\omega=0,$$ for all $f\in\mathcal{C}^\infty(M)$.

\bigskip

\begin{remark}
Let $(M,g)$ be a Riemannian manifold with Levi-Civita connection $\nabla$ and $\Lambda$ be a bivector on $M$. If there exists a tensor field $T^{i}_{jk}$ on $M$  such that $T^{i}_{jk}=T^{i}_{kj}$ and $$\nabla_k \Lambda^{ij}+T^{i}_{hk}\Lambda^{hj}+T^{j}_{hk}\Lambda^{ih}=0,$$ then the bivector $\Lambda$ is a Poisson bivector, i.e. $\Lambda$ defines a Poisson structure on the Riemannian manifold.
\end{remark}

\bigskip

If $(M,\omega)$ is a symplectic manifold, then $\omega$ is a closed non-degenerate 2-form. This determines on $M$ a Poisson structure through the prescription
 $$\{f,g\}:=\omega({\bf X}_f,{\bf X}_g),$$ where ${\bf X}_f$ and ${\bf X}_g$ are defined by \eqref{contract3}.

\bigskip
\bigskip

\subsection{General distributions and Poisson cohomology}

\bigskip

\begin{definition}
Let $(M, \Lambda)$ be a Poisson manifold. If the set $\mathcal{D}_p(M)$ is defined as $$\mathcal{D}_p(M):=\{v\in T_p(M)/\exists f\in \mathcal{C}^\infty (M), {\bf X}_f(p)=v\},$$ then the set of linear subspaces $\mathcal{D}(M)=\{\mathcal{D}_p(M)\}$ is called a general distribution. Moreover, it is called the characteristic distribution of the Poisson structure whenever for all $p\in M$ there are vector fields ${\bf X}_{f_1},\dots,{\bf X}_{f_k}\in \mathcal{D}(M)$ such that for each point in $M$ they span the subspace $\mathcal{D}_p(M)$.
\end{definition}

\bigskip
\bigskip

\begin{example}
Let $\{f_1,f_2,f_3\}$ be the set of Hamiltonian functions \begin{equation*}
f_1=x,\quad f_2=x^2+zy, \quad f_3=1,
\end{equation*} 
and $\omega=dx\wedge dy$ the canonical symplectic form in the plane. Then the Hamiltonian vector fields are given by 
 \begin{equation*}
{\bf X}_{f_1}=-\frac{\partial}{\partial y}, \quad {\bf X}_{f_2}=-2x\frac{\partial}{\partial y}+z\frac{\partial}{\partial x}, \quad {\bf X}_{f_3}=0.
\end{equation*}
In this case, the distribution $\mathcal{D}=<{\bf X}_{f_1},{\bf X}_{f_2},{\bf X}_{f_3}>$ has rank 2 if and only if $z\neq 0$, as
 \begin{equation*}
\det\begin{pmatrix}
0 & -1  \\
z & -2x \\
\end{pmatrix}
=z.
\end{equation*}
A point $(x,y,z)\in M$ such that $z\neq 0$ is called \textit{regular}, otherwise it is called  \textit{singular}.
\end{example}

\bigskip
\bigskip

In this context, it is possible to establish a simple criterion that ensures the complete integrability of a distribution (the proof can be found e.g. in \cite{Bhas, Va94}).

\bigskip

\begin{definition}
Let $\mathcal{D}(M)$ be a general distribution on a Poisson manifold, $M$. The distribution is invariant if there are vector fields on $M$ such that for all $p\in M$, these vector fields span $\mathcal{D}_p$ at $p$, and such that for all $t\in \mathbb{R}$ and $p\in M$ the exponential map
 \begin{equation*}
    (\exp t{\bf X})_*(\mathcal{D}_p(M))=\mathcal{D}_{\exp t{\bf X}(p)}(M). 
\end{equation*}
is defined.
\end{definition}

\bigskip

\begin{theorem}[Stefan--Sussmann Frobenius theorem]
A general distribution $\mathcal{D}(M)$ is completely integrable if  and only if  it is invariant.
\end{theorem}

\bigskip

If the manifold $M$ is additionally a Poisson manifold, the leaves of $\mathcal{D}(M)$ are called \textit{symplectic leaves}, with $\mathcal{D}(M)$ being referred to as a \textit{symplectic foliation} on $M$.

\bigskip

\begin{theorem}
Let $(M,\Lambda)$ be a Poisson manifold and $\mathcal{D}(M)$ its characteristic distribution. Then $\mathcal{D}(M)$ is completely integrable and the Poisson structure defines a symplectic structure on each leaf of the characteristic distribution.
\end{theorem}

\bigskip



\noindent For completeness, we briefly recall some elementary notions about cohomology groups. 

\medskip 

\begin{definition}
Let $(M,\Lambda)$ be a Poisson manifold, and $\delta_\Lambda$ the coboundary operator such that $\delta_\Lambda(X)=[\Lambda,X]\in \mathfrak{X}^{k+1}(M)$ for all $X\in \mathfrak{X}^k(M)$. Then, the cohomology groups are defined as \begin{equation}
H^k_\Lambda(M):=\frac{\ker(\delta_\Lambda:\mathfrak{X}^{k}(M))\rightarrow \mathfrak{X}^{k+1}(M)}{ \mathrm{im}(\delta_\Lambda:\mathfrak{X}^{k-1}(M))\rightarrow \mathfrak{X}^{k}(M))}.
\end{equation}
\end{definition}

\bigskip

\begin{remark}
For low values of $k$, the cohomology groups have a concise geometrical interpretation:  
\begin{description}
    \item[k=0:] If ${\bf X}_f\in \mathfrak{X}(M)$ is a Hamiltonian vector field, then $\delta_\Lambda(f)={\bf X}_f$. Hence, $H^0_\Lambda(M)$ is the set of functions $f\in \mathcal{C}^\infty(M)$ such that $\{f,\cdot\}=0$, i.e., the Casimir functions of the Poisson structure. Hence, $H^1_\Lambda(M)$ coincides with the center of $\mathcal{C}^\infty(M)$.
    \item[k=1:] Let ${\bf X}\in \mathfrak{X}(M)$ be a vector field. If ${\bf X}$ is a infinitesimal automorphism of the Poisson structure, then $\mathcal{L}_{\bf X} \Lambda=\delta_\Lambda {\bf X}=0$. Therefore, $H^1_\Lambda(M)$ is the quotient of the space of infinitesimal automorphisms of the Poisson manifold by the space of the Hamiltonian vector fields.
    \item[k=2:] This group has the class of the element $\Lambda$ defined by $[\Lambda,\Lambda]=0$. If this class is 0, there is a vector field ${\bf X}$ such that $\delta_\Lambda({\bf X})=\mathcal{L}_{\bf X}\Lambda$. Poisson manifolds with this property are called exact Poisson manifolds.
\end{description}
If $\omega$ is a $k-$form on $M$, then the identity $\delta_\Lambda(\sharp(\omega))=\sharp(d\omega)$ is satisfied\footnote{The map $\sharp:T^*(M)\rightarrow T(M)$ is a homomorphism, such that for all $\beta,\alpha\in T^*(M)$ is defined by $\beta(\sharp(\alpha))=\Lambda(\alpha,\beta)$. If the bivector $\Lambda$ is non-degenerate then the map $\sharp$ is an isomorphism.}. It follows that the map $\sharp$ induces a homomorphism of $H^k_{dR}(M)\rightarrow H^k_\Lambda(M)$, where $H^k_{dR}(M)$ denotes the de Rham cohomology. If the bivector is non-degenerate, then the map is an isomorphism.

\end{remark}

\bigskip
\bigskip


\section{Poisson--Lie groups and Lie bialgebras}

\bigskip

\begin{definition}
Let $G$ be a Lie group. If $(\mathcal{C}^{\infty}(G),\Delta, \{\cdot,\cdot \})$ is a Poisson--Hopf algebra, then $G$ is a Poisson--Lie group.
\end{definition}

\bigskip

Let $G$ be a Poisson--Lie group with its Poisson structure defined by the bivector $\Lambda$. For all $f\in \mathcal{C}^\infty(G)$ we have a vector field locally determined by 
 \begin{equation*}
{\bf X}_f:=\{f,\cdot\}=X^{ij}\frac{\partial f}{\partial x_i}\frac{\partial}{\partial x_j}.
\end{equation*}
As the right-invariant vector fields $\{{\bf R}_i\}$ are a basis for $T_g(G)$ for arbitrary elements $g\in G$, we conclude that there exist functions $\alpha^i$ such that ${\bf X}_f=\alpha^i{\bf R}_i$. Hence, all Poisson structures of a Poisson--Lie group can be written in local coordinates as \begin{equation*}
\Lambda=\lambda^{ij}{\bf R}_i\wedge {\bf R}_j,
\end{equation*} 
where $\lambda^{ij} \in \mathcal{C}^\infty(G)$ and $i<j$.

\bigskip

Let $\mathfrak{g}$ be a Lie algebra of a Poisson--Lie group $G$. Then, the Lie algebra has a Lie bracket associated with the group structure. Moreover, there is a Lie structure in $\mathfrak{g}^*$ due to the linearization of the Poisson structure in $G$. The Lie structure in $\mathfrak{g}^*$ is well-defined as \begin{equation}
    [v_1,v_2]:=(d\{f_1,f_2\})\arrowvert _e
\end{equation} for all $f_1,f_2 \in \mathcal{C}^\infty (G)$, where $\{\cdot,\cdot\}$ is the Poisson bracket and $v_i:=df_i\arrowvert _e$. Hence, the cocommutator is defined  by means of the following relation 
\begin{equation}
    \delta([X,Y])=[\delta(X),1\otimes Y+Y\otimes 1]+[1\otimes X+X\otimes 1,\delta(Y)]
\end{equation} for all $X,Y \in \mathfrak{g}$, such that if $\Lambda_R:G\rightarrow \mathfrak{g}\otimes \mathfrak{g}$ is the right-translation of the Poisson bivector on $G$, then $\delta:\mathfrak{g}\rightarrow \mathfrak{g} \otimes \mathfrak{g}$ is the tangent map $\delta:=T_e\Lambda_R$. Thus, $\delta^*$ defines a Lie bracket on $\mathfrak{g}^*$, such that $\delta^*(v \otimes v')=[v,v']$, and $v,v' \in \mathfrak{g}^*$. In this context, the Lie algebra $(\mathfrak{g},\delta)$ is called  a \textit{Lie bialgebra}.

\bigskip

If there is an element $\rho \in \mathfrak{g}\otimes \mathfrak{g}$, such that $\delta(\rho)= [ 1 \otimes X+X\otimes 1,\rho]$, then the Lie bialgebra $(\mathfrak{g},\delta, \rho)$ is called a \textit{coboundary Lie bialgebra}. The element $\rho$ is the \textit{classical r-matrix} of $\mathfrak{g}$.

\bigskip
\bigskip

Let $(\mathfrak{g},\delta, \rho)$ be a coboundary Lie bialgebra. If $\{X_1,\dots, X_n\}$ is a basis of $\mathfrak{g}$ then the classical r-matrix turns out to be $\rho=\rho^{ij}X_i\otimes X_j$. By virtue of the elements \begin{align}
    \rho_+ :=\frac{1}{2}(\rho^{i,j}+\rho^{j,i})X_i\otimes X_j,\  \
    \rho_{12}:=\rho^{ij}X_i\otimes X_j\otimes 1, \  \\
    \rho_{13}:=\rho^{ij}X_i\otimes 1\otimes X_j, \  \
    \rho_{23}:=\rho^{ij}1\otimes X_i\otimes X_j,
\end{align}
$(\mathfrak{g},\delta, \rho)$ is a Lie bialgebra if, and only if, the following identities are satisfied: \begin{equation}
    (ad_\mathfrak{g}\otimes ad_\mathfrak{g})(\rho _+)=0,
\end{equation}
and \textit{the modified classical Yang-Baxter equation} holds
\begin{equation}
    (ad_\mathfrak{g}\otimes ad_\mathfrak{g}\otimes ad_\mathfrak{g})[[\rho,\rho]]=0,
\end{equation}
where $[[\cdot,\cdot]]$ is the Schouten-–Nijenhuis bracket $$[[\rho,\rho]]:=[\rho_{12},\rho_{13}]+[\rho_{12},\rho_{23}]+[\rho_{13},\rho_{23}].$$ The first condition guarantees the antisymmetry of $\delta^*$, while the last equation ensures that the Jacobi identity in $\mathfrak{g}^*$ is satisfied.

\bigskip
\bigskip

\begin{definition}
Two Lie bialgebras, $(\mathfrak{g},\delta_1)$ and $(\mathfrak{g},\delta_2)$ are equivalent if there is a Lie automorphism $\Theta$ of $\mathfrak{g}$ such that $$\delta_2=(\Theta^{-1}\otimes \Theta^{-1})\circ\delta_1\circ \Theta,$$ that is, the next diagram  \begin{equation*}
     \xymatrix{{\mathfrak{g}} \ar[r]^{\delta_1}  \ar[d]_{\Theta} & {\mathfrak{g}\otimes \mathfrak{g}}\ar[d]_{\Theta \otimes \Theta} \\
{\mathfrak{g} } \ar[r]_{\delta_2} &
{\mathfrak{g} \otimes \mathfrak{g}}} 
 \end{equation*} is commutative.
\end{definition}

\bigskip

\begin{remark} (see \cite{tjin}) \
\begin{itemize}
    \item If $\mathfrak{g}$ is a semisimple Lie algebra, then all its Lie bialgebras are coboundary Lie bialgebras.
    \item If $\mathfrak{g}$ is an abelian Lie algebra, then there is a non-coboundary Lie bialgebra $(\mathfrak{g}, \delta)$, and if the Lie algebra $\mathfrak{g}$ is not abelian, then there is a non-trivial coboundary Lie bialgebra $(\mathfrak{g}, \delta)$.
    \item In general, the Lie bialgebra $(\mathfrak{g}, \delta)$ can be a coboundary Lie bialgebra, while its dual bialgebra, $(\mathfrak{g}^*, \delta^*)$ is not a coboundary Lie bialgebra.
    \item Let $\mathfrak{g}$ be a Lie algebra, the commutation relations 
    \begin{equation}
        [X_i,X_j]=c^{k}_{ij}X_k, \quad    [v_i,v_j]=\lambda^{k}_{ij}v_k,\quad
        [v_i,X_j]=c^{i}_{jk}v_k-\lambda^{j}_{ik}X_k, \nonumber
    \end{equation}
    span a Lie algebra over $\mathfrak{g}\otimes\mathfrak{g}^*$ called a \textit{Manin--Lie algebra}, $\mathfrak{g}\bowtie \mathfrak{g}^*$.\end{itemize} 

\end{remark}

\bigskip
\bigskip


\subsection{Quantum deformations and quantum algebras}

\bigskip

In this section, we shall consider the algebra $A=C^\infty(M)$ with $M$ being a Poisson manifold. Albeit quantum deformations can be defined in a rather general context,  we restrict ourselves to the syudy on Poisson manifolds. For the general case, see e.g. \cite{Balltesis, Chari,Dri,  Va94}.

\bigskip

\begin{definition}
The algebra $A_z$ is a \textit{$z$-parametric deformation} of the algebra $A$, if $A_z$ is a formal series algebra $A[[z]]$ such that the quotient $A/zA_z$ is isomorphic to $A$.
\end{definition}

\bigskip

Furthermore, in terms of the product, $A_z$ is a \textit{quantization} of the Poisson algebra $(A,m,\{\cdot,\cdot\})$ if there is an associative $*_z$-product such that it is a deformation of $m$ given  by $$f*_z g:=fg+\frac{1}{z}\{f,g\}+o(z^2),$$  $f*_z a=a*_z f$ for all $a\in \mathbb{C}$ and $f,g\in A$. If there is a homomorphism of Poisson algebras $\Phi$ such that  $$\Phi(f)*_z\Phi(g)=\Phi(f*_z g),$$ then the $*_z$-product is invariant. It follows that the limit  $$\{f,g\}=\lim_{z \to 0}\frac{1}{z}(f*_z g-g*_z f)$$ 
recovers the classic bracket, implying the relation  
\begin{equation}
    [f,g]:=f*_z g-g*_z f=z \{f,g\}+o(z^2).
\end{equation}

\bigskip

Let $(A,m,\Delta)$ be a Hopf algebra and $(A, \{\cdot,\cdot \})$ be a Poisson structure for $A$. Then $(A_z\Delta_z)$ is a quantum deformation of $(A,m,\Delta, \{\cdot,\cdot \})$ if there is a coproduct $\Delta_z$ such that 
\begin{equation}
    \Delta_z (f*_z g)=\Delta_z (f)*_z \Delta_z (g),
\end{equation} for all $f,g\in A.$

\bigskip

\begin{remark}
Let $\omega$ be a bilinear form on the vector space $\mathcal{C}^\infty (M)$ of a Poisson Manifold $M$ and let $z$ be a parameter. Hence, a deformation of $\omega$ is a formal power series. If $\omega$ determines a Poisson bracket, a deformation  of the Poisson--Lie bracket will be denoted as $\omega_z(f,g)=\{f,g\}_z$.
\end{remark}

\bigskip

\begin{definition}
A Hopf algebra $(A_z,m_z,\Delta_z)$ is a quantization of a co-Poisson--Hopf algebra $(A,m,\Delta, \delta)$ if $A_z$ is a deformation of a Hopf algebra $(A,m,\Delta)$ and there is a $*_z$-co-product defined by \begin{equation}
    \Delta_z (X)=\Delta(X)+\frac{z}{2}\delta(X)+o(z^2)
\end{equation} compatible with a product $m_z$.
\end{definition}

\bigskip

\begin{remark}
If $\mathfrak{g}$ is a Lie bialgebra over $\mathbb{R}$ or $\mathbb{C}$, then admits a quantization.
\end{remark}

\bigskip

Within the Hopf algebras, quantum groups represent a specially interesting class that have shown to be of capital importance in Geometry \cite{Chari,Majid}. For example, as algebraic groups are well described by its Hopf algebra of functions,  the deformed version of the latter Hopf algebra describes a quantized version of the algebraic group, which generally does not correspond anymore to an algebraic group.
\bigskip

\begin{definition}
Let $\mathfrak{g}$ be a Lie algebra,  $(\mathcal{U}_z(\mathfrak{g}),m_z,\Delta_z)$ is a quantum algebra if it is a quantization of a co-Poisson--Hopf algebra $(\mathcal{U}(\mathfrak{g}),m,\Delta_0, \delta)$.
\end{definition}

\bigskip
\bigskip

\begin{example}
Let us consider $\mathfrak{sl}(2)$ with the standard basis $\{J_3, J_+,J_-\}$ satisfying the commutation relations
$$
[ J_3,J_\pm  ]  = \pm  2J_\pm     ,\qquad  [J_+ , J_- ] = J_3.
$$
In this basis, the Casimir operator reads 
\be
C=\frac{1}{2}  J_3^2+(J_+ J_- + J_- J_+).
\label{cas1}
\ee
 Considering the non-standard (triangular or Jordanian) quantum deformation $\mathcal{U}_z(\mathfrak{sl}(2))$ of $\mathfrak{sl}(2)$~\cite{Ohn} (see also \cite{non} and references therein), we are led to the following deformed coproduct 
\bea
&& \Delta_z(J_{+})=  J_+ \otimes 1+
1\otimes J_+ , \nonumber\\
&&\Delta_z(J_l)=J_l \otimes \eee^{2 z J_+} + \eee^{-2 z J_+} \otimes
J_l ,\qquad l\in\left\{-,3\right\}
\nonumber
\eea
and the commutation rules 
\bea
&& \left[J_3,J_+\right]_z= 2\,{{\rm shc}(2z J_+)}  J_{+}, \qquad\left[J_+,J_-\right]_z=J_3, \nonumber\\
&& \left[J_3,J_-\right]_z=-J_{-} \,{\rm ch}(2z J_+)-{\rm ch}(2z J_+)J_{-}.
\nonumber
\eea
Here ${\rm shc}$ denotes the cardinal hyperbolic sinus function defined by 
$$
\shc( \xi):=  \left\{ 
\begin{array}{ll}
\displaystyle \frac{\sinh(\xi)}{\xi} , & \ \  \mbox{for}\  \xi\ne 0, \\
1,& \ \  \mbox{for}\  \xi=0.
\end{array}
 \right. 
$$

\bigskip

It is known that every quantum algebra $\mathcal{U}_z(\mathfrak{g})$ related to a semi-simple Lie algebra $\mathfrak{g}$ admits an isomorphism of algebras $\mathcal{U}_z(\mathfrak{g})\rightarrow \mathcal{U}(\mathfrak{g})$ (see \cite[Theorem 6.1.8]{Chari}). This allows us to obtain a Casimir operator of $U_z(\mathfrak{sl}(2))$ from $C$ in (\ref{cas1}) as (see \cite{Chari} for details)
$$
C_z=\frac{1}{2}J_3^{2}+  {{\rm shc}( 2z J_{+})}{J_+}J_{-} + J_{-}{J_+} \, {{\rm shc}( 2z J_{+})} +\frac{1}{2}\,{\rm ch}^{2}(2zJ_{+}),
\label{gx}
$$
As expected, this coincides with the expression  formerly given in ~\cite{non}.
\end{example}

\bigskip
\bigskip
 
If $G$ is a Lie group with Lie algebra $\mathfrak{g}$, then we know that there is a duality between the universal enveloping algebra $\mathcal{U}(\mathfrak{g})$ and $\mathcal{C}^\infty(G)$ \cite{tjin}. The function $\rho: \mathfrak{g}\rightarrow End \mathcal{C}^\infty(G)$ defined by $$(\rho (X)\phi)(g):=\frac{d}{dt}\phi(exp^{tX}g)\shortmid_{t=0},$$ where $X\in \mathfrak{g}$, $\phi \in \mathcal{C}^\infty(G)$ and $g\in G$, can be extended to a homomorphism of $\mathcal{U}(\mathfrak{g})$ into $End_\mathbb{C} \mathcal{C}^\infty(G)$. There is an right invariant action $R:G\rightarrow \mathcal{C}^\infty(G)$ such that $R_g(\phi)(h)=\phi(gh)$, and such that the Leibniz rule is fulfilled. Then, it is possible to define a bilinear map $\langle \cdot,\cdot \rangle$ of $\mathcal{C}^\infty(G)\otimes \mathcal{U}(\mathfrak{g})$ onto $\mathbb{C}$ 
  \begin{equation}
    \langle \phi, x \rangle=(p(x)\phi)(e),
\end{equation}
where $e$ is the neutral element of $G$. Hence, the map defined by 
\begin{align}
    \mathcal{C}^\infty(G)&\longrightarrow  \mathcal{U}(\mathfrak{g})^*\\
    \phi &\longmapsto  \langle \phi, \cdot \rangle 
\end{align} is an immersion, according to the Gelfand-–Naimark theorem \cite{GelNeu}. We conclude that the ring of functions $\mathcal{C}^\infty(G)$ can be considered as a dual of $\mathcal{U}(\mathfrak{g})$ \footnote{It is not true that both Hopf algebras, $\mathcal{C}^\infty(G)$ and $\mathcal{U}(\mathfrak{g})$, are duals of each other, due to some problems that arise in  infinite dimensional Lie algebras \cite{Abe,Chari}.}.

\bigskip

\begin{remark}[Quantum group]
It ought to be observed that currently there is no universal and unified definition of quantum groups \cite{Balltesis, Chari}, albeit all definitions explicitly refer to the Hopf algebra structure. Alternative approaches are e.g. given by the following: 

\begin{itemize}
\item   Approach in Noncommutative Geometry: as a deformation of algebraic groups. Here matrix groups are subjected to satisfy certain algebraic identities.

\item The Faddeev theory \cite{Fade}: it uses solutions of the quantum Yang--Baxter equation (YBE). This approach is the preferred one in Quantum Field Theory.
\end{itemize} 

In the following we will use the notion of quantum group as introduced by Drinfel'd in \cite{Dri}: A quantum group is a non-commutative Hopf algebra (deformation of the universal enveloping algebra of a Lie algebra) that gives rise to a Poisson--Hopf algebra $(\mathcal{C}^\infty(G),\{\cdot,\cdot\},\Delta)$.
\end{remark}

\bigskip

\chapter{Poisson--Hopf Algebra Deformations of Lie Systems} 

\label{Chapter3} 
\renewcommand{\theequation}{3.\arabic{equation}}



\sect{Formalism}

\bigskip

For  the sake of  simplicity we consider explicit computations merely on $\mathbb{R}^2$, but we stress that this approach can be applied, mutatis mutandis, to construct Poisson--Hopf algebra deformations of Lie--Hamilton systems defined on any manifold.

\bigskip
\bigskip


\subsect{Lie--Hamilton systems}

\bigskip

Let us consider the local coordinates  $\{x_{1},\dots, x_{n}\}$ on an  $n$-dimensional manifold $M$. Geometrically, every non-autonomous system of first-order differential equations on $M$ of the form
\begin{equation}
 \frac{{\rm d} x_{i}}{{\rm d} t  }=f_{i}(t,x_{1},\dots,x_{n}), \qquad i=1,\dots, n,
 \label{system}
\end{equation}
where $f_{i}:\mathbb{R}^{n+1}\rightarrow \mathbb{R}$ are arbitrary functions, amounts to a $t$-dependent vector field  ${\bf X}:\mathbb{R}\times M\rightarrow {\rm T}M$  given by 
\begin{equation}\label{Vect}
{\bf X}_t :\mathbb{R}\times M\ni (t,x_{1},\dots,x_{n})\mapsto \sum_{i=1}^{n} f_{i}(t,x_{1},\dots,x_{n})\frac{\partial}{\partial x_{i}}\in {\rm T}M .
\end{equation}
This justifies to represent (\ref{Vect}) and its related system of differential equations (\ref{system}) by ${\bf X}_t$ (cf. \cite{Dissertations}).

\bigskip

According to the \textit{Lie--Scheffers Theorem}~\cite{CGM00,CGM07,LS}, see the theorem \ref{LSTheorem}, a system $\bf X$ is a Lie system if, and only if, 
\be
{\bf X}_t(x_{1},\dots,x_{n}):= {\bf X}(t,x_{1},\dots,x_{n})=\sum_{i=1}^\ell b_i(t){\bf X}_i(t,x_{1},\dots,x_{n}) ,
\label{aabb}
\ee
  for some $t$-dependent functions $b_1(t),\ldots,b_\ell(t)$ and vector fields ${\bf X}_1,\ldots,{\bf X}_\ell$ on $M$ that span  an $\ell$-dimensional real Lie algebra $V$ of vector fields, i.e. the Vessiot--Guldberg Lie algebra of ${\bf X}$.  

  A Lie system ${\bf X}$ is, furthermore, a LH one~\cite{BBHLS,BCHLS13Ham,BHLS, Dissertations,CLS13,HLT} if it admits a Vessiot--Guldberg Lie algebra $V$ of Hamiltonian vector fields relative to a Poisson structure. This amounts to the existence, around each generic point of $M$, of a symplectic form, $\omega$, such that:
\be
\mathcal{L}_{{\bf X}_i}\omega=0 ,
\label{der}
\ee
for a basis ${\bf X}_1,\ldots,{\bf X}_\ell$ of $V$ (cf.~Lemma 4.1 in \cite{BBHLS}). To avoid minor technical details and to highlight our main ideas, hereafter it will be  assumed, unless otherwise stated, that the symplectic form and remaining structures are defined globally. More accurately, a  local description around a generic point in $M$ could easily  be  carried out.

Each vector field ${\bf X}_i$  admits a Hamiltonian function $h_i$ given by the rule:
\be
\iota_{{\bf X}_i}\omega={\rm d}h_i,
\label{contract}
\ee
where $\iota_{{\bf X}_i}\omega$ stands for the contraction of the vector field ${\bf X}_i$ with the symplectic form $\omega$. Since $\omega$ is non-degenerate, every function $h$ induces a unique associated Hamiltonian vector field ${\bf X}_h$ (Chapter \ref{Chapter1}).

\bigskip
\bigskip

\subsect{Poisson--Hopf algebras}

\bigskip

 The core in what follows is the fact that the space $\mathcal{C}^\infty\left(\lh^* \right)$ can be endowed with a {\it Poisson--Hopf algebra} structure. 
  We recall that an associative algebra $A$  with a {\it product} $m$ and a {\it unit} $\eta$  is said to be a {\em Hopf algebra} over $\mathbb{R}$
\cite{Abe,Chari, Majid} if there exist two homomorphisms called {\em coproduct}  $(\Delta :
A\longrightarrow A\otimes A )$ and {\em counit} $(\epsilon : A\longrightarrow
\Bbb R)$, along with an  antihomomorphism, the {\em antipode} $\gamma :
A\longrightarrow A$,  such that the following diagram \eqref{diagramHopf} is commutative, section \ref{PHALGRAsection}.

If $A$ is a commutative Poisson algebra and $\Delta$ is a Poisson algebra morphism, then  $(A,m, \eta,\Delta,\epsilon,\gamma)$ is a {\it Poisson--Hopf algebra}  over $\Bbb R$.
We recall that the Poisson bracket on $A\otimes A$ reads
$$
\{ a\otimes b, c\otimes d\}_{A\otimes A}=\{ a, c\}\otimes  b d + a c\otimes \{ b, d\} ,\qquad \forall a,b,c,d\in A .
$$

\bigskip

In our particular case, $\mathcal{C}^\infty\left(\lh^* \right)$ becomes a Hopf algebra relative to its natural associative algebra with unit provided that
$$
\begin{gathered}
 \Delta (f)(x_1,x_2):=f(x_1+x_2),\qquad m(h\otimes g)(x):=h(x)g(x),\\
 \epsilon (f):=f(0),\qquad \eta(1)(x):=1,\qquad \gamma(f)(x):=f(-x),
 \end{gathered}
 $$
 for every $x,x_1,x_2\in \lh$ and $f,g,h\in \mathcal{C}^\infty(\lh^*)$. Therefore, the space $\mathcal{C}^\infty\left(\lh^* \right)$ becomes a Poisson--Hopf algebra by endowing it with the Poisson structure defined by the Kirillov--Kostant--Souriau bracket related to a Lie algebra structure on $\lh$.

 \bigskip
\bigskip


\subsect{Deformations of Lie--Hamilton systems and generalized distributions} \label{SecDefLHsystem}

\bigskip

In this section we propose a systematic procedure to obtain deformations of LH systems by using  LH algebras and deformed Poisson--Hopf algebras that lead to appropriate extensions of the theory of LH systems. Explicitly, the construction is based upon the following  {four} steps:
  
  \bigskip

\begin{enumerate}

\item Consider a LH  system ${\bf X}$ (\ref{aabb}) on  $\mathbb{R}^{2n}$ with respect to a symplectic form $\omega$ and admitting a LH algebra $\lh$ spanned by a basis of functions $h_1,\ldots, h_\ell\in \mathcal{C}^\infty(\mathbb{R}^{2n})$ with structure constants $c_{ij}^k$, i.e.
\be
\{h_i,h_j\}_{\omega}= \sum_{{k=1}}^{\ell} c_{ij}^{k}h_{k},\qquad i,j=1,\ldots,\ell.
 \nonumber
\ee

\item Introduce a Poisson--Hopf algebra deformation $\qlhz$ of $\mathcal{C}^\infty(\lh^*)$ with deformation parameter $z\in\mathbb R$ (in a quantum group setting we would have $q:={\rm e}^z$) as the space of smooth functions $F(h_{z,1},\ldots,h_{z,\ell})$ with fundamental  Poisson bracket given by
\be
\{h_{z,i},h_{z,j}\}_{\omega}= F_{z,ij}(h_{z,1},\dots,h_{z,\ell }),
\label{zab}
\ee
where $F_{z,ij}$ are certain smooth functions also depending smoothly on the deformation parameter $z$ and  such that
\be
\lim_{z\to 0} h_{z,i}=h_i ,\qquad \lim_{z\to 0}\nabla h_{z,i}=\nabla h_i,  \qquad   \lim_{z\to 0}   F_{z,ij}(h_{z,1},\dots,h_{z,\ell }) =\sum_{k=1}^\ell c_{ij}^k h_k, 
\label{zac}
\ee
where $\nabla$ stands for the gradient relative to the Euclidean metric on $\mathbb{R}^{2n}$.
Hence,
\be
  \lim_{z\to 0} \{h_{z,i},h_{z,j}\}_{\omega}=\{h_i,h_j\}_\omega .
\label{zad}
\ee

\item Define the deformed vector fields ${\bf X}_{z,i}$ by the rule
\be
\iota_{{\bf X}_{z,i}}\omega :={\rm d}h_{z,i},
\label{contract2}
\ee
so that
\be
\lim_{z\to 0} {\bf X}_{z,i}= {\bf X}_i.
\label{zae}
\ee

\item Define the deformed LH system of the initial   system ${\bf X}$  (\ref{aabb}) by
\be
{\bf X}_z:=\sum_{i=1}^\ell b_i(t){\bf X}_{z,i} .
\label{aabbc}
\ee 
\end{enumerate}

\bigskip

Now some remarks are in order. First, note that for a given LH algebra $\lh$ there exist as many Poisson--Hopf algebra deformations as non-equivalent Lie bialgebra structures $\delta$ on $\lh$~\cite{Chari}, where the 1-cocycle $\delta$ essentially provides the first-order deformation in $z$ of the coproduct map $\Delta$. For three-dimensional real Lie algebras the full classification of Lie bialgebra structures is known, and some classification results are also known for certain higher-dimensional Lie algebras (see~\cite{dualPL, BBM3d} and references therein). Once a specific Lie bialgebra $(\lh,\delta)$ is chosen, the full Poisson--Hopf algebra deformation can be systematically obtained by making use of the Poisson version of the quantum duality principle for Hopf algebras, as we will explicitly see in the next section for an $(\mathfrak{sl}(2),\delta)$ Lie bialgebra.

Second,  the deformed vector fields ${\bf X}_{z,i}$ (\ref{contract2}) will not, in general, span a finite-dimensional Lie algebra, which implies that  (\ref{aabbc}) is not a Lie system. In fact, the 
sequence of Lie algebra morphisms (\ref{seq}) and the properties of Hamiltonian vector fields \cite{Va94} lead to 
\be
[{\bf X}_{z,i},{\bf X}_{z,j}]= [
\varphi(h_{z,i}),\varphi(h_{z,j})]= \varphi(
\{h_{z,i},h_{z,j}\}_{\omega})= \varphi( F_{z,ij}(h_{z,1},\dots,h_{z,l }))=-\sum_{k=1}^\ell\frac{\partial F_{z,ij}}{\partial h_{z,k}}{\bf X}_{z,k}.
\nonumber
\ee
In other words,
\be
\left[{\bf X}_{z,i},{\bf X}_{z,j}\right]= \sum_{k=1}^{\ell} G_{z,ij}^{k} (x,y ) {\bf X}_{z,k},      \label{FRO}
\ee
where the $G_{z,ij}^{k} (x,y)$ are smooth functions relative to the coordinates $x,y$ and the deformation parameter $z$. 
Despite this, the relations (\ref{zad}) and the continuity of $\varphi$ imply that
\be 
[{\bf X}_i,{\bf X}_j]=\varphi(\{h_i,h_j\})_\omega=\varphi\left(\lim_{z\rightarrow 0}\{h_{z,i},h_{z,j}\}_\omega\right)=\lim_{z\to 0}\varphi\{h_{z,i},h_{z,j}\}_\omega=\lim_{z\to 0}[{\bf X}_{z,i},{\bf X}_{z,j}] .
\nonumber
\ee
Hence
\be  \lim_{z\to 0}G_{z,ij}^{k} (x,y ) ={\rm constant} 
\nonumber
\ee
holds for all indices.
Geometrically, the conditions (\ref{FRO}) establish that the vector fields ${\bf X}_{z,i}$ span an involutive smooth generalized distribution $\mathcal{D}_z$. In particular, the distribution $\mathcal{D}_0$ is spanned by the Vessiot--Guldberg Lie algebra $\langle {\bf X}_1,\dots,{\bf X}_l\rangle$. This causes $\mathcal{D}_0$ to be integrable on the whole $\mathbb{R}^{2n}$ in the sense of Stefan--Sussman ~\cite{Va94,Pa57,WA}. The integrability of $\mathcal{D}_z$, for $z\neq 0$, can only be ensured on open connected subsets of $\mathbb{R}^{2n}$ where $\mathcal{D}_z$ has constant rank~\cite{Va94}.

Third, although the vector fields ${\bf X}_{z,i}$  depend smoothly on $z$, the distribution $\mathcal{D}_z$ may change abruptly. For instance, consider the case given by the LH system ${\bf X}=\partial_x+ty\partial_x$ relative to the symplectic form $\omega=\dd x\wedge \dd y$ and admitting a LH algebra $\lh=\langle h_1:=y,\ h_2:=y^2/2\rangle$. Let us define  $h_{z,1}:=y$ and $h_{z,2}:=y^2/2+zx$. Then ${\bf X}_z=\partial_x+t(y\partial_x-z\partial_y)$ and $\dim \mathcal{D}_0(x,y)=1$, but $\dim \mathcal{D}_z(x,y)=2$ for $z\neq 0$. Hence, the deformation of LH systems may change in an abrupt way the dynamical and geometrical properties of the systems ${\bf X}_z$ (cycles, periodic solutions, etc).
 
Fourth, the deformation parameter $z$ provides an additional degree of freedom that enables the control or modification of the deformed system ${\bf X}_z$. In fact, as $z$ can be taken small, perturbations of the initial Lie system ${\bf X}$ can be obtained from the deformed one ${\bf X}_z$ in a natural way. 

And, finally, we stress that, by construction, the very same procedure can be applied to other $2n-$dimensional manifolds different to $\mathbb{R}^{2n}$, to higher dimensions as well as to multiparameter Poisson--Hopf algebra deformations of Lie algebras endowed with two or more deformation parameters. 

\bigskip

\begin{remark}
The coalgebra method employed in \cite{BCHLS13Ham} to obtain superposition rules and constants of motion for LH systems on a manifold $M$ relies almost uniquely in the Poisson--Hopf algebra structure related to $\mathcal{C}^\infty(\mathfrak{g}^*)$ and a Poisson map 
$$
D:\mathcal{C}^\infty(\mathfrak{g}^*)\rightarrow \mathcal{C}^\infty(M),$$ 
where we recall that $\mathfrak{g}$ is a Lie algebra isomorphic to a LH algebra, $\mathcal{H}_\omega$, of the LH system.  

Relevantly, quantum deformations allow us to repeat this scheme by substituting the Poisson algebra $\mathcal{C}^\infty(\mathfrak{g}^*)$ with a quantum deformation $\mathcal{C}^\infty(\mathfrak{g}_z^*)$, where $z\in \mathbb{R}$, and obtaining an adequate Poisson map 
$$
D_z:\mathcal{C}^\infty(\mathfrak{g}_z^*)\rightarrow \mathcal{C}^\infty(M).
$$
The above procedure enables us to deform the LH system into a $z$-parametric family of Hamiltonian systems whose dynamics is determined by a Steffan--Sussmann distribution and a family of Poisson algebras. If  $z$ tends to zero, then the properties of the (classical) LH system are recovered by a limiting process, hence enabling to construct new deformations exhibiting physically relevant 
properties. 
\end{remark}

\bigskip
\bigskip

\subsect{Constants of the motion}

\bigskip

The fact that $ \qlhz$ we are handling Poisson--Hopf algebra allows us to apply the coalgebra formalism established in ~\cite{BCHLS13Ham}  in order to obtain $t$-independent constants of the motion for ${\bf X}_z$.

Let $S\left(\lh\right)$ be the {\it symmetric algebra} of $\lh$, i.e. the associative unital algebra of polynomials on the elements of $\lh$. The Lie algebra structure on $\lh$ can be extended to a Poisson algebra structure in $S\left(\lh\right)$ by requiring $[v,\cdot ]$ to be a derivation on the second entry for every $v\in \lh$. Then, $S\left(\lh\right)$  can be  endowed with a Hopf algebra structure with a  non-deformed (trivial) coproduct map $\Delta$ defined  by
\begin{equation}
 {\Delta} :S\left(\lh \right)\rightarrow
S\left(\lh\right) \otimes S\left(\lh\right)    ,\qquad      {\Delta}(v_i):=v_i\otimes 1+1\otimes v_i,  \qquad    i=1,\dots, \ell,
\label{baa}
\end{equation}
which is a Poisson algebra homomorphism relative to the Poisson structure on $S(\lh)$ and the one induced in $S(\lh)\otimes S(\lh)$. Recall that every element of $S(\lh)$ can be understood as a function on $\lh^*$. Moreover, as $S(\mathcal{H}_\omega)$ is dense in the space
$\mathcal{C}^\infty(\mathcal{H}^*_\omega)$ of smooth functions on the dual $\mathcal{H}^*_\omega$ of the LH algebra $\mathcal{H}_\omega$, the coproduct in $S(\mathcal{H}_\omega)$ can be extended in a unique way to  
\begin{equation}
 {\Delta} :\mathcal{C}^\infty\left(\lh^*\right)\rightarrow
\mathcal{C}^\infty\left(\lh^*\right) \otimes \mathcal{C}^\infty\left(\lh^*\right).
\nonumber
\end{equation}
Similarly, all structures on $S(\lh)$ can be extended turning  $\mathcal{C}^\infty(\mathcal{H}^*_\omega)$ into a Poisson--Hopf algebra. Indeed, the resulting structure is the natural one in $\mathcal{C}^\infty(\mathcal{H}^*_\omega)$ given in section 2.2.

\bigskip

Let us assume  now that   $\mathcal{C}^\infty\left(\lh^*\right)$  
has a   Casimir  invariant 
\be
C=C(v_1,\dots,v_\ell),
\nonumber
\ee
where $v_1,\ldots,v_\ell$ is a basis for $\lh$. The initial LH system allows us to define a Lie algebra morphism $\phi:\lh\rightarrow \mathcal{C}^\infty(M)$, where  $M$ is a submanifold of $\mathbb R^{2n}$ where all functions $h_i:=\phi(v_i)$, for $i=1,\ldots,\ell$, are well defined.
 Then, the Poisson algebra morphisms 
\be
D: \mathcal{C}^\infty\left( \lh^* \right) \rightarrow \mathcal{C}^\infty(M),\qquad  D^{(2)} :   \mathcal{C}^\infty\left(   \lh^* \right)\otimes \mathcal{C}^\infty\left(  \lh^* \right)\rightarrow \mathcal{C}^\infty(M)\otimes \mathcal{C}^\infty(M),
\label{morphisms}
\ee
defined respectively by
\be
D( v_i):= h_i(x_1,y_1), \qquad
 D^{(2)} \left( {\Delta}(v_i) \right):= h_i(x_1,y_1)+h_i(x_2,y_2)   ,\qquad i=1,\dots, \ell,
\label{bb}
\ee
lead to the $t$-independent constants of motion  $F^{(1)}:= F$ and $F^{(2)}$ for the Lie system  ${\bf X}$ given in (\ref{aabb}) where
\be
  F:= D(C),\qquad F^{(2)}:=  D^{(2)} \left( {\Delta}(C) \right).
\label{bc}
\ee

The very same procedure can also be applied to any   Poisson--Hopf algebra $\qlhz$ with deformed coproduct $\Delta_z$ and Casimir invariant
$C_z=C_z(v_{1},\dots,v_{\ell})$, where $\{ v_{1},\dots,v_{\ell} \}$      fulfill the same   Poisson brackets ({\ref{zab}), and  such that
  \be
\lim_{z\to 0} \Delta_{z}=\Delta , \qquad \lim_{z\to 0} C_{z}= C.
\nonumber
\ee
Following~\cite{BCHLS13Ham}, the element $C_z$ turns out to be the cornerstone in the construction of the deformed constants of the motion for the `generalized' LH system ${\bf X}_z $.

\part{Applications to the Theory of Quantum Poisson--Hopf Algebras} \label{Part3}

\chapter{Poisson--Hopf Algebra Deformations of $\mathfrak{sl}(2)$-related Systems} 

\label{Chapter4} 
\renewcommand{\theequation}{4.\arabic{equation}}

\section{Poisson--Hopf algebra deformations of $\mathfrak{sl}(2)$}

\bigskip

Once the general description of our approach has been  introduced, we present in this section the general properties of the Poisson analogue of the so-called non-standard quantum deformation of the simple real Lie algebra $\mathfrak{sl}(2)$. This deformation will be applied in  the sequel to get deformations of  the    Milne--Pinney equation or Ermakov system and of some Riccati equations, since all these systems are known to be endowed with a LH algebra $\lh$ isomorphic to  $\mathfrak{sl}(2)$~\cite{BBHLS,BCHLS13Ham,   BHLS}. 

\bigskip

Let us consider the basis $\{J_3, J_+,J_-\}$ for $\mathfrak{sl}(2)$ with Lie brackets and Casimir operator given by
\be
[ J_3,J_\pm  ]  = \pm  2J_\pm     ,\qquad  [J_+ , J_- ] = J_3,\qquad C=\tfrac 12  J_3^2+(J_+ J_- + J_- J_+).
\label{crules}
\ee
Amongst the three possible quantum deformations of $\mathfrak{sl}(2)$, we shall hereafter consider the   non-standard (triangular or Jordanian) quantum deformation, $U_z(\mathfrak{sl}(2))$ (see~\cite{non,Ohn,  Shariati} for further details). The Hopf algebra structure of $U_{z}(\mathfrak{sl}(2))$ has the following deformed coproduct and compatible deformed commutation rules  
\be
\Delta_z(J_+)=J_+\otimes 1 + 1 \otimes J_+,\qquad 
\Delta_z(J_j)=J_j\otimes {\rm e}^{2z J_+} + {\rm e}^{- 2z J_+}  \otimes J_j ,\qquad j \in \{-, 3\},
\label{codef}\nonumber
\ee
\be
[ J_3,J_+ ] _z=   \frac{\sinh (  2z J_+)}{  z}    ,\qquad [ J_3, J_-]_z=- J_- \cosh(2zJ_+)  - \cosh(2zJ_+) J_-   ,\qquad  [J_+ , J_- ]_z= J_3.
\label{corudef}\nonumber
\ee
The counit and antipode can be explicitly found in~\cite{non,Ohn}, and the deformed Casimir reads~\cite{non}  
\be
 C_z=\frac 12 J_3^2+\frac{\sinh(2z J_+) }{2z} \,  J_- + J_-  \,\frac{\sinh(2z J_+) }{2z} + \frac 12 \cosh^2( 2z J_+) .
\label{bf}\nonumber
\ee

\bigskip

Let $\mathfrak{g}$ be the Lie algebra of $G$.  
It is well known (see~\cite{Chari,Majid}) that quantum algebras $\mathcal{U}_z(\mathfrak{g})$ are Hopf algebra duals of quantum groups $G_z$. On the other hand, quantum groups $G_z$ are just quantizations of Poisson--Lie groups, which are Lie groups endowed with a multiplicative Poisson structure, {i.e.}~a Poisson structure for which the Lie group multiplication is a Poisson map. In the case of $\mathcal{U}_z(\mathfrak{sl}(2))$, such Poisson structure on $SL(2)$ is explicitly given by the Sklyanin bracket coming from the classical $r$-matrix
 \be
 r=z J_3\wedge J_+,
 \label{rmns}
 \ee
which is a solution of the (constant) classical Yang--Baxter equation.
 
 \bigskip

 Moreover, the `quantum duality principle`~\cite{Dri,STS} states that quantum algebras can be thought of as `quantum dual groups' $G_z^\ast$, which means that any quantum algebra can be obtained as the Hopf algebra quantization of the dual Poisson--Lie group $G^\ast$. The usefulness of this approach to construct explicitly the Poisson analogue of quantum algebras was developed in~\cite{dualPL}.
 
 In the case of $\mathcal{U}_z(\mathfrak{sl}(2))$, the Lie algebra $\mathfrak{g}^\ast$ of the dual Lie group $G^\ast$ is given by the dual of the cocommutator map $\delta$ that is obtained from the classical $r$-matrix as
  \begin{equation}
\delta(x)=[ x\otimes 1+1\otimes x , r],\qquad \forall x\in \mathfrak{g}.
\label{rmatrix}
\end{equation}
In our case, from~\eqref{crules} and~\eqref{rmns} we explicitly obtain
\be
\delta(J_3)=2z \, J_3 \wedge J_+ ,\qquad
\delta(J_+)= 0,\qquad
\delta(J_-)= 2z \, J_- \wedge J_+ ,
\nonumber
\ee
and the dual Lie algebra $\mathfrak{g}^\ast$ reads
\begin{equation}
[j^+,j^3]=-2 z \, j^3, \qquad [j^+,j^-]=-2 z \, j^-, \qquad [j^3,j^-]=0,
\label{book}
\end{equation}
where $\{j^3,j^+,j^-\}$ is the basis of $\mathfrak{g}^\ast$,  and $\{J_3,J_+,J_-\}$ can now be  interpreted as local coordinates on the dual Lie group $G^\ast$. The dual Lie algebra~\eqref{book} is the so-called `book' Lie algebra, and the complete set of its Poisson--Lie structures was explicitly obtained in~\cite{BBM3d} (see also~\cite{LV}, where book Poisson--Hopf algebras were used to construct integrable deformations of Lotka--Volterra systems). In particular, if we consider the  coordinates on $G^\ast$ given by
\be
v_1=   J_+,\qquad v_2 =  \tfrac 12 J_3,\qquad v_3= -  J_-,
\nonumber
\ee
the Poisson--Lie structure on the book group whose Hopf algebra quantization gives rise to the quantum algebra $\mathcal{U}_z(\mathfrak{sl}(2))$ is given by the fundamental Poisson brackets~\cite{BBM3d}
\be 
\{v_1,v_2\}_z=-\shc (2z v_1)v_1,\qquad 
 \{v_1,v_3\}_z=-2 v_2,\qquad
\{v_2,v_3\}_z= -  \cosh(2 z v_1) v_3,
\label{gb}
\ee
together with the coproduct map
\begin{equation}
 \Delta_z(v_1)=  v_1 \otimes 1+
1\otimes v_1 , \qquad
\Delta_z(v_k)=v_k \otimes \eee^{2 z v_1} + \eee^{-2 z v_1} \otimes
v_k   ,\qquad  k=2,3,
 \label{ga}
 \end{equation}
which is nothing but the group law for the book Lie group $G^\ast$  in the chosen coordinates (see~\cite{LV,BBM3d,dualPL} for a detailed explanation).  
Therefore, ~\eqref{gb} and~\eqref{ga} define a Poisson--Hopf algebra structure on $\mathcal{C}^\infty(G^\ast)$, which can be thought of as a Poisson--Hopf algebra deformation of the Poisson algebra $\mathcal{C}^\infty(\mathfrak{sl}(2)^\ast)$, since we have identified the local coordinates on $\mathcal{C}^\infty(G^\ast)$ with the generators of the Lie--Poisson algebra  $\mathfrak{sl}(2)^\ast$. 

\bigskip

Notice that we have introduced in  (\ref{gb})  the hereafter called {\it cardinal hyperbolic sinus  function} (see Appendix \ref{AppendixB}) defined by\footnote{Some   properties of this function along with its relationship with Lie systems are given in the Appendix \ref{AppendixA} }
\be
\shc(x):=\frac {\sinh (x)}{x}.
\label{shc}
\ee

Summarizing, the Poisson--Hopf algebra given by~\eqref{gb} and~\eqref{ga}, together with its Casimir function
 \be
 {C}_z=\shc( 2z v_1)\, v_1v_3-v_2^2   ,
\label{gc}
\ee
will be the deformed Poisson--Hopf algebra that we will use in the sequel in order to construct deformations of LH systems based on $\mathfrak{sl}(2)$. Note that the usual  Poisson--Hopf algebra $\mathcal{C}^\infty(\mathfrak{sl}(2)^\ast)$ is smoothly recovered under the $z\to 0$ limit leading to the non-deformed Lie--Poisson coalgebra 
\be
\{ v_1,v_2\}=- v_1,\qquad \{ v_1,v_3\} =- 2v_2,\qquad  \{ v_2,v_3\} =-v_3,
\label{brack2}
\ee 
 with undeformed coproduct (\ref{baa}) and Casimir
\be
C=v_1 v_3 - v_2^2 .
\label{ai}
\ee

\bigskip

We stress that this application of the `quantum duality principle' would allow one to obtain the Poisson analogue of any quantum algebra $\mathcal{U}_z(\mathfrak{g})$, which by following the method here presented could be further applied in order to construct the corresponding deformation of the LH systems associated to the Lie--Poisson algebra $\mathfrak{g}$. In particular, the Poisson versions of the other quantum algebra deformations of $\mathfrak{sl}(2)$ can be obtained in the same manner with no technical obstructions (for instance, see~\cite{dualPL} for the explicit construction of the `standard' or Drinfel'd--Jimbo deformation).

\bigskip

\begin{remark}
 Since $\mathcal{U}_{z}(\mathfrak{sl}(2))$ is a Poisson algebra, one can define a Lie algebra representation $v\in\mathfrak{sl}(2)\mapsto[v,\cdot]_z\in{\rm End}(\mathcal{U}_{z}(\mathfrak{sl}(2)))$, which makes $\mathcal{U}_{z}(\mathfrak{sl}(2))$ into a $\mathfrak{sl}(2)$-space.
Similarly, $\mathcal{C}_z^{\infty}(\mathfrak{sl}(2)^\ast)$ is also a $\mathfrak{sl}(2)$-space relative to the Lie algebra representation induced by the Poisson
structure on $\mathcal{C}_z^{\infty}(\mathfrak{sl}(2)^\ast)$, i.e.
$$
\rho_{2}:v\in\mathfrak{sl}(2)\mapsto\{v,\cdot\}_{\mathfrak{sl}(2),z}\in{\rm End}(\mathcal{C}_z^{\infty}(\mathfrak{sl}(2)^\ast)).
$$
There exists a $z$-parametrized family of linear morphisms of the form
$$
\phi_z:P\in \mathcal{U}_{z}(\mathfrak{sl}(2))\mapsto f_P\in \mathcal{C}_z^{\infty}(\mathfrak{sl}(2)^*),\qquad \forall z\in \mathbb{R},
$$
satisfying that $\phi_z([v,\cdot])=\{v,\phi_z(\cdot)\}_{\mathfrak{sl}(2)^*,z}$ for every $v\in \mathfrak{sl}(2)$, i.e. $\phi_z$ is a morphism of $\mathfrak{sl}(2)$-spaces. 

There is a canonical way of constructing $\phi_0$ by setting $\phi_0([P])$ to be the unique symmetric polynomial in the equivalence class $[P]$. This construction is no longer available for $\mathcal{U}_z(\mathfrak{sl}(2))$. To define $\phi_z$ is enough to use that every class of equivalence $[P]$ in $\mathcal{U}_z(\mathfrak{sl}(2))$  has a unique decomposition as a linear combination of the elements of every {\it Poincar\'e--Birkhoff--Witt}    basis for $\mathcal{U}_z(\mathfrak{sl}(2))$. Then, $\phi_z$ is the linear morphism on $\mathcal{U}_z(\mathfrak{sl}(2))$ that acts as the identity on the elements of the chosen Poincar\'e--Birkhoff--Witt basis.

To illustrate the above point, let us recall that $\mathcal{U}_z(\mathfrak{sl}(2))$ can be defined as the algebra generated by the operators 
$$J_- , \qquad J_3, \qquad K:=e^{2zJ_+},\qquad K^{-1}:=e^{-2zJ_+}.
$$
 Then, a  Poincar\'e--Birkhoff--Witt  basis is given by the polynomials $J_-^mK^pJ^l_3$, where $m,l\geq 0$ and $p\in \mathbb{Z}$. In other words, every element in $\mathcal{U}_z(\mathfrak{sl}(2))$ admits a unique representation as a polynomial in this basis. This allows us to define a morphism of $\mathfrak{sl}(2)$-spaces:
$$
\phi_z:P(J_-,K,J_3)\in \mathcal{U}_{z}(\mathfrak{sl}(2))\mapsto f_P(J_-,K,J_3)\in \mathcal{C}_z^{\infty}(\mathfrak{sl}(2)^\ast),\qquad \forall z\in \mathbb{R}.
$$
Hence,  any Casimir element of the Poisson--Hopf algebra
$\mathcal{U}_{z}(\mathfrak{sl}(2))$ gives rise to a Casimir of $\mathcal{C}_z^{\infty}(\mathfrak{sl}(2)^\ast)$. For instance,  ${C}_z$ (\ref{gc})  is the Poisson analog of $-\tfrac 12 {C}_z$ (\ref{bf}).

\end{remark}

\bigskip
 \bigskip

\section{Poisson--Hopf deformations of $\mathfrak{sl}(2)$ Lie--Hamilton systems}\label{PHDefLHSstem}
 
 \bigskip

 This section concerns the analysis of Poisson--Hopf deformations of LH systems on a manifold $M$  with a Vessiot--Guldberg algebra isomorphic to $\mathfrak{sl}(2)$. Our geometric analysis will allow us to introduce the notion of a Poisson--Hopf Lie system that,  roughly speaking, is a family of non-autonomous Hamiltonian systems of first-order differential equations constructed as a deformation of a LH system by means of the representation of the deformation of a Poisson--Hopf algebra in a Poisson manifold. 
 
 \bigskip

Let us endow a manifold $M$ with a symplectic structure $\omega$ and consider a Hamiltonian Lie group action $\Phi:SL(2,\mathbb{R})\times M\rightarrow M$. A basis of fundamental vector fields of  $\Phi$, let us say $\left\{{\bf X}_1,{\bf X}_2,{\bf X}_3\right\}$, enable us to define a Lie system 
$$
{\bf X}_t=\sum_{i=1}^3b_i(t){\bf X}_i,
$$
for arbitrary $t$-dependent functions $b_1(t),b_2(t),b_3(t)$, and $\{{\bf X}_1,{\bf X}_2,{\bf X}_3\}$ spanning a Lie algebra isomorphic to $\mathfrak{sl}(2)$. As is well known, there are only three non-diffeomorphic classes of Lie algebras of Hamiltonian vector fields isomorphic to $\mathfrak{sl}(2)$ on the plane \cite{BBHLS}. 
Since ${\bf X}_1,{\bf X}_2,{\bf X}_3$ admit Hamiltonian functions $h_1,h_2,h_3$, the $t$-dependent vector field ${\bf X}$ admits a $t$-dependent Hamiltonian function 
$$
h=\sum_{i=1}^3b_i(t)h_i.
$$
Due to the cohomological properties of $\mathfrak{sl}(2)$ (see e.g. \cite{Va94}), the Hamiltonian functions $h_1,h_2,h_3$ can always be chosen so that the space $\langle h_1,h_2,h_3\rangle$ spans a Lie algebra isomorphic to $\mathfrak{sl}(2)$ with respect to $\{\cdot,\cdot\}_\omega$.

Let $\{v_1,v_2,v_3\}$ be the basis for $\mathfrak{sl}(2)$ given in (\ref{book}) and let $M$ be a manifold where the functions $h_1,h_2,h_3$ are smooth. Further, the Poisson--Hopf algebra structure of  $\mathcal{C}^\infty(\mathfrak{sl}^*(2))$ is given by (\ref{brack2}). In these conditions, there exists a Poisson algebra morphism $D:\mathcal{C}^\infty(\mathfrak{sl}^*(2))\rightarrow \mathcal{C}^\infty(M)$ satisfying
$$
D(f(v_1,v_2,v_3))=f(h_1,h_2,h_3),\quad \forall f\in \mathcal{C}^\infty(\mathfrak{sl}^*(2)).
$$
Recall that the deformation $\mathcal{C}^\infty(\mathfrak{sl}_z^*(2))$ of $\mathcal{C}^\infty(\mathfrak{sl}^*(2))$ is a Poisson--Hopf algebra with the new Poisson structure induced by the relations (\ref{gb}). Let us define the submanifold $\mathcal{O}=:\{\theta\in \mathfrak{sl}^*(2): v_1(\theta)\neq0\}$ of $\mathfrak{sl}^*(2)$. Then, the Poisson structure on $\mathfrak{sl}^*(2)$ can be restricted to the space $\mathcal{C}^\infty(\mathcal{O})$. In turn, this enables us to expand the Poisson--Hopf algebra structure in $\mathcal{C}^\infty(\mathfrak{sl}^*(2))$ to $\mathcal{C}^\infty(\mathcal{O})$. Within the latter space, the elements 
\bea
&& v_{z,1}:=v_1,\qquad v_{z,2}:={\shc}(2z v_{1})v_{2},\nonumber\\
&&v_{z,3}:={\shc}(2zv_1)\frac{v^2_2}{v_1}+\frac{c}{4{\shc}(2zv_1)v_1} ,
\label{nf}
\eea
are easily verified to satisfy the same commutation relations with respect to $\{\cdot,\cdot\}$ as the elements $v_1,v_2,v_3$ in $\mathcal{C}^\infty(\mathfrak{sl}_z^*(2))$ with respect to $\{\cdot,\cdot\}_z$ (\ref{gb}), i.e.
\bea
&&
\{v_{z,1},v_{z,2}\}=-{\shc}(2zv_{z,1})v_{z,1}, \qquad \{v_{z,1},v_{z,3}\}=-2v_{z,2}, \nonumber\\
&&\{v_{z,2},v_{z,3}\}=-\,{\rm ch}(2zv_{z,1}) v_{z,3}   . 
\label{ggbb}
\eea
In particular, from (\ref{nf}) with $z=0$ we find that
\bea
&&
\{v_{0,1},v_{0,2}\}=-v_{0,1},\qquad
\{v_{0,1},v_{0,3}\}=\frac{2\{v_{0,1},v_{0,2}\} v_{0,2}}{v_{0,1}}=-2v_{0,2},  \nonumber\\
&&\{v_{0,2},v_{0,3}\}=-\frac{v_{0,2}^2}{v_{0,1}^2}\{v_{0,2},v_{0,1}\}-\frac{c}{4v_{0,1}^2}\{v_{0,2},v_{0,1}\}=-\frac{v_{0,2}^2}{v_{0,1}}-\frac{c}{4v_{0,1}}=-v_{0,3}. 
\nonumber
\eea
The functions $v_{z,1},v_{z,2},v_{z,3}$ are not functionally independent, as they satisfy the constraint 
\begin{equation}
{\shc}(2zv_{1,z})v_{1,z}v_{3,z}-v^2_{2,z}=c/4.\label{kas}
\end{equation}

The existence of the functions $v_{z,1},v_{z,2},v_{z,3}$ and the relation (\ref{kas}) with the Casimir of the deformed Poisson--Hopf algebra is by no means casual. Let us explain why $v_{z,1},v_{z,2},v_{z,3}$ 
exist and how to obtain them easily. 

\bigskip

Around a generic point $p\in \mathfrak{sl}^*(2)$, there always exists  an open $U_p$ containing $p$ where both Poisson structures give a symplectic foliation by surfaces. Examples of symplectic leaves for $\{\cdot,\cdot\}$ and $\{\cdot,\cdot\}_z$ are displayed in Fig. \ref{Figure1}.

\begin{figure}[t]
\begin{center}
\includegraphics[scale=0.5]{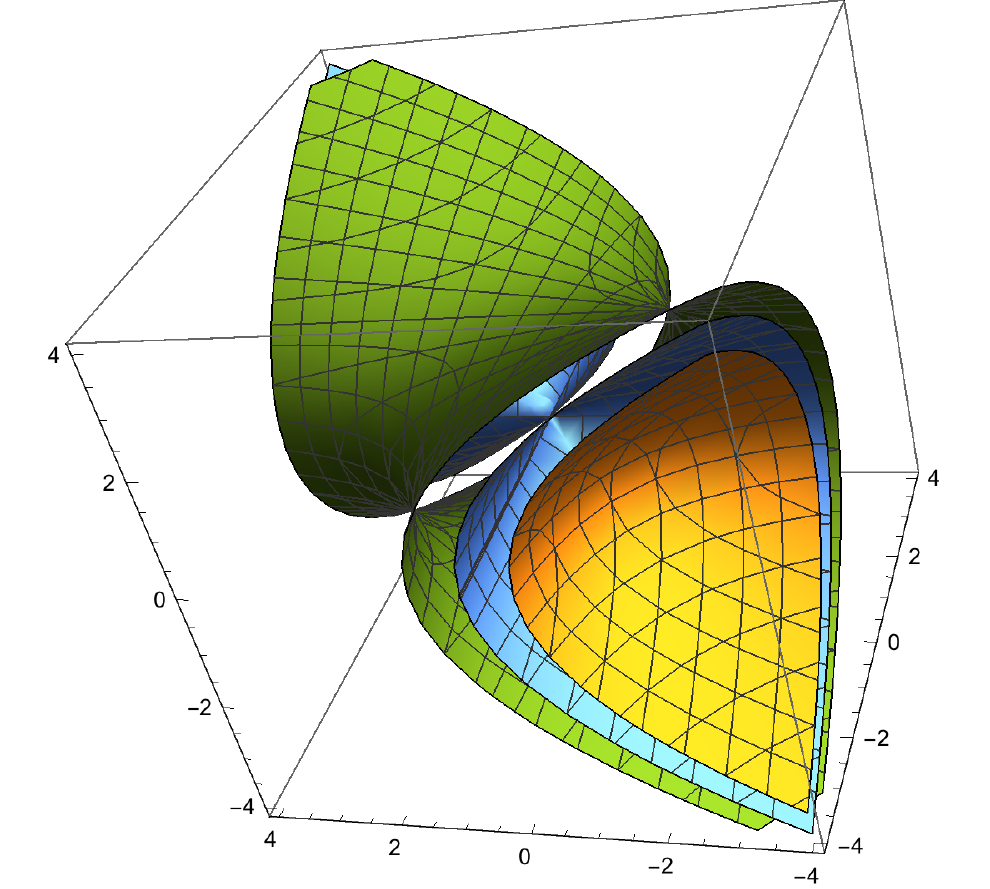}\qquad \qquad
\includegraphics[scale=0.5]{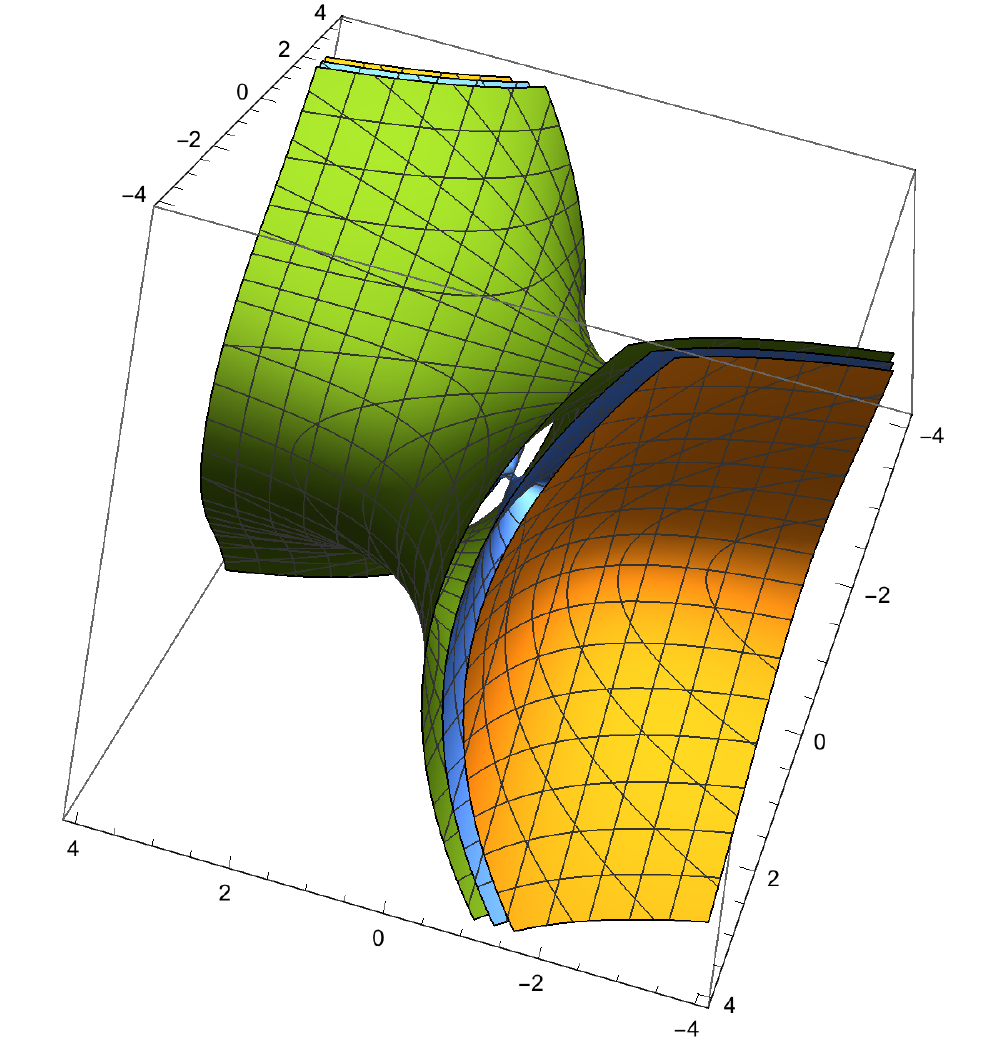}
\end{center}
\caption{Representatives of the submanifolds in $\mathfrak{sl}^*(2)$ given by the surfaces with constant value of the Casimir for the Poisson structure in $\mathfrak{sl}^*(2)$ (left) and its deformation (right). Such submanifolds are symplectic submanifolds where the Poisson bivectors $\Lambda$ and $\Lambda_z$ admit a canonical form.}
\label{Figure1}        
\end{figure}

The splitting theorem on 
Poisson manifolds  \cite{Va94} ensure that if $U_p$ is small enough, then there exist two different coordinate systems $\{x,y,C\}$ and $\{x_z,y_z,C_z\}$ where the Poisson bivectors related to $\{\cdot,\cdot\}$ and $\{\cdot,\cdot\}_z$ read $\Lambda=\partial_{x}\wedge \partial_y$ and $\Lambda_z=\partial_{x_z}\wedge 
\partial_{y_z}$.   Hence, $C_z$ and $C$ are Casimir functions for $\Lambda_z$ and $\Lambda$, respectively. Moreover, $x_z=x_z(x,y,C),y_z=y_z(x,y,C), C_z=C_z(x,y,C)$. 
It follows from this that 
$$
\Phi:f(x_z,y_z,C_z)\in C^\infty_z(U_p)\mapsto f(x,y,C)\in C^\infty(U_p)
$$
 is a Poisson algebra morphism. 

\bigskip

If $\left\{v_1,v_2,v_3\right\}$ are the standard coordinates on $\mathfrak{sl}^*
(2)$ and the relations (\ref{gb}) are satisfied, then $v_i=\xi_i(x_z,y_z,C_z)$ holds for certain functions $\xi_1,\xi_2,\xi_3:\mathbb{R}^3\rightarrow \mathbb{R}$. Hence, the $\hat v_{z,i}=\xi_i(x,y,C)$ close the same commutation relations relative to $\{\cdot,\cdot\}$ as the $v_i$ do with respect to $\{\cdot,\cdot\}_z$. As $C$ is a Casimir invariant, the functions $v_{z,i}:=\xi_i(x,y,c)$, with a constant value of $c$, still close the same commutation relations among themselves as the $v_i$. Moreover, the functions $v_{z,i}$ become functionally dependent. Indeed,
$$
C_z=C_z(v_1,v_2,v_3)=C_z(\xi_1(x_z,y_z,C_z),\xi_2(x_z,y_z,C_z),\xi_3(x_z,y_z,C_z)).
$$
Hence,
$c=C_z(\xi_1(x_z,y_z,c),\xi_2(x_z,y_z,c),\xi_3(x_z,y_z,c))$ and we conclude that $c=C_z(v_{z,1},v_{z,2},v_{z,3}).$ 

\bigskip

The previous argument allows us to recover the functions (\ref{nf}) in an algorithmic way. Actually, the functions $x_z,y_z,C_z$ and $x,y,C$ can be easily chosen to be 
$$
x_z:=v_1,\qquad y_z:=-\frac{v_2}{{\shc}(2zv_1)v_1},\qquad C_z:={\shc}(2zv_1)v_1v_3-v_2^2,
$$
 as well as 
 $$
 x=v_1,\qquad y=-v_2/v_1,\qquad C=v_1v_3-v_2^2.
 $$
Therefore, 
\bea
&&
\xi_1(x_z,y_z,C_z)=x_z,\qquad \xi_2(x_z,y_z,C_z)=-y_z{\shc}(2zx_z)x_z, \nonumber\\
&& \xi_3(x_z,y_z,C_z)=\frac{ C_z+x_z^2y_z^2{\shc}^2(2zx_z)}{ {\shc}(2zx_z)x_z}.
\nonumber
\eea
Assuming that $C_z=c/4$,  
replacing $x_z,y_z$ by $x=v_1,y=-v_2/v_1$, respectively, and taking into account that $v_{z,i}:=\xi_i(x,y,c)$, one retrieves (\ref{nf}).  

\bigskip

It is worth mentioning that due to the simple form of the Poisson bivectors in splitting form for three-dimensional Lie algebras, this method can be easily applied to such a type of Lie algebras.

Next, the above relations enable us to construct a Poisson algebra morphism 
$$
D_z:f(v_1,v_2,v_3)\in \mathcal{C}^\infty(\mathfrak{sl}_z^*(2))\mapsto D(f(v_{1,z},v_{2,z},v_{3,z}))\in \mathcal{C}^\infty(M)
$$
for every value of $z$ allowing us to pass the structure of the Poisson--Hopf algebra $\mathcal{C}^\infty(\mathfrak{sl}_z^*(2))$ to $C^\infty(M)$. As a consequence, $D_z( {C}_z)$ satisfies the relations 
$$
\{D_z( {C}_z),h_{z,i}\}_\omega=0,\qquad i=1,2,3.
$$
Using the symplectic structure on $M$ and the functions $h_{z,i}$ written in terms of $\left\{h_1,h_2,h_3\right\}$, one can easily obtain the deformed vector fields ${\bf X}_{z,i}$ in terms of the vector fields ${\bf X}_i$. Finally, as ${\bf X}_{z,t}=\sum_{i=1}^3b_i(t){\bf X}_{z,i}$ holds, it is straightforward to verify that the brackets  
$$
{\bf X}_{z,i}D(C_z)=\{D(C_z),h_{z,i}\}=0,
$$
imply that the function $D(C_z)$ is a $t$-independent constant of the motion for each of the deformed LH system ${\bf X}_{z,t}$. 

\bigskip
 
Consequently, deformations of LH-systems based on $\mathfrak{sl}(2)$ can be treated simultaneously, starting from their classical 
LH counterpart. The final result  is summarized in the following statement.

\bigskip

\begin{theorem}\label{MT} If $\phi:\mathfrak{sl}(2)\rightarrow \mathcal{C}^\infty(M)$ is a morphism of Lie algebras with respect to the Lie bracket in $\mathfrak{sl}(2)$ and a Poisson bracket in $\mathcal{C}^\infty(M)$, then for each $z\in \mathbb{R}$ there exists a Poisson algebra morphism $D_z:\mathcal{C}^\infty(\mathfrak{sl}_z^*(2))\rightarrow \mathcal{C}^\infty(M)$ such that for a basis $\{v_1,v_2,v_3\}$ satisfying the commutation relations (\ref{brack2}) is given by
\bea
 \!  \! && D_z(f(v_1,v_2,v_3))\nonumber\\
 \!  \! &&=f\! \left( \!  \phi(v_1),{\shc}(2z \phi(v_1))\phi(v_{2}), {\shc}(2z\phi(v_1))\frac{\phi^2(v_2)}{\phi(v_1)}+\frac{c}{4{\shc}(2z\phi(v_1))\phi(v_1)}\right). 
\nonumber
\eea
Provided that $h_i:=\phi(v_i)$, the deformed Hamiltonian functions $h_{z,i}:=D_z(v_i)$ adopt the form
\bea
&&h_{z,1}=    h_1 ,\qquad  h_{z,2}=  {\shc}(2z  h_1 ) h_{2}    ,  \nonumber\\
&&h_{z,3}=    {\shc}(2z h_1)\frac{h_2^2 }{ h_1}+\frac{c}{4{\shc}(2z h_1) h_1}  ,
\nonumber
\eea
which satisfy the commutation relations (\ref{ggbb}).

The Hamiltonian vector fields ${\bf X}_{z,i}$ associated with  $h_{z,i}$ through (\ref{contract2}) turn out to be
\bea
&&
{\bf X}_{z,1}={\bf X}_{1},\qquad {\bf X}_{z,2}=\frac{h_2}{h_1}\bigl(\!{\cosh(2zh_1)}-{{\shc}(2zh_1)}\bigr){\bf X}_1+{\shc}(2zh_1){\bf X}_2, \nonumber\\
&&
{\bf X}_{z,3}= \left[\frac{h_2^2}{h_1^2}\bigl( {\rm ch}(2z h_1)-2 \,{\shc}(2z h_1)\bigr)- \frac{c\; {\rm ch}(2z h_1)}{4h_1^2 {\shc}^2(2z h_1)}\right] {\bf X}_1\nonumber\\
&&\qquad \qquad\qquad +2\,\frac{h_2}{h_1}\,{\shc}(2zh_1){\bf X}_2 ,\nonumber
\eea
and satisfy the following commutation relations coming from 
\bea
&&
[{\bf X}_{z,1},{\bf X}_{z,2}]={\rm ch}(2zh_{z,1}){\bf X}_{z,1},\qquad [{\bf X}_{z,1},{\bf X}_{z,3}]=2\,{\bf X}_{z,2},\nonumber\\
&&
[{\bf X}_{z,2},{\bf X}_{z,3}]={\rm ch}(2zh_{z,1}){\bf X}_{z,3}+4 z^2{\shc}^2(2z h_{z,1}) h_{z,1}h_{z,3}\,{\bf X}_{z,1} .
\nonumber
\eea
\end{theorem}

 \bigskip

As a consequence, the deformed Poisson--Hopf system can be generically described in terms of the Vessiot--Guldberg Lie algebra corresponding to the non-deformed LH system as follows: 
\bea
&&{\bf X}_{z,t}=\sum_{i=1}^3 b_i(t) {\bf X}_{z,i}=
 \left[ b_1(t)+ b_2(t)\,\frac{h_2}{h_1}\bigl( \!{\cosh(2zh_1)}-{{\shc}(2zh_1)}\bigr)\right] {\bf X}_1\nonumber\\
&&\qquad\qquad\qquad  +b_3(t)\left[\frac{h_2^2}{h_1^2}\bigl( {\rm ch}(2z h_1)-2\, {\shc}(2z h_1)\bigr)- \frac{c\; {\rm ch}(2z h_1)}{4 h_1^2 {\shc}^2(2z h_1)}\right]{\bf X}_1 \nonumber\\
&&\qquad\qquad\qquad   +\,{\shc}(2zh_1)\left(b_2(t)+ 2b_3(t)\frac{ h_2}{h_1}\right) {\bf X}_2. \label{DYF}\nonumber
\eea

This unified approach to nonequivalent deformations of LH systems possessing a common underlying Lie algebra suggests the following definition.

\bigskip

\begin{definition}\label{defPHsistem} Let $(\mathcal{C}^\infty(M),\{\cdot,\cdot\})$ be a Poisson algebra. A {\it Poisson--Hopf Lie system} is pair consisting of a Poisson--Hopf algebra $\mathcal{C}^\infty(\mathfrak{g}_z^*)$ and a $z$-parametrized family of Poisson algebra representations $D_z:\mathcal{C}^\infty(\mathfrak{g}_z^*)\rightarrow \mathcal{C}^\infty(M)$ with $z\in \mathbb{R}$.
\end{definition}

\bigskip

Next, constants of the motion for ${\bf X}_{z,t}$ can be deduced by applying the coalgebra approach introduced in~\cite{BCHLS13Ham} in the way  briefly described in Section 3. In the deformed case, we consider the Poisson algebra morphisms 
\bea
&& D_z: \mathcal{C}^\infty\left( \mathfrak{sl}_z^*(2) \right) \rightarrow \mathcal{C}^\infty(M),   \nonumber\\
 && D_z^{(2)} :   \mathcal{C}^\infty\left(  \mathfrak{sl}_z^*(2)  \right)\otimes \mathcal{C}^\infty\left( \mathfrak{sl}_z^*(2)\right)\rightarrow \mathcal{C}^\infty(M)\otimes \mathcal{C}^\infty(M),
 \nonumber
\eea 
which by taking into account the coproduct (\ref{ga}) are defined by
\bea 
&& D_z( v_i):= h_{z,i}({\bf x}_1) \equiv   h_{z,i}^{(1)}  , \quad i=1,2,3, \nonumber\\ 
&&  D_z^{(2)} \left( {\Delta}_z(v_1) \right) =h_{z,1}({\bf x}_1)+h_{z,1}({\bf x}_2) \equiv  h_{z,1}^{(2)}   ,  \nonumber \\
&& 
D_z^{(2)} \left( {\Delta}_z(v_k) \right) = h_{z,k}({\bf x}_1)  {\rm e}^{2 z h_{z,1}({\bf x}_2)}  + {\rm e}^{-2 z h_{z,1}({\bf x}_1)} h_{z,k}({\bf x}_2)\equiv  h_{z,k}^{(2)}   ,  \quad k= 2,3,
\nonumber
\eea
where
  ${\bf x}_s$ $(s=1,2)$ are global coordinates in $M$.
We remark that, by construction,  the functions $h_{z,i}^{(2)}$   satisfy the  same Poisson brackets (\ref{ggbb}). 
 Then 
$t$-independent  constants of motion are given by (see (\ref{bc}))
$$
  F_z\equiv F_z^{(1)}:= D_z(C_z),\qquad F_z^{(2)}:=  D_z^{(2)} \left( {\Delta_z}(C_z) \right),
$$
where  $C_z$ is     the deformed Casimir (\ref{gc}). Explicilty, they read
\bea
&&  F_z=  \shc\!\left(2 z h_{z,1}^{(1)}  \right)  h_{z,1}^{(1)}   h_{z,3}^{(1)} - \left( h_{z,2}^{(1)}  \right)^2 = \frac c    4 \,  ,\nonumber\\[2pt]
&& F_z^{(2)}=  \shc\!\left(2 z h_{z,1}^{(2)} \right)  h_{z,1}^{(2)}   h_{z,3}^{(2)} - \left( h_{z,2}^{(2)}  \right)^2  \label{amz} .
\nonumber
\eea

\bigskip
\bigskip

 \section{Deformations of $\mathfrak{sl}(2)$ Lie--Hamilton systems in $\mathbb{R}^2$}\label{SecDefsl2}

\bigskip

We now apply last theorem to the three classes of LH systems in the plane with a Vessiot--Guldberg Lie algebra isomorphic to $\mathfrak{sl}(2)$ according to the local classification performed in \cite{BBHLS}, which was based on the previous results ~\cite{GKO}. Consider the manifold $M=\mathbb R^2$ and the coordinates $\mathbf{x}=(x,y)$. According to \cite{BBHLS,BHLS}, there are only thee classes of LH systems, denoted by P$_2$, I$_4$ and I$_5$ and they correspond to a positive, negative and zero value of the Casimir constant $c$, respectively. Recall that these are non-diffeomorphic, so that there does not exist any local $t$-independent change of variables mapping one into another.


\begin{table}[t]
\caption{The three  classes of LH systems on the plane with underlying Vessiot--Guldberg Lie algebra isomorphic to $\mathfrak{sl}(2)$. For each class, it is displayed, in this order,  a basis of vector fields ${\bf X}_{i}$, Hamiltonian functions $h_i$, symplectic form $\omega$,   the constants of motion $F$ and $F^{(2)}$  as well as the corresponding specific LH systems.}
\begin{tabular}
[l]{l}
\hline
 \noalign{\smallskip}
$\bullet$ Class P$_2$ with $c= 4>0$\\[4pt]
$\displaystyle{\quad  {\bf X}_{1}=
\frac{\partial}{\partial x} \qquad  {\bf X}_{2}=x\frac{\partial}{\partial x}+y\frac{\partial
}{\partial y} \qquad  {\bf X}_{3}=
 (  x^{2}-y^{2} )  \frac{\partial}{\partial x}+2xy\frac{\partial
}{\partial y}}$\\[6pt]
$\displaystyle{\quad  h_{1}= -\frac{1}{y}  \qquad   h_2= -\frac{x}{y}  \qquad  h_3=  -\frac{x^{2}+y^{2}}{y}  \qquad    \omega=\frac{{\rm d}x\wedge
{\rm d}y}{y^{2}} }$  \\[6pt]
$\displaystyle{\quad F=1 \qquad F^{\left(  2\right)  }=\frac{ (  x_{1}-x_{2} )  ^{2}+ (
y_{1}+y_{2} )  ^{2}}{y_{1}y_{2}} }$  \\ 
 \noalign{\smallskip}
-- Complex Riccati equation\\
--  Ermakov system, Milne--Pinney  and Kummer--Schwarz equations with
$c>0$\\
 \noalign{\smallskip}
 \hline
 \noalign{\smallskip}
$\bullet$ Class I$_4$ with $c= -1<0$\\[4pt]
 $\displaystyle{\quad  {\bf X}_{1}= \frac{\partial}{\partial x}+\frac{\partial}{\partial y}\qquad  {\bf X}_{2}= x\frac{\partial
}{\partial x}+y\frac{\partial}{\partial y} \qquad  {\bf X}_{3}=
x^{2}\frac{\partial}{\partial x}+y^{2}\frac{\partial}{\partial y}}$\\[8pt]
$\displaystyle{ \quad  h_{1}=  \frac{1}{x-y}  \qquad  h_2=\frac{x+y}{2 (  x-y )  } \qquad  h_3=\frac{xy}{x-y}\qquad   
\omega=\frac{{\rm d} x\wedge {\rm d} y}{ (  x-y )^{2}  }}$ \\[8pt]   
$\displaystyle{ \quad F=-\frac{1}{4}   \qquad   F^{ (  2 )  }=-\frac{ (  x_{2}-y_{1} )
 (  x_{1}-y_{2} )  }{ (  x_{1}-y_{1} )   (  x_{2} 
-y_{2} )  }  }$\\
     \noalign{\smallskip}
 -- Split-complex Riccati equation\\
--  Ermakov system, Milne--Pinney  and Kummer--Schwarz equations with
$c<0$\\
-- Coupled Riccati equations\\
     \noalign{\smallskip}
 \hline
 \noalign{\smallskip}
$\bullet$ Class I$_5$ with $c=  0$\\[4pt]
 $\displaystyle{\quad  {\bf X}_{1}=  
\frac{\partial}{\partial x}\qquad   {\bf X}_{2}=  x\frac{\partial}{\partial x}+\frac{y}{2} 
\frac{\partial}{\partial y} \qquad   {\bf X}_{3}=   x^{2}\frac{\partial}{\partial x}+xy\frac{\partial}{\partial y}   }$\\[8pt]
$\displaystyle{ \quad  h_{1}=  -\frac{1}{2y^{2}} \qquad  h_{2}=  -\frac{x}{2y^{2}}  \qquad  h_{3}= -\frac{x^{2}}{2y^{2}}   \qquad
\omega = \frac{{\rm d} x\wedge{\rm d}y}{y^{3}}  }$ \\[8pt] 
$\displaystyle{ \quad F=0\qquad F^{\left(  2\right)  }=\frac{ (  x_{1}-x_{2} )  ^{2}}{  
4y_{1}^2y_{2}^2 }  }$\\
     \noalign{\smallskip}
 -- Dual-Study Riccati equation\\
 --  Ermakov system, Milne--Pinney  and Kummer--Schwarz equations with
$c=0$\\
--  Harmonic oscillator\\
-- Planar diffusion Riccati system\\
  \noalign{\smallskip}
\hline
\end{tabular}\label{TABLE4.1}
\end{table}



\begin{table}[t]
\caption{Poisson--Hopf deformations  of the   three  classes of  $\mathfrak{sl}(2)$-LH systems written in   Table 1.   The symplectic form $\omega$ is the same given in Table 1 and $F\equiv F_z$.}
\begin{tabular}
[l]{l}
\hline
 \noalign{\smallskip}
$\bullet$ Class P$_2$ with $c= 4>0$\\[4pt]
$\displaystyle{   {\bf X}_{z,1}= \frac{\partial}{\partial x} \qquad   {\bf X}_{z,2}=  x\,\mathrm{ch} (  2z/y )
\frac{\partial}{\partial x}+y\,\mathrm{shc}(2z/y)\frac{\partial}{\partial y}     }$  \\[8pt]
$\displaystyle{    {\bf X}_{z,3}=
\left(  x^{2}-\frac{y^{2}}{\mathrm{shc}^{2}(2z/y)}\right)  \mathrm{ch} 
(2z/y)\frac{\partial}{\partial x}+2xy\, \mathrm{shc}(2z/y)\frac{\partial
}{\partial y} }$ \\[8pt]
$\displaystyle{   {\bf h}_{z,1}=  -\frac{1}{y}    \qquad  {\bf h}_{z,2}=  -\frac{x}{y}\,\mathrm{shc} (   {2z}/{y} ) 
\qquad  {\bf h}_{z,3}= -\frac{x^{2} \,\mathrm{shc}^2 (   {2z}/{y} )  +y^{2}}{y\,\mathrm{shc} 
 (   {2z}/{y} ) }}$\\[6pt]
$\displaystyle{   F_z^{(2)}=    \frac{( x_1- x_2)^2 }{ y_1  y_2}\, \shc (2z/  y_1)   \shc (2z/  y_2)   \, \eee^{2z \!/\! y_1}  \eee^{-2z \! /\! y_2} }$\\[8pt]
 $\displaystyle{\qquad\qquad \quad    + \frac{ ( y_1+  y_2)^2}{ y_1  y_2} \,  \frac{\shc^2 (2z/  y_1+ 2z/  y_2) }{  \shc (2z/y_1)   \shc (2z/y_2)  } \,\eee^{2z \! / \! y_1}  \eee^{-2z \!/ \! y_2} }$\\[6pt]
  \noalign{\smallskip}
 \hline
 \noalign{\smallskip}
$\bullet$ Class I$_4$ with $c= -1<0$\\[4pt]
 $\displaystyle{   {\bf X}_{z,1} = \frac{\partial}{\partial x}+\frac{\partial}{\partial y}     }$  \\[8pt]
 $\displaystyle{   
   {\bf X}_{z,2} = 
 \frac{x+y}{2}\,\mathrm{ch}\!\left(\frac{2z}{x-y}\right)\left(  \frac{\partial}{\partial
x}+\frac{\partial}{\partial y}\right)  +\frac{x-y}{2}\,\mathrm{shc}\!\left(\frac
{2z}{x-y}\right)\left(  \frac{\partial}{\partial x}-\frac{\partial}{\partial
y}\right)       }$  \\[8pt]
    $\displaystyle{     {\bf X}_{z,3} = \frac{1}{4} \,\mathrm{ch} \!\left(\frac{2z}{x-y} \right)\left[  (  x+y )
^{2}+ (  x-y )^{2}\mathrm{shc}^{-2}\!\left(\frac{2z}{x-y}\right)\right]  \left(
\frac{\partial}{\partial x}+\frac{\partial}{\partial y}\right)    }$\\[8pt]
 $\displaystyle{\qquad\qquad \quad + \frac{1}{2}\left(  x-y\right)  ^{2}\mathrm{shc}\!\left(\frac{2z}{x-y}\right)\left(
\frac{\partial}{\partial x}-\frac{\partial}{\partial y}\right)  }$\\[8pt]
$\displaystyle{   {\bf h}_{z,1}=   \frac{1}{x-y} \quad  {\bf h}_{z,2}= \frac{(x+y) \,\mathrm{shc}\!\left(  \frac{2z}{x-y}\right)  }{2 (  x-y )  } 
\quad  {\bf h}_{z,3}= \frac{ (  x+y )
^{2}\mathrm{shc}^2\! \left(  \! \frac{2z}{x-y} \right)  - (  x-y )  ^{2} 
}{4 (  x-y ) \, \mathrm{shc}\!\left(\!   \frac{2z}{x-y}\right)  }}$\\[10pt]
$\displaystyle{   F_z^{(2)}= \frac{  ( x_1- x_2+ y_1- y_2)^2   }{4( x_1- y_1)( x_2- y_2)} \, \shc\! \left(\frac{2z} { x_1- y_1}\right)      \shc \!\left(\frac{2z} { x_2- y_2}\right)      \eee^{-\frac{2z}{ x_1- y_1}} \eee^{\frac{2z}{ x_2- y_2}}   }$\\[8pt]
 $\displaystyle{     -  \frac{( x_1+ x_2- y_1- y_2)  \shc\! \left( \!  \frac{2z} { x_1- y_1}+\frac{2z} { x_2- y_2}\right)   }{4( x_1- y_1)( x_2- y_2)}   \left[ \frac{  \eee^{\frac{2z}{ x_2- y_2}}  ( x_1- y_1) }{  \shc \bigl(\frac{2z} { x_1- y_1}\bigr) }   + \frac{    \eee^{-\frac{2z}{ x_1- y_1}} ( x_2- y_2) }{  \shc \bigl(\frac{2z} { x_2- y_2}\bigr) }   \right] }$\\[10pt]
     \noalign{\smallskip}
 \hline
 \noalign{\smallskip}
$\bullet$ Class I$_5$ with $c=  0$\\[4pt]
 $\displaystyle{    {\bf X}_{z,1} =    \frac{\partial}{\partial x}  \qquad  
   {\bf X}_{z,2} =    x\,\mathrm{ch}\!\left(   {z}/{y^{2} 
}\right)  \frac{\partial}{\partial x}+\frac{y}{2}\,\mathrm{shc}\!\left(   
{z}/{y^{2}}\right)  \frac{\partial}{\partial y}  }$   \\[8pt]
    $\displaystyle{     {\bf X}_{z,3} =   x^{2}\,\mathrm{ch}\!\left(   {z}/{y^{2}}\right)  \frac{\partial}{\partial
x}+x y\, \mathrm{shc}\!\left(   {z}/{y^{2}}\right)  \frac{\partial}{\partial
y}      }$\\[6pt]
$\displaystyle{   {\bf h}_{z,1}=  -\frac{1}{2y^{2}}
\qquad  {\bf h}_{z,2}= -\frac{x}{2y^{2}}\,\mathrm{shc}\!\left(   {z}/{y^{2}}\right)  \qquad  {\bf h}_{z,3}=   -\frac
{x^{2}}{2y^{2}} \,\mathrm{shc}\!\left(   {z}/{y^{2}}\right) } $\\[6pt]
$\displaystyle{   F_z^{(2)}=    \frac{( x_1- x_2)^2 }{4  y_1^2  y_2^2} \,\shc \!\left(z/  y_1^2\right)   \shc \!\left(z/  y_2^2\right)  \eee^{z \!/\! y_1^2}  \eee^{-z \! /\! y_2^2} }$ 
     \\[6pt]
   \noalign{\smallskip}
\hline
\end{tabular}\label{TABLE4.2}
\end{table}

Table \ref{TABLE4.1} summarizes  the three cases, covering vector fields, Hamiltonian functions, symplectic structure and $t$-independent constants of motion. The particular LH systems which are diffeormorphic within each class are also mentioned~\cite{BHLS}. Notice that for all of them  it is satisfy the following commutation relations for the vector fields and Hamiltonian functions (the latter with respect to corresponding $\omega$):
\bea
&&
[{\bf X}_{1},{\bf X}_{2}]={\bf X}_{1},\qquad [{\bf X}_{1},{\bf X}_{3}]=2 {\bf X}_{2},\qquad [{\bf X}_{2},{\bf X}_{3}]={\bf X}_{3},\nonumber\\
&& \{ h_1,h_2\}_\omega = - h_1,\qquad \{ h_1,h_3\}_\omega = -2  h_2,\qquad \{ h_2,h_3\}_\omega = - h_3.
\nonumber
\eea

\bigskip

By applying the theorem \ref{MT} with  the results of Table \ref{TABLE4.1} we obtain the corresponding deformations which are displayed in Table \ref{TABLE4.2}. 
It is straightforward to verify that the classical 
limit $z\rightarrow 0$ in Table \ref{TABLE4.2} recovers the corresponding starting LH systems and related structures of Table \ref{TABLE4.1}, in agreement with   the relations (\ref{zac}) and (\ref{zae}).

\bigskip
\bigskip

\subsection{A method to construct Lie--Hamilton systems}\label{SeccMEthodConst}

\bigskip

Last chapter showed that deformations of a LH system with a fixed LH algebra $\mathcal{H}_\omega\simeq \mathfrak{g}$ can be obtained through a Poisson algebra $C^\infty(\mathfrak{g}^*)$, a given deformation and a certain Poisson morphism $D:C^\infty(\mathfrak{g}^*)\rightarrow C^\infty(M)$. This section presents a simple method to obtain $D$ from an arbitrary $\mathfrak{g}^*$ onto a symplectic manifold $\mathbb{R}^{2n}$. 

\bigskip

\begin{theorem}\label{MT2} Let $\mathfrak{g}$ be a Lie algebra whose Kostant--Kirillov--Souriau Poisson bracket admits a symplectic foliation in $\mathfrak{g}^*$ with a $2n$-dimensional $\mathcal{S}\subset \mathfrak{g}^*$. Then, there exists a LH algebra on the plane  given by
$$
\Phi:\mathfrak{g}\rightarrow C^\infty\bigl(\mathbb{R}^{2n}\bigr)
$$
relative to the canonical Poisson bracket on the plane.
\end{theorem}

\bigskip

\begin{proof}
The Lie algebra $\mathfrak{g}$ gives rise to a Poisson structure on $\mathfrak{g}^*$ through the Kostant--Kirillov--Souriau bracket $\{\cdot,\cdot\}$. This induces a symplectic foliation on $\mathfrak{g}^*$, whose leaves are symplectic manifolds relative to the restriction of the Poisson bracket. Such leaves are characterized by means of the Casimir functions of the Poisson bracket. By assumption, one of these leaves is $2n$-dimensional. In such a case, the Darboux Theorem warrants that the Poisson bracket on each leave is locally symplectomorphic to the Poisson bracket of the canonical symplectic form on $\mathbb{R}^{2n}\simeq T^*\mathbb{R}^n$. In particular, there exist some Darboux coordinates mapping the Poisson bracket on such a leaf into the canonical symplectic bracket on $T^*\mathbb{R}^n$. The corresponding change of variables into the canonical form in Darboux coordinates can be understood as a local diffeomorphism $h:\mathcal{S}_k\rightarrow \mathbb{R}^{2n}$ mapping the Poisson bracket $\Lambda_k$ on the leaf $\mathcal{S}_k$ into the canonical Poisson bracket on $T^*\mathbb{R}^n$. Hence, $h$ gives rise to a canonical Poisson algebra morphism $\phi_h:C^\infty(\mathcal{S}_k)\rightarrow C^\infty(T^*\mathbb{R}^n)$.  

As usual, a basis $\{v_1,\ldots, v_r\}$ of $\mathfrak{g}$ can be considered as a coordinate system on $\mathfrak{g}^*$. In view of the definition of the Kostant--Kirillov--Souriau bracket, they span an $r$-dimensional Lie algebra. In fact, if $[v_i,v_j]=\sum_{k=1}^rc_{ij}^kv_k$ for certain constants $c_{ij}^k$, then $\{v_i,v_j\}=\sum_{k=1}^rc_{ij}^kv_k$. Since $\mathcal{S}_k$ is a symplectic submanifold, there is a local immersion $\iota:\mathcal{S}_k\hookrightarrow \mathfrak{g}^*$ which is a Poisson manifold morphism. In consequence, 
$$
\{\iota^*v_i,\iota^*v_j\}=\sum_{k=1}^rc_{ij}^k\iota^*v_k.
$$
Hence, the functions $\iota^*v_i$ span a finite-dimensional Lie algebra of functions on $\mathcal{S}$. Since $\mathcal{S}$ is $2n$-dimensional, there exists a local diffeomorphism $\phi:\mathcal{S}\rightarrow \mathbb{R}^{2n}$ and
$$
\Phi:v\in \mathfrak{g}\mapsto \phi\circ\iota^*v\in C^\infty\bigl(\mathbb{R}^{2n} \bigr)
$$
is a Lie algebra morphism.
\end{proof}

\bigskip

{\rm

Let us apply the above mechanism to explain the existence of three types of LH systems on the plane. We already know that the Lie algebra $\mathfrak{sl}(2)$ gives rise to a Poisson algebra in $C^\infty(\mathfrak{g}^*)$. In the standard basis $v_1,v_2,v_3$ with commutation relations (\ref{brack2}), the Casimir is (\ref{ai}).
It turns out that the symplectic leaves of this Casimir are of three types: 
\begin{itemize}
 \item A one-sheeted hyperboloid when $v_1v_3-v_2^2=k<0$.
 
 \item A conical hyperboloid surface when $v_1v_3-v_2^2=0$.
 
\item A two-sheeted hyperboloid when $v_1v_3-v_2^2=k>0$.
\end{itemize}

In each of the three cases we have the Poisson bivector
$$
\Lambda=-v_1\frac{\partial}{\partial v_1}\wedge\frac{\partial}{\partial v_2}-2v_2\frac{\partial}{\partial v_1}\wedge\frac{\partial}{\partial v_3}-v_3\frac{\partial}{\partial v_2}\wedge\frac{\partial}{\partial v_3}.
$$
Then, we have a changes of variables passing from the above form into Darboux coordinates
$$
\bar v_1=v_1,\qquad \bar v_2=-v_2/v_1,\qquad  C= v_1v_3-v_2^2.
$$
Then,
$$
v_1=v_1,\qquad v_2=-\bar v_1 \bar v_2,\qquad  v_3=(C+\bar v^2_1\bar v^2_2)/\bar v_1.
$$ 
On a symplectic leaf, the value of $C$ is constant, say $C=c/4$, and the restrictions of the previous functions to the leaf read
$$
\iota ^*v_1=v_1,\qquad \iota^*v_2=-\bar v_1\bar v_2,\qquad  \iota^*v_3=c/(4\bar v_1)+\bar v_1\bar v^2_2.
$$
This can be viewed as a mapping $\Phi:\mathfrak{sl}(2)\rightarrow C^\infty(\mathbb{R}^2)$ such that
$$
\phi(v_1)=x,\qquad \phi(v_2)=-xy,\qquad \phi(v_3)=c/(4x)+xy^2,
$$
which is obviously a Lie algebra morphism relative to the standard Poisson bracket in the plane. It is simple to proof that when $c$ is positive,  negative  or zero, one obtains three different types of Lie algebras of functions and their associated vector fields span the Lie algebras P$_2$, I$_4$ and I$_5$  as enunciated in \cite{BBHLS}. Observe that since $\phi(v_1)\phi(v_3)-\phi(v_2)^2=c/4$, there exists no change of variables on $\mathbb{R}^2$ mapping one set of variables into another for different values of $c$. Hence, Theorem \ref{MT2} ultimately explains the real origin of all the  $\mathfrak{sl}(2)$-LH systems on the plane.

\bigskip

It is known that $\mathfrak{su}(2)$ admits a unique Casimir, up to  a proportional constant, and the
symplectic leaves induced in $\mathfrak{su}^*(2)$ are spheres. The application of the previous method originates a unique Lie algebra representation, which gives rise to the unique LH system on the plane related to $\mathfrak{so}(3)$. All the remaining LH systems on the plane can be generated in a similar fashion. The deformations of such Lie algebras will generate all the possible deformations of LH systems on the plane.

}


 

\chapter{Deformation of ODEs} 

\label{Chapter5} 
\renewcommand{\theequation}{5.\arabic{equation}}


\section{Deformed Milne--Pinney  equation and oscillator systems}\label{DefMPOSCILATORSYS}

\bigskip
{\rm

In this section we construct the non-standard deformation of the well-known Milne--Pinney (MP) equation \cite{Mi30,PIN}, which is known to be a LH system~\cite{BBHLS, BHLS}. Recall that the MP equation corresponds  to the equation of motion of the isotropic oscillator with a time-dependent frequency and a `centrifugal' term. As we will show in the sequel, the main feature of this deformation is that the new oscillator system has both a position-dependent mass and a time-dependent frequency.

\bigskip
\bigskip

\subsect{Non-deformed  system}

\bigskip

The MP equation \cite{Mi30,PIN} has the following expression
\begin{equation}\label{mp}
\frac{\dd^2x}{\dd t^2}=-\Omega^2(t)x+\frac{c}{x^3},
\end{equation}
where $\Omega(t)$ is any $t$-dependent function and $c\in \mathbb{R}$. 
By introducing a new variable $y:= \dd x/\dd t$, the system \eqref{mp} becomes a first-order system of differential equations on ${\rm T}\mathbb{R}_0$, where $\mathbb{R}_0:=\mathbb{R}\backslash\{0\}$, of the form
\be
\frac{\dd x}{\dd t}=y,\qquad \frac{\dd y}{\dd t}=-\Omega^2(t)x+\frac{c}{x^3}.
\label{FirstLie}
\ee
This system is indeed part of the one-dimensional Ermakov  system~\cite{Dissertations,Er08,Le91,LA08} and diffeomorphic to the one-dimensional $t$-dependent frequency counterpart~\cite{BBHLS, BCHLS13Ham, BHLS} of the Smorodinsky--Winternitz oscillator \cite{WSUF65}.

\bigskip

The system (\ref{FirstLie})  determines a Lie system  with  associated  $t$-dependent vector field~\cite{BHLS}
\be
{\bf X} ={\bf X}_3+\Omega^2(t){\bf X}_1,
\label{MP}
\ee
where
\begin{equation}\label{FirstLieA}  
{\bf X}_1:=-x\frac{\partial}{\partial y},\qquad {\bf X}_2:=\frac 12 \left(y\frac{\partial}{\partial y}-x\frac{\partial}{\partial x}\right),\qquad {\bf X}_3:=y\frac{\partial}{\partial x}+\frac{c}{x^3}\frac{\partial}{\partial y},
\end{equation}
span a Vessiot--Guldberg Lie algebra $V^{\rm MP}$ of vector fields isomorphic to $\mathfrak{sl}(2)$  (for any value of $c$) with commutation relations given by  
\begin{equation}\label{aa}
[{\bf X}_1,{\bf X}_2]={\bf X}_1,\qquad [{\bf X}_1,{\bf X}_3]=2{\bf X}_2,\qquad [{\bf X}_2,{\bf X}_3]={\bf X}_3 .
\end{equation}
The vector fields of  $V^{\rm MP}$ are defined on $\mathbb R^2_{x\ne 0}$, where they span a regular distribution of order two.

\bigskip

Furthermore,  ${\bf X} $  is a LH system with respect to the symplectic form $\omega={\rm d}x\wedge {\rm d}y$ and the vector fields (\ref{FirstLieA}) admit Hamiltonian functions given by
\be
h_1=\frac 12 x^2 ,\qquad h_2=-\frac 12 xy ,\qquad h_3=\frac 12 \left(y^2 +\frac{c}{x^2} \right),
\label{ham2}
\ee 
that fulfill the following commutation relations with respect  to the Poisson bracket induced by $\omega$:
\be
\{ h_1,h_2\}_\omega=- h_1,\qquad \{ h_1,h_3\}_\omega =- 2h_2,\qquad  \{ h_2,h_3\}_\omega =-h_3 .
\label{brack}
\ee
Then, the functions $h_1,h_2,h_3$ span a LH algebra  ${\cal H}_{\omega}^{\rm {MP}} \simeq \mathfrak{sl}(2)$ of functions on $\mathbb R^2_{x\ne 0}$; the $t$-dependent Hamiltonian associated with the $t$-dependent vector field (\ref{MP}) reads
\be
h =h_3+\Omega^2(t)h_1   .
\label{hMP}
\ee
We recall that this Hamiltonian  is a natural one, that is, it can be written in terms of a kinetic energy $T$ and potential $U$
by identifying the variable $y$ as the conjugate momentum $p$ of the coordinate $x$:
\be
h =T+U= \frac 12\, p^2 + \frac 12 \Omega^2(t) x^2 + \frac{c}{2x^2}.
\label{ham}
\ee
Hence $h$  determines the composition  of a one-dimensional  oscillator with  a time-dependent frequency  $ \Omega(t)$ and unit mass  with a   Rosochatius  or Winternitz potential; the latter  is  just a centrifugal barrier whenever $c>0$ (see~\cite{nonlinear} and references therein).   The LH system  (\ref{FirstLie})  thus comes from the Hamilton equations of $h$ and, obviously, when $c$ vanishes, these reduce  to the equations of motion of a harmonic oscillator with a
time-dependent frequency. 

\bigskip

We stress that  it has been already proved in ~\cite{BBHLS,BHLS} that the MP equations (\ref{FirstLie}) comprise the {\em three} different types of possible $\mathfrak{sl}(2)$-LH systems according to the value of the constant $c$:   class  {\rm P}$_2$ for $c>0$;  class {\rm I}$_4$ for $c<0$; and class {\rm I}$_5$ for $c=0$. This means that any other LH system related to a Vessiot--Guldberg Lie algebra of Hamiltonian vector fields isomorphic to $\mathfrak{sl}(2)$ must be, up to a $t$-independent change of variables, of the form (\ref{FirstLie}) for a positive, zero or negative value of $c$.

This implies that the second-order Kummer--Schwarz equations~\cite{CGL11, LS12} and several  types of  Riccati equations~\cite{CGLS, Eg07,pilar,Mariton, SSVG14, SSVG11,  Wi08} are comprised within  ${\cal H}_{\omega}^{\rm {MP}}$ (depending on the sign of $c$). The relationships amongst all of these systems are  ensured by construction and these can be explicitly obtained  through either diffeomorphisms or changes of variables (see~\cite{BBHLS,BHLS} for details).

The constants of motion for the MP equations can be obtained by applying the coalgebra formalism introduced in~\cite{BCHLS13Ham} and briefly summarized in section 2.4. Explicitly, 
let us consider the   Poisson--Hopf algebra ${\cal C}^\infty({\cal H}_{\omega}^{\rm {MP}*} )$   with basis  $\{ v_1,v_2,v_3\}$,  coproduct    (\ref{baa}), fundamental Poisson brackets  (\ref{brack2}) and Casimir (\ref{ai}).
The Poisson algebra morphisms  (\ref{morphisms}) 
 \be
 D: {\cal C}^\infty({\cal H}_{\omega}^{\rm {MP}*}) \rightarrow {\cal C}^\infty(\mathbb R^2_{x\ne 0}) ,\quad D^{(2)} :    {\cal C}^\infty( {\cal H}_{\omega}^{\rm {MP}*} ) \otimes  {\cal C}^\infty ( {\cal H}_{\omega}^{ \rm {MP}*} )\rightarrow {\cal C}^\infty(\mathbb R^2_{x\ne 0})\otimes {\cal C}^\infty(\mathbb R^2_{x\ne 0}) ,
 \nonumber
 \ee
  defined by (\ref{bb}),  where $h_i$ are the Hamiltonian functions   (\ref{ham2}), lead to the   $t$-independent  constants of the motion  $F^{(1)}:=F$ and $F^{(2)}$  given by (\ref{bc}), through   the Casimir (\ref{ai}),   for the Lie system  ${\bf X}$ (\ref{FirstLie}); namely~\cite{BCHLS13Ham}
\bea
&& F= h_1(x_1,y_1) h_3(x_1,y_1)- h_2^2(x_1,y_1)=\frac c 4 ,\nonumber\\[2pt]
&&F^{(2)}=\bigl( \left[ h_1(x_1,y_1)+h_1(x_2,y_2)\right] \left[ h_3(x_1,y_1)+h_3(x_2,y_2)\right]  \bigr) -\bigl( h_2(x_1,y_1)+h_2(x_2,y_2) \bigl)^2\nonumber\\[2pt]
&&\qquad\, =  \frac 14 ({x_1}{y_2} -{x_2} {y_1})^2 
+\frac c 4\,  \frac{(x_1^2+x_2^2)^2}{x_1^2 x_2^2} .
\label{am}
\eea

\bigskip

We observe that $F^{(2)}$ is just a Ray--Reid invariant for generalized Ermakov systems \cite{Le91,RR79} and that  it is related to  the one obtained in~\cite{coalgebra2,letterBH} from a coalgebra approach  applied to superintegrable systems. 

By permutation of the indices corresponding to the variables of  the non-trivial invariant $F^{(2)}$,  we find two other constants of the motion:
\be 
    F_{13}^{(2)}=S_{13} ( F^{(2)}   ) ,\qquad    F_{23}^{(2)}=S_{23} ( F^{(2)}   ) ,
\label{an}
\ee
 where $S_{ij}$ is the permutation of variables $(x_{i},y_i)\leftrightarrow
(x_j,y_j)$. Since    $\partial(F^{(2)},F^{(2)}_{23})/\partial(x_1,y_1)\neq 0$, both constants of motion are functionally independent (note that the pair $(F^{(2)},F^{(2)}_{13})$  is functionally independent as well). From these two invariants, the corresponding superposition rule can be  derived in a straightforward manner. Its explicit expression can be found in~\cite{BCHLS13Ham}.

\bigskip
\bigskip

\subsect{Deformed Milne--Pinney  equation}

 \bigskip

In order to apply the non-standard deformation of $\mathfrak{sl}(2)$ described in chapter \ref{Chapter4}  to the MP equation, we need to find  the deformed counterpart $h_{z,i}$ $(i=1,2,3)$ of the Hamiltonian functions $h_i$ (\ref{ham2}), so fulfilling the   Poisson brackets  (\ref{gb}), by keeping the canonical symplectic form $\omega$. 

\bigskip

This problem can be rephrased as the one consistent in finding symplectic realizations of a given Poisson algebra, which can be solved once a particular symplectic leave is fixed as a level set for the Casimir functions of the algebra, where the generators of the algebra can be  expressed in terms of the corresponding Darboux coordinates. 
In the particular case of the ${\cal U}_z(\mathfrak{sl}(2))$ algebra, the explicit solution (modulo canonical transformations) was obtained in~\cite{chains} where the algebra~\eqref{gb} was found to be generated by the functions
\bea 
&&v_1(q,p)=\frac 12\,q^2,\cr
&&v_2(q,p)=-\frac 12\frac {\sinh z q^2}{z q^2} \, q p , \cr
&&v_3(q,p)=\frac 12\frac {\sinh z q^2}{z q^2}\,  p^2 +
\frac 12\frac{z c}{\sinh z q^2},
\nonumber
\eea
where $\omega={\rm d} q\wedge {\rm d} p$, and the Casimir function~\eqref{gc} reads ${C}_z=c/4$. In practical terms, such a solution can easily be  found by solving firstly the non-deformed case $z\to 0$ and, afterwards, by deforming the $v_i(q,p)$ functions under the constraint that the Casimir ${C}_z$ has to take a constant value.
With this result at hand, the corresponding deformed vector fields ${\bf X}_{z,i}$ can   be computed by imposing the relationship (\ref{contract2}) and the final result is summarized in the following statement. 

\bigskip
  
\begin{proposition}
\label{proposition1} (i) The Hamiltonian functions defined by
\be
h_{z,1}:=\frac 12 x^2 ,  \qquad
 h_{z,2}:= -      \frac 12\shc (z x^2)\, x  y    , \qquad
 h_{z,3}:= \frac 12 \left(\! \shc (z x^2) \, y^2 + \frac 1{ \shc (zx^2)}\,
\frac{c}{x^2}  \right)  ,
\label{gd}
\ee
 close the Poisson brackets (\ref{gb}) with respect to the symplectic form  $\omega={\rm d}x\wedge {\rm d}y$ on $\mathbb R^2_{x\ne 0}$, namely
\begin{equation}\label{gb2}
\begin{gathered}
\{h_{z,1},h_{z,2}\}_\omega=-\shc (2z h_{z,1} )h_{z,1},\qquad 
 \{h_{z,1},h_{z,3}\}_\omega=-2 h_{z,2},\\[2pt]
\{h_{z,2},h_{z,3}\}_\omega= - \cosh(2 z h_{z,1})  h_{z,3},
\end{gathered}
\end{equation}
where $\shc(x)$ is defined in (\ref{shc}).
 Relations~\eqref{gb2} define  the deformed Poisson  algebra $C^\infty({\cal H}_{z,\omega}^{\rm {MP}*})$. 

\noindent
 (ii)  The vector fields ${\bf X}_{z,i}$ corresponding to $h_{z,i}$  read
\begin{equation}
\begin{gathered}
{\bf X}_{z,1}   =   -x\frac{\partial}{\partial y},\qquad {\bf X}_{z,2}=\left(\cosh(zx^{2})-\frac 12 \shc(zx^{2})\right)y\frac{\partial}{\partial y}-\frac 12 \shc(zx^{2})\,x\frac{\partial}{\partial x}, \\
{\bf X}_{z,3}   =   \shc(zx^{2})\,y\frac{\partial}{\partial x}+\left[\frac{c}{x^{3}}\, \frac{\cosh(zx^{2})}{\shc^{2}(zx^{2})} +\frac{\shc(zx^{2})-\cosh(zx^{2})}{x}\,y^{2}\right]\frac{\partial}{\partial y},\nonumber
\end{gathered}
\end{equation}
which satisfy 
\begin{equation}
\begin{gathered}
\left[{\bf X}_{z,1},{\bf X}_{z,2}\right]=\cosh (z x^2) \, {\bf X}_{z,1},\qquad [{\bf X}_{z,1},{\bf X}_{z,3}]=2 {\bf X}_{z,2}, \\[2pt] 
[{\bf X}_{z,2},{\bf X}_{z,3}]=\cosh (z x^2) \, {\bf X}_{z,3}+ z^2\left(    c  + x^2  y^2\,  \shc^2 (z x^2)  \right)  {\bf X}_{z,1}. 
\label{com2}
\end{gathered}
\end{equation}
\end{proposition}

\smallskip

Since $\lim_{z\to 0}\shc(z x^2)=1$ and $\lim_{z\to 0}\cosh(z x^2)=1$, it can directly  be checked that
all the classical limits   (\ref{zac}), (\ref{zad}) and (\ref{zae}) are fulfilled. As expected,  the Lie derivative of $\omega$  with respect to each ${\bf X}_{z,i}$  vanishes.

At this stage, it is important to realize that, albeit (\ref{gb2}) are genuine  Poisson brackets defining the Poisson algebra   $C^\infty({\cal H}_{z,\omega}^{\rm {MP}*})$, the commutators (\ref{com2}) show that ${\bf X}_{z,i}$ do not span a new Vessiot--Guldberg Lie algebra; in fact, the commutators give rise to linear combinations of the vector fields ${\bf X}_{z,i}$ with coefficients that are functions depending on the coordinates and the deformation parameter.

Consequently, proposition~\ref{proposition1} leads to  a deformation of the initial Lie system (\ref{MP}) and of the LH one (\ref{hMP})  defined by
   \be
{\bf X}_z:={\bf  X}_{z,3}+\Omega^2(t){\bf X}_{z,1},\qquad h_z:=h_{z,3}+\Omega^2(t)h_{z,1}.
\label{MPz}
\ee
Thus we    obtain  the following $z$-parametric   system of differential equations that generalizes (\ref{FirstLie}):
\bea
&&\frac{\dd x}{\dd t}=\shc (z x^2)\, y,\nonumber\\[2pt]
&& \frac{\dd y}{\dd t}=-\Omega^2(t)x+    
\frac{c}{x^3} \, \frac{\cosh (z x^2)  }{ \shc^2(z x^2) }+ \frac{\shc (z x^2)- \cosh (z x^2)}{x}  \, y^2 .
\label{FirstLie2}
\eea
From the first equation, we can write   $$y=\frac 1{   \shc (z x^2) }\, \frac{\dd x }{\dd t},
$$
and by substituting this expression into the second equation in (\ref{FirstLie2}), we obtain   
a deformation of the MP equation~\eqref{mp}  in the form
\be
\frac{\dd^2 x}{\dd t^2}  + \left(\frac{1}{x}- \frac{z x}{ \tanh (z x^2)} \right)   \biggl(\frac{\dd x}{\dd t} \biggr)^2 =-\Omega^2(t) \, x  \shc (z x^2)+ \,\frac{c\,z }{x \tanh (z x^2)} .
\nonumber
\ee
Note that this really is    a deformation of the MP equation in the sense that   the limit $z\to 0$   recovers  the standard one (\ref{mp}).
 
\bigskip
\bigskip


\subsect{Constants of motion for the deformed  Milne--Pinney system}

\bigskip

 An essential feature of the formalism here presented is the fact that  $t$-independent constants of motion for the deformed system ${\bf X}_z$ (\ref{MPz}) can be deduced by using the coalgebra structure of $C^\infty( {\cal H}_{z,\omega}^{\rm {MP}*})$.  Thus we start with  the Poisson--Hopf algebra    $C^\infty( {\cal H}_{z,\omega}^{\rm {MP}*} )$ with deformed coproduct $\Delta_z$ given by (\ref{ga}) and, following section 2.4~\cite{BCHLS13Ham}, we consider the Poisson algebra morphisms 
\be
D_z: C^\infty( {\cal H}_{z,\omega}^{\rm {MP}*} )\rightarrow C^\infty( \mathbb R^2_{x\ne 0}),\quad  D_z^{(2)} : C^\infty( {\cal H}_{z,\omega}^{\rm {MP}*} )\otimes   C^\infty( {\cal H}_{z,\omega}^{\rm {MP}*} )\rightarrow C^\infty(\mathbb R^2_{x\ne 0})\otimes C^\infty(\mathbb R^2_{x\ne 0}),
\nonumber
\ee
which are  defined by   
\bea 
&& D_z( v_i)= h_{z,i}(x_1,y_1):=  h_{z,i}^{(1)}  , \quad i=1,2,3, \nonumber\\ 
&&  D_z^{(2)} \left( {\Delta}_z(v_1) \right) = h_{z,1}(x_1,y_1)+h_{z,1}(x_2,y_2):= h_{z,1}^{(2)} \,  ,  \nonumber \\
&& 
D_z^{(2)} \left( {\Delta}_z(v_j) \right) = h_{z,j}(x_1,y_1)  {\rm e}^{2 z h_{z,1}(x_2,y_2)}  + {\rm e}^{-2 z h_{z,1}(x_1,y_1)} h_{z,j}(x_2,y_2):=  h_{z,j}^{(2)} \,  ,\quad j= 2,3,
\nonumber
\eea
where $h_{z,i}$ are the Hamiltonian functions (\ref{gd}), so fulfilling (\ref{gb2}). Hence (see \cite{chains})
\bea
&&  h_{z,1}^{(2)} =  \frac 12(x_1^2+x_2^2)  ,\nonumber\\[2pt]
&&  h_{z,2}^{(2)} =  -      \frac 12 \left(  \! {\shc (z x_1^2)}  \, x_1  y_1  {\rm e}^{z x_2^2}   +   {\rm e}^{- z x_1^2}      {\shc (z x_2^2)}  \, x_2  y_2 \right)    , \nonumber\\[2pt]
&&  h_{z,3}^{(2)} =\frac  12   \left(  \!  {\shc (z x_1^2)}\,  y_1^2 +
\frac{c}{ x_1^2 \shc (zx_1^2)} \right) {\rm e}^{z x_2^2}   + \frac 12\,   {\rm e}^{- z x_1^2}   \left(  \! {\shc (z x_2^2)} \, y_2^2 +
\frac{c}{ x_2^2\, \shc (zx_2^2)} \right). 
\nonumber
\eea
Recall that, by construction, the functions $h_{z,i}^{(2)}$ fulfill the Poisson brackets (\ref{gb2}). 
The $t$-independent constants of motion are then obtained through 
$$
  F_z= D_z(C_z),\qquad F_z^{(2)}=  D_z^{(2)} \left( {\Delta_z}(C_z) \right),
$$
where $C_z$ is the Casimir (\ref{gc}); these are 
\bea
&& \!\!\!\!  \!\!\!\!  \!\!\!\! \!  F_z=  \shc\bigl(2 z h_{z,1}^{(1)}  \bigr)  h_{z,1}^{(1)}   h_{z,3}^{(1)} - \bigl( h_{z,2}^{(1)}  \bigr)^2 = \frac c    4 \,  ,\nonumber\\[2pt]
&& \!\!\!\!  \!\!\!\!  \!\!\!\! \!  F_z^{(2)}=  \shc\bigl(2 z h_{z,1}^{(2)} \bigr)  h_{z,1}^{(2)}   h_{z,3}^{(2)} - \bigl( h_{z,2}^{(2)}  \bigr)^2  \label{amz}\\[2pt]
&& = 
 \frac 14 \left[  \shc(z x_1^2)      \shc(z x_2^2) \,  ({x_1}{y_2} -{x_2} {y_1})^2     +c\,  \frac{ \shc^2\bigl( z(x_1^2+x_2^2) \bigr)}{\shc( z x_1^2) \shc( z x_2^2)}\,  \frac{(x_1^2+x_2^2)^2}{x_1^2 x_2^2} \right]  {\rm e}^{- z x_1^2} {\rm e}^{z x_2^2} ,
\nonumber
\eea
so providing the corresponding deformed  Ray--Reid invariant, being (\ref{am}) its non-deformed counterpart  with $z=0$. Notice that this invariant is related  to the so-called `universal   constant of the motion'  coming from  ${\cal U}_z(\mathfrak{sl}(2))$  and given in~\cite{letterBH}.
As in (\ref{an}), other equivalent constants of motion can be deduced from $ F_z^{(2)}$ by permutation of the variables.


\bigskip
\bigskip

\subsect{A new  oscillator system with position-dependent mass}\label{ANOscliatorPDM}

\bigskip

If we set $p:= y$, the $t$-dependent Hamiltonian $h_z$  in (\ref{MPz})  can be written, through (\ref{gd}),  as:
\be
h_z= T_z+U_z= \frac 12  {\shc (z x^2)}\,  p^2+ \frac 12\Omega^2(t) x^2  +
\frac{ c}{2 x^2  {\shc (z x^2)} }  \,  , 
\nonumber
\ee
so deforming $h$ given in  (\ref{ham}). The corresponding Hamilton equations are just   (\ref{FirstLie2}). 

  It is worth mentioning that $h_z$ can be interpreted naturally within the framework of 
 position-dependent mass oscillators (see~\cite{ BurgosAnnPh11,CrNN07,  CrR09, GhoshRoy15, MustJpa15, Quesne07,Quesne15Jmp, Ran14Jmp} and references therein). The above Hamiltonian naturally suggests the definition of a position-dependent mass function in the form
\be
m_z(x) :=\frac 1{  \shc (z x^2) }= \frac{z x^2}{ \sinh (z x^2)}  \, ,\qquad \lim_{z\to 0} m_z(x) =1,  \qquad \lim_{x\to \pm\infty } m_z(x) =0.
\label{masa}
\ee
Then $h_z$ can be rewritten as
\be
h_z=  \frac {p^2}{2m_z(x)}  + \frac 12m_z(x) \Omega^2(t)\left[ x^2  \shc (z x^2)   \right]  +
\frac{  c}{2m_z(x)} \left[  \frac{1}{x ^2\shc^2(z x^2)} \right]   .
\nonumber
\ee
Thus  the Hamiltonian $h_z$ can be regarded as a system corresponding to   a particle with  position-dependent mass $m_z(x)$ under a deformed oscillator potential  $U_{z,{\rm osc}}(x)$ with time-dependent frequency  $\Omega(t)$ and   a deformed potential $U_{z,{\rm RW}}(x)$ given by
\bea
&& 
U_{z,{\rm osc}}(x): =  x^2 \shc (z x^2)=  \frac{\sinh (z x^2)}{z} \, ,\label{oscz}\\
&& 
 U_{z,{\rm RW}}(x):=\frac{1}{x ^2\shc^2(z x^2)}= \left( \frac{z x}{\sinh (z x^2)}\right)^2,
\nonumber
\eea
such that
\bea
&&  \lim_{z\to 0} U_{z,{\rm osc}}(x) =x^2,  \qquad\  \lim_{x\to \pm\infty }U_{z,{\rm osc}}(x)  =+\infty, \nonumber\\[2pt]
&&  \lim_{z\to 0}  U_{z,{\rm RW}}(x)=\frac 1{x^2},  \qquad\  \lim_{x\to \pm\infty } U_{z,{\rm RW}}(x) =0 .
\nonumber 
\eea
The deformed mass and the oscillator potential functions are represented in figures \ref{pdm} and  \ref{osc}

The Hamilton equations (\ref{FirstLie2}) can easily be  expressed in terms of $m_z(x)$ as
\bea
&&\!\!\!\!\!\!\!\!\!\!   \dot x = \frac{\partial h^{\rm MP}_z}{\partial p}= \frac p{m_z(x)} ,\nonumber\\
&&\!\!\!\!\!\!\!\!\!\!   \dot p= -  \frac{\partial h^{\rm MP}_z}{\partial x} = -m_z(x)  \Omega^2(t)  \, x \shc (z x^2) + \frac{c}{m_z(x)} \, \frac {\cosh (z x^2)} {x^3 \shc^3 (z x^2)}   +p^2  \frac{ m^\prime_z(x)}{2 m^2_z(x)},
\nonumber
\eea
and the constant of the motion (\ref{amz})   turns out to be
\be
 F_z^{(2)}=    \frac 14 \left[  \frac {({x_1}{p_2} -{x_2} {p_1})^2 }{m_z(x_1)m_z(x_2) }    +c\, m_z(x_1)m_z(x_2)    { \shc^2\bigl( z(x_1^2+x_2^2) \bigr)} \,  \frac{(x_1^2+x_2^2)^2}{x_1^2 x_2^2} \right]  {\rm e}^{- z x_1^2} {\rm e}^{z x_2^2}.
\nonumber
\ee


\begin{figure}[h!]
\begin{center}
\includegraphics[height=6.0cm]{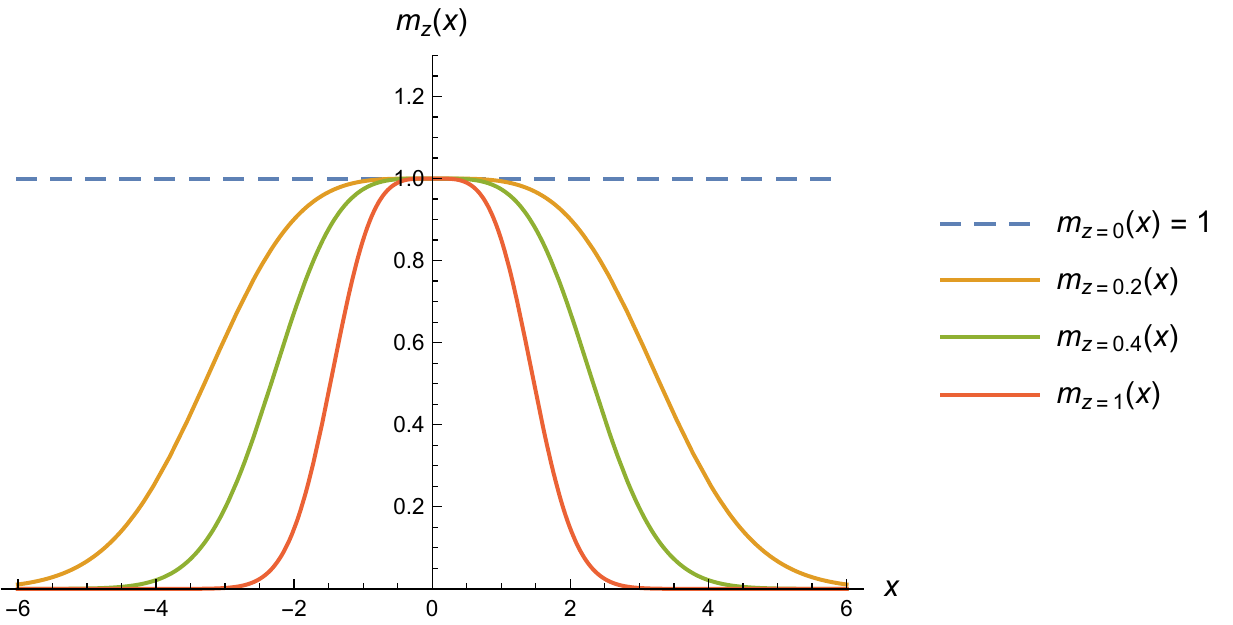}
\caption{\small The position-dependent mass (\ref{masa}) for different values of the deformation parameter $z$.}
 \label{pdm}
\end{center}
\end{figure}



\begin{figure}[h!]
\begin{center}
\includegraphics[height=6.0cm]{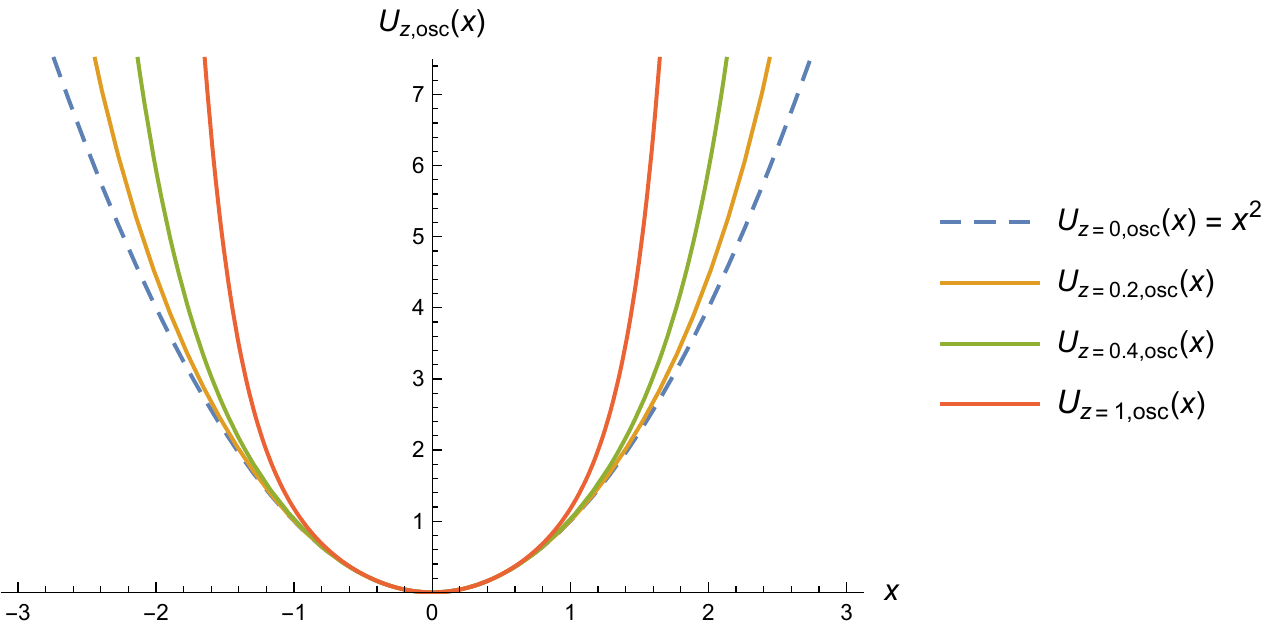}
\caption{\small The deformed oscillator potential  (\ref{oscz}) for different values of the deformation parameter $z$.}
 \label{osc}
\end{center}
\end{figure}


\bigskip
\bigskip


\sect{Deformed Riccati equation}\label{RiccatiSeccion}

\bigskip

\subsect{Deformed complex Riccati  equation}

\bigskip

 In this section we  consider the   complex Riccati equation given by
\begin{equation}
\frac{{\rm d} z}{{\rm d} t}=b_1(t)+b_2(t)z+b_3(t)z^2,\qquad z\in\mathbb{C},
\label{da}
\end{equation}
where $b_i(t)$  are arbitrary $t$-dependent real coefficients. We recall that (\ref{da}) is related to certain planar Riccati equations~\cite{Eg07,Wi08} and that several mathematical and physical applications can be found in~\cite{Or12,Sc12}.

By writing $z= \xx+i \yy$, we find that (\ref{da}) gives rise to a system of the type (\ref{system}), namely 
\begin{equation}
\frac{{\rm d} \xx}{{\rm d} t}=b_1(t)+b_2(t)\xx+b_3(t)(\xx^2- \yy^2),\qquad \frac{{\rm d} \yy}{{\rm d} t}=b_2(t)\yy+2b_3(t)\xx \yy .
\label{db}
\end{equation}
Thus the  associated  $t$-dependent vector field  reads
\be
 {\bf X}=b_1(t){\bf X}_1+b_2(t){\bf X}_2+b_3(t){\bf X}_3,
 \label{dc}
\ee
 where
\begin{equation}
{\bf X}_1= \frac{\partial}{\partial \xx},\qquad {\bf X}_2= \xx\frac{\partial}{\partial \xx}+\yy\frac{\partial}{\partial \yy} ,\qquad {\bf X}_3= (\xx^2- \yy^2)\frac{\partial}{\partial \xx}+2\xx \yy\frac{\partial}{\partial \yy} ,
\label{vectRiccati2}
\end{equation}
span a Vessiot--Guldberg Lie algebra $V^{\rm CR}\simeq \mathfrak{sl}(2)$ with the same commutation relations (\ref{aa}). It has already be proven that the system ${\bf X}$  is a LH one belonging to the  class  {\rm P}$_2$~\cite{ BBHLS, BHLS} and  that their vector fields span a regular distribution on $\mathbb{R}^2_{\yy\neq 0}$. The  symplectic form, coming from (\ref{der}),  and    the corresponding Hamiltonian functions (\ref{contract})   turn out to be
 \be
\omega=\frac{\dd \xx\wedge \dd \yy}{\yy^2},\qquad h_1= -\frac 1{\yy},\qquad h_2= -\frac \xx \yy , \qquad h_3=- \frac{\xx^2+\yy^2}{\yy} ,
\label{de}
\ee
 which fulfill the commutation rules (\ref{brack}) so defining   a LH algebra  ${\cal H}_{\omega}^{\rm {CR}}$.  
A $t$-dependent Hamiltonian associated with ${\bf X}$ reads
\be
h=b_1(t)h_1+b_2(t)h_2+b_3(t)h_3   .
\label{de2}
\ee
In this case, the constants of the motion     (\ref{bc}) are found to be $F=1$ and~\cite{BHLS}
\be
 F^{(2)}= \frac{(\xx_1-\xx_2)^2+(\yy_1+ \yy_2)^2}{\yy_1 \yy_2} \, .
 \label{df}
 \ee

As commented above,  the  Riccati system (\ref{db}) is locally diffeomorphic  to the MP equations (\ref{FirstLie})  with  $c>0$, both belonging to the same   class  {\rm P}$_2$~\cite{BBHLS}. Explicitly,  the change of variables
\be
x=\pm \frac{c^{1/4}}{ \sqrt{|v| }},\qquad  y=\mp \frac{c^{1/4} \, u}{ \sqrt{|v|}},\qquad u=-\frac{y}x,\qquad |v|=  \frac{c^{1/2}}{x^2},\qquad c>0,
\label{dg}
\ee
map, in this order,  the vector fields (\ref{FirstLieA}) on $\mathbb{R}^2_{x\neq 0}$, the symplectic form $\omega={\rm d}x\wedge {\rm d}y$, Hamiltonian functions (\ref{ham2}) and the constant of motion (\ref{am}) 
onto the vector fields (\ref{vectRiccati2}) on $\mathbb{R}^2_{\yy\neq 0}$, (\ref{de}) and (\ref{df}) (up to a multiplicative constant $\pm\frac 12 c^{1/2}$). 

 \bigskip

To obtain the corresponding   (non-standard) deformation of the complex Riccati  system (\ref{db}),   the very same change of variables (\ref{dg}) can be considered since, in our approach, the symplectic form (\ref{de}) is kept non-deformed.  Thus, by starting from proposition~\ref{proposition1}  and applying (\ref{dg}) (with $c=4$ for simplicity), we get the following result.

\bigskip

\begin{proposition}
\label{proposition2}
(i) The Hamiltonian functions given by
\be
h_{z,1}=- \frac 1 {\yy} \, ,  \qquad
 h_{z,2}= - \shc (2z /\yy)  \,    \frac \xx\yy  \,  , \qquad
 h_{z,3}=-  \frac{ \shc^2 (2z /\yy) \, \xx^2 + \yy^2 }{\shc (2z /\yy)\,  \yy}    \, ,
\nonumber
\ee
fulfill the  commutation rules (\ref{gb2}) with respect to the Poisson bracket induced by the symplectic form  $\omega$ (\ref{de})   defining the deformed Poisson algebra ${\cal C}^\infty({\cal H}_{z,\omega}^{\rm {CR}*})$. 

\noindent
(ii) The   corresponding   vector fields ${\bf X}_{z,i}$ read
\begin{eqnarray}
&& {\bf X}_{z,1}   =  \frac{\partial}{\partial \xx},\qquad {\bf X}_{z,2}=\xx \cosh(2z/ \yy)\frac{\partial}{\partial \xx} + \yy  \shc(2z/  \yy) \frac{\partial}{\partial \yy} , \nonumber\\
&& {\bf X}_{z,3}   =    \left(\xx^2 -\frac{\yy^2}{ \shc^2(2z/ \yy)} \right)   \cosh(2z /\yy)\frac{\partial}{\partial \xx}+2\xx\yy  \shc(2z/ \yy)\frac{\partial}{\partial \yy} ,\nonumber
\end{eqnarray}
which satisfy 
\bea
&& [{\bf X}_{z,1},{\bf X}_{z,2}]=\cosh (2z/ \yy) \, {\bf X}_{z,1},\qquad [{\bf X}_{z,1},{\bf X}_{z,3}]=2 {\bf X}_{z,2},\nonumber\\[2pt] 
&& [{\bf X}_{z,2},{\bf X}_{z,3}]=\cosh (2z/ \yy) \, {\bf X}_{z,3}+ 4 z^2\left( 1  + \frac{\xx^2}{\yy^2}  \shc^2 (2z/ \yy)  \right)  {\bf X}_{z,1}. 
\nonumber
\eea
\end{proposition}

\bigskip

Next the deformed counterpart of the Riccati Lie system (\ref{dc}) and of the   LH one (\ref{de2}) is defined by
 \be
{\bf X}_z:=b_1(t){\bf X}_{z,1}+b_2(t){\bf X}_{z,2}+b_3(t){\bf X}_{z,3},\qquad h_z:=b_1(t)h_{z,1}+b_2(t)h_{z,2}+b_3(t)h_{z,3} .
\label{dl}
\ee
And the   $t$-independent constants of motion turn out to be $F_z=1$ and
\be
 F_z^{(2)}=\left(   \shc (2z/ \yy_1)   \shc (2z/ \yy_2)   \frac{(\xx_1-\xx_2)^2 }{\yy_1 \yy_2} + \frac{\shc^2 (2z/ \yy_1+ 2z/ \yy_2) }{  \shc (2z/ \yy_1)   \shc (2z/ \yy_2)  } \frac{ (\yy_1+ \yy_2)^2}{\yy_1 \yy_2}  \right) \eee^{2z/\yy_1}  \eee^{-2z/\yy_2} \, .
\nonumber
\ee
Therefore  the   deformation of the system (\ref{db}), defined by ${\bf X}_z$ (\ref{dl}),  reads
\bea
&& \frac{{\rm d} \xx}{{\rm d} t}=b_1(t)+b_2(t) \xx \cosh(2z/ \yy)  +b_3(t)  \left(\xx^2 -\frac{\yy^2}{ \shc^2(2z/ \yy)} \right)   \cosh(2z /\yy), \nonumber\\[2pt]
&& \frac{{\rm d} \yy}{{\rm d} t}=b_2(t)  \yy  \shc(2z/  \yy)+2b_3(t) \xx\yy  \shc(2z/ \yy) .
\nonumber
\eea

\bigskip
\bigskip

\subsect{Deformed coupled Riccati  equations}

\bigskip

As a last application, let us consider two coupled Riccati equations given by
 \cite{Mariton}
\begin{equation}
\frac{{\rm d}\xx}{{\rm d}t}=a_0(t)+a_1(t)\xx+a_2(t)\xx^2,\qquad \frac{{\rm d}\yy}{{\rm d}t}=a_0(t)+a_1(t)\yy+a_2(t)\yy^2,
\label{ea}
\end{equation}
constituting a particular case of the systems of Riccati equations studied in~\cite{BCHLS13Ham,CGLS}.  

Clearly, the system (\ref{ea}) is a Lie system associated with a $t$-dependent vector field
\be
 {\bf X}=a_0(t){\bf X}_1+a_1(t){\bf X}_2+a_2(t){\bf X}_3,
 \label{eb}
\ee
where
\begin{equation}
{\bf X}_1= \frac{\partial}{\partial \xx}+ \frac{\partial}{\partial \yy},  \qquad {\bf X}_2= \xx\frac{\partial}{\partial \xx}+\yy\frac{\partial}{\partial \yy} ,\qquad {\bf X}_3= \xx^2\frac{\partial}{\partial \xx}+\yy^2\frac{\partial}{\partial \yy} ,
\label{ec}
\end{equation}
close on the commutation rules (\ref{aa}), so  spanning  a Vessiot--Guldberg Lie algebra $V^{\rm 2R}\simeq \mathfrak{sl}(2)$. 
Furthermore, ${\bf X}$  is a LH system which belongs to the class  {\rm I}$_4$~\cite{ BBHLS, BHLS} restricted to $\mathbb{R}^2_{\xx\neq \yy}$.
The symplectic form and Hamiltonian functions for ${\bf X}_1,{\bf X}_2, {\bf X}_3$ read   
 \be
\omega=\frac{\dd \xx\wedge \dd \yy}{(\xx-\yy)^2} ,\qquad h_1= \frac 1{\xx-\yy},\qquad h_2= \frac 12\left( \frac {\xx+\yy}  {\xx-\yy}\right) , \qquad h_3= \frac{\xx \yy}{\xx-\yy}.
\label{ee}
\ee
The functions $h_1,h_2,h_3$ satisfy the   commutation rules (\ref{brack}),  thus spanning a LH algebra  ${\cal H}_{\omega}^{\rm {2R}}$.  
Hence, the $t$-dependent Hamiltonian associated with ${\bf X}$  is given by
\be
h=a_0(t)h_1+a_1(t)h_2+a_2(t)h_3  .
\label{ee2}
\ee
The   constants of the motion     (\ref{bc}) are  now $F=-1/4$ and~\cite{BHLS}
\be
 F^{(2)}=-  \frac{ (\xx_2- \yy_1  ) (\xx_1- \yy_2  )}    { (\xx_1- \yy_1  ) (\xx_2- \yy_2  )} \, .
 \label{ef}
 \ee

The LH system    (\ref{ea}) is locally diffeomorphic  to the MP equations (\ref{FirstLie}) but now  with  $c<0$~\cite{BBHLS}. 
Such a diffeomorphism is achieved through the     change of variables given by
\bea
&& x=\pm \frac{ (4|c|)^{1/4}}{ \sqrt{|\xx-\yy| }},\qquad  y=\mp \frac { (4|c|)^{1/4} (\xx+\yy)}{2 \sqrt{|\xx-\yy| }},\qquad  c<0, \nonumber\\[2pt]
&& \xx=\pm \frac{ |c|^{1/2}}{x^2}-\frac y x ,\qquad \yy=\mp \frac{ |c|^{1/2}}{x^2}-\frac y x  ,
\label{eg}
\eea
which map  the MP vector fields (\ref{FirstLieA}) with domain   $\mathbb{R}^2_{x\neq 0}$, symplectic form $\omega={\rm d}x\wedge {\rm d}y$, Hamiltonian functions (\ref{ham2}) and constant of motion (\ref{am}) 
  onto (\ref{ec}) with domain $\mathbb{R}^2_{\xx\neq \yy}$, (\ref{ee}) and (\ref{ef}) (up to a multiplicative constant $\pm |c|^{1/2}$), respectively.

As in the previous section, the   (non-standard) deformation of the coupled Riccati    system (\ref{ea})  is obtained by   starting again from proposition~\ref{proposition1}  and now    applying the change of variables (\ref{eg}) with $c=-1$ (without loss of generality) finding the following result. 

 \bigskip

\begin{proposition}
\label{proposition3}
(i)  The Hamiltonian functions given by
\bea
h_{z,1}=  \frac 1 {\xx-\yy} \, ,  \qquad
 h_{z,2}= \frac 12 \shc \bigl(\tfrac{2z} {\xx-\yy}\bigr) \biggl(   \frac{ \xx+\yy}{ \xx-\yy} \biggr) ,\\
 \qquad h_{z,3}=   \frac{   \shc^2 \bigl(\frac{2z} {\xx-\yy}\bigr)( \xx+\yy)^2 -( \xx-\yy)^2 }{ 4   \shc \bigl(\frac{2z} {\xx-\yy}\bigr)  (\xx-\yy) }  \, ,
\nonumber
\eea
satisfy the  commutation relations (\ref{gb2})   with respect to the symplectic form  $\omega$ (\ref{ee})   and define  the deformed Poisson algebra  ${\cal C}^\infty({\cal H}_{z,\omega}^{\rm {2R}*})$.\\
(ii) Their corresponding    deformed   vector fields   turn out to be
\begin{eqnarray}
&& {\bf X}_{z,1}   =  \frac{\partial}{\partial \xx}+  \frac{\partial}{\partial \yy},\nonumber\\[2pt]
&& {\bf X}_{z,2}=\frac 12  (\xx+\yy)  \cosh \bigl(\tfrac{2z} {\xx-\yy}\bigr)     \left(  \frac{\partial}{\partial \xx}+  \frac{\partial}{\partial \yy} \right)
+\frac 12  (\xx-\yy)  \shc \bigl(\tfrac{2z} {\xx-\yy}\bigr)     \left(  \frac{\partial}{\partial \xx}-  \frac{\partial}{\partial \yy} \right), \nonumber\\
&& {\bf X}_{z,3}   =  \frac 14 \left[  (\xx+\yy)^2+\frac{ (\xx-\yy)^2 }{ \shc^2 \bigl(\tfrac{2z} {\xx-\yy}\bigr)}  \right] \cosh \bigl(\tfrac{2z} {\xx-\yy}\bigr)     \left(  \frac{\partial}{\partial \xx}+  \frac{\partial}{\partial \yy} \right)+\frac 12  (\xx^2-\yy^2)  \shc \bigl(\tfrac{2z} {\xx-\yy}\bigr)     \left(  \frac{\partial}{\partial \xx}-  \frac{\partial}{\partial \yy} \right) ,\nonumber
\end{eqnarray}
which fulfill 
\bea
&& [{\bf X}_{z,1},{\bf X}_{z,2}]=\cosh \bigl(\tfrac{2z} {\xx-\yy}\bigr)  {\bf X}_{z,1},\qquad [{\bf X}_{z,1},{\bf X}_{z,3}]=2 {\bf X}_{z,2},\nonumber\\[2pt] 
&& [{\bf X}_{z,2},{\bf X}_{z,3}]=\cosh \bigl(\tfrac{2z} {\xx-\yy}\bigr)   {\bf X}_{z,3}- z^2\left[ 1  - \biggl(\frac{\xx+\yy}{\xx-\yy} \biggr)^2\!  \shc^2 \bigl(\tfrac{2z} {\xx-\yy}\bigr)  \right]  {\bf X}_{z,1}. 
\nonumber
\eea
\end{proposition}

\bigskip

 The deformed counterpart of the coupled Ricatti Lie system (\ref{eb}) and of the   LH one (\ref{ee2}) is defined by
 \be
{\bf X}_z:=a_0(t){\bf X}_{z,1}+a_1(t){\bf X}_{z,2}+a_2(t){\bf X}_{z,3},\qquad h_z:=a_0(t)h_{z,1}+a_1(t)h_{z,2}+a_2(t)h_{z,3} .
\label{el}
\ee
And the $t$-independent constants of motion are $F_z=-1/4$ and
\bea
&&  F_z^{(2)}=\frac{   \eee^{-\frac{2z}{\xx_1-\yy_1}} \eee^{\frac{2z}{\xx_2-\yy_2}}   }{4(\xx_1-\yy_1)(\xx_2-\yy_2)} \left[   \shc \bigl(\tfrac{2z} {\xx_1-\yy_1}\bigr)      \shc \bigl(\tfrac{2z} {\xx_2-\yy_2}\bigr)  (\xx_1-\xx_2+\yy_1-\yy_2)^2  \right. \nonumber\\[2pt]
&&  \qquad  \left. -  \left(  \frac{  \eee^{\frac{2z}{\xx_1-\yy_1}}  (\xx_1-\yy_1) }{  \shc \bigl(\tfrac{2z} {\xx_1-\yy_1}\bigr) }   + \frac{  \eee^{-\frac{2z}{\xx_2-\yy_2}}  (\xx_2-\yy_2) }{  \shc \bigl(\tfrac{2z} {\xx_2-\yy_2}\bigr) }   \right)  \shc \bigl(\tfrac{2z} {\xx_1-\yy_1}+\tfrac{2z} {\xx_2-\yy_2}\bigr) (\xx_1+\xx_2-\yy_1-\yy_2)      \right] .
 \label{em}\nonumber
\eea
Therefore, the  deformation of the system (\ref{ea}) is determined  by ${\bf X}_z$ (\ref{el}).  Note that the resulting system  presents a strong interaction amongst the variables $(u,v)$ through $z$,
which goes far beyond the initial (naive) coupling corresponding to set the same $t$-dependent parameters $a_i(t)$ in both one-dimensional Riccati equations; namely 
 \bea
&& \frac{{\rm d} \xx}{{\rm d} t}=a_0(t)+\frac{a_1(t)}{2}\left[  (\xx+\yy)  \cosh \bigl(\tfrac{2z} {\xx-\yy}\bigr)    + (\xx-\yy)  \shc \bigl(\tfrac{2z} {\xx-\yy}\bigr)     \right]
  \nonumber\\[2pt]
&&\qquad\quad +\frac{a_2(t)}4 \left[    \left(  (\xx+\yy)^2+\frac{ (\xx-\yy)^2 }{ \shc^2 \bigl(\tfrac{2z} {\xx-\yy}\bigr)}  \right) \cosh \bigl(\tfrac{2z} {\xx-\yy}\bigr)       + 2 (\xx^2-\yy^2)  \shc \bigl(\tfrac{2z} {\xx-\yy}\bigr)    \right]   , \nonumber\\[2pt]
&& \frac{{\rm d} \yy}{{\rm d} t}=a_0(t)+\frac{a_1(t)}{2}\left[  (\xx+\yy)  \cosh \bigl(\tfrac{2z} {\xx-\yy}\bigr)   - (\xx-\yy)  \shc \bigl(\tfrac{2z} {\xx-\yy}\bigr)     \right]
  \nonumber\\[2pt]
&&\qquad\quad +\frac{a_2(t)}4 \left[    \left(  (\xx+\yy)^2+\frac{ (\xx-\yy)^2 }{ \shc^2 \bigl(\tfrac{2z} {\xx-\yy}\bigr)}  \right) \cosh \bigl(\tfrac{2z} {\xx-\yy}\bigr)       - 2 (\xx^2-\yy^2)  \shc \bigl(\tfrac{2z} {\xx-\yy}\bigr)    \right]   .
\label{en}\nonumber
\eea

}

\chapter{Oscillator Systems from $\mathfrak{h}_4$} 

\label{Chapter6} 
\renewcommand{\theequation}{6.\arabic{equation}}

\section{Deformed oscillators from the oscillator algebra $\mathfrak{h}_4$}

\bigskip

\noindent In this paragraph we analyze a quite different type of deformation based on the Poincar\'e algebra $\mathfrak{h}_4$, which is
the second of the relevant subalgebras of the two-photon Lie algebra $\mathfrak{h}_6$, corresponding to the highest dimensional Lie algebra of vector fields on the real plane that appears in the context of planar Lie--Hamilton systems admits two non-equivalent quantum deformations leading to correspondingly non-equivalent Lie--Poisson systems. The Lie algebra  $\mathfrak{h}_4$ is amidst the notable non-semisimple Lie algebras used in physics, where it provides  
a unified description of coherent, squeezed and intelligent states of light (\cite{Zha} and references therein), also having applications in the theory of integrable systems \cite{BH01}  and the description of damped harmonic oscillators \cite{BHLS}, among others. 

\bigskip

\noindent The two-photon Lie algebra $\mathfrak{h}_6$, as considered in \cite{Zha}, is spanned by the six operators 
\begin{equation}
N=a_{+}a_{-},\quad A_{+}=a_{+},\quad A_{i}=a_{i},\quad B_{+}=a_{+}^{2},\quad B_{i}=a_{i}^{2},\quad M={\rm I},\label{bas1}
\end{equation}
where $a_{+}$ and $a_{-}$ are the generators of the boson algebra $\left[a_{-},a_{+}\right]={\rm I}$. Over this basis, the commutation relations are given by (see \cite{BHP97}): 
\begin{eqnarray}
&& \left[A_{\pm},B_{\pm}\right]=0,\quad \left[A_{-},A_{+}\right]=M,\quad  \left[M,\cdot\right]=0,\quad
\left[B_{-},B_{+}\right]=  4N+2M, \nonumber\\[2pt]  
&& \left[A_{\pm},B_{\mp}\right]=\mp 2A_{\mp},\quad \left[N,A_{\pm}\right]=  \pm A_{\pm},\quad \left[N,B_{\pm}\right]=\pm2B_{\pm}.\label{bas2}
\end{eqnarray}
It follows at once from these relations that the operators $\left\{A_{+},A_{-},M\right\}$ span a subalgebra isomorphic to the Heisenberg-Weyl algebra $\mathfrak{h}_3$, which can be extended to the oscillator algebra $\mathfrak{h}_4$ spanned by these operators along with the number operator $N$.  

\medskip

\noindent The two-photon Lie algebra admits two independent Casimir operators, one corresponding to the centre generator $M$ and a second one having degree four in the generators and given by 
\begin{equation}
C_{4}=\left(MB_{+}-A_{+}^2\right)\left(MB_{-}-A_{-}^2\right)-\left(MN-A_{-}A_{+}+\frac{1}{2}M^2\right)^2.               \label{casi}
\end{equation} 

\medskip

\noindent For computational convenience, let us consider the change of basis $D=-(N+\frac{1}{2}M)$. Now, if we understand the generators $\left\{M, A_{-},-A_{+},D, \frac{B_{-}}{2}, -\frac{B_{+}}{2}\right\}$ of the two-photon algebra as functions on $\mathfrak{h}_{6}^{\ast}$ in the natural way \cite{BH01}, we can consider the coordinates $\left\{v_0,\cdots , v_5\right\}$ on $\mathfrak{h}_6$. Taking into account the induced Kirillov--Konstant--Souriau Poisson structure on ${\cal C}^{\infty}\left(\mathfrak{h}_{6}^{\ast}\right)$ with respect to the canonical symplectic 2-form in the plane $\omega={\rm d}x\wedge {\rm d}y$, we are led to the brackets
\begin{eqnarray}
&&  \left\{v_{1},v_2\right\}=v_0, \quad \left\{v_{1},v_3\right\}=-v_1,\quad \left\{v_{1},v_4\right\}=0,\quad \left\{v_{1},v_5\right\}=-v_2,\quad \left\{v_{2},v_3\right\}=v_2,\nonumber \\[2pt]
&&  \left\{v_{2},v_4\right\}=-v_1,\quad \left\{v_{2},v_5\right\}=0,\quad \left\{v_{3},v_4\right\}=2v_4,\quad\left\{v_{3},v_5\right\}=-2v_5,\quad \left\{v_{4},v_5\right\}=v_3.\label{hf1}
\end{eqnarray}
The Casimir functions are found to $v_0$ and  
\begin{equation}
C_3=2\left(v_1^2v_5-v_2^2v_4-v_1v_2v_3\right)-v_0\left(v_3^2+4v_4v_5\right).\label{inv}
\end{equation}

\bigskip

We observe that the cubic invariant is a consequence of the Lie algebra $\mathfrak{h}_6$ having the structure of a semi-direct product of $\mathfrak{su}(1,1)$ and a Heisenberg algebra \cite{C45}. Defining 
\begin{equation}
V_3=v_0v_3+v_1v_2,\quad V_2=v_0v_4-\frac{1}{2}v_1^2,\quad V_3=v_0v_5+\frac{1}{2}v_2^2
\end{equation}
it follows at once that $v_0C_3=V_3^2+4 V_1V_2$, showing its relation to the fourth-order Casimir operator (\ref{casi}). 

\medskip \noindent Using now the identity $\iota_{{\bf X}_i}\omega={\rm d}v_i$, the corresponding Hamiltonian vector fields are given by 
\begin{equation}
{\bf X}_1=\frac{\partial}{\partial x},\quad {\bf X}_2=\frac{\partial}{\partial y},\quad {\bf X}_3=x\frac{\partial}{\partial x}-y\frac{\partial}{\partial y},\quad {\bf X}_4=y\frac{\partial}{\partial x},\quad {\bf X}_5=x\frac{\partial}{\partial y},\label{HVF}
\end{equation}
with the nontrivial commutation relations 
\begin{eqnarray}\label{2pb}
&& \left[{\bf X}_1,{\bf X}_3\right]={\bf X}_1,\quad   \left[{\bf X}_1,{\bf X}_5\right]={\bf X}_2,\quad  \left[{\bf X}_2,{\bf X}_3\right]=-{\bf X}_2,\quad 
 \left[{\bf X}_2,{\bf X}_4\right]={\bf X}_1,\nonumber\\[2pt]
&& \left[{\bf X}_3,{\bf X}_4\right]=-2{\bf X}_4, \quad
 \left[{\bf X}_3,{\bf X}_5\right]=2{\bf X}_5,\quad  \left[{\bf X}_4,{\bf X}_5\right]=-{\bf X}_3.
\end{eqnarray}

\bigskip
\bigskip

\section{Poisson--Hopf deformation of   $\mathfrak{h}_4$ Lie--Hamilton systems }

\bigskip

\noindent  As observed earlier, the subalgebra of $\mathfrak{h}_6$ generated by $\left\{N,M,A_{+},A_{-}\right\}$ is isomorphic to the four-dimensional Poincar\'e algebra $\mathfrak{h}_4$. This Lie algebra possesses two Casimir operators, given respectively by $M$ and $C_2=2MN-A_{+}A_{-}-A_{-}A_{+}$. The quantum deformation of $\mathfrak{h}_4$ (see \cite{BHN97}) is determined by the coassociative coproduct 
\begin{eqnarray}\label{dex1}
&& \Delta(A_+)= 1\otimes A_{+}+A_{+}\otimes 1,\quad \Delta(M)= 1\otimes M+ M\otimes 1,\nonumber\\[2pt]   &&
 \Delta(N)=1\otimes N+ N\otimes {\rm e}^{zA_{+}},\quad 
\Delta(A_{-})=1\otimes A_{-}+ A_{-}\otimes {\rm e}^{zA_{+}}+ z N \otimes {\rm e}^{z A_{+}} M,\label{baX}\end{eqnarray}
from which the deformed commutation relations result as: 
\begin{eqnarray}\label{dex2}
\left[ N, A_{+}\right]= \frac{{\rm e}^{z A_{+}}-1}{z},\quad \left[ N, A_{-}\right]= -A_{-},\quad \left[ A_{-}, A_{+}\right]= M {\rm e}^{z A_{+}}.\label{bax2}
\end{eqnarray}
Al already used, we consider the change of basis $D=-N-\frac{1}{2}M$.  By (\ref{bax2}), the deformed Hamiltonian functions for $\mathfrak{h}_4$ are given by  
\begin{eqnarray}
v_{0,z}=v_{0},\quad v_{1,z}={\rm e}^{z v_2}v_1,\, v_{2,z}=v_2,\quad v_{3,z}=\frac{1-{\rm e}^{z v_2}}{z}v_1,\label{bax3}
\end{eqnarray}
corresponding to the classical Hamiltonian functions $v_i$ in the limit $z\rightarrow 0$:
\begin{equation}
\lim_{z\rightarrow 0} v_{j,z}= v_{j},\quad 0\leq j\leq 3
\end{equation}
and Poisson brackets  
\begin{eqnarray}
 \left\{v_{1,z},v_{2,z}\right\}_z=v_{0,z} {\rm e}^{zv_{2,z}},\quad  
\left\{v_{1,z},v_{3,z}\right\}_z=-v_{1,z},\quad  \left\{v_2,v_3\right\}_z=\frac{e^{zv_{2,z}}-1}{z}.\label{bax4}
\end{eqnarray}
With respect to the canonical 2-form $\omega={\rm d}x\wedge {\rm d}y$, the Poisson structure of ${\cal U}(\mathfrak{h}_4)$ is given by the Hamiltonian functions $v_{j}$ ($0\leq j\leq 3$) with Poisson brackets  (\ref{bax4}) and Casimir invariants $v_0$ and $C_2=v_0v_3-v_1v_2$. 

  \bigskip

 \noindent The $t$-dependent vector field ${\bf X}_t=\sum_{j=1}^{3} f_{i}(t){\bf X}_{j,z}$ leads to the Poisson--Hopf system 
  \begin{eqnarray}
&&\frac{dx}{dt}={\rm e}^{-xz} f_{1}(t)+\frac{1-{\rm e}^{-xz}}{z}f_{3}(t),\nonumber\\[2pt]
&& \frac{dy}{dt}=yz{\rm e}^{-xz}f_{1}(t)+f_{2}(t)-y {\rm e}^{-xz}f_{3}(t),\label{dep}
\end{eqnarray}
for arbitrary functions $f_j(t)$ ($1\leq j\leq 3$). For the choice $f_3(t)=0$, the preceding system corresponds to the quantum deformation of the Lie--Hamilton system based on the Heisenberg Lie algebra $\mathfrak{h}_3$. We further observe that, under this latter assumption, the classical and deformed systems essentially have the same form, as the system consists in this case of a separable equation and a linear first-order inhomogeneous equation.  In this sense, the Poisson--Hopf system associated to $\mathfrak{h}_3$ does not lead to new types of systems.  

\bigskip

\noindent
For the generic deformation based on the Poincar\'e algebra, the deformed noncentral invariant is given by 
\[
C_{2,z}=v_{0,z}v_{3,z}-v_{1,z}\frac{{\rm e}^{-z v_{2,z}}-1}{z}.
\]
The constants of the motion of the deformed system (\ref{dep}), computed by means of the coalgebra, are easily verified to take the form 
\begin{eqnarray}
&& F_z(C_{2,z})=\frac{{\rm e}^{-zx}}{z}\left(zx+{\rm e}^{-zx}-1\right)y,\nonumber\\[2pt]
&& F_z^{(2)}(C_{2,z})=\frac{{2(y_1+y_2)- \rm e}^{-z(x_1+x_2)}\left(2+z(x_1+x_2)\right)\left(y_1{\rm e}^{zx_2}+y_2{\rm e}^{zx_1}\right)    }{z}
\end{eqnarray}
with the expected classical limits (see Table 2 in \cite{BHLS}) 
\begin{equation}
\lim_{z\rightarrow 0} F_z(C_{2,z})=0,\quad  \lim_{z\rightarrow 0} F_z^{(2)}(C_{2,z})=(x_1-x_2)(y_1-y_2).
\end{equation}
 
 \bigskip
\bigskip


\section{Poisson--Hopf deformation of $\mathfrak{h}_6$ Lie--Hamilton systems from $\mathfrak{h}_4$}\label{h4 damped oscilator}

\bigskip

The non-standard (also called Jordanian) quantum deformation of the two photon algebra $\mathfrak{h}_6$ is a natural extension of the preceding quantum deformation 
of the Poincar\'e algebra $\mathfrak{h}_4$ seen previously. It has been extensively studied, for which reason we omit the details (these can be found e.g. in \cite{BHP97,BH01}) and merely indicate the coproduct and commutation relations. The coassociative coproduct for the quantum two-photon algebra ${\cal U}_{z}(\mathfrak{h}_6)$ is given by:
\begin{eqnarray}
&& \Delta(A_+)= 1\otimes A_{+}+A_{+}\otimes 1,\quad \Delta(M)= 1\otimes M+ M\otimes 1,\nonumber\\[2pt]   
&& \Delta(N)=1\otimes N+ N\otimes {\rm e}^{zA_{+}},\quad \Delta(B_+)=1\otimes B_{+}+ B_{+}\otimes {\rm e}^{-2zA_{+}},\nonumber\\[2pt] 
&& \Delta(A_{-})=1\otimes A_{-}+ A_{-}\otimes {\rm e}^{zA_{+}}+ z N \otimes {\rm e}^{z A_{+}} M,\label{bas3} \\[2pt] 
&& \Delta(B_{-})=1\otimes B_{-}+ B_{-}\otimes {\rm e}^{2zA_{+}}+ z N \otimes {\rm e}^{z A_{+}} \left(A_{-}- z M N\right) -z A_{-}\otimes {\rm e}^{z A_{+}}N.\nonumber
\end{eqnarray}
The corresponding compatible deformed commutation relations are thus given by: 
\begin{eqnarray}
&& \left[ N, A_{+}\right]= \frac{{\rm e}^{z A_{+}}-1}{z},\quad \left[ N, A_{-}\right]= -A_{-},\quad \left[ A_{-}, A_{+}\right]= M {\rm e}^{z A_{+}},\nonumber\\[2pt] 
&& \left[ N, B_{+}\right]=2 B_{+}, \quad  \left[ N, B_{-}\right]=-2B_{-}-z A_{-}N,\quad \left[ A_{+},B_{+}\right]=0,\nonumber\\[2pt]   
&& \left[ A_{-}, B_{+}\right]=2\frac{1-{\rm e}^{-zA_{+}}}{z}, \quad \left[ A_{-}, B_{-}\right]=-z A_{-}^2, \quad \left[ \cdot , M\right]=0,\label{bas4}\\[2pt]  
&& \left[ A_{+}, B_{-}\right]=-\left(1+{\rm e}^{zA_{+}}\right)A_{-} +z {\rm e}^{zA_{+}}MN,\nonumber\\[2pt]  && 
\left[ B_{-}, B_{+}\right]= 2\left(1+{\rm e}^{-zA_{+}}\right)N+2M- 2z A_{-}B_{+}.\nonumber
\end{eqnarray}
It follows from (\ref{bas4}) that the deformed Hamiltonian functions adopt the form 
have the form 
\begin{eqnarray}
&&  v_{0,z}=v_{0},\quad v_{1,z}={\rm e}^{z v_2}v_1,\quad v_{2,z}=v_2,\quad v_{3,z}=\frac{1-{\rm e}^{z v_2}}{z}v_1,\nonumber\\[2pt] 
&&  v_{4,z}=\frac{1}{2}{\rm e}^{z v_2}v_1^2,\quad v_{5,z}=-\frac{1}{2}\left(\frac{{\rm e}^{-z v_2}-1}{z}\right)^2,\label{bas5}
\end{eqnarray}
which faithfully reproduce the Hamiltonian functions $v_i$ in the limit $z\rightarrow 0$:
\begin{equation}
\lim_{z\rightarrow 0} v_{j,z}= v_{j},\quad 0\leq j\leq 5.
\end{equation}
The explicit Poisson brackets for the deformation are thus 
\begin{eqnarray}
&&  \left\{v_{1,z},v_{2,z}\right\}_z=v_{0,z} {\rm e}^{zv_{2,z}},\quad \left\{v_{1,z},v_{3,z}\right\}_z=-v_{1,z},\quad
\left\{ v_{1,z},v_{4,z}\right\}_{z} =  -\frac{zv_{1,z}^{2}}{2},\quad \nonumber\\[2pt] &&  \left\{ v_{1,z},v_{5,z}\right\}_{z} =  \frac{e^{-zv_{2,z}}-1}{z},\quad
 \left\{v_2,v_3\right\}_z=\frac{e^{zv_{2,z}}-1}{z},\quad \nonumber\\[2pt] && \left\{ v_{2,z},v_{4,z}\right\}_{z} =  -\frac{v_{1,z}}{2}(1+e^{zv_{2,z}})-\frac{ze^{zv_{2,z}}v_{0,z}}{4}\left(2v_{3,z}+v_{0,z}-1\right),\nonumber\\[2pt]
&&  \left\{ v_{2,z},v_{5,z}\right\}_{z}=0,\quad  \left\{ v_{3,z},v_{4,z}\right\}_{z} =  2v_{4,z}-\frac{zv_{1,z}}{4}\left(2v_{3,z}+v_{0,z}-1\right),\quad  \left\{ v_{3,z},v_{5,z}\right\}_{z} =  -2v_{5,z},\nonumber\\[2pt]
&&   \left\{ v_{4,z},v_{5,z}\right\}_{z} =v_3\frac{1+e^{-zv_{2,z}}}{2}+\left(\frac{v_{0,z}}4-\frac 14\right)(e^{-zv_{2,z}}-1)-zv_1v_5,\quad \left\{v_{0,z},\cdot\right\}=0. \label{defo1}
\end{eqnarray}

\bigskip

\noindent As follows from inspection of the coassociative coproduct (\ref{bas3}) of the quantum two-photon algebra, the deformed generators of $\mathfrak{h}_4$ give rise to a subalgebra of the quantum deformation of $\mathfrak{h}_6$, hence the deformed Hamiltonian functions can be identified with the $v_{j,z}$ in equation (\ref{bas5}) for $0\leq j\leq 3$, while the corresponding Hamiltonian vector fields can be identified with the fields ${\bf X}_{j,z}$ given in (\ref{dhvf}) for these indices. We observe that, in this case, the Hamiltonian functions $\left\{v_{3,z},v_{4,z},v_{5,z}\right\}$ do not close as a subalgebra, as they involve the remaining generators, indicating that we are dealing with a different quantum deformation of $\mathfrak{h}_6$ as that considered earlier extending the deformation on the subalgebra $\mathfrak{sl}(2,\mathbb{R})\oplus\mathbb{R}$.  A cumbersome but routine computation shows that the Casimir invariants  for (\ref{defo1}) are $v_{0,z}$ for being a central element and the function 
\begin{eqnarray}
 C_{3,z}=z v_{0,z}v_{1,z}v_{5,z}\left(2v_{3,z}+v_{0,z}-1\right)-\frac{2v_{4,z}\left({\rm e}^{z v_{2,z}}-1\right)^2}{z^2}{\rm e}^{-2z v_{2,z}}+2v_{1,z}^2v_{5,z}-v_{0,z}v_{3,z}^2\nonumber\\[2pt]
 -4v_{0,z}v_{4,z}v_{5,z}-\frac{\left({\rm e}^{z v_{2,z}}-1\right)\left(2v_{3,z}\left({\rm e}^{z v_{2,z}}+1\right)+(v_{0,z}-1)\left(1-{\rm e}^{z v_{2,z}}\right) \right)}{z {\rm e}^{2z v_{2,z}}}.\label{invd}
\end{eqnarray}
It is straightforward to verify that in the limit we obtain the cubic invariant (\ref{inv})
\begin{equation}
\lim_{z\rightarrow 0} C_{3,z}=C_3.
\end{equation}

\bigskip

\noindent The Hamiltonian vector fields ${\bf X}_{j,z}$ associated to the functions $v_{j,z}$ are given by 
\begin{eqnarray}
&& {\bf X}_{0,z}={\bf X}_0,\quad {\bf X}_{1,z}={\rm e}^{z\, v_{2,z}}{\bf X}_{1}+ z v_{1,z} {\bf X}_{2},\quad {\bf X}_{2,z}={\bf X}_2,\quad {\bf X}_{3,z}=\frac{1-{\rm e}^{z\, v_{2,z}}}{z}{\bf X}_1-v_{1,z}{\bf X}_{2}\nonumber\\[2pt]
&& {\bf X}_{4,z}=v_{1,z}{\bf X}_1+z v_{4,z}{\bf X}_2,\quad {\bf X}_{5,z}={\rm e}^{-z\, v_{2,z}}\frac{{\rm e}^{-z\, v_{2,z}}-1}{z}{\bf X}_2.\label{dhvf}
\end{eqnarray}
with non-vanishing commutators  
\begin{eqnarray}
&&  \left[{\bf X}_{1,z},{\bf X}_{2,z}\right]=-{\rm e}^{z\, v_{2,z}}{\bf X}_{2,z},\quad \left[{\bf X}_{1,z},{\bf X}_{3,z}\right]={\bf X}_{1,z},\quad \left[{\bf X}_{1,z},{\bf X}_{4,z}\right]=z v_{1,z}{\bf X}_{1,z},\nonumber\\[4pt]
&&  \left[{\bf X}_{1,z},{\bf X}_{5,z}\right]={\rm e}^{-z\, v_{2,z}}{\bf X}_{2,z},\quad \left[{\bf X}_{2,z},{\bf X}_{3,z}\right]=-{\rm e}^{z\, v_{2,z}}{\bf X}_{2,z},\quad \left[{\bf X}_{2,z},{\bf X}_{4,z}\right]={\bf X}_{1,z},\nonumber\\[4pt]
&&  \left[{\bf X}_{3,z},{\bf X}_{4,z}\right]=-\left(1+{\rm e}^{z\, v_{2,z}}\right){\bf X}_{4,z}-z v_{4,z}{\rm e}^{z\, v_{2,z}}{\bf X}_{2,z},\quad \left[{\bf X}_{3,z},{\bf X}_{5,z}\right]=2 {\bf X}_{5,z},\nonumber\\[4pt]
&&  \left[{\bf X}_{4,z},{\bf X}_{5,z}\right]={\rm e}^{-z\, v_{2,z}} v_{1,z}\left({\rm e}^{-z\, v_{2,z}}-1\right){\bf X}_{2,z}-{\rm e}^{-z\, v_{2,z}}{\bf X}_{3,z}.
\end{eqnarray}
Hence, for both the vector fields and the commutators the limit is found to recover the classical Lie-Hamilton system: 
\begin{equation}
\lim_{z\rightarrow 0} {\bf X}_{j,z}={\bf X}_j,\quad \lim_{z\rightarrow 0}\left[{\bf X}_{j,z},{\bf X}_{k,z}\right]= \left[{\bf X}_{j},{\bf X}_{k}\right],\quad  0\leq j\leq 5.
\end{equation}

\noindent For generic functions $f_{i}(t)$, the deformed Poisson--Hopf system determined by the $t$-dependent vector field ${\bf X}_t=\sum_{j=1}^{5} f_{i}(t){\bf X}_{i,z}$ is explicitly given by 
\begin{eqnarray}
&& \frac{dx}{dt}={\rm e}^{-xz} f_{1}(t)+\frac{1-{\rm e}^{-xz}}{z}f_{3}(t)+ y {\rm e}^{-xz}f_{4}(t),\nonumber\\[2pt]
&&  \frac{dy}{dt}=yz{\rm e}^{-xz}f_{1}(t)+f_{2}(t)-y {\rm e}^{-xz}f_{3}(t)+\frac{y^2z}{2}{\rm e}^{-xz}f_{4}(t)+\frac{{\rm e}^{xz}\left({\rm e}^{xz}-1\right)}{z}f_{5}(t).\label{QDF}
\end{eqnarray}

\bigskip

\noindent The $t$-independent constants of the motion of this system are obtained by application of the coalgebra formalism (see \cite{coalgebra1}), from which the following expressions are obtained:
\begin{eqnarray}
&& F(C_{3,z})=0,\qquad F^{(2)}(C_{3,z})=0,\nonumber \\[2pt]
&&  F^{(3)}(C_{3,z})=\frac{{\rm e}^{-z(x_1+x_2)}}{2z}\left(4 {\rm e}^{zx_1}-2 {\rm e}^{2zx_1}+4 {\rm e}^{zx_2}-2 {\rm e}^{-2zx_2}+ {\rm e}^{2z(x_1+x)}-2 {\rm e}^{z(x_1+x_2)}-3\right)\times \nonumber \\[2pt]
&&\qquad\qquad \qquad \left(y_1  {\rm e}^{zx_2}+y_2 {\rm e}^{zx_1}\right)-\frac{y_1^2}{z^2}{\rm e}^{-2zx_1}\left({\rm e}^{2zx_1}-1\right)\left({\rm e}^{zx_2}-1\right)-\frac{y_2^2}{z^2}\left({\rm e}^{2zx_2}-1\right) \times \nonumber \\[2pt]
&&\qquad\qquad \qquad \left({\rm e}^{zx_1}-1\right){\rm e}^{-2zx_2}+\frac{y_1y_2}{z^2}{\rm e}^{-z(x_1+x_2)}  \left( {\rm e}^{z(2x_1+3x_2)} +{\rm e}^{z(3x_1+2x_2)} -2{\rm e}^{z(x_1+2x_2)} \right.\nonumber\\[2pt]
&&\qquad\qquad \qquad \left.-2{\rm e}^{z(2x_1+x_2)} -2{\rm e}^{2z(x_1+x_2)}+4{\rm e}^{z(x_1+x_2)} -2 {\rm e}^{3zx_2}+6{\rm e}^{2zx_2}-5{\rm e}^{zx_2}-2{\rm e}^{3zx_1}\right.\nonumber\\[2pt]
&&\qquad\qquad \qquad\left.+6{\rm e}^{2zx_1} -5{\rm e}^{zx_1} +2\right).\label{indd}
\end{eqnarray} 
It can be routinely verified that $\lim_{z\rightarrow 0}F^{(2)}(C_{3,z})=0$, which is in agreement with the well known fact that $F^{(2)}(C_{3})=0$ holds for the classical counterpart.  Applying again the same procedure, another deformed invariant $F^{(3)}(C_{3,z})$ can be constructed so that in the limit we recover the first nonvanishing $t$-independent constant of the motion 
$$F^{(3)}(C_3)=\left(x_1(y_2-y_3)+x_2(y_3-y_1)+x_3(y_1-y_2)\right)^2$$ of the classical Lie--Hamilton system \cite{BHLS}.

\bigskip
\bigskip

\subsection{Deformation of the damped oscillator}

\bigskip

\noindent As a physically interesting application of the preceding deformation, let us consider a one-dimensional damped oscillator of the form
\begin{eqnarray}
&& \frac{dx}{dt}= a(t) x+ b(t) y +f(t),\nonumber\\[2pt]
&& \frac{dp}{dt}=- c(t) x- a(t) p -g(t),\label{gho}
\end{eqnarray}
where $a(t),b(t),c(t),f(t)$ and $g(t)$ are arbitrary functions. The system (\ref{gho}) is actually a Lie--Hamilton system associated to the $t$-dependent vector field $${\bf X}_t=f(t){\bf X}_1-g(t){\bf X}_2+a(t){\bf X}_3+b(t){\bf X}_4-c(t){\bf X}_5,$$ where 
\begin{equation}
{\bf X}_1=\frac{\partial}{\partial x},\quad {\bf X}_2=\frac{\partial}{\partial p},\quad {\bf X}_3=x\frac{\partial}{\partial x}-p\frac{\partial}{\partial p},\quad {\bf X}_4=p\frac{\partial}{\partial x},\quad {\bf X}_5=x\frac{\partial}{\partial p}\label{VF2} 
\end{equation}
are Hamiltonian vector fields
with respect to the symplectic form $\omega={\rm d}x\wedge {\rm d}p$. The Hamiltonian functions $h_{i}$ associated to the vector fields (\ref{VF2}) are respectively
\begin{equation}
h_{1}=p,\quad h_2=-x,\quad h_3=xp,\quad h_4=\frac{1}{2}p^2,\quad h_5=-\frac{1}{2}x^2,
\end{equation}
showing that the resulting Vessiot--Guldberg Lie algebra  of the Lie--Hamilton system is isomorphic to the two-photon Lie algebra $\mathfrak{h}_6$ as given in Equation (\ref{HVF}).\footnote{
The invariants and nonlinear superposition rule for this type of Lie--Hamilton system have been analyzed in detail in \cite{BHLS}, for which reason we skip their detailed expressions.} 
The quantum deformation of (\ref{gho}) therefore corresponds to a system of type (\ref{QDF}), for the choice of functions $f_{1}(t)=f(t),f_{2}(t)=-g(t),f_{3}(t)=a(t),f_{4}(t)=b(t)$ and $f_5(t)=c(t)$. As follows from (\ref{indd}), the deformed system possesses a non-vanishing $z$-dependent constant of the motion $F^{(2)}(C_{3,z})$. 

\bigskip

\noindent We observe that the operators $\left\{B_{+},B_{-},N+\frac{1}{2}M\right\}$ 
span a simple non-compact Lie algebra isomorphic to $\mathfrak{su}(1,1)\simeq \mathfrak{sl}(2,\mathbb{R})$. As both of these subalgebras admit quantum deformations, it is natural to ask whether these can be extended to the whole two-photon algebra in a consistent way, such that all Poisson--Hopf deformations of Lie--Hamilton systems in the plane can be described uniformly in terms of subalgebras of the two photon algebra. In this sense, an extension of the $\mathfrak{sl}(2,\mathbb{R})$-related deformed systems remain to be described. Various preliminary results in this direction have been obtained, with a complete description currently in progress. 
 

\chapter{Applications to Biology} 

\label{Chapter8} 
\renewcommand{\theequation}{7.\arabic{equation}}

\section{An epidemic SIS model}

\bigskip

If one wants to study the evolution of a SIS epidemic exposed to a constant heat source, like centrally heated buildings; one can make use of quantum stochastic differential equations.
It was shown that the stochastic SIS-epidemic model can be interpreted as a Hamiltonian system. Lie--Hamiltonian systems admit a quantum deformation, so does the stochastic SIS-epidemic model, because it is a Lie--Hamilton system.

This chapter \cite{Covid} proposes a quantum version of a stochastic SIS-epidemic model without using stochastic calculus, but using the proper Hamiltonian approximation for the mean and the variance.

Epidemic models try to predict the spread of an infectious disease afflicting a specific population, see \cite{Brauer,Mi83}. These models are rooted in the works of Bernoulli in the 18th century, when he proposed a mathematical model to defend the practice of inoculating against smallpox \cite{smallpox}. This was the start of germ theory.

At the beginning of the 20th century, the emergence of compartmental models was starting to develop. Compartmental models are deterministic models in which the population is divided into compartments, each representing a specific stage of the epidemic. For example, $S$ represents the susceptible individuals to the disease, $I$ designates the infected individuals, whilst $R$ stands for the recovered ones. The evolution of these variables in time is represented by a system of ordinary differential equations whose independent variable, the time, is denoted by $t$. 
Some of these first models are the Kermack--McKendrick \cite{KK} and the Reed--Frost \cite{Ab52} epidemic models, both describing the dynamics of healthy and infected individuals among other possibilities. There are several types of compartmental models \cite{Miller}, as it can be the SIS model, in which after the infection the individuals do not acquire immunity, the SIR model, in which after the infection the individuals acquire immunity, 
the SIRS model, for which immunity only lasts for a short period of time, the MSIR model, in which infants are born with immunity, etc. In this present work our focus is on the SIS model. 

\bigskip
\bigskip

\section{The SIS model}

\bigskip

The susceptible-infectious-susceptible (SIS) epidemic model assumes a population of size $N$ and one single disease disseminating.
The infectious period extends throughout the whole course of the disease until the recovery of the patient with two possible states, either infected or susceptible. This implies that there is no immunization in this model. 
In this approach the only relevant variable is the instantaneous density of infected individuals $\rho=\rho(\tau)$ depending on the time parameter $\tau$, and taking values in $[0,1]$. The density of infected individuals decreases with rate $\gamma \rho$, where $\gamma$ is the recovery rate, and the rate of growth of new infections is proportional to $\alpha \rho(1-\rho)$, where the intensity of contagion is given by the transmission rate $\alpha$. These two processes are modelled through the compartmental equation
\begin{equation}\label{compsis1}
\frac{d \rho}{d\tau}=\alpha \rho(1-\rho)-\gamma \rho.
\end{equation}
One can redefine the timescale as $t := \alpha \tau$ and introduce $\rho_0:=1-\gamma/\alpha$, so we can rewrite \eqref{compsis1} as
\begin{equation}\label{sismodel0}
\frac{d\rho}{dt}= \rho(\rho_0-\rho).
\end{equation}

\bigskip

Clearly, the equilibrium density is reached if $\rho=0$ or $\rho=\rho_0$.
Although compartmental equations have proven their efficiency for centuries, they are still based on strong hypotheses. For example, the SIS model works more efficiently under the random mixing and large population assumptions. The first assumption is asking homogeneous mixing of the population, that is, individuals contact with each other randomly and do not gather in smaller groups, as abstaining themselves from certain communities. This assumption is nevertheless rarely justified. The second assumption is the rectangular and stationary age distribution, which means that everyone in the population lives to an age $L$, and for each age up to $L$, which is the oldest age, there is the same number of people in each subpopulation. This assumption seems feasible in developed countries where there exists very low infant mortality, for example, and a long live expectancy. Nonetheless, 
it looks reasonable to implement probability at some point to permit random variation in one or more inputs over time. Some recent experiments provide evidence that temporal fluctuations can drastically alter
the prevalence of pathogens and spatial heterogeneity also introduces an extra layer of complexity as it can delay the pathogen transmission.

\bigskip
\bigskip

\subsection{The SIS model with fluctuations} 

\bigskip

It is needless to point out that fluctuations should be considered in order to capture the spread of infectious diseases more closely. Nonetheless, the introduction of these fluctuations is not trivial \cite{Covid}. One way to account for fluctuations is to consider stochastic variables. On the other hand, it seems that in the case of SIS models there exist improved differential equations for the mean and variance of infected individuals. 

Recently, in \cite{NakamuraMartinez}, the model assumes the spreading of the disease as a Markov chain in discrete time in which at most one single recovery or transmission occurs in the duration of this infinitesimal interval.

\bigskip

As a result \cite{kiss}, the first two equations for instantaneous mean density of infected people $\langle\rho\rangle$ and the variance $\sigma^2=\langle\rho^2\rangle-\langle\rho\rangle^2$ are 
\begin{equation}\label{sismodel1}
\begin{split}
\frac{d\langle\rho\rangle}{dt}&=\langle\rho\rangle \left( \rho_0-\langle\rho\rangle \right)-\sigma^2 ,\\
\frac{d\sigma^2}{dt}&=2\sigma^2 \left(\rho_0-\langle\rho\rangle\right)-\Delta_3 -\frac{1}{N}\langle\rho (1-\rho)\rangle+\frac{\gamma}{N\alpha}\langle\rho\rangle ,
\end{split}
\end{equation}
where $\Delta_3 =\langle\rho^3\rangle-\langle\rho\rangle^3$. This system  finds excellent agreement with empirical data \cite{NakamuraMartinez}. 
Equations \eqref{sismodel0} and \eqref{sismodel1} are equivalent when $\sigma$ becomes irrelevant compared to $\langle\rho\rangle$. Therefore, a generalization of compartmental equations only requires mean and variance, neglecting higher statistical moments. The skewness coefficient vanishes as a direct consequence of this assumption, so that $\Delta_3 := 3\sigma^2\langle\rho \rangle$. For a big number of individuals $(N\gg 1)$, the resulting equations are
\begin{equation}\label{sislabellog}
\begin{split}
\frac{d \ln{\langle\rho\rangle}}{d t}&=\rho_0-\langle\rho\rangle-\frac{\sigma^2}{\langle\rho\rangle},\\
\frac{1}{2}\frac{d \ln{\sigma^2}}{d t}&=\rho_0-2\langle\rho\rangle.
\end{split}
\end{equation}
The system right above can be obtained from a stochastic expansion as it is given in \cite{vilar}, as well.  

\bigskip
\bigskip

\subsection{Hamiltonian character of the model}

\bigskip

The investigation of the geometric and the algebraic foundations of a system permits to employ several powerful techniques of geometry and algebra while performing the qualitative analysis of the system. This even results in an analytical/general solution of the system in our case. For example \cite{LeAn03} and \cite{NuLe04} for Lie symmetry approach to solve the classical SIS model. The Hamiltonian analysis of a system plays an important role in the geometrical analysis of a given system.

In \cite{NakamuraMartinez}, the SIS system \eqref{sislabellog} involving fluctuations has been recasted in Hamiltonian form in the following way: the dependent variables are the mean $\langle\rho\rangle$ and the variance $\sigma^2$, and they both depend on time. Then, we define the dynamical variables $q=\langle\rho\rangle$ and  $p=1/\sigma$, so the system \eqref{sislabellog} turns out to be 
\begin{equation}\label{sismodel3}
\begin{split}
\frac{dq}{dt}&=q\rho_0-q^2-\frac{1}{p^2},
\\
\frac{dp}{dt}&=-p\rho_0+2pq.
\end{split}
\end{equation}
We employ the abbreviation SISf for system \eqref{sismodel3} to differentiate it from the classical SIS model in \eqref{sismodel0}. The letter ``f" accounts for ``fluctuations".

We have computed the general solution to this system, finding a more general solution than the one provided by Nakamura and Mart\'inez in \cite{NakamuraMartinez}. Indeed, we have found obstructions in their model solution. We shall comment this in the last section gathering all our new results.

\bigskip

The general solution for this system reads:
\begin{equation} \label{solsismodel3}
\begin{split}
q(t)&=\frac{\rho e^{\rho t}(C_1\rho^2-4)e^{\rho t}+2C_1C_2\rho^2}{(C_1^2\rho^2-4)e^{2\rho t}+4C_2\rho^2(C_1e^{\rho t}+C_2)},\\[2pt]
p(t)&=C_1+\frac{C_1^2\rho^2-4}{4\rho^2-C_2}+C_2e^{-\rho t}.
\end{split}
\end{equation}

\bigskip

In order to develop a geometric theory for this system of differential equations, we need to choose certain particular solutions that we shall make use of. Here we present three different choices and their corresponding graphs according to the change of variables $q=\langle \rho\rangle $ and $p=1/\sigma$.

\begin{figure}[htb]
\centering

\includegraphics[height=7.5cm]{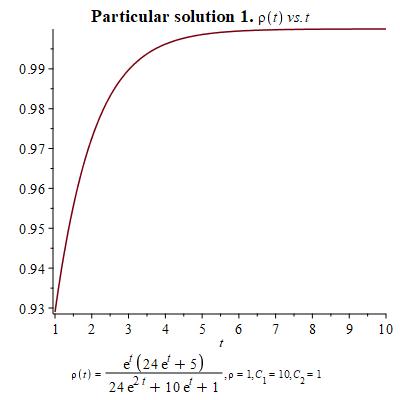}

\includegraphics[height=7.5cm]{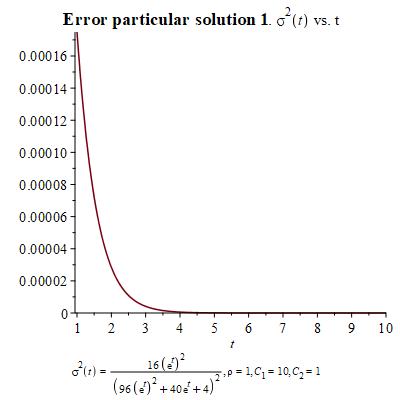}

\caption{The first particular solution}
\label{PS1}
\end{figure}

\newpage

\begin{figure}[htb]
\centering

\includegraphics[height=7.5cm]{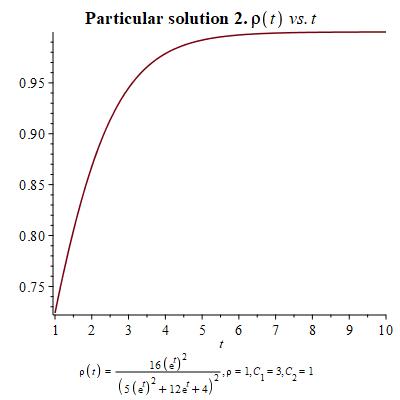}

\includegraphics[height=7.5cm]{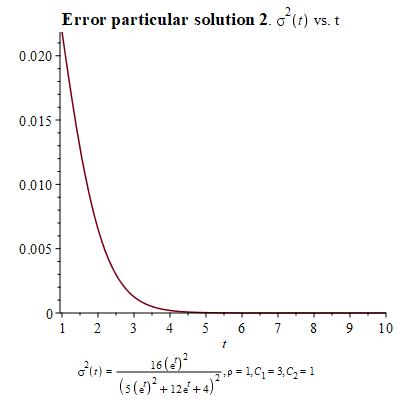}

\caption{The second particular solution}
\label{PS2}
\end{figure}

\begin{figure}[htb]
\centering

\includegraphics[height=8cm]{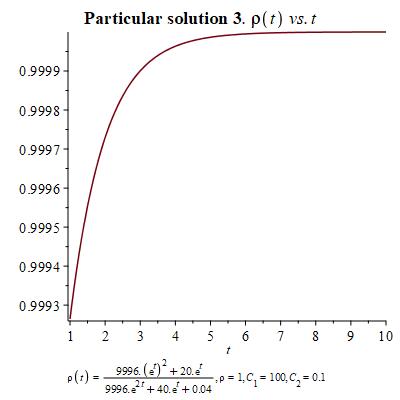}

\includegraphics[height=8cm]{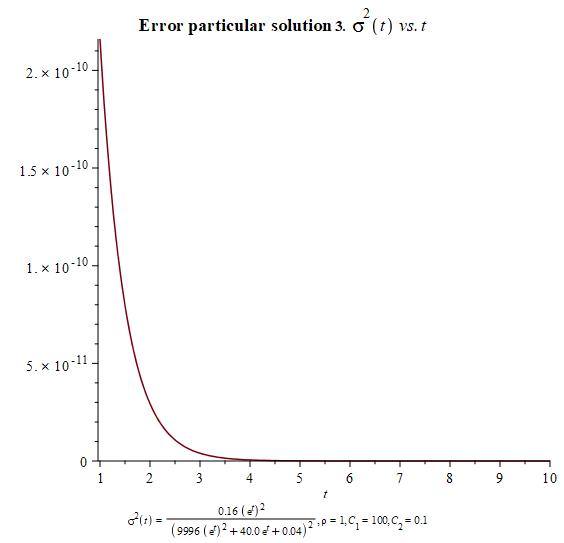}

\caption{The second particular solution}
\label{PS3}
\end{figure}

\bigskip

Let us turn now to interpret these equations geometrically on a symplectic manifold.
The symplectic two-form $\omega=dq\wedge dp$ is a canonical skew-symmetric tensorial object in two-dimensions. For a chosen (real-valued) Hamiltonian function $h=h(q,p)$, the dynamics is governed by a Hamiltonian vector field $X_h$ defined through the Hamilton equation
\begin{equation} \label{Ham-Eq}
\iota_{X_h}\omega=dh.
\end{equation}
 In terms of the coordinates $(q,p)$, the Hamilton equations \eqref{Ham-Eq} become
\begin{equation} \label{loc-Ham-Eq}
\frac{d q} {dt}=\frac{\partial h}{\partial p},\qquad \frac{d p} {dt}=-\frac{\partial h}{\partial q}.
\end{equation}
It is possible to realize that the SISf system (\ref{sismodel3}) is a Hamiltonian system since it fulfills the Hamilton equations \eqref{Ham-Eq}. To see this, consider the Hamiltonian function
\begin{equation}\label{hamiltoniansis}
h=qp\left(\rho_0-q\right)+\frac{1}{p}.
\end{equation}
and substitute it into \eqref{loc-Ham-Eq}. A direct calculation will lead us to (\ref{sismodel3}). The skew-symmetry of the symplectic two-form implies that the Hamiltonian function is constant all along the motion. In holonomic Classical Mechanics, where the Hamiltonian is taken to be the total energy, this corresponds to the conservation of energy. 

\bigskip
\bigskip

\subsection{Lie Analysis of the SISf model}

\bigskip

The model \eqref{sismodel3} can be generalized to a model represented by a time-dependent vector field 
\begin{equation}
X_t=\rho_0(t)X_1+X_2,
\end{equation}
where the constitutive vector fields are computed to be
\begin{equation} \label{sisliesystem-}
X_1=q\frac{\partial}{\partial q}-p\frac{\partial}{\partial p},\quad X_2=\left(-q^2-\frac{1}{p^2}\right)\frac{\partial}{\partial q}+2qp\frac{\partial}{\partial p}.
\end{equation}
The generalization comes from the fact that $\rho_0(t)$ is no longer a constant, but it can evolve in time. 

First, for the vector fields in \eqref{sisliesystem-}, a direct calculation shows that the Lie bracket 
\begin{equation}\label{sisliesystem}
[X_1,X_2]=X_2
\end{equation}
is closed within the Lie algebra. This implies that the SISf model \eqref{sismodel3} is a Lie system. The Vessiot-Guldberg algebra spanned by $X_1, X_2$ is an imprimitive Lie algebra of type $I_{14}$ according to the classification presented in \cite{Ballesteros2}. 

\bigskip

If we copy the configuration space twice, we will have four degrees of freedom $(q_1,p_1,q_2,p_2)$ and we will archieve precisely two first-integrals as a consequence of the Fr\"obenius theorem.
A first-integral for $X_t$ has to be a first-integral for $X_1$ and $X_2$ simultaneously. We define the diagonal prolongation $\widetilde{X}_1$ of the  vector field $X_1$ in the decomposition \eqref{sisliesystem}. Then we look for a first integral $F_1$ such that $\widetilde{X}_1[F_1]$ vanishes identically. Notice that if $F_1$ is a first-integral of the vector field $\widetilde{X}_1$ then it is a first integral of $\widetilde{X}_2$ due to the commutation relation.
For this reason, we start by integrating the prolonged vector field 
\begin{equation}
\widetilde{X}_1=q_1\frac{\partial}{\partial q_1}+q_2\frac{\partial}{\partial q_2}-p_1\frac{\partial}{\partial p_1}-p_2\frac{\partial}{\partial p_2},
\end{equation}
through the following characteristic system
\begin{equation}
\frac{dq_1}{q_1}=\frac{dq_2}{q_2}=\frac{dp_1}{-p_1}=\frac{dp_2}{-p_2}.
\end{equation}
Fix the dependent variable $q_1$ and obtain a new set of dependent variables, say $(K_1,K_2,K_3)$, which are computed to be 
\begin{equation}\label{kchange}
K_1=\frac{q_1}{q_2},\qquad K_2=q_1p_1,\qquad K_3=q_1p_2.
\end{equation}

Now, this induces the following basis in the tangent space
\begin{equation}
\begin{split}
\frac{\partial}{\partial K_1}=q_2 \frac{\partial }{\partial q_1}-\frac{q_2 p_1}{q_1} \frac{\partial }{\partial p_1}- \frac{q_2 p_2}{q_1} \frac{\partial }{\partial p_1}, \qquad \frac{\partial}{\partial K_2}=\frac{1}{q_1}\frac{\partial}{\partial p_1}, \qquad
\frac{\partial}{\partial K_3}=
\frac{1}{q_1}\frac{\partial}{\partial p_2}.
\end{split}
\end{equation}
provided that $q_1$ is not zero. Introducing the coordinate changes exhibited in \eqref{kchange} into the diagonal projection  $\widetilde{X}_2$ of the vector field $X_2$, we arrive at the following expression
\begin{equation}
\begin{split}
\widetilde{X}_2=&\left(2K_1-\left(1+\frac{1}{K_1^2}\right)\right)\frac{\partial}{\partial K_1}+\left(\left(\frac{1}{K_2^2}+\frac{1}{K_3^2}\right)K_2^2-\left(1+\frac{1}{K_1^2}\right)K_2\right)\frac{\partial}{\partial K_2}\nonumber\\
&+\left(2\frac{K_3}{K_2}-\left(1+\frac{1}{K_1^2}\right)K_3\right)\frac{\partial}{\partial K_3}.
\end{split}
\end{equation}
To integrate the system once more, we use the method of characteristics again and obtain
\begin{equation}\label{logsys}
\frac{d\ln{|K_1|}}{1-\frac{1}{K_1^2}}=\frac{d\ln{|K_2|}}{\frac{1}{K_2}+\frac{K_2}{K_3}-1-\frac{1}{K_1^2}}=\frac{d\ln{|K_3|}}{\frac{2}{K_2}-\left(1+\frac{1}{K_1^2}\right)}.
\end{equation}

We obtain two first integrals by integrating in pairs $(K_1,K_2)$ and $(K_1,K_3)$, where $K_1$ is fixed. After some cumbersome calculations we obtain
\begin{equation} \label{KKK}
K_2=\frac{K_1\left(4k_2^2 K_1^2+4k_1k_2K_1+k_1^2-4\right)}{2(K_1+1)(K_1-1)k_2(2k_2K_1+k_1)}, \qquad 
K_3=\frac{K_1\left(k_2K_1^2+k_1K_1+\frac{k_1^2-4}{4k_2}\right)}{(K_1+1)(K_1-1)}.
\end{equation}
By substituting back the coordinate transformation  \eqref{kchange} into the solution \eqref{KKK}, we arrive at the following implicit equations
\begin{equation}\label{exactsol}
\begin{split}
q_1&=-\frac{q_2\left(k_1k_2\pm \sqrt{4k_2^2p_2^2q_2^2+k_1^2k_2p_2q_2-4k_2^3p_2q_2-4k_2p_2q_2+4k_2^2}\right)}{2k_2(-p_2q_2+k_2)},\\[2pt]
p_1&=\frac{4q_1^2k_2^2+4q_1q_2k_1k_2+q_2^2k_1^2-4q_2^2}{2k_2(2q_1^3k_2+q_1^2k_1q_2-2q_1k_2q_2^2-k_1q_2^3)}.
\end{split}
\end{equation}
Let us notice that the equations \eqref{exactsol} depend on a particular solution $(q_2,p_2)$ and two constants of integration $(k_1,k_2)$ which are related to initial conditions. 

\bigskip

Let us show now the graphs and values of the initial conditions for which the solution reminds us of sigmoid behavior, which is the expected growth of $\rho(t)$. As particular solution for $(q_2,p_2)$, we have made use of particular solution 2 given in Figure \ref{PS2} through its corresponding values of $q,p$ through the change of variables $q=\langle \rho \rangle $ and $p=1/\sigma$.

\begin{figure}[h!]
\centering
\includegraphics[height=9.5cm]{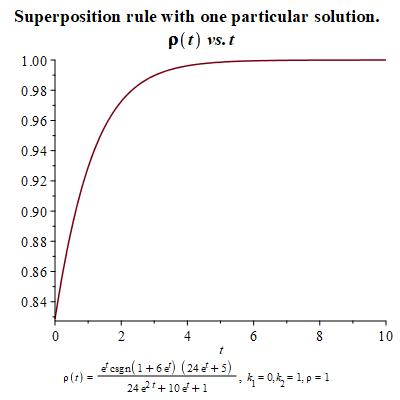}
\label{SR1notlin}
\caption{Superposition rule for exact solution}
\end{figure}

\bigskip

Since the solution \eqref{exactsol} is quite complicated,  one may look for a solution of a linearized model. We first employ the following change of coordinates
\begin{equation}\label{changelog}
\{u=\ln{|K_1|}, v=\ln{|K_2|}, w=\ln{|K_3|}\}. 
\end{equation}
In terms of these new variables, the system \eqref{logsys} reads
\begin{equation} \label{sis-s--s}
\frac{du}{1-e^{-2u}}=\frac{dv}{e^{-v}+e^{v-w}-1-e^{-2u}}=\frac{dw}{2e^{-v}-(1+e^{-2u})}.
\end{equation}
One can solve the system above by introducing a linear approximation
\begin{equation}
\begin{split}
1-e^{-2u}&\simeq  2u,  \\
e^{-v}+e^{v-w}-1-e^{-2u}&\simeq   2u-w,\\
2e^{-v}-(1+e^{-2u})&\simeq  2u-2v,
\end{split}
\end{equation}
after which \eqref{sis-s--s} reads
\begin{equation}\label{linearsys}
\frac{du}{2u}=\frac{dv}{2u-w}=\frac{dw}{2u-2v}.
\end{equation}
We can solve now $v$ and $w$ in terms of $u$ and obtain
\begin{equation}
v(u)=k_1u^{-\sqrt{2}/2}+k_2u^{\sqrt{2}/2}+u, \qquad
w(u)=\sqrt{2}\left(k_1u^{-\sqrt{2}/2}-k_2u^{\sqrt{2}/2}\right).
\end{equation}
We need to isolate the constants of integration $k_1$ and $k_2$. Hence, the two first integrals read now
\begin{equation}
k_1= u^{\frac{\sqrt{2}}{2}}\big(\sqrt{2}v-\sqrt{2}u+w\big)/{2\sqrt{2}}, \qquad 
k_2=u^{-\frac{\sqrt{2}}{2}}
\big(\sqrt{2}v-\sqrt{2}u-w\big)/2\sqrt{2}.
\end{equation}
Now, if we substitute the coordinate changes in \eqref{changelog} and in \eqref{kchange}, we arrive at the following general solution
\begin{equation}\label{ppiosuprule}
q_1=q_2\exp\Big(-\frac{\ln{(q_2p_2)}}{1+k_1+k_2}\Big),\qquad 
p_1=\frac{1}{q_2}\exp\Big(\frac{\sqrt{2}}{2}\frac{(k_1-k_2)\ln{(q_2p_2)}}{1+k_1+k_2}\Big).
\end{equation}
which can be written as 
\begin{equation}\label{ppiosuprule2}
q_1=q_2\Big(q_2p_2\Big)^{\frac{-1}{1+k_1+k_2}},\qquad  p_1=\frac{1}{q_2}\Big(q_2p_2 \Big)^{\frac{\sqrt{2}}{2}\frac{k_1-k_2}{1+k_1+k_2}}. 
\end{equation}

\bigskip

Notice that the solution depends on a particular solution $(q_2,p_2)$ and two constants of integration $(k_1,k_2)$, as in \eqref{exactsol}.

Let us show now the graphs and values of the initial conditions for which the solution reminds us of sigmoid behavior, which is the expected growth of $\rho(t)$. As particular solution for $(q_2,p_2)$, we have made use of particular solution 2 given in Figure \ref{PS2} through its corresponding values of $q,p$ through the change of variables $q=<\rho>$ and $p=1/\sigma$.

\begin{figure}[htb]
\centering

\includegraphics[height=9.5cm]{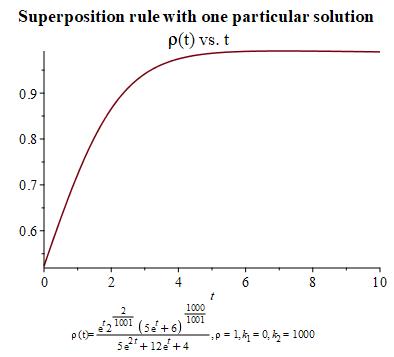}

\includegraphics[height=9.5cm]{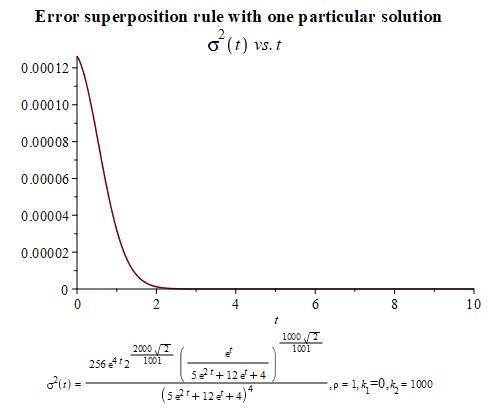}

\caption{Superposition rule for linear approximation}
\label{SR1}
\end{figure}


\bigskip
\bigskip

\section{Lie--Hamilton analysis of the SISf model} \label{Lie-Ham-Sec}

\bigskip

In this section, we shall show that the SISf model \eqref{sismodel3} is a Lie-Hamilton system \cite{LS}. Among the developed methods for Lie--Hamilton systems, we consider a very important recent method for the obtainance of solutions as superposition principles through the Poisson coalgebra method \cite{Ballesteros2}. 

\bigskip

We have already proven in \eqref{sisliesystem} that \eqref{sismodel3} defines a Lie system. In order to see if it is a Lie--Hamilton system, we first need to check  whether the vector fields in \eqref{sisliesystem} are Hamiltonian vector fields. Consider now the canonical symplectic form $\omega=dq\wedge dp$. It is easy to check that
the vector fields $X_1$ and $X_2$ in \eqref{sisliesystem} are Hamiltonian with respect to the Hamiltonian functions
\begin{equation}\label{hamfunctsis}
h_1=-qp,\quad \quad h_2=-q^2p+\frac{1}{p},
\end{equation}
respectively. It is easy to see that the Poisson bracket of these two functions reads $\{h_1,h_2\}=h_2$. It means that the Hamiltonian functions form a finite dimensional Lie algebra, denoted in the literature as $I^{r=1}_{14A}\simeq \mathbb{R}\ltimes \mathbb{R}$, and it is isomorphic to the one defined by vector fields $X_1, X_2$. The Hamiltonian function for the total system is 
\begin{equation}\label{hamcovid}
h=\rho_0(t)h_1+h_2=-q^2p+\frac{1}{p}-\rho_0(t)qp
\end{equation}
and it is exactly the Hamiltonian function \eqref{hamiltoniansis} proposed in \cite{NakamuraMartinez}.

\bigskip

Lie-Hamilton systems can also be integrated in terms of a superposition rule. We need to find a Casimir function for the Poisson algebra, but unfortunately, there exists no nontrivial Casimir in this particular case.
 It is interesting to see how a symmetry of the Lie algebra $\{X_1,X_2\}$ commutes with the Lie bracket, i.e. the vector field
\begin{equation}
Z=-\frac{1}{2}\frac{p(C_2p^2q^2+4C_1pq+C_2)}{(pq-1)(pq+1)}\frac{\partial}{\partial p}+\frac{C_1p^4q^4+C_2p^3q^3-C_2pq-C_1}{p(pq-1)^2(pq+1)^2}\frac{\partial}{\partial q}
\end{equation}
fulfills $[X_1,Z]=0,\quad [X_2,Z]=0.$ Notice too that $Z$ is a conformal vector field, that is, 
\begin{equation}
\mathcal{L}_Z\omega=-(C_2/2)\omega.
\end{equation}

\bigskip

Since it is a Hamiltonian system, one would expect that a first integral for $Z$, let us say $f$, would Poisson commute with the Poisson algebra $\{h_1,h_2\}$, since $Z=-\hat{\Lambda}(df)$. Nonetheless, this is not the case unless $f=\text{constant}$. 
This implies that the Casimir is a constant, hence trivial and the coalgebra method can not be directly applied. However, there is a way in which we can circumvent this problem by considering an inclusion of the algebra $I^{r=1}_{14A}$ as a Lie subalgebra of a Lie algebra to another class admitting a Lie--Hamiltonian algebra with a non-trivial Casimir. In this case, we will consider the algebra, denoted by $I_8\simeq \mathfrak{iso}(1,1)$, due to the simple form of its Casimir. If we obtain the superposition rule for $I_8$, we  simultaneously obtain the superposition for $I_{14A}^{r=1}$ as a byproduct. 

\bigskip
\bigskip


\subsection{Superposition rules for the SISf model: $\mathfrak{iso}(1,1)$}

\bigskip

The Lie--Hamilton algebra $\mathfrak{iso}(1,1)$ has the commutation relations
\begin{equation}\label{basisiso}
\{h_1,h_2\}=h_0,\quad \{h_1,h_3\}=-h_1,\quad  \{h_2,h_3\}=h_2, \quad \{h_0,\cdot\}=0,
\end{equation}
with respect to $\omega=dx\wedge dy$ in the basis 
$\{h_1=y,h_2=-x,h_3=xy,h_0=1\}$. The Casimir associated to this Lie--Hamilton algebra is
$C=h_1h_2+h_3h_0$. Let us apply
the coalgebra method to this case. Mapping the representation without coproduct, the first
iteration is trivial, i.e., $F=0$. We could use the second-order coproduct and third-order coproduct $\Delta^{(2)}$ and $\Delta^{(3)}$, or the second-order coproduct $\Delta^{(2)}$ together with the permuting sub-indices property.
We need three constants of motion, this would be equivalent to integrating 
the diagonal prolongation $\widetilde X$ on $(\mathbb{R}^2)^3$. Using the coalgebra method
and sub-index permutation, one obtains
\begin{equation}
\begin{split}
  F^{(2)}&=  
(x_{1}-x_{2})   (y_{1}-y_{2}) =   k_1 , \\
F_{23}^{(2)}&=(x_{1}-x_{3})   (y_{1}-y_{3}) =   k_2 , \\
F_{13}^{(2)}& =(x_{3}-x_{2})   (y_{3}-y_{2}) =   k_3 .
\end{split}
\end{equation}
From them, we can choose two functionally independent constants of motion. Our choice is $F^{(2)}=k_1,F^{(2)}_{23}=k_2$. The introduction of  $k_3$  simplifies the final result, with expression 
\begin{equation}\label{sup2}
\begin{split}
x_1(x_2,y_2,x_3,y_3,k_1,k_2,k_3)&=\frac 12(x_2+x_3) +\frac{k_2-k_1\pm B}{2(y_2-y_3)}   , \\
y_1(x_2,y_2,x_3,y_3,k_1,k_2,k_3)  &=\frac 12(y_2+y_3) +\frac{k_2-k_1\mp B}{2(x_2-x_3)} ,
\end{split}
\end{equation}
where
\begin{equation}
B= \sqrt{ k_1^2+k_2^2+k_3^2-2(k_1k_2+k_1k_3+k_2k_3) }.
\end{equation}

In the case that matters to us, $I_{14A}^{r=1}$, the third constant $k_3$ is a function $k_3=k_3(x_2,y_2,x_3,y_3)$ and $B\geq 0$. Notice though that this superposition rule is expressed in the basis \eqref{basisiso}, therefore, we need the change of coordinates between the present $\mathfrak{iso}(1,1)$ and our problem \eqref{hamfunctsis}.
See that the commutation relation $\{h_1,h_3\}=-h_1$ in \eqref{basisiso} coincides with the commutation relation $\{h_1,h_2\}=h_2$ of our pandemic system \eqref{hamfunctsis}. 
So, by comparison, we see there is a change of coordinates
\begin{equation}\label{changecoord}
x=-qp,\qquad y=q-\frac{1}{qp^2}.
\end{equation}

\bigskip

This way, introducing this change \eqref{changecoord} in \eqref{sup2}, the superposition principle for our Hamiltonian pandemic system reads
\begin{equation}\label{sassaa}
\begin{split}
q&=A\times C^{-1}
\\
p&=-C\times D^{-1}
\end{split}
\end{equation}
where $$A:=\left(\frac{\frac 12 q_2+\frac 12 q_3+(k_2-k_1\pm B)}{(2p_2-2p_3)}\right)^2\left(\frac 12 p_2+\frac 12 p_3+\frac{(k_2-k_1\mp B)}{(2q_2-2q_3)}\right),$$ $$C:=\left(\frac{\frac 12 q_2+\frac 12 q_3+(k_2-k_1\pm B)}{(2p_2-2p_3)}\right)^2-1$$ and $$D:=\frac 12 q_2+\frac 12 q_3+\frac{(k_2-k_1\pm B)}{(2p_2-2p_3)}\left(\frac 12 p_2+\frac 12 p_3+\frac{(k_2-k_1\mp B)}{(2q_2-2q_3)}\right).$$
Here, $(q_2,p_2)$ and $(q_3,p_3)$ are two particular solutions and $k_1,k_2,k_3$ are constants of integration.

\bigskip

Now, we show the graphics for $<\rho>=q(t)$ and $\sigma^2=1/p^2$ using the two particular solutions in Figure \ref{PS2} and Figure \ref{PS3} provided in the introduction. Notice that we have renamed $c=(k_2-k_1\pm B)$ and $k=(k_2-k_1\mp B)$.

\begin{figure}[htb]
\centering

\includegraphics[height=10.5cm]{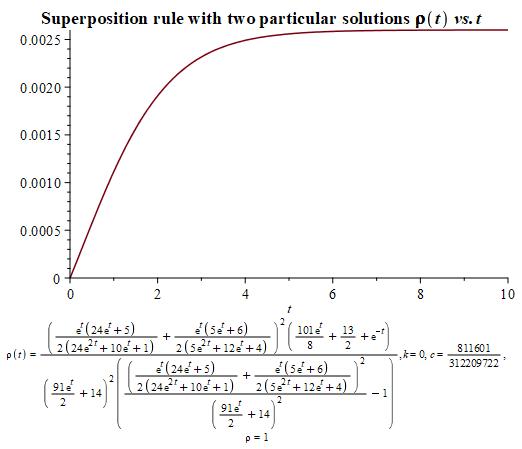}

\includegraphics[height=10.5cm]{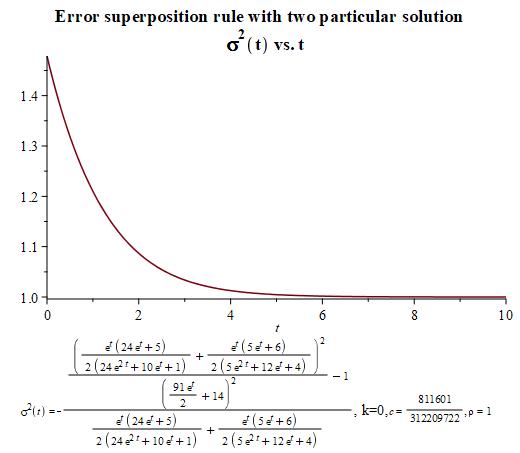}

\caption{Superposition rule with two particular solutions}
\label{SR2}
\end{figure}

\bigskip
\bigskip

\section{A deformed SISf model}

\bigskip

For the SISf model \eqref{sismodel3}, we start with the Vessiot--Guldberg algebra \eqref{sisliesystem} labelled as $I_{14A}^{r=1}$. To obtain a deformation of a Lie algebra $I_{14A}^{r=1}$, we need to rely on a bigger Lie algebra, in this case, we make use of $\mathfrak{sl}(2)$. To this end, consider the vector fields $X_1$ and $X_2$ in \eqref{sisliesystem-}, and let $X_3$ be a vector field given by \begin{equation}
    X_3:=\frac{p^2q^2(-2p^2q^2+c+6)+c}{2(p^2q^2-1)^2}\frac{\partial}{\partial q}-\frac{p^3q(c+2)}{(p^2q^2-1)^2}\frac{\partial}{\partial p},
\end{equation} where $c\in \mathbb{R}$. Then, $\{X_1,X_2,X_3\}$ span a Vessiot--Lie algebra $V$ isomorphic to $\mathfrak{sl}(2)$ that satisfies the following commutation relations \begin{equation}\label{sl2commrule}
[X_1,X_2]=X_2,\qquad [X_1,X_3]=-X_3,\qquad [X_2,X_3]=2X_1 .
\end{equation}

This vector field $X_3$ admits a Hamiltonian function, say $h_3$,  with respect to the canonical symplectic form on $\mathbb{R}^2$, so that we have the family
 \begin{equation}
    h_1=-qp,\qquad \quad h_2=\frac{1}{p}-q^2p, \qquad h_3=\frac{2p^3q^2+c}{2-2p^2q^2}.
\end{equation} 

\bigskip

Hence, $\{h_1,h_2,h_3\}$ span a Lie--Hamilton algebra $\mathcal{H}_\omega$; isomorphic to $\mathfrak{sl}(2)$ where the commutation relations with respect to the Poisson bracket induced by the canonical symplectic form $\omega$ on $\mathbb{R}^2$ are given by 
\begin{equation}\label{sl2Poissoncommrule}
\{h_1,h_2\}_\omega=h_2,\qquad \{h_1,h_3\}_\omega=-h_3,\qquad \{h_2,h_3\}_\omega=2h_1.
\end{equation}

\textbf{Step 1.}
Applying the non-standard deformation of $\mathfrak{sl}(2)$ in \cite{Ballesteros5} we  arrive at the Hamiltonian functions 
\begin{equation}\label{gd1} 
   h_{z;1}=-shc(2zh_{z;2})qp,  \qquad
 h_{z;2}=\frac{1}{p}-q^2p, \qquad
 h_{z;3}=-\frac{p\left(shc(2zh_{z;2})2q^2p^2+c\right)}{2shc(2zh_{z;2})(q^2p^2-1)} ,
\end{equation}
Accordingly, the Poisson brackets are computed to be
\begin{equation}\label{gb22}
\begin{gathered}
\{h_{z;1},h_{z;2}\}_\omega=shc(2zh_{z;2})h_{z;2},\qquad 
 \{h_{z;2},h_{z;3}\}_\omega=2h_{z;1},\\[2pt]
 \{h_{z;1},h_{z;3}\}_\omega=-cosh(2zh_{z;2})h_{z;3},
\end{gathered}
\end{equation}

\textbf{Step 2.} The vector fields $X_{z;1}$ and $X_{z;2}$ associated to the Hamiltonian functions $h_{z;1}$ and $h_{z;2}$ exhibited in \eqref{gd1} are
\begin{equation}
\begin{split}
X_{z,1}&=\frac{cosh\left(2z(\frac{1}{p}-q^2p)\right)}{(p^2q^2-1)^2}\left[(1-p^4q^4) \frac{\partial}{\partial q}+(2p^5q^4-p^3q^2)\frac{\partial}{\partial p}\right]\\[4pt] &\qquad \qquad +\frac{shc\left(2z(\frac{1}{p}-q^2p)\right)}{(p^2q^2-1)}\left[ q \frac{\partial}{\partial q}-p(p^2q^2+1)\frac{\partial}{\partial p}\right],\\[4pt] 
X_{z,2}&=\left(-q^2-\frac{1}{p^2}\right)\frac{\partial}{\partial q}+2qp\frac{\partial}{\partial p}.
\end{split}
\end{equation}
We do not write explicitly the expression of the vector field $X_{z;3}$ because it does not play a relevant role in our system. The deformed vector fields keep the commutation relations
\begin{equation}\label{com22}
\left[X_{z;1},X_{z;2}\right]=cosh\left(2z\left(\frac{1}{p}-q^2p\right)\right) \, X_{z;2}.
\end{equation}

\textbf{Step 3.}  The total Hamiltonian function for the deformed model is
\begin{equation}\label{defham}
h_z=\rho(t)h_{z;1}+h_{z;2}=-\rho(t) shc(2zh_{z;2})qp+\frac{1}{p}-q^2p.
\end{equation} 
so that the deformed dynamics is computed to be
\begin{equation}\label{dssis}
\begin{split}
    \frac{dq}{dt}&=\left(\frac{cosh\left(2z(\frac{1}{p}-q^2p)\right)}{(p^2q^2-1)^2}(1-p^4q^4) +\frac{shc\left(2z(\frac{1}{p}-q^2p)\right)}{(p^2q^2-1)}\ q \right)\rho_0(t)-q^2-\frac{1}{p^2},\\[4pt]
    \frac{dp}{dt}&=\left(\frac{cosh\left(2z(\frac{1}{p}-q^2p)\right)}{(p^2q^2-1)^2}(2p^5q^4-p^3q^2)-p\frac{shc\left(2z(\frac{1}{p}-q^2p)\right)}{(p^2q^2-1)}(p^2q^2+1)\right)\rho_0(t)-2qp.
\end{split}
\end{equation}
This system describes a family of z-parametric differential equations that generalizes the SISf model \eqref{sismodel3}, where the demographic interaction and both rates allow a more realistic representation of the epidemic evolution. According to the kind of deformation, this may be called a quantum family SISf model. Note that the SISf model can be recovered in the limit when $z$ tends to zero.  

\bigskip
\bigskip


\subsection{Constants of motion} 

\bigskip

For the present case, the constants of motion are computed to be \begin{equation}
    F^{(1)}=\frac{c}{4},\qquad F^{(2)}=\left(h_{2}^{(1)}+h_{2}^{(2)}\right)\left(h_{3}^{(1)}+h_{3}^{(2)}\right)-\left(h_{1}^{(1)}+h_{1}^{(2)}\right)^2,
\end{equation}
after the quantization, the latter one becomes 
\begin{equation}\label{constant1}
   F_z^{(2)}= shc\left(2zh_{z;2}^{(2)}\right)h_{z;2}^{(2)}h_{z;3}^{(2)}-\left(h_{z;1}^{(2)}\right)^2,
\end{equation}  where $h_{z;j}^{(2)}:=D_z^{(2)}(\Delta_z(v_j))$. This coproduct $\Delta_z$ can be described as a follows \begin{equation*}
\Delta_z(v_2)=  v_2 \otimes 1+1\otimes v_2 , \qquad
\Delta_z(v_j)=v_j \otimes e^{2 z v_2} + e^{-2 z v_2} \otimes v_j   ,\qquad  j=1,3.
\end{equation*} 
More explictly, using the expressions given in \eqref{gd}, we have
\begin{equation}
\begin{split}
h_{z;j}^{(2)}=&h_{z;j}(q_1,p_1)e^{2zh_{z;2}(q_2,p_2)}+h_{z;j}(q_2,p_2)e^{-2zh_{z;2}(q_1,p_1)}, \qquad j=1,3 \\[2pt] h_{z;2}^{(2)}=&h_{z;2}(q_1,p_1)+h_{z;2}(q_2,p_2).
\end{split}
\end{equation}

So, to retrieve another constant of motion we can apply the trick of permuting indices.
Then, here we have a second constant of motion, writing it implicitly,
\begin{equation}\label{constant2}
  F_{z,(23)}^{(2)}= shc\left(2zh_{z;2 (23)}^{(2)}\right)h_{z;2(23)}^{(2)}h_{z;3(23)}^{(2)}-\left(h_{z;1 (23)}^{(2)}\right)^2,
  \end{equation}
  where the sub-index $(23)$ means that the variables $(q_2,p_2)$ are interchanged with $(q_3,p_3)$ when they appear in the deformed Hamiltonian functions $h_{z;j}$ and
  \begin{align*}
h_{z;j (23)}^{(2)}=&h_{z;j}(q_1,p_1)e^{2zh_{z;2}(q_3,p_3)}+h_{z;j}(q_3,p_3)e^{-2zh_{z;2}(q_1,p_1)}, \qquad j=1,3 \\[2pt] h_{z;2 (23)}^{(2)}=&h_{z;2}(q_1,p_1)+h_{z;2}(q_3,p_3).
\end{align*}
In \eqref{constant1}, we have
\begin{equation}
\begin{split}
h_{z;2}(q_1,p_1)&=-shc(2zh_{z;2})q_1p_1,\quad h_{z;2}=\frac{1}{p_1}-q_1^2p_1,\quad h_{z;3}=-\frac{p_1\left(shc(2zh_{z;2})2q_1^2p_1^2+c\right)}{2shc(2zh_{z;2})(q_1^2p_1^2-1)}, \\[4pt]
h_{z;2}(q_2,p_2)&=-shc(2zh_{z;2})q_2p_2,\quad h_{z;2}=\frac{1}{p_2}-q_2^2p_2,\quad h_{z;3}=-\frac{p_2\left(shc(2zh_{z;2})2q_2^2p_2^2+c\right)}{2shc(2zh_{z;2})(q_2^2p_2^2-1)} ,
\end{split}
\end{equation}
whilst in \eqref{constant2}
\begin{equation}
\begin{split}
h_{z;2}(q_1,p_1)&=-shc(2zh_{z;2})q_1p_1,\quad h_{z;2}=\frac{1}{p_1}-q_1^2p_1,\quad h_{z;3}=-\frac{p_1\left(shc(2zh_{z;2})2q_1^2p_1^2+c\right)}{2shc(2zh_{z;2})(q_1^2p_1^2-1)}, \\[4pt]
h_{z;2}(q_3,p_3)&=-shc(2zh_{z;2})q_3p_3,\quad h_{z;2}=\frac{1}{p_3}-q_3^2p_3,\quad h_{z;3}=-\frac{p_3\left(shc(2zh_{z;2})2q_3^2p_3^2+c\right)}{2shc(2zh_{z;2})(q_3^2p_3^2-1)}. 
\end{split}
\end{equation}

\bigskip

If we set these two first integrals equal to a constant, $F_{z,(23)}^{(2)}=k_{23}$ and $F_z^{(2)}=k_{12}$, with $k_{23},k_{12}\in \mathbb{R}$, one is able to retrieve a superposition principle for $q_1=q_1(q_2,q_3,p_2,p_3,k_{12},k_{23})$ and $p_1=p_1(q_2,q_3,p_2,p_3,k_{12},k_{23})$.
Notice that here $(q_2,p_2)$ and $(q_3,p_3)$ are two pairs of particular solutions and
$k_{12},k_{23}$ are two constants over the plane to be related to initial conditions \cite{Covid}. 
\renewcommand{\theequation}{5.\arabic{equation}}
\chapter{Conclusions} 

\label{Conclusiones} 

\bigskip

In this work, the notion of Poisson--Hopf deformation of  LH systems has been proposed. This  framework differs radically from other approaches to the LH systems theory~\cite{BCHLS13Ham,CGL10,Dissertations,CLS13,PW}, as our resulting deformations do not formally correspond to LH systems, but to an extended notion of them that requires a (non-trivial) Hopf structure and is related with the non-deformed LH system by means of a limiting process in which the deformation parameter $z$ vanishes.  Moreover, the introduction of Poisson--Hopf structures  allows for the generalization of the type of systems under inspection, since the finite-dimensional Vessiot--Guldberg Lie algebra is replaced by an involutive distribution in the Stefan--Sussman sense (Chapter 3). 

In Chapter 4, the Poisson analogue of the non-standard quantum deformation of $\mathfrak{sl}(2)$ has been studied, establishing explicitly the constants of the motion for the quantum deformed systems. The three non-equivalent LH systems in the plane based on the Lie algebra $\mathfrak{sl}(2)$ have been described in unified form, which provides a nice geometrical interpretation of both these systems and their corresponding quantum deformations. Chapter 5 is devoted to the analysis of specific systems of differential equations and their deformed counterpart. We first consider the Milne--Pinney equation, the deformations of which provide us with new oscillator systems with the particularity that the mass of the particle  is dependent on the position, and where the constants of the motion are explicitly obtained. In particular, the Schr\"odinger problem for position-dependent mass Hamiltonians is directly connected with the quantum dynamics of charge carriers in semiconductor heterostructures and nanostructures (see, for instance,~\cite{Bastard, qDWW,Roos}). In this context, it is worthy to be observed that the standard or Drinfel'd--Jimbo deformation of $\mathfrak{sl}(2)$ would not lead to an oscillator with a position-dependent mass as, in that case, the deformation function would be $\!\shc(z q p)$ instead of $\!\shc(z q^2)$; this can clearly be  seen in the corresponding symplectic realization given in~\cite{BCFHL}. This fact explains that  we have chosen the non-standard deformation of $\mathfrak{sl}(2)$ due to its physical applications.
 In spite of this, the Drinfel'd--Jimbo deformation could provide additional deformation for the Milne--Pinney  equation, leading to systems that are non-equivalent  to those studied here. In any case, these examples suggest an alternative approach to dynamical systems with a nonconstant mass, for which the classical tools are of limited applicability. The second type of LH systems that has been studied corresponds to the complex and coupled Riccati equations, which have been extensively studied in the literature. For them, the deformed versions of the corresponding LH systems and their constants of the motion have been obtained. The main results of Chapters 3-5 have been published in \cite{Ballesteros5} and \cite{BCFHL}.   
In Chapter 6 we focus on oscillator systems obtained as a deformation of LH systems based on the oscillator algebra $\frak{h}_4$, seen as a subalgebra of the 2-photon algebra $\mathfrak{h}_6$. In particular, these deformations can be obtained as the restriction of the corresponding deformed $\mathfrak{h}_6$ Poisson--Hopf  systems. An illustrative example of this type of deformations is given by a generalization of the damped oscillator. It remains to explicitly determine a superposition rule for such systems, a problem that it is currently in progress. In Chapter 7 the affine Lie algebra $\mathfrak{b}_2$, seen as a subalgebra of $\mathfrak{sl}(2)$,  is used to obtain quantum deformed systems applicable in the context of epidemiological models. This approach constitutes a novelty, as the techniques usually employed for this type of models are essentially of stochastic nature. The results of this chapter have recently been submitted for publication.

\medskip

There is a plethora of problems and applications that emerge from the Poisson--Hopf  algebra deformation formalism. Although the results have been principally focused on the two-dimensional case, for which an explicit classification of LH algebras exists \cite{BBHLS,BCHLS13Ham}, the results are valid for arbitrary manifolds and higher-dimensional Vessiot--Guldberg Lie algebras. A systematic analysis of the known systems would certainly lead to a richer spectrum of properties for the deformed systems that deserve further investigation. In particular, the dynamical properties of specific systems of differential equations can be studied with these techniques, and it is expected that some new and intriguing features will emerge from this analysis. 

\medskip

As a byproduct, and related to the current COVID-19 pandemic,  one may wonder whether a description of the pandemic could be related to a SISf-pandemic model. The SISf model is a very first approximation for a trivial infection process, in which there are only two possible states for an individual in the population: they are either infected or susceptible to the infection. Hence, this model does not provide the possibility of acquiring immunity at any point. It seems that COVID-19 provides some certain types of immunity, but only   a thirty percent of the infected individuals, hence, a SIR model that considers ``R" for recovered individuals (not susceptible anymore, i.e., immune) is not a proper model for the current situation. It would be interesting to have a model contemplating immune and nonimmunized individuals simultaneously. Currently we are still in search of a stochastic Hamiltonian model that includes potential immunity and nonimmunity. 

\section{Future work}
One of the most important questions to be addressed is whether the Poisson--Hopf algebra  approach can provide an effective procedure to  derive a  deformed analogue of superposition principles for deformed LH systems. It would also be interesting to know whether such a description is simultaneously applicable to the various non-equivalent deformations, like an extrapolation of the notion of Lie algebra contraction to Lie systems. Another open problem worthy to be considered  is the possibility of getting a unified description of such systems in terms of a certain amount of fixed `elementary' systems, thus implying a first rough systematization of LH-related systems from a more general perspective than that of finite-dimensional Lie algebras. Some possible future work in this direction can be summarized as follows: 

\medskip

\begin{itemize}

\medskip

\item In the classification of LH systems on the plane, the so-called 2-photon algebra plays a central role, as it is the highest dimensional algebra that can appear with the properties of a LH algebra. The study of their quantum deformations is, therefore, a fundamental question to complete the analysis of the deformations of the LH systems on the plane. It should be noted that there are essentially two different possibilities for these deformations, depending on the structure of two prominent subalgebras, the algebra $ h_4 $ and $\mathfrak{sl}(2)$, which gives rise to systems and deformations with different properties. The first case, based on the oscillator algebra $\mathfrak{h}_4$, has partially been considered in Chapter 6. However, there still remains the problem of obtaining an effective superposition rule, the implementation of which is currently being studied. The analysis must be completed identifying particular classes of systems of differential equations that can be deformed by this procedure, and that can be interpreted as perturbations of the initial system. The second  case, based on the extension of the results for $\mathfrak{sl}(2)$ to the 2-photon algebra,  is structurally quite different due to the corresponding quantum deformation. It is expected that new systems with diverging properties will emerge from this analysis. From the point of view of applications, these systems have a multitude of interesting properties, such as new systems of the Lotka--Volterra type or oscillator systems with variable frequencies or masses dependent on one or more parameters, but whose dynamics can be characterized by the existence of a procedure for the systematic and explicit construction of the constants of motion and the superposition rules. The complete analysis of the Poisson--Hopf deformations of the LH systems based on the 2-photon algebra is currently in progress, to be sent soon for publication.

 \medskip

\item On the other hand, it should be noted that, currently, there is no classification of LH systems for dimension $n \geq 3$. A problem of interest that arises in this context is to analyze the possibility of generating new LH systems, both classical and deformed, by means of the extension of the systems on the plane, in combination with the projections of the realizations of vector fields. In this context, it is known that projections of Lie algebra realizations associated with a linear representation give rise to non-linear realizations. Analyzing the question from the perspective of functional algebras (Hamiltonians), it is conceivable that there exist compatible symplectic structures that give rise to LH systems in higher dimensions, as well as a dependence of the symplectic forms of the representation. Criteria of this type can be combined with quantum deformations of known Lie algebras, in order to obtain new applications of these in the context of differential equations. 

 \medskip
 
\item As a complement to the SISf models based on $\mathfrak{b}_2$ as a subalgebra of $\mathfrak{sl}(2)$, it is natural to construct the corresponding model but considering $\mathfrak{b}_2$ as a subalgebra of the oscillator algebra $\mathfrak{h}_4$. Again, the quite distinct quantum deformation leads to systems with different properties, and both approaches should be compared in detail, analyzing the numerical solutions deduced from both approaches. The first steps in this direction are also currently under scrutiny. 

 \medskip

\item We would also like to extend our study to more complicated compartmental models, although at a first glance we have not been able to identify more Lie systems, at least in their current   form. We suspect that the Hamiltonian description of these compartmental models could nonetheless behave as a Lie system, as it has happened in our presented case. This shall be part of our future endeavors. Moreover, one could inspect in more meticulous detail how the solutions of the quantum-deformed system \eqref{dssis} recover the nondeformed solutions when the introduced parameter tends to zero. We need to further study how this precisely models a heat bath, and if this new integrable system could correspond to other models apart from infectious models. We would like to figure out whether it is possible to modelize subatomic dynamics with the resulting deformed Hamiltonian \eqref{defham}.
 There exists a stochastic theory of Lie systems developed in \cite{Ortega} that could be another starting point to deal with compartmental systems. In the present work we were lucky to find a theory with fluctuations that happened to match a stochastic expansion, but this is rather more of an exception than a rule. Indeed, it seems that the most feasible way to propose stochastic models is using the stochastic Lie theory instead of expecting a glimpse of luck with fluctuations. As stated, finding particular solutions is by no means trivial. The analytic search is a very intricate task. We think that in order to fit particular solutions in the superposition principle, one may need to compute these particular solutions numerically. Some specific numerical methods for particular solutions of Lie systems can be devised in \cite{piet}.  

 \medskip

\item \noindent Finally, starting from the Chebyshev equation, it has been shown that the point of Noether symmetries of this equation can be expressed for arbitrary $n$ in terms of the Chebyshev polynomials $T_{n}(x), U_{n}(x)$ of first and second kind, respectively. 
Moreover, it has been observed that the generic realization of the Lie point symmetry algebra $\frak{sl}(3,\mathbb{R})$ can be enlarged to more general linear homogeneous second-order ODEs, the solutions of which are expressible in terms of trigonometric or hyperbolic functions. In particular, the commutators of the generic point symmetries show that various of the algebraic relations of the general solutions actually arise as a consequence of the symmetry. The same conclusions hold for the structure of the five-dimensional subalgebra of Noether symmetries. The realization of the symmetry generators has been shown to remain valid for differential equations of hypergeometric type, enabling us to obtain realizations of $\frak{sl}(3,\mathbb{R})$ in terms of hypergeometric functions in general and various orthogonal polynomials in particular, such as the Chebyshev or Jacobi polynomials.
 Another remarkable fact emerges from this analysis; namely, that the forcing terms are always independent on the ``velocities" $y_{1}^{\prime},y_{2}^{\prime}$. This is again a consequence of the chosen generic realization, and the question whether other generic realizations in terms of the general solution of the ODE (or system) enable to determine forcing terms that explicitly depend on the derivatives, and even lead to autonomous differential equations (systems), arises naturally. In this context, it would be desirable to obtain a realization of $\frak{sl}(3,\mathbb{R})$ that not only enables to describe generically the point and Noether symmetries of the Jacobi polynomials, but also applies to the differential equations associated to the remaining families of orthogonal polynomials, specifically the Laguerre and Hermite polynomials. This would allow to construct further non-linear equations and systems possessing a subalgebra of Noether symmetries, the generators of which are given in terms of these orthogonal polynomials.

\end{itemize}


\appendix 


\part{Appendix}
\chapter{Lie algebras: Elementary properties} 

\label{AppendixA}
\setcounter{equation}{0}
\renewcommand{\theequation}{A.\arabic{equation}}

\section{Lie algebras}

\noindent In this Appendix, we recall the main structural
properties of Lie algebras used in this work. Details can be found
in \cite{Ste,WY}.

\bigskip

\noindent Given a Lie algebra $\frak{g}$ of dimension $n$ over a
basis $\left\{ X_{1},\cdots,X_{n}\right\}  $ with commutators
\begin{equation}
\left[  X_{i},X_{j}\right]  =C_{ij}^{k}X_{k},\label{la1}%
\end{equation}
the structure constants $\left\{  C_{ij}^{k}\right\}  $ correspond
to the coefficients of a skew-symmetric 2-covariant
1-contravariant tensor $\mu$ defined on the linear space
underlying $\frak{g}$ and satisfying the Jacobi identity. If
$\left\{ \omega^{1},\cdots,\omega^{n}\right\}  $ denotes the dual
basis of 1-forms to $\left\{  X_{1},\cdots,X_{n}\right\}  $, the
Maurer-Cartan equations of $\frak{g}$ are defined as
\begin{equation}
d\omega^{k}=-\frac{1}{2}C_{ij}^{k}\,\omega^{i}\wedge\omega^{j},\;1\leq
i,j,k\leq n.\label{la2}%
\end{equation}
In particular, the Jacobi identity is satisfied if and only of the
2-forms $d\omega^{k}$ are closed \cite{Sat}.

\begin{definition}
The adjoint representation $ad:\frak{g}\rightarrow End\left(
\frak{g}\right)
$ of a Lie algebra $\frak{g}$ is given by%
\begin{equation}
ad\left(  X\right)  \left(  Y\right)  =\left[  X,Y\right]
,\;X,Y\in
\frak{g.}\label{la3}%
\end{equation}
\end{definition}

\begin{definition}
The Killing form $\kappa$ of $\frak{g}$ is the bilinear symmetric
form $\kappa:\frak{g}\times\frak{g}\rightarrow\mathbb{R}$ defined
by
\begin{equation}
\kappa\left(  X,Y\right)  ={\rm Tr}\left(  ad\left(  X\right)
\cdot ad\left(
Y\right)  \right)  ,\;X,Y\in\frak{g.}\label{la4}%
\end{equation}
\end{definition}
\medskip
\noindent In particular, it follows that a Lie algebra $\frak{g}$
is semisimple if and only if $\kappa$ is non-degenerate, i.e.,
$\det\left(  \frak{\kappa}\right) \neq0$. For real Lie algebras,
the signature of the Killing form further determines the
isomorphism class \cite{WY}.

\bigskip

\begin{proposition}
[Levi decomposition] Any Lie algebra $\frak{g}$ admits a
decomposition
\begin{equation}
\frak{g}=\frak{s}\overrightarrow{\oplus}\frak{r},\label{la4a}%
\end{equation}
where $\frak{r}$ is a maximal solvable ideal of $\frak{g}$ and
$\frak{s}\simeq \frak{g}/\frak{r}$ is the maximal semisimple
subalgebra.
\end{proposition}

\bigskip

\noindent The semisimple algebra $\frak{s}$ of (\ref{la4a}) is
usually called the Levi subalgebra of $\frak{g}$.

\medskip
\noindent The Lie algebras $\frak{sl}(2,\mathbb{R})$ and
$\frak{sl}(3,\mathbb{R})$ are semisimple by the preceding
criterion. Moreover, they are simple Lie algebras \cite{SNO}.

\chapter{The hyperbolic sinc function} 

\label{AppendixB} 

\section{ The hyperbolic sinc function}

\setcounter{equation}{0}
\renewcommand{\theequation}{B.\arabic{equation}}

 \bigskip
 
   The hyperbolic counterpart of the  well-known sinc  function is defined by
\be
\shc( x):= \frac{\sinh (x)}{x} = \left\{ 
\begin{array}{ll}
\frac{\sinh (x)}{x}, &\mbox{for}\ x\ne 0, \\
1,&\mbox{for}\  x=0.
\end{array}
 \right. 
 \nonumber
\ee
 The power series around $x=0$ reads
 \be
 \shc( x) = \sum_{n=0}^\infty \frac{x^{2n}}{(2n+1)!} \, .
 \nonumber
 \ee
And its derivative is given by
\be
\frac{\rm d}{{\rm d}x} \shc(x)= \frac{\cosh (x)}{x}- \frac{\sinh (x)}{x^2} =  \frac {\cosh( x) - \shc(x)}{x} \, .
\nonumber
\ee
 Hence the behaviour of $\shc( x)$ and its derivative remind that of the hyperbolic cosine and sine functions, respectively.  
  We represent them  in figure~\ref{fig3}.


\begin{figure}[t]
\begin{center}
\includegraphics[height=6.0cm]{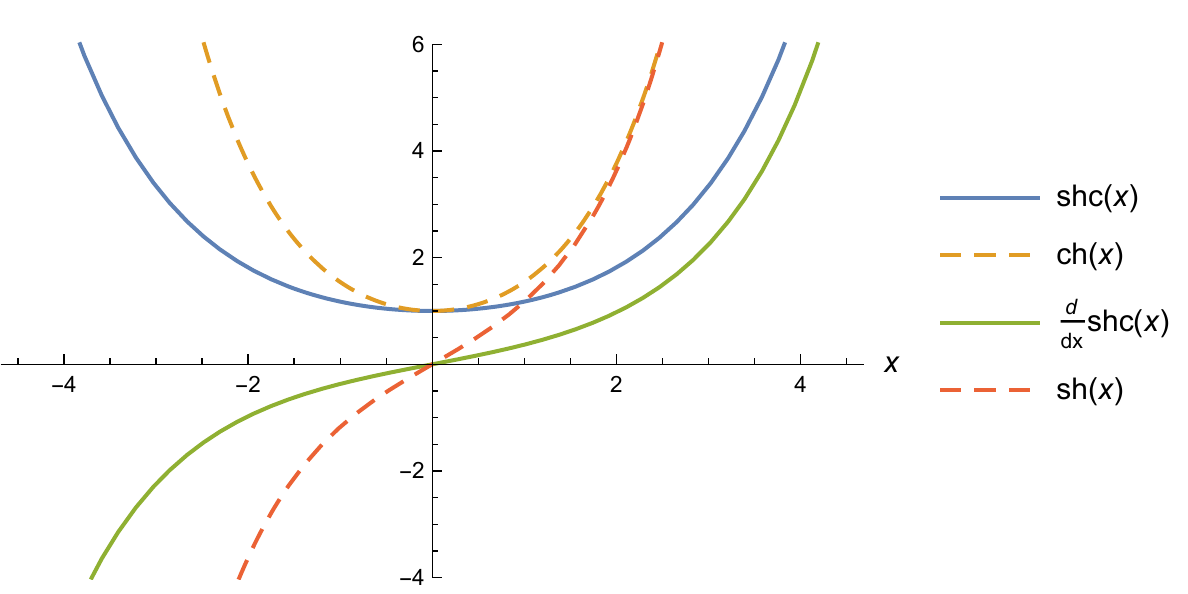}
\caption{\small The hyperbolic sinc function versus the hyperbolic cosine function and the derivative of the former versus the hyperbolic sine function.}
 \label{fig3}
\end{center}
\end{figure}


  A novel relationship of the  $\shc$ function  (and also of the $\sinc$ one) with Lie systems can be established   by considering the following second-order ordinary differential equation
  \be
t\,  \frac{\dd^2 x}{\dd t^2}+ 2 \,  \frac{\dd x}{\dd t}-\eta^2  t\, x =0 ,
\label{diff}
  \ee
  where $\eta$ is a non-zero real parameter. Its general solution can be written as
  \be
  x(t)=A \shc(\eta t)+B\, \frac{\cosh(\eta t)}{t} \,  ,\qquad A,B\in \mathbb{R}.
  \nonumber
  \ee
Notice that if we set $\eta=i \lambda$ with $\lambda\in \mathbb{R}^\ast$ we recover the known result for the sinc function:
  \be
  t\,  \frac{\dd^2 x}{\dd t^2}+ 2 \,  \frac{\dd x}{\dd t}+\lambda^2  t\, x =0 , \qquad x(t)=A \sinc(\lambda t)+B\, \frac{\cos(\lambda t)}{t} \,  .
  \label{Ad}
  \ee
   Next the differential equation (\ref{diff})  can be written as a system of two first-order differential equations by setting $y={\dd x}/{\dd t}$, namely
\be
\frac{\dd x}{\dd t}=y,\qquad \frac{\dd y}{\dd t}=- \frac{2}{t}\, y+ \eta^2 x.
\label{eqLie}
\nonumber
\ee
These equations  determine a Lie system  with  associated  $t$-dependent vector field 
\be
{\bf X}=- \frac 2 t\, {\bf X}_1+ {\bf X}_2 + \eta^2 {\bf X}_3,
\label{Aa}
\ee
where
\begin{equation} 
{\bf X}_1=y\frac{\partial}{\partial y},\qquad {\bf X}_2= y\frac{\partial}{\partial x} ,\qquad {\bf X}_3=  x\frac{\partial}{\partial y} ,\qquad {\bf X}_4=x \frac{\partial}{\partial x}+ y\frac{\partial}{\partial y},
\nonumber
\end{equation}
   fulfill the commutation relations
\begin{equation}
[{\bf X}_1,{\bf X}_2]={\bf X}_2,\qquad [{\bf X}_1,{\bf X}_3]=-{\bf X}_3,\qquad [{\bf X}_2,{\bf X}_3]=2 {\bf X}_1- {\bf X}_4 ,\qquad [{\bf X}_4, \, \cdot \, ]=0 .
\nonumber
\end{equation}
Hence, these  vector fields span a Vessiot--Guldberg Lie algebra  $V$   isomorphic to $\mathfrak{gl}(2)$ with domain   $\mathbb R^2_{x\ne 0}$. 
In fact,  $V$ is diffeomorphic to the class ${\rm I}_7 \simeq \mathfrak{gl}(2)$ of the classification given in~\cite{BBHLS}. The diffemorphism can be explictly performed by means of the change of variables
$u=y/x$ and $v=1/x$,  leading to the vector fields of class ${\rm I}_7$ with domain $\mathbb R^2_{v\ne 0}$ given in~\cite{BBHLS}
$$
{\bf X}_1= u\frac{\partial}{\partial u},\qquad  {\bf X}_2= -u^2\frac{\partial}{\partial u}- u  v \frac{\partial}{\partial v},\qquad
{\bf X}_3=  \frac{\partial}{\partial u},\qquad  {\bf X}_4= - v \frac{\partial}{\partial v} .
$$
Therefore ${\bf X}$ (\ref{Aa}) is a Lie system but not a  LH one since there does not exist any compatible symplectic form  satisfying  (\ref{der}) for  class ${\rm I}_7$  as shown in~\cite{BBHLS}.

\bigskip

Finally, we   point out that the very same result follows  by starting from the differential equation (\ref{Ad}) associated with the sinc function.


\chapter{Orthogonal systems and symmetries of ODEs} 

\label{AppendixC}
\setcounter{equation}{0}
\renewcommand{\theequation}{C.\arabic{equation}}

\section{Point symmetries of ordinary differential equations}

\bigskip

\noindent Symmetries of differential equations can be formulated
by various different approaches, from the classical one by means
of vector fields and their $k^{th}$-order prolongations to their
reformulation in terms of differential forms (see e.g.
\cite{AA,HY,Ol,Lie,WA} and references therein). From the
computational point of view, a rather convenient approach to
determine symmetries of differential equations is based on the
reformulation of the symmetry condition in terms of differential
operators \cite{Ol}. As in the following we adopt this method for
the computation of symmetries, we briefly review the main facts of
the procedure (see e.g. \cite{Ste} for details). It is well known
that a scalar second-order ordinary differential equation
\begin{equation}
y^{\prime\prime}=\omega\left(  x,y,y^{\prime}\right)  \label{SYS}%
\end{equation}
can be formulated in equivalent form in terms of the partial
differential equation (PDE)
\begin{equation}
\mathbf{A}f=\left(  \frac{\partial}{\partial
x}+y^{\prime}\frac{\partial }{\partial y}+\omega\left(
x,y,y^{\prime}\right)  \frac{\partial}{\partial
y^{\prime}}\right)  f=0. \label{VFA}%
\end{equation}
A vector field $X=\xi\left(  x,y\right) \frac{\partial }{\partial
x}+\eta\left(  x,y\right) \frac{\partial}{\partial y}\in
\frak{X}\left( \mathbb{R}^{2}\right)  $ is called a (Lie) point
symmetry generator of the equation (\ref{SYS}) if the prolonged
vector field
\begin{equation}
\dot {X}=X+\left(\frac{d\eta}{dx}-\frac{d\xi}{dx}y^{\prime}\right)
\frac{\partial}{\partial y^{\prime}}\label{SX}
\end{equation}
satisfies the commutator
\begin{equation}
\left[  \dot{X},\mathbf{A}\right]  =-\frac{d\xi}{dx}\mathbf{A}. \label{SB1}%
\end{equation}
We observe in particular that the condition on the prolongation of
the symmetry generator $X$ is automatically given by the
commutator. If we expand the latter, it follows that the only
non-vanishing component is that related to the basic vector field
$\frac{\partial}{\partial y^{\prime}}$. From this we extract the
equation defining the symmetry condition
\begin{align}
& \left(  y^{\prime}\right)^{3}\frac{\partial^{2}\xi}{\partial y^{2}%
}+\left(  y^{\prime}\right)^{2}\left(
2\frac{\partial^{2}\xi}{\partial x\partial
y}-\frac{\partial\xi}{\partial y}\frac{\partial\omega}{\partial
y^{\prime}}-\frac{\partial^{2}\eta}{\partial
y^{2}}\right)+\xi\frac{\partial\omega}{dx}+\eta\frac{\partial\omega}{\partial
y}+\frac{\partial\eta}{\partial x}\frac{\partial\omega}{\partial y^{\prime}%
}-\frac{\partial^{2}\eta}{\partial x^{2}}\nonumber\\
& +\omega\left( 2\frac{\partial\xi}{\partial
x}-\frac{\partial\eta}{\partial y}\right)+y^{\prime}\left(
3\omega\frac{\partial\xi}{\partial
x}+\frac{\partial\omega}{\partial y^{\prime}}\left(
\frac{\partial\eta}{\partial y}-\frac{\partial\xi}{\partial
x}\right)  +\frac{\partial^{2}\xi}{\partial
x^{2}}-2\frac{\partial^{2}\eta }{\partial x\partial
y}\right)=0.\label{SBE}
\end{align}
As the components $\xi$ and $\eta$ of the symmetry generator $X$
do not depend on $y^{\prime}$, the latter equation can be
separated into a system of partial differential equations. In
particular, for a second-order linear homogeneous differential
equation
\begin{equation}
y^{\prime\prime}+g_{1}\left(  x\right)  y^{\prime}+g_{2}\left(
x\right)  y=0,
\label{edo2}%
\end{equation}
we have $\omega=-g_{1}\left(  x\right) y^{\prime}-g_{2}y\left(
x\right) $, and the preceding symmetry condition separates into
the following four partial differential equations (PDEs in short):
\begin{align}
& \frac{\partial^{2}\xi}{\partial y^{2}}=0;\;\left(
2\frac{\partial^{2}\xi }{\partial x\partial y}-2g_{1}\left(
x\right)  \frac{\partial\xi}{\partial
y}-\frac{\partial^{2}\eta}{\partial x\partial y}\right)  =0;\nonumber\\
& \frac{\partial^{2}\xi}{\partial x^{2}}-g_{1}\left(  x\right)
\frac {\partial\xi}{\partial x}-3y\,g_{2}\left(  x\right)
\frac{\partial\xi }{\partial y} -2\frac{\partial^{2}\eta}{\partial
x\partial y}+\frac
{dg_{1}\left(  x\right)  }{dx}\xi=0;\label{SB}\\
& g_{2}\left(  x\right)  \left(  \frac{\partial\eta}{\partial
y}-2\frac {\partial\xi}{\partial x}-\eta\right)  -g_{1}\left(
x\right)  \frac
{\partial\eta}{\partial x}-\frac{\partial^{2}\eta}{\partial x^{2}}%
-y\frac{dg_{2}\left(  x\right)  }{dx}\xi  =0.\nonumber
\end{align}
As the equation (\ref{edo2}) is linear and homogeneous, it is well
known that the symmetry algebra is maximal and isomorphic to the
rank two simple Lie algebra $\frak{sl}\left(  3,\mathbb{R}\right)
$ \cite{AA,Ol,Lie}. For linear homogeneous ordinary differential
equations, the vector field $X_{1}=y\frac{\partial}{\partial y}$
is always a point symmetry. Moreover, if
\begin{equation}
y\left(  x\right)  =\lambda_{1}T\left(  x\right)
+\lambda_{2}U\left(
x\right)  ;\;\lambda_{1},\lambda_{2}\in\mathbb{R} \label{sol1}%
\end{equation}
denotes the general solution of (\ref{edo2}), two additional
independent symmetries of the equation can be chosen as
\begin{equation}
X_{2}=T\left(  x\right)  \frac{\partial}{\partial
y},\;X_{3}=U\left(
x\right)  \frac{\partial}{\partial y}. \label{sol2}%
\end{equation}
These three symmetries satisfy the commutators
\begin{equation}
\left[  X_{1},X_{i}\right]  =-X_{i},\;i=2,3;\;\left[
X_{2},X_{3}\right]  =0,\label{lee}
\end{equation}
therefore span a solvable Lie algebra of type $A_{3,3}$
\cite{Le,SNO}. It is well known that a scalar second-order ODE
admits a Lie algebra of point symmetries of dimension
$n=0,1,2,3,8$ \cite{Lie,Ma}. From the various studies concerning
the structure of linearizable differential equations (see e.g.
\cite{Bag,C13a,Ma,Le} and references therein), it follows that
invariance with respect to the solvable algebra $A_{3,3}$ implies
that the symmetry algebra of the ODE (\ref{edo2}) is maximal,
hence isomorphic to $\frak{sl}\left( 3,\mathbb{R}\right)$. By
means of a local transformation, the ODE can be reduced to the
free particle equation $z^{\prime\prime}=0$.

\bigskip

\noindent For later use, for the general solution (\ref{sol1}) of
equation (\ref{edo2}) we denote the Wronskian $W\left\{ T\left(
x\right) ,U\left( x\right) \right\}  $ as
\begin{equation}
\mathbf{W}=\det\left(
\begin{array}
[c]{cc}%
T\left(  x\right)  & U\left(  x\right) \\
\frac{d}{dt}T\left(  x\right)  & \frac{d}{dt}U\left(  x\right)
\end{array}
\right)  .\label{WR}
\end{equation}

\subsection{Symmetries of the Chebyshev equation}

\noindent Chebyshev polynomials possibly constitute the simplest
case of orthogonal polynomials, and possess various interesting
structural properties that have found extensive application in
numerical analysis \cite{RIV}. In contrast to the other classical
orthogonal polynomials, the Laguerre, Legendre and Hermite
polynomials, which appear in a wide variety of physical problems
and therefore are of considerable importance in the description of
natural phenomena \cite{KAM}, Chebyshev polynomials appear
only marginally (e.g. in connection with the Lissajous figures
\cite{GOL}, although they have found extensive application in
approximation theory and numerical methods \cite{RIV}.

\bigskip

\noindent The Chebyshev polynomials of first and second kind are
defined by
\begin{align}
T_{n}\left(  x\right)   &  =\cos\left(  n\,\arccos x\right)  =\frac{1}%
{2}\left[  \left(  x+i\sqrt{1-x^{2}}\right)  ^{n}+\left(  x-i\sqrt{1-x^{2}%
}\right)  ^{n}\right]  ,\label{TS1}\\
U_{n}\left(  x\right)   &  =\sin\left(  n\,\arccos x\right)  =\frac{1}%
{2i}\left[  \left(  x+i\sqrt{1-x^{2}}\right)  ^{n}-\left(  x-i\sqrt{1-x^{2}%
}\right)  ^{n}\right]  .\label{TS1b}
\end{align}
An alternative formulation of Chebyshev polynomials is given by
means of the functions
\begin{equation}
V_{n}\left(  x\right)  =\cos\left(  n\,\arcsin x\right)
,\;W_{n}\left(
x\right)  =\sin\left(  n\,\arcsin x\right)  \label{TS2}%
\end{equation}
satisfying the following identities:
\begin{equation}%
\begin{array}
[c]{cc}%
V_{2m+1}\left(  x\right)  =\left(  -1\right)  ^{m}U_{2m+1}\left(
x\right) , & V_{2m}\left(  x\right)  =\left(  -1\right)
^{m}T_{2m}\left(  x\right)
,\\
W_{2m+1}\left(  x\right)  =\left(  -1\right)  ^{m}T_{2m+1}\left(
x\right) , & W_{2m}\left(  x\right)  =\left(  -1\right)
^{m}U_{2m}\left(  x\right)  .
\end{array}
\label{TS3}%
\end{equation}
For the weight function $\varphi\left(  x\right)  =\left(  \sqrt{1-x^{2}%
}\right)  ^{-1}$ and the closed interval $\left[-1,1\right]$, the
Chebyshev polynomials satisfy the orthogonality relations
\begin{equation}
\int_{-1}^{1}T_{n}\left(  x\right)  T_{m}\left(  x\right)  \frac{dx}%
{\sqrt{1-x^{2}}}=\left\{
\begin{array}
[c]{cl}%
0 & m\neq n\\
\frac{\pi}{2} & m=n\neq0\\
\pi &  m=n=0
\end{array}
\right.  \label{TS4}%
\end{equation}
and
\begin{equation}
\int_{-1}^{1}U_{n}\left(  x\right)  U_{m}\left(  x\right)  \frac{dx}%
{\sqrt{1-x^{2}}}=\left\{
\begin{array}
[c]{cl}%
0 & m\neq n\\
\frac{\pi}{2} & m=n\neq0\\
0 & m=n=0
\end{array}
\right.  .\label{TS5}%
\end{equation}
\smallskip
\noindent The Chebyshev polynomials can also be constructed by
means of an iterative process using the Rodrigues formula (see
e.g. \cite{CHI,SZ}). For any $n\geq 0$, $T_{n}(x)$ and $U_{n}(x)$
are respectively given by a $n^{th}$-order differential operator
\begin{align}
T_{n}\left(  x\right)   &  =\left(  -1\right)
^{n}2^{n}\frac{n!}{\left( 2n\right)
!}\sqrt{1-x^{2}}\frac{\mathrm{d}^{n}}{\mathrm{d\,x}^{n}}\left[
1-x^{2}\right]  ^{n-\frac{1}{2}},\label{TSu1}\\
U_{n}\left(  x\right)   &  =\left(  -1\right)
^{n-1}2^{n}\frac{n!\,n}{\left( 2n\right)
!}\frac{\mathrm{d}^{n-1}}{\mathrm{d\,x}^{n-1}}\left[
1-x^{2}\right]  ^{n-\frac{1}{2}}.\label{TSu2}
\end{align}
From the latter identities we can easily deduce the relations
\begin{equation}
\frac{d}{dx}T_{n}\left(  x\right)
=\frac{n}{\sqrt{1-x^{2}}}U_{n}\left(
x\right)  ;\;\frac{d}{dx}U_{n}\left(  x\right)  =-\frac{n}{\sqrt{1-x^{2}}%
}T_{n}\left(  x\right)  ,\label{TS11}%
\end{equation}
from which we get
\begin{equation}
\frac{d}{dx}T_{n}\left(  x\right)  \frac{d}{dx}U_{n}\left(
x\right)
+n\frac{T_{n}\left(  x\right)  U_{n}\left(  x\right)  }{1-x^{2}}=0\label{TS12}%
\end{equation}
for any $n$. Using these relations, it is straightforward to
verify that the polynomials $T_{n}\left( x\right) $ and
$U_{n}\left( x\right)  $ are independent solutions of the linear
homogeneous second-order ODE
\begin{equation}
\left(  1-x^{2}\right)  y^{\prime\prime}-x\,y^{\prime}+n^{2}y=0.\label{TS6}%
\end{equation}
As observed in the previously, three of the symmetry generators
are immediate:
\begin{equation}
X_{1}=y\frac{\partial}{\partial y},\;X_{2}=T_{n}\left(  x\right)
\frac{\partial}{\partial y},\;X_{3}=U_{n}\left(  x\right)
\frac{\partial
}{\partial y}.\label{TS7}%
\end{equation}

\bigskip

\noindent  We conclude that the Chebyshev equation (\ref{TS6}) has
maximal symmetry $\mathcal{L}\simeq \frak{sl}\left(
3,\mathbb{R}\right)$, and hence constitutes a linearizable
equation. Discarding the case $n=0$ for being reducible, if we
compute the point symmetries for equation (\ref{TS6}) and $n=1$,
we find that the symmetries (\ref{TS7}), together with the
following vector fields, form a basis of
$\mathcal{L}$:\footnote{This specific realization differs from
that considered in \cite{EDU}.}
\begin{align}
X_{4} &  =\sqrt{1-x^{2}}y\left(  x\frac{\partial}{\partial
x}+y\frac{\partial }{\partial y}\right)  ;\,X_{5}=\left(
x^{2}-1\right)  y\frac{\partial }{\partial
x}+y^{2}x\frac{\partial}{\partial y};\,X_{6}=x\sqrt{1-x^{2}}\left(
x\frac{\partial}{\partial x}+y\frac{\partial}{\partial y}\right);\nonumber\\
X_{7} &  =x\left(  x^{2}-1\right)  \frac{\partial}{\partial
x}+\left( x^{2}+1\right)  y\frac{\partial}{\partial
y};\;X_{8}=\sqrt{1-x^{2}}\left( \left(  x^{2}-1\right)
\frac{\partial}{\partial x}+yx\frac{\partial}{\partial
y}\right).\label{TARF}
\end{align}

\noindent As $T_{1}(x)=x$ and $U_{1}(x)=\sqrt{1-x^{2}}$, it
follows at once that the symmetry generators (\ref{TARF}) can all
be expressed in terms of the Chebyshev polynomials for $n=1$,
using appropriately the relations (\ref{TSu1})-(\ref{TSu2}) and
those derived from them. It is therefore natural to ask whether
this realization for the Lie algebra $\frak{sl}(3,\mathbb{R})$ can
be modified in order to describe the point symmetries of the
Chebyshev equation for arbitrary $n$. The answer, which is in the
affirmative, will be proven to remain valid for differential
equations having solutions of trigonometric and hyperbolic types.

\bigskip
\bigskip

\section{Functional realization of $\frak{sl}(3,\mathbb{R})$}

\bigskip

\noindent As follows from equation (\ref{TS1}), the Chebyshev
polynomials constitute a particular case of trigonometric
functions of the type
\begin{equation}
y\left(  x\right)  =\lambda_{1}\sin H\left(  x\right)
+\lambda_{2}\cos H\left(  x\right),\label{GE1}%
\end{equation}
with $H\left(  x\right)  $ being an arbitrary differentiable
function and $\lambda_{1},\lambda_{2}\in\mathbb{R}$. Functions of
the form (\ref{GE1}) can be shown to be solutions to the linear
second-order homogeneous equation
\begin{equation}
\frac{dy^{2}}{dx^{2}}-\left(  \frac{\frac{d^{2}H}{dx^{2}}}{\frac{dH}{dx}%
}\right)  \frac{dy}{dx}+\left(  \frac{dH}{dx}\right)  ^{2}y=0.\label{GE2}%
\end{equation}
In analogy with the previous example, it is reasonable to ask
whether for this equation, that also exhibits maximal
$\frak{sl}\left( 3,\mathbb{R}\right) $-symmetry, the symmetry
generators can be described generically in terms of the
fundamental solutions $T\left( x\right)  =\sin H\left(  x\right) $
and $U\left( x\right) =\cos H\left( x\right) $. Making the
substitution $g\left( x\right) =\left( \frac{dH}{dx}\right) ^{2}$,
the ODE\ (\ref{GE2}) transforms onto
\begin{equation}
\frac{dy^{2}}{dx^{2}}-\frac{g^{\prime}\left(  x\right)  }{2g\left(
x\right)
}\frac{dy}{dx}+g\left(  x\right)  y=0.\label{GE3}%
\end{equation}
Skipping the assumption that $g\left(  x\right)  $ is obtained
from the derivative of $H\left(  x\right)  $, we can formulate the
symmetry problem for the more general ODE (\ref{GE3}). Without
loss of generality, we can suppose that the general solution of
this equation is given by  $y\left(  x\right) =\lambda_{1}T\left(
x\right)  +\lambda_{2}U\left(  x\right)  ,$ where $T\left(
x\right)  $ and $U\left(  x\right)  $ are two independent
solutions.

\bigskip

\begin{proposition}
For arbitrary functions $g\left(  x\right)\neq 0$, the vector fields%
\begin{equation}
X_{4}=-\frac{T^{\prime}\left(  x\right)  }{g\left(  x\right)
}y\frac {\partial}{\partial x}+T\left(  x\right)
y^{2}\frac{\partial}{\partial
y};\quad X_{5}=-\frac{U^{\prime}\left(  x\right)  }{g\left(  x\right)  }%
y\frac{\partial}{\partial x}+U\left(  x\right)
y^{2}\frac{\partial}{\partial
y}\label{GE4}%
\end{equation}
are point symmetries of (\ref{GE3}).
\end{proposition}

\bigskip

\begin{proof}
Let $T\left(  x\right)  $ be a solution of the ODE (\ref{GE3}).
Denoting $\frac{dT}{dx}=T^{\prime}\left(  x\right)  $, the
prolongation $\dot{X}_{4}$ of the vector field $X_{4}$ is
explicitly given by
\begin{flalign}
\dot{X}_{4}=-\frac{T^{\prime}\left(  x\right)  }{g\left(  x\right)  }%
y\frac{\partial}{\partial x}+T\left(  x\right)
y^{2}\frac{\partial}{\partial y}+\left(  y^{2}T^{\prime}\left(
x\right)  +yy^{\prime}\frac{T\left( x\right)  g\left(  x\right)
-T^{\prime}\left(  x\right)  g^{\prime}\left( x\right)
}{g^{2}\left(  x\right)  }+y^{\prime2}\frac{T^{\prime}\left(
x\right)  }{g\left(  x\right)  }\right)  \frac{\partial}{\partial y^{\prime}%
}.
\end{flalign}
We further define the quantity $R=\left(
\frac{d^{2}T}{dx^{2}}-\frac{g^{\prime}\left(  x\right)
}{2\,g\left(  x\right)  }\frac{dT}{dx}+g\left(  x\right)  T\left(
x\right) \right)  $, which reduces to zero as $T\left( x\right) $
solves the equation. If we now evaluate the commutator
(\ref{SB1}), after some simplification we obtain the following
expression for the symmetry condition:
\begin{equation}
\left(  \frac{-2y^{\prime2}}{g\left(  x\right)
}+\frac{yy^{\prime}g^{\prime }\left(  x\right)  }{g^{2}\left(
x\right)  }+y^{2}\right)  \,R-\frac {yy^{\prime}}{g\left(
x\right)  }\frac{d\,R}{dx}=0,
\end{equation}
showing that $X_{4}$ is a point symmetry. Permuting $T\left(
x\right)  $ and $U\left(  x\right)  $, the same argument shows
that $X_{5}$ is also a symmetry of the ODE.
\end{proof}

\bigskip

\noindent Clearly the vector fields $\left\{X_{1},\cdots
,X_{5}\right\}$ are independent, as their component in
$\frac{\partial}{\partial y}$ depends on different powers of $y$
and $U(x),T(x)$ are independent solutions of the ODE. In order to
complete a basis of symmetries, we use the fact that the
commutator of two point symmetries is a point symmetry
\cite{AA,Dik}. To this extent, we consider the additional vector
fields
\begin{eqnarray}
X_{6}  =\left[  X_{2},X_{4}\right]  = & -\frac{T\left(  x\right)
U^{\prime
}\left(  x\right)  }{g\left(  x\right)  }\frac{\partial}{\partial x}%
+\frac{2T\left(  x\right)  U\left(  x\right)g(x)
+T^{\prime}\left( x\right) U^{\prime}\left(  x\right)  }{g\left(
x\right) }y\frac{\partial}{\partial
y},\nonumber\\
X_{7} =\left[  X_{2},X_{5}\right]  = & -\frac{T\left(  x\right)
T^{\prime
}\left(  x\right)  }{g\left(  x\right)  }\frac{\partial}{\partial x}%
+\frac{2T^{2}\left(  x\right)g(x)  +\left(  T^{\prime}\left(
x\right) \right)
^{2}}{g\left(  x\right)  }y\frac{\partial}{\partial y},\label{GE5}\\
X_{8} =\left[  X_{3},X_{5}\right]  = & -\frac{T^{\prime}\left(
x\right)
U\left(  x\right)  }{g\left(  x\right)  }\frac{\partial}{\partial x}%
+\frac{2T\left(  x\right)  U\left(  x\right)g(x)
+T^{\prime}\left( x\right) U^{\prime}\left(  x\right)  }{g\left(
x\right) }y\frac{\partial}{\partial y}\nonumber
\end{eqnarray}
We observe that, according to the ODE (\ref{GE3}), the solutions
$T(x)$ and $U(x)$ are functionally related through $g(x)$, i.e.,
\begin{equation}
g(x)=\frac{T^{\prime}(x)}{C_{1}-T^{2}(x)}=\frac{U^{\prime}(x)}{C_{2}-U^{2}(x)}
\end{equation}
for some constants $C_{1},C_{2}$. As $T(x)$ and $U(x)$ do not
themselves reduce to constants, a routine but cumbersome
computation shows that $\left\{X_{6},X_{7},X_{8}\right\}$ are
linearly independent vector fields. As a consequence, $\left\{
X_{1},\cdots,X_{8}\right\}$ are linearly independent and can be
taken as a basis of the symmetry algebra $\mathcal{L}$ of the
differential equation (\ref{GE3}). We will see that this
realization describes the symmetries of the ODE for arbitrary
choices of $g(x)$, always yielding the same commutators.

\bigskip

\noindent We now proceed to compute the structure constants of
$\frak{sl}(3,\mathbb{R})$ over the preceding basis. Up to now, the
only known commutators are the following:
\begin{equation}%
\begin{array}
[c]{llll}%
\left[  X_{1},X_{2}\right]  =-X_{2}, & \left[  X_{1},X_{3}\right]
=-X_{3}, &
\left[  X_{1},X_{4}\right]  =X_{4}, & \left[  X_{1},X_{5}\right]  =X_{5},\\
\left[  X_{1},X_{6}\right]  =0, & \left[  X_{1},X_{7}\right]  =0,
& \left[
X_{1},X_{8}\right]  =0, & \left[  X_{2},X_{3}\right]  =0,\\
\left[  X_{2},X_{4}\right]  =X_{6}, & \left[  X_{2},X_{5}\right]
=X_{7}, & \left[  X_{3},X_{5}\right]  =X_{8}. &
\end{array}
\label{LB1}%
\end{equation}
We observe in particular that $X_{1}$ acts diagonally on the
remaining generators, hence it must belong to the Cartan
subalgebra $\frak{h}$ of $\frak{s}l\left(
3,\mathbb{R}\right)$.\footnote{For the basic definitions on Lie
algebras, see \cite{SNO,WY}.} Further, as $\frak{h}$ is an Abelian
subalgebra of $\frak{sl}\left(  3,\mathbb{R}\right)  $, the
brackets (\ref{LB1}) imply that a second generator of $\frak{h}$
must be a linear combination of $X_{6},X_{7}$ and $X_{8}$.
However, as the functions $g(x)$, $U\left(  x\right)  $ and
$T\left(  x\right)  $ are unknown, we ignore the coefficients of
the remaining commutators. One possibility to circumvent this
difficulty is to forget provisionally that $\left\{
X_{1},\cdots,X_{8}\right\} $ arise as symmetries of an ODE and
focus only on the algebraic problem. We suppose that $\left\{
X_{1},\cdots,X_{8}\right\}  $ are independent generators of an
eight-dimensional Lie algebra, and that the commutators
(\ref{LB1}) hold. Using the Jacobi condition, we can derive a
parameterized expression for the remaining brackets, or
alternatively we can compute the Maurer--Cartan equations
associated to this basis \cite{Sat}. Proceeding like this, it
follows that $\left\{ X_{1},\cdots,X_{8}\right\}  $ define a Lie
algebra if the
following conditions hold:%

\begin{flalign}%
\begin{tabular}
[c]{lll}%
$\left[  X_{2},X_{6}\right]  =2a_{1}a_{2}X_{2},$ & $\left[  X_{2}%
,X_{7}\right]  =2a_{1}X_{2},$ & $\left[  X_{2},X_{8}\right]  =a_{1}a_{2}%
X_{2}+a_{1}X_{3},$\\
\multicolumn{2}{l}{$\left[  X_{3},X_{4}\right]  =\alpha X_{1}+a_{2}X_{6}%
+a_{3}X_{7}+a_{2}X_{8},$} & $\left[  X_{3},X_{6}\right]
=a_{1}\left(
a_{2}X_{3}-a_{3}X_{2}\right)  ,$\\
\multicolumn{2}{l}{$\left[  X_{3},X_{7}\right]  =a_{1}\left(  a_{2}X_{2}%
+X_{3}\right)  ,$} & $\left[  X_{3},X_{8}\right]  =2a_{1}a_{2}X_{3},$\\
$\left[  X_{4},X_{5}\right]  =0,\;$ & $\left[  X_{4},X_{6}\right]
=-2a_{1}a_{2}X_{4},$ & $\left[  X_{4},X_{7}\right]  =-a_{1}\left(  X_{4}%
+a_{2}X_{5}\right)  ,$\\
\multicolumn{2}{l}{$\left[  X_{4},X_{8}\right]  =a_{1}\left(  a_{3}X_{5}%
-a_{2}X_{4}\right)  ,$} & $\left[  X_{5},X_{6}\right]
=-a_{1}\left(
X_{4}+a_{2}X_{5}\right)  ,$\\
$\left[  X_{5},X_{7}\right]  =-2X_{5},$ & $\left[
X_{5},X_{8}\right]
=-2a_{1}a_{2}X_{5},$ & $\left[  X_{6},X_{7}\right]  =a_{1}\left(  X_{6}%
-a_{2}X_{7}\right)  ,$\\
\multicolumn{2}{l}{$\left[  X_{6},X_{8}\right]  =a_{1}\left(
\alpha
X_{1}+a_{2}X_{6}+2a_{3}X_{7}+a_{2}X_{8}\right)  ,$} & $\left[  X_{7}%
,X_{8}\right]  =a_{1}\left(  X_{8}-a_{2}X_{7}\right)  ,$%
\end{tabular}
\label{LB2}%
\end{flalign}
where $a_{1},a_{2},a_{3}\in\mathbb{R}$ are arbitrary constants and
$\alpha=-3a_{1}\left(  a_{3}+a_{2}^{2}\right)  $. As the Lie
algebra must be isomorphic to $\frak{s}l\left( 3,\mathbb{R}\right)
$, its Killing form $\kappa$ must be non-degenerate \cite{Sat}. A
routine computation shows that $\kappa$ has the following
determinant:
\begin{equation}
\det\kappa=-559872\,a_{1}\left(  a_{3}+a_{2}^{2}\right)  ^{4}.\label{LB3}%
\end{equation}
Thus, if $\kappa$ is non-degenerate, then we must always have
$a_{1}\neq0$ and $a_{3}\neq-a_{2}^{2}$. As $a_{1}\neq0$, a change
of scale always allows us to suppose that $a_{1}=1$.

\bigskip

\noindent We now return to the interpretation of $\left\{
X_{1},\cdots,X_{8}\right\}  $ as point symmetries of (\ref{GE3})
in the realization (\ref{GE4})-(\ref{GE5}). In order to satisfy
the commutators (\ref{LB2}), the functions $T\left(  x\right) ,
U\left(  x\right)  $ and their derivatives $T^{\prime}\left(
x\right), U^{\prime}\left(  x\right)  $ will have to satisfy
certain supplementary constraints that depend on the specific
values of $a_{2}$ and $a_{3}$. One one hand, such constraints will
enable us to deduce relations between $T(x)$, $U(x)$ and further
$g(x)$, in order to appear as the solutions of the differential
equation (\ref{GE3}). On the other hand, we will derive the
admissible values of $a_{2}$ and $a_{3}$ for which the commutators
are compatible with the generators being realized as vector
fields.\footnote{In other words, the values for which the vector
fields define a realization of the Lie algebra
$\frak{sl}(3,\mathbb{R})$. Details can be found in \cite{GKO}.}

\bigskip

\noindent Developing formally the commutator of $X_{2}$ and
$X_{6}$, imposition of the identity
$\left[X_{2},X_{6}\right]+2a_{2}X_{2}=0$ forces the functional
relation
\begin{equation}
a_{2}T^{\prime}(x)+T^{2}(x)U^{\prime}(x)-U^{\prime}(x)-T(x)U(x)T^{\prime}(x)=0.\label{rey1}
\end{equation}
It is not difficult to see that this equation admits an
integrating factor, enabling us to rewrite (\ref{rey1}) as
\begin{equation}
\frac{d}{d\;x}\left(\left(U(x)-a_{2}T(x)\right)\left(T^{2}(x)-1\right)^{-\frac{1}{2}}+\beta\right)=0\label{rey2}
\end{equation}
for some constant $\beta$. As a consequence, we obtain that
\begin{equation}
U(x)=a_{2}T(x)-\beta\left(T^{2}(x)-1\right)^{\frac{1}{2}},\label{rey3}
\end{equation}
where $\beta\neq 0$ as $\mathbf{W}\neq 0$. Now the commutator
$\left[X_{2},X_{7}\right]+X_{2}=0$ is satisfied if
\begin{equation}
(T^{\prime}(x))^2+(T^{2}(x)-1)g(x)=0\label{rey4}
\end{equation}
holds. Analyzing now the bracket $\left[X_{3},X_{4}\right]$ we
obtain, after some simplification, the numerical relation
\begin{equation}
\beta^2-a_{3}-a_{2}^{2}=0.\label{KRE}
\end{equation}
With these conditions, the only remaining commutator that still
imposes a constraint is $\left[X_{6},X_{8}\right]$. Developing the
latter leads to the condition
\begin{equation}
a_{2}\left(T(x)^2-1\right)\left(a_{2}T(x)-\sqrt{a_{3}+a_{2}^{2}}\sqrt{T^{2}(x)-1}\right)=0.\label{rey5}
\end{equation}
As the solution $T(x)={\rm cons.}\in\mathbb{R}$ is excluded by the
previous conditions (as otherwise the Wronskian is
$\mathbf{W}=0$), we necessarily have that $a_{2}=0$. In
particular, we have that the admissible functions for which the
realization (\ref{GE4})-(\ref{GE5}) are point symmetries of the
ODE (\ref{GE3}) are given by
\begin{eqnarray}
g(x) = &
\left(T^{\prime}(x)\right)^{2}\left(1-T^{2}(x)\right)^{-1},
\\\label{rey6a}
U(x) = & \sqrt{a_{3}}\sqrt{T^{2}-1}. \label{rey6b}
\end{eqnarray}
As a consequence, the squares of the functions $U(x)$ and $T(x)$
are related by
\begin{equation}
U^{2}(x)-a_{3}T^{2}(x)+a_{3}=0.\label{rey7}
\end{equation}
We observe in particular that, in addition, the identity
\begin{equation}
g(x)T(x)U(x)+T^{\prime}(x)U^{\prime}(x)=0\label{rey8}
\end{equation}
linking $g(x)$ with the solution of the ODE is satisfied.

\bigskip

\noindent In view of this result, we will essentially obtain two
types of functions (trigonometric and hyperbolic) for which the
symmetries are given by (\ref{GE4})-(\ref{GE5}), depending on the
sign of the parameter $a_{3}$. From (\ref{LB2}), we obtain that
the commutator table for the point symmetry algebra $\mathcal{L}$
and the given realization is the following:\footnote{As the Lie
bracket is skew-symmetric, we only display the commutators
$\left[X_{i},X_{j}\right]$ for $i<j$.}

\begin{flalign}%
\begin{tabular}
[c]{c|cccccccc}%
$\left[  \cdot,\cdot\right]  $ & $X_{1}$ & $X_{2}$ & $X_{3}$ &
$X_{4}$ & $X_{5}$ & $X_{6}$ & $X_{7}$ & $X_{8}$\\\hline
$X_{1}$ & $0$ & $-X_{2}$ & $-X_{3}$ & $X_{4}$ & $X_{5}$ & $0$ & $0$ & $0$\\
$X_{2}$ &  & $0$ & $0$ & $X_{6}$ & $X_{7}$ & $0$ & $2X_{2}$ & $X_{3}$\\
$X_{3}$ &  &  & $0$ & $3a_{3}X_{1}+a_{3}X_{7}$ & $X_{8}$ & $-a_{3}X_{2}$ & $X_{3}$ & $0$\\
$X_{4}$ &  &  &  & $0$ & $0$ & $0$ & $-X_{4}$ & $a_{3}X_{5}$\\
$X_{5}$ &  &  &  &  & $0$ & $-X_{4}$ & $-2X_{5}$ & $0$\\
$X_{6}$ &  &  &  &  &  & $0$ & $X_{6}$ & $-3a_{3}X_{1}+2a_{3}X_{7}$\\
$X_{7}$ &  &  &  &  &  &  & $0$ & $X_{8}$\\
$X_{8}$ &  &  &  &  &  &  &  & $0$%
\end{tabular}
\label{TCALO}%
\end{flalign}

\bigskip
\bigskip

\subsection{Noether symmetries}

\bigskip

\noindent The differential equation (\ref{GE3}) can be seen as the
equation of motion of a particle in one dimension. As such systems
are always integrable and conservative (see e.g.
\cite{HAV,PER,SAN}), it follows that there exists a Lagrangian
$L(x,y,y^{\prime})$ such that (\ref{GE3}) arises as the Lagrange
equation of second kind
\begin{equation}
\frac{d}{d\;x}\left(\frac{\partial\; L}{\partial\;
y^{\prime}}\right)-\frac{\partial\; L}{\partial\;
y}=0.\label{rey11}
\end{equation}
In particular, as the ODE (\ref{GE3}) can be reduced to the free
particle equation $z^{\prime\prime}(s)=0$ by means of a local
transformation \cite{AA,OV}, it follows that the symmetry algebra
$\mathcal{L}$ must contain a five-dimensional subalgebra
$\mathcal{L}_{NS}$ corresponding to Noether symmetries
\cite{STEE}.

\bigskip

\noindent Recall that a  point symmetry $X=\xi(x,y)\frac{\partial
}{\partial x}+\eta(x,y)\frac{\partial }{\partial y}$ is called a
Noether symmetry if there exists a function $V\left(  x,y\right) $
such that the condition
\begin{equation}
\dot{X}\left(  L\right)  +A\left(  \xi\right)  L-A\left(  V\right)
=0\label{Noe3}%
\end{equation}
is satisfied. As a consequence, the function
\begin{equation}
\psi=\xi(x,y)\left[  y^{\prime}\frac{\partial L}{\partial
y^{\prime}}-L\right]
-\eta(x,y)\frac{\partial L}{\partial y^{\prime}}+V(x,y) \label{Noe4}%
\end{equation}
will be a constant of the motion of the system \cite{GOL,SAN}.

\bigskip

\noindent For the Lagrangian defined as
\begin{equation}
L\left(x,y,y^{\prime}\right)=\frac{1}{2\sqrt{g(x)}}\left((y^{\prime})^{2}-g(x)y^{2}\right),\label{rey12}
\end{equation}
the equation of motion is equivalent to the differential equation
(\ref{GE3}), so that, without loss of generality, we can suppose
that $L$ is the Lagrangian of the system.

\bigskip

\noindent Evaluating the symmetry condition (\ref{Noe3}) for $L$
and separating the resulting expression into powers of
$y^{\prime}$, we obtain the system of PDEs for the components of a
Noether symmetry:
\begin{flalign}
\frac{\partial\xi}{\partial y}   =0,\quad \xi\,\left(  x,y\right)
g^{\prime }\left(  x\right)  -4g\left(  x\right)
\frac{\partial\eta}{\partial
y}+2g\left(  x\right)  \frac{\partial\xi}{\partial x}=0,\\
2g^{2}\left(  x\right)  \frac{\partial\xi}{\partial y}+-4g\left(
x\right)
\frac{\partial\eta}{\partial x}+4g\left(  x\right)  ^{\frac{3}{2}}%
\frac{\partial V}{\partial y}  =0,\\
\xi\left(  x,y\right)  g\left(  x\right)  g^{\prime}\left(
x\right) y^{2}+4g^{2}\left(  x\right)  y\,\eta\left(  x,y\right)
+2g^{2}\left( x\right)  y^{2}\frac{\partial\xi}{\partial
x}+4g\left(  x\right)  ^{\frac {3}{2}}\frac{\partial V}{\partial
x}  =0.\label{4gl}
\end{flalign}
The first condition implies that $\xi\left(  x,y\right)
=\varphi\left( x\right)  $. Inserting this into the second
equation further shows that $\eta\left(  x,y\right)  $ satisfies
the equation
\begin{equation}
\frac{\partial\eta}{\partial y}=\frac{1}{4}\left(
\frac{\varphi^{\prime
}\left(  x\right)  g^{\prime}\left(  x\right)  }{g\left(  x\right)  }%
+2\varphi^{\prime}\left(  x\right)  \right)
\end{equation}
with solution $\eta\left(  x,y\right)  =\frac{1}{4}\left(  \frac
{\varphi^{\prime}\left(  x\right)  g^{\prime}\left(  x\right)
}{g\left( x\right)  }+2\varphi^{\prime}\left(  x\right)  \right)
y+\theta\left( x\right)  $. Therefore, the generic form of a
Noether symmetry is given by
\begin{equation}
X=\varphi\left(  x\right)  \frac{\partial}{\partial x}+\left(  \frac{1}%
{4}\left(  \frac{\varphi\left(  x\right)  g^{\prime}\left(
x\right) }{g\left(  x\right)  }+2\varphi^{\prime}\left(  x\right)
\right)
y+\theta\left(  x\right)  \right)  \frac{\partial}{\partial y}.\label{la7}%
\end{equation}

\noindent Reordering the terms and simplifying, the third equation
can be brought to the form
\begin{align}
& 4g\left(  x\right)  ^{\frac{5}{2}}\frac{\partial V}{\partial y}
+y\left[  \varphi\left(  x\right) \left(  g^{\prime}\left(
x\right)  ^{2}-g\left(  x\right)  g^{\prime\prime }\left( x\right)
\right)  -g\left(  x\right)  \left(  2g\left(  x\right)
\varphi^{\prime\prime}\left(  x\right)  +g^{\prime}\left( x\right)
\varphi^{\prime}\left(  x\right)  \right)  \right]  \nonumber\\
& -4g^{2}\left( x\right)  \theta^{\prime}\left(  x\right)=0, \label{la9}%
\end{align}
from which the expression for $V(x,y)$ is obtained as
\begin{align}
& V\left(  x,y\right)  =-\frac{1}{8}\frac{\left[ \varphi\left(
x\right) \left(  g^{\prime}\left(  x\right)  ^{2}-g\left( x\right)
g^{\prime\prime}\left(  x\right)  \right)  -g\left(  x\right)
\left(  2g\left(  x\right)  \varphi^{\prime\prime}\left(  x\right)
+g^{\prime}\left(  x\right)  \varphi^{\prime}\left(  x\right)
\right) \right]  }{g\left(  x\right)  ^{\frac{5}{2}}}y^{2}+h\left(
x\right)\nonumber \\
& +\frac{g^{2}\left(  x\right) \theta^{\prime}\left( x\right)
}{g\left(  x\right) ^{\frac{5}{2}}}y.\label{la10}%
\end{align}
Inserting $\xi\left(  x,y\right)  $, $\eta\left(  x,y\right)  $
and $V\left( x,y\right)  $ into the equation (\ref{4gl}) and
simplifying the resulting expression, we finally obtain the
conditions to be satisfied by $\varphi\left(  x\right)  $ and
$\theta\left( x\right)  $ in order to define a Noether symmetry:
\begin{flalign}
\theta\left(  x\right)  g\left(  x\right)
-\frac{1}{2}\frac{g^{\prime}\left( x\right)  }{g\left(  x\right)
}\theta^{\prime}\left(  x\right)
+\theta^{\prime\prime}\left(  x\right)   &  =0,\\
4g\left(  x\right)  ^{2}\varphi^{\prime\prime\prime}\left(
x\right)
+\varphi^{\prime}\left(  x\right)  \left(  4g\left(  x\right)  ^{2}%
g^{\prime\prime}\left(  x\right)  +16g\left(  x\right)
^{4}-5g\left(
x\right)  g^{\prime}\left(  x\right)  ^{2}\right)  + &\nonumber\\
\varphi\left(  x\right)  \left(  2g\left(  x\right)
^{2}g^{\prime\prime \prime}\left(  x\right)  +5g^{\prime}\left(
x\right)  ^{3}-7g\left( x\right)  g^{\prime}\left(  x\right)
g^{\prime\prime}\left(  x\right) +8g\left(  x\right)
^{3}g^{\prime}\left(  x\right)  \right)   &  =0.\label{eds}
\end{flalign}

\noindent Now, as any Noether symmetry of (\ref{GE3}) must be a
linear combination of $\left\{ X_{1},\cdots,X_{8}\right\}  $, we
conclude from (\ref{GE4}) and (\ref{GE5}) that
\begin{equation}
X=k_{1}X_{1}+k_{2}X_{2}+k_{3}X_{3}+k_{6}X_{6}+k_{7}X_{7}+k_{8}X_{8}%
,\label{la8}%
\end{equation}
as only these symmetries are at most linear in the variable $y$,
being thus compatible with the form (\ref{la7}). Now $X_{2}$ and
$X_{3}$ are clearly Noether symmetries if we take either
$\theta\left( x\right) =T\left(  x\right)  $ or $\theta\left(
x\right)  =U\left( x\right)  $. As a consequence, the three
remaining Noether symmetries must be a linear combination of
$X_{1},X_{6},X_{7}$ and $X_{8}$. Instead of inspecting the
preceding equation (\ref{eds}) for any arbitrary linear
combinations, we check whether $X_{6}$ or $X_{8}$ as given in
(\ref{GE5}) satisfy the condition. Take for instance
$X_{6}$.\footnote{Again, permuting $T(x)$ and $U(x)$, the analysis
is extensible to the vector field $X_{8}$.}

\noindent Using the constraint (\ref{rey8}) and the fact that
$T\left( x\right), U\left(  x\right)  $ are independent solutions
of the differential equation (\ref{GE3}), the expansion of
condition (\ref{Noe3}) applied to $X_{6}$ reduces to
\begin{align}
\dot{X}_{6}\left(  L\right)  +A\left(  \xi\right)  L-A\left(
V\right)    & =y^{\prime}\left(  \frac{U\left(  x\right)
T^{\prime}\left(  x\right)
+T\left(  x\right)  U^{\prime}\left(  x\right)  }{\sqrt{g\left(  x\right)  }%
}y-\frac{\partial V}{\partial y}\right) \nonumber\\
& -\frac{y^{2}}{2\sqrt{g\left(  x\right)  }}\left(
T^{\prime}\left(  x\right) U^{\prime}\left(  x\right)  -5g\left(
x\right)  T\left(  x\right)  U\left( x\right)  \right)
-\frac{\partial V}{\partial x}.\label{rey13b}
\end{align}
From the term in $y^{\prime}$ we obtain the auxiliary function
\begin{equation}
V\left(  x,y\right)  =\frac{U\left(  x\right)  T^{\prime}\left(
x\right)
+T\left(  x\right)  U^{\prime}\left(  x\right)  }{2\sqrt{g\left(  x\right)  }%
}y^{2}+h\left(  x\right)  .\label{rey14}%
\end{equation}
Inserting this into the last term of  (\ref{rey13b}) and
manipulating algebraically the expression we obtain that
\begin{flalign}
\frac{y^{2}\left(  5g\left(  x\right)  T\left(  x\right)  U\left(
x\right)
-T^{\prime}\left(  x\right)  U^{\prime}\left(  x\right)  \right)  }%
{2\sqrt{g\left(  x\right)  }}-\frac{\partial V}{\partial
x}=-\frac{y^{2}\left[  U\left( x\right)  \,\Lambda _{1}+T\left(
x\right) \,\Lambda_{2}+3\,\Lambda_{3}\right]}{2\sqrt{g\left(
x\right)  }}
+h^{\prime }\left(  x\right),\label{rey15}%
\end{flalign}
where the $\Lambda_{i}$ ($1\leq i\leq 3$) are defined as
\begin{align*}
\Lambda_{1}  & =\left(  T^{\prime\prime}\left(  x\right)
-\frac{g^{\prime }\left(  x\right)  }{2g\left(  x\right)
}T^{\prime}\left(  x\right)
+g\left(  x\right)  T\left(  x\right)  \right)  ,\\
\Lambda_{2}  & =\left(  U^{\prime\prime}\left(  x\right)
-\frac{g^{\prime }\left(  x\right)  }{2g\left(  x\right)
}U^{\prime}\left(  x\right)
+g\left(  x\right)  U\left(  x\right)  \right)  ,\\
\Lambda_{3}  & =\left(  T\left(  x\right)  U\left(  x\right)
g\left( x\right)  +T^{\prime}\left(  x\right)  U^{\prime}\left(
x\right)  \right)  .
\end{align*}
As $T\left(  x\right)  $ and $U\left(  x\right)  $ are solutions
of (\ref{GE3}), it is immediate that $\Lambda_{1}=\Lambda_{2}=0$,
while $\Lambda_{3}=0$ follows from the constraint  (\ref{rey8}).
We conclude that for $h\left(  x\right)  =\alpha\in\mathbb{R}$,
the point symmetry $X_{6}$ is also a Noether symmetry. Changing
$T\left(  x\right)  $ for $U\left( x\right)  $ further shows that
$X_{8}$ also constitutes a Noether symmetry of the equation. Using
that Noether symmetries are preserved by commutators \cite{STEE},
it follows that $\left[  X_{6},X_{8}\right]
=-3a_{3}X_{1}+2a_{3}X_{7}$ is also a Noether symmetry.

\bigskip

\begin{proposition}
Any Noether symmetry $X$ of the ODE (\ref{GE3}) has the form
\begin{equation}
X=\lambda_{1}X_{2}+\lambda_{2}X_{3}+\lambda_{3}X_{6}+\lambda_{4}X_{8}%
+\lambda_{5}\left[  X_{6},X_{8}\right]  \label{rey16}%
\end{equation}
for some scalars $\lambda_{1},\cdots,\lambda_{5}\in\mathbb{R}$. In
particular, the Lie algebra $\mathcal{L}_{NS}$ of Noether
symmetries admits the following
Levi decomposition%
\begin{equation}
\mathcal{L}_{NS}=\frak{sl}\left(  2,\mathbb{R}\right)
\overrightarrow{\oplus
}_{V_{2}}\mathbb{R}^{2},\label{rey17}%
\end{equation}
where the Levi subalgebra $\frak{s}=\frak{sl}\left(
2,\mathbb{R}\right)  $ is generated by $X_{6},X_{8}$ and $Y=\left[
X_{6},X_{8}\right]$. The symmetries $X_{2},X_{3}$ transform
according to the 2-dimensional irreducible representation $V_{2}$
of $\frak{s}$.
\end{proposition}

\bigskip

\noindent The first part follows from the previous computations.
Now, using Table (\ref{TCALO}), the commutators of the Noether
symmetries are the following:
\begin{equation}%
\begin{tabular}
[c]{c|ccccc}%
$\left[  \cdot\,,\,\cdot\right]  $ & $Y$ & $X_{6}$ & $X_{8}$ &
$X_{2}$ & $X_{3}$\\\hline
$Y$ & $0$ & $-2a_{3}X_{6}$ & $2a_{3}X_{8}$ & $-a_{3}X_{2}$ & $a_{3}X_{3}$\\
$X_{6}$ &  & $0$ & $Y$ & $0$ & $a_{3}X_{2}$\\
$X_{8}$ &  &  & $0$ & $-X_{3}$ & $0$\\
$X_{2}$ &  &  &  & $0$ & $0$\\
$X_{3}$ &  &  &  &  & $0$%
\end{tabular}
\label{rey21}%
\end{equation}
showing that the Levi subalgebra is isomorphic to $\frak{sl}\left(
2,\mathbb{R}\right)  $ and generated by $X_{6},X_{8}$ and $Y$. The
symmetries $X_{2}$ and $X_{3}$ are easily seen to form a maximal
solvable ideal of $\mathcal{L}_{NS}$, hence the Levi decomposition
of the Lie algebra is given by (\ref{rey17}).

\bigskip

\noindent From (\ref{Noe4}) it is immediate to verify that the
constants of the motion associated to the symmetries $X_{2}$ and
$X_{3}$ are
\begin{flalign}
\psi_{1}=\frac{y^{\prime}T(x)-y\;T^{\prime}(x)}{\sqrt{g(x)}},\quad
\psi_{2}=\frac{y^{\prime}U(x)-y\;U^{\prime}(x)}{\sqrt{g(x)}}.\label{KdB}
\end{flalign}
For the symmetries in the Levi subalgebra
$\frak{sl}(2,\mathbb{R})$, the constants of the motion are
quadratic in $y^{\prime}$ and easily seen to be functionally
dependent on $\psi_{1}$ and $\psi_{2}$, just as it is expected
from the free particle equation $z^{\prime\prime}=0$
\cite{HAV,GOL}.

\bigskip
\bigskip

\section{Orthogonal functions as solutions to the ODE (\ref{GE3})}

\bigskip

\noindent Among the many interesting questions arising in the
theory of special functions, considerable attention has been
devoted to the problem of obtaining and characterizing orthogonal
polynomials by means of differential equations
\cite{ACZ,BRE,FEL,HAH,Ince,LES}. In this context, it is well known
that the so-called classical orthogonal polynomials constitute
essentially the only class to be determined by a second-order
differential equation of Sturm-Liouville type \cite{ACZ,HAH}.
\newline An important structural result in the theory of classical
orthogonal polynomials states that the Rodrigues formula
\begin{equation}
F_{n}\left(  x\right)  =\frac{1}{p\left(  x\right)  }D^{n}\left[
p\left(
x\right)  Q\left(  x\right)  ^{n}\right]  ,\label{orp1}%
\end{equation}
for quadratic polynomials $Q\left(  x\right)  =\left(  b-x\right)
\left( x-a\right)  $, where $p\left(  x\right)  $ is a weight
function in the finite interval $\left(  a,b\right)  $, provides a
polynomial $F_{n}\left(  x\right) $ of degree $n$ in $x$ for any
$n\geq0$ only if the weight function has the form
\begin{equation}
p\left(  x\right)  =\left(  b-x\right)  ^{\alpha}\left(
x-a\right)  ^{\beta
};\;\alpha>-1,\;\beta>-1.\label{orp2}%
\end{equation}
The $F_{n}\left(  x\right)  $ correspond to the class of Jacobi
polynomials. For polynomials $Q\left(  x\right)  $ of degrees
$d\leq1$ two other cases are given, corresponding to the Laguerre
and Hermite polynomials (see e.g. \cite{TRI}). The possible orders
of linear differential equations satisfied by non-classical or
generalized orthogonal polynomials have been analyzed by various
authors, albeit no generically valid analogue of the Rodrigues
formula has been found for these generalizations
\cite{HAH,KRA}. The formula (\ref{orp1}) is interesting in its
own right, as it allows to derive the second-order linear ordinary
differential equation satisfied by the polynomial $F_{n}\left(
x\right)  $ for any $n\geq0$. For the case of a weight function
(\ref{orp2}), the ODE has the following form:\footnote{For the
general ODE for the classical
orthogonal polynomials, see e.g. \cite{TRI}.}%
\begin{align}
& \left(  x-a\right)  \left(  b-x\right)
F_{n}^{\prime\prime}\left( x\right) +\left(  a\left(
1+\alpha\right)  +b\left( 1+\beta\right)  -\left(
2+\alpha+\beta\right)  x\right) F_{n}^{\prime}\left(
x\right)\nonumber\\
& +n\left(n+1+\alpha+\beta\right)  F_{n}\left(  x\right)  =0.\label{orp3}%
\end{align}
As seen before, the Chebychev polynomials $T_{n}\left(  x\right) $
and $U_{n}\left(  x\right)  $ arise from (\ref{orp1}) for
$a=-1,b=1$ and $\alpha=\beta=-\frac{1}{2}$.\footnote{This case is
generally seen as a special case of ultraspherical functions. See
\cite{ABS} for details.} In this context, the question arises
whether, besides the Chebyshev case previously considered, there
are other possible values of $\alpha,\beta,a$ and $b$ such that
the ODE (\ref{orp3}) has the form (\ref{GE3}). Starting from the
function
\[
g\left(  x\right)  =\frac{n\left(  n+1+\alpha+\beta\right)
}{\left( x-a\right)  \left(  b-x\right)  },
\]
the constraint
\begin{equation}
-\frac{g^{\prime}\left(  x\right)  }{2g\left(  x\right)
}=\frac{\left( a\left(  1+\alpha\right)  +b\left(  1+\beta\right)
-\left(  2+\alpha +\beta\right)  x\right)  }{\left(  x-a\right)
\left(  b-x\right)
}\label{orp4}%
\end{equation}
leads, after comparison of the differential equations (\ref{GE3})
and (\ref{orp3}), to the equation
\begin{equation}
a+b-2x=2\left(  a\left(  1+\alpha\right)  +b\left(  1+\beta\right)
-\left(
2+\alpha+\beta\right)  x\right)  .\label{orp5}%
\end{equation}
It follows at once that this identity holds only if $\alpha=\beta=-\frac{1}%
{2}$, without any conditions on $a$ and $b$. As a consequence, the
orthogonal
polynomials of the form%
\begin{equation}
P_{n}\left(  x\right)  =\left(  b-x\right)  ^{\frac{1}{2}}\left(
x-a\right) ^{\frac{1}{2}}\,D^{n}\left[  \left(  b-x\right)
^{n-\frac{1}{2}}\left(
x-a\right)  ^{n-\frac{1}{2}}\right]  ,\;n\geq0\label{orp6}%
\end{equation}
are always solutions of the ODE\
\begin{equation}
\left(  x-a\right)  \left(  b-x\right)  F_{n}^{\prime\prime}\left(
x\right) +\left(  \frac{a}{2}+\frac{b}{2}-x\right)
F_{n}^{\prime}\left(  x\right)
+n^{2}F_{n}\left(  x\right)  =0\label{orp6b}%
\end{equation}
of type (\ref{GE3}), with $g\left(  x\right)  $ given by
\begin{equation}
g\left(  x\right)  =\frac{n^{2}}{\left(  x-a\right)  \left(
b-x\right)
}.\label{orp7}%
\end{equation}
The ODE\ (\ref{orp6b}) is clearly of hypergeometric type (see \cite{ABS,CHI,KAM}%
), hence the orthogonal functions obtained will always be
expressible in terms of hypergeometric functions.

\bigskip
\bigskip

\subsection{Solutions of trigonometric type}

\bigskip

\noindent If we consider the values $b=-a$, the ODE\ (\ref{orp6b})
admits the general solution of trigonometric type
\begin{equation}
F_{n}\left(  x\right)  =C_{1}\cos\left(  n\,\arctan\frac{x}{\sqrt{a^{2}-x^{2}%
}}\right)  +C_{2}\sin\left(
n\,\arctan\frac{x}{\sqrt{a^{2}-x^{2}}}\right)
.\label{orp8}%
\end{equation}
With $H\left(  x\right)  =n\,\arctan\left(  x\,\left(
a^{2}-x^{2}\right) ^{-\frac{1}{2}}\right)  $, taking $T\left(
x\right)  =\sin H\left(  x\right) $ and $U(x)=\cos H\left(
x\right)  $, the constraints (\ref{rey6a}) and (\ref{rey6b}) are
satisfied for $a_{3}=-1$, while the relations (\ref{rey7}) and
(\ref{rey8}) follow at once. In this case, the symmetry generators
of the differential equation are given by
\begin{align*}
X_{1}  & =y\frac{\partial}{\partial y};\;X_{2}=\sin H\left(
x\right) \frac{\partial}{\partial y};\;X_{3}=\cos H\left(
x\right)  \frac{\partial }{\partial y};\;X_{4}=\frac{\sin H\left(
x\right)  }{H^{\prime}\left( x\right)  }y\frac{\partial}{\partial
x}+y^{2}\cos H\left(  x\right)
\frac{\partial}{\partial y};\\
X_{5}  & =-\frac{\cos H\left(  x\right)  }{H^{\prime}\left(  x\right)  }%
y\frac{\partial}{\partial x}+y^{2}\sin H\left(  x\right)
\frac{\partial }{\partial y};\;X_{6}=\frac{\sin^{2}H\left(
x\right)  }{H^{\prime}\left( x\right)  }\frac{\partial}{\partial
x}+\frac{y}{2}\sin\left(  2H\left(
x\right)  \right)  \frac{\partial}{\partial y};\\
X_{7}  & =-\frac{\sin\left(  2H\left(  x\right)  \right)
}{2H^{\prime}\left( x\right)  }\frac{\partial}{\partial x}-y\left(
1+\sin^{2}H\left(  x\right) \right)  \frac{\partial}{\partial
y};\;X_{8}=-\frac{\cos^{2}H\left(  x\right)
}{H^{\prime}\left(  x\right)  }y\frac{\partial}{\partial x}+\frac{y}{2}%
\sin\left(  2H\left(  x\right)  \right)  \frac{\partial}{\partial
y}.
\end{align*}
For this basis of generators, the commutator table of
$\frak{sl}(3,\mathbb{R})$ is explicitly given by

\begin{equation}%
\begin{tabular}
[c]{c|cccccccc}%
$\left[  \cdot,\cdot\right]  $ & $X_{1}$ & $X_{2}$ & $X_{3}$ &
$X_{4}$ & $X_{5}$ & $X_{6}$ & $X_{7}$ & $X_{8}$\\\hline
$X_{1}$ & $0$ & $-X_{2}$ & $-X_{3}$ & $X_{4}$ & $X_{5}$ & $0$ & $0$ & $0$\\
$X_{2}$ &  & $0$ & $0$ & $X_{6}$ & $X_{7}$ & $0$ & $2X_{2}$ & $X_{3}$\\
$X_{3}$ &  &  & $0$ & $3X_{1}-X_{7}$ & $X_{8}$ & $X_{2}$ & $X_{3}$ & $0$\\
$X_{4}$ &  &  &  & $0$ & $0$ & $0$ & $-X_{4}$ & $-X_{5}$\\
$X_{5}$ &  &  &  &  & $0$ & $-X_{4}$ & $-2X_{5}$ & $0$\\
$X_{6}$ &  &  &  &  &  & $0$ & $X_{6}$ & $3X_{1}-2X_{7}$\\
$X_{7}$ &  &  &  &  &  &  & $0$ & $X_{8}$\\
$X_{8}$ &  &  &  &  &  &  &  & $0$%
\end{tabular}
\label{TCAL}%
\end{equation}

\bigskip

\noindent We observe that for $a=1$, the preceding solution
(\ref{orp8}) can be simplified by means of the trigonometric
identity
\[
\arcsin x=\arctan\frac{x}{\sqrt{1-x^{2}}},
\]
and thus we recover the classical Chebyshev polynomials $T_{n}(x)$
and $U_{n}(x)$. The basis of symmetries of (\ref{TS6}) is
explicitly given by
\begin{align}
X_{1} &  =y\frac{\partial}{\partial y};\;X_{2}=T_{n}\left(
x\right) \frac{\partial}{\partial y};\;X_{3}=U_{n}\left(  x\right)
\frac{\partial
}{\partial y};\;X_{4}=\frac{\sqrt{1-x^{2}}T_{n}\left(  x\right)  }{n}%
y\frac{\partial}{\partial x}+y^{2}U_{n}\left(  x\right)
\frac{\partial
}{\partial y};\nonumber\\
X_{5} &  =-\frac{\sqrt{1-x^{2}}U_{n}^{\prime}\left(  x\right)  }{n}%
y\frac{\partial}{\partial x}+y^{2}T_{n}\left(  x\right)
\frac{\partial
}{\partial y};\,X_{6}=\frac{\sqrt{1-x^{2}}T_{n}^{2}\left(  x\right)  }{n}%
\frac{\partial}{\partial x}+yT_{n}\left(  x\right)  U_{n}\left(
x\right)
\frac{\partial}{\partial y};\nonumber\\
X_{7} &  =-\frac{\sqrt{1-x^{2}}T_{n}\left(  x\right)  U_{n}\left(
x\right) }{n}\frac{\partial}{\partial x}+y\left(
1+T_{n}^{2}\left(  x\right)  \right)
\frac{\partial}{\partial y};\;\nonumber\\
X_{8} &  =-\frac{\sqrt{1-x^{2}}U_{n}^{2}\left(  x\right)
}{n}\frac{\partial }{\partial x}+yT_{n}\left(  x\right)
U_{n}\left(  x\right)  \frac{\partial }{\partial y}.\label{KRASW}
\end{align}

\noindent For $n=1$, we recover exactly the vector fields in
(\ref{TARF}), showing that the generic realization describes
naturally the basis of symmetries of the Chebyshev equation
(\ref{TS6}) for arbitrary values of $n$.

\bigskip

\noindent For $a\neq1$, the orthogonal polynomials deduced from
the Rodrigues formula are still deeply related to the Chebyshev
case, and by means of a new scaled variable $z=x\,a^{-1}$, it can
be shown that the functions
\begin{equation}
T_{n}\left(  \frac{x}{a}\right)  ,\;U_{n}\left(
\frac{x}{a}\right)
,\;n\geq0\label{orp9}%
\end{equation}
solve the differential equation (\ref{orp6b}). This case hence
does not add essentially new variants.

\bigskip

\noindent For $a\neq0$ and $b+a\neq0$, the general solution of
(\ref{orp6b}) can be written as
\begin{flalign}
F_{n}\left(  x\right)  =C_{1}\cos\left(  n\,\arctan\frac{a+b+2x}%
{2\sqrt{\left(  a-x\right)  \left(  x-b\right)  }}\right)
+C_{2}\sin\left(
n\,\arctan\frac{a+b+2x}{2\sqrt{\left(  a-x\right)  \left(  x-b\right)  }%
}\right)  .\label{orp10}%
\end{flalign}
In this case, the polynomials $P_{n}\left(  0\right)  $ have
nonzero terms in any even and odd order, hence they will be
expressible in terms of linear combinations of Jacobi polynomials
\cite{ABS,SZ}. As an example, we enumerate the first five
polynomials that arise from this choice of the parameters:
\[
\begin{tabular}
[c]{l|l}%
$n$ & $P_{n}\left(  x\right)  $\\\hline
$0$ & $1$\\
&\\
$1$ & $\frac{1}{2}\left(  \left(  a+b\right)  -2x\right)  $\\
$2$ & $\frac{3}{4}\left(  \left(  a^{2}+6ab+b^{2}\right)  -8\left(
a+b\right)  x+8x^{2}\right)  $\\
&\\
$3$ & $\frac{15}{8}\left(  \left(
a^{3}+15a^{2}b+15ab^{2}+b^{3}\right) -6\left(  a+3b\right)  \left(
3a+b\right)  x+48\left(  a+b\right)
x^{2}-32x^{3}\right)  $\\
&\\
$4$ & $\frac{105}{16}\left(  \left(  a^{4}+28a^{3}b+70a^{2}b^{2}%
+28ab^{3}+b^{4}\right)  -32\left(  a+b\right)  \left(
a^{2}+6ab+b^{2}\right)
x\right)  +$\\
& $\frac{105}{16}\left(  32\left(  5a^{2}+14ab+5b^{2}\right)
x^{2}-256\left(
a+b\right)  x^{3}+128x^{4}\right)  $\\
&\\
$5$ & $\frac{945}{32}\left(  a+b\right)  \left(  a^{4}+44a^{3}b+166a^{2}%
b^{2}+44ab^{3}+b^{4}\right)  -\frac{4725}{16}\left(  5a^{2}+10ab+5b^{2}%
\right)  \times$\\
& $\left(  a^{2}+10ab+5b^{2}\right)  x+\frac{4725}{2}\left(
a+b\right)
\left(  5a^{2}+22ab+5b^{2}\right)  x^{2}-4725\left(  7a^{2}+18ab+\right.  $\\
& $\left.  7b^{2}\right)  x^{3}+37800\left(  a+b\right)  x^{4}-15120x^{5}$\\
&\\\hline
\end{tabular}
\]
The orthogonality relation for these polynomials is given by the
formula
\begin{equation}
\int_{a}^{b}P_{n}\left(  x\right)  P_{m}\left(  x\right)  \left(
b-x\right) ^{-\frac{1}{2}}\left(  x-a\right)
^{-\frac{1}{2}}dx=\delta_{n}^{m}\frac
{\prod_{l=0}^{n-1}\left(  2l+1\right)  }{2^{2n+1}}.\label{orp12}%
\end{equation}

\bigskip

\noindent Let us finally observe that for the case $a=0$, $b=1$,
the differential equation
\begin{equation}
x\left(  1-x\right)  F_{n}^{\prime\prime}\left(  x\right)  +\left(
\frac {1}{2}-x\right)  F_{n}^{\prime}\left(  x\right)
+n^{2}F_{n}\left(  x\right)
=0\label{opr13}%
\end{equation}
is of the classical Jacobi type \cite{KAM}. Its general solution
can be conveniently expressed as\footnote{The general solution can
also be expressed in terms of hyperbolic functions. }
\begin{flalign}
F_{n}\left(  x\right)  =C_{1}\cos\left(
2n\,\arctan\frac{\sqrt{x}}{\sqrt
{b-x}}\right)  +C_{2}\sin\left(  2n\,\arctan\frac{\sqrt{x}}{\sqrt{b-x}%
}\right)  .\label{orp14}%
\end{flalign}
The Jacobi polynomials $\frak{F}_{n}\left(  0,\frac{1}{2},x\right)
$ are obviously a solution to this equation, hence considering the
transformation $x=1-2\,z$, the orthogonal polynomials obtained
from the Rodrigues formula can be easily related to the Chebyshev
polynomials $T_{n}\left(  1-2z\right)  $.

\bigskip
\bigskip

\subsection{Solutions of hyperbolic type}

\bigskip

\noindent Functions of hyperbolic type can be obtained as
solutions of the ODE (\ref{GE3}) choosing the auxiliary function
$g\left( x\right) =-\left( \frac {dH}{dx}\right) ^{2}$. In this
case, we write the general solution as
\begin{equation}
y\left(  x\right)  =\lambda_{1}\sinh H\left(  x\right)
+\lambda_{2}\cosh
H\left(  x\right)  ,\label{HY1}%
\end{equation}
and the relations (\ref{rey7})-(\ref{rey8}) are satisfied for
$a_{3}=1$ taking $U(x)=\cosh H\left(  x\right)$, $T(x)=\sinh
H\left(  x\right)$.\footnote{Observe that interchanging the role
of $T(x)$ and $U(x)$ always changes the sign of $a_{3}$.} The
basis of point symmetries is explicitly given by
\begin{align*}
X_{1} &  =y\frac{\partial}{\partial y};\;X_{2}=\cosh H\left(
x\right) \frac{\partial}{\partial y};\;X_{3}=\sinh H\left(
x\right)  \frac{\partial }{\partial y};\;X_{4}=\frac{\cosh H\left(
x\right)  }{H^{\prime}\left( x\right)  }y\frac{\partial}{\partial
x}+y^{2}\sinh H\left(  x\right)
\frac{\partial}{\partial y};\\
X_{5} &  =\frac{\sinh H\left(  x\right)  }{H^{\prime}\left(  x\right)  }%
y\frac{\partial}{\partial x}+y^{2}\cosh H\left(  x\right)
\frac{\partial }{\partial y};\;X_{6}=\frac{\cosh^{2}H\left(
x\right)  }{H^{\prime}\left( x\right)  }\frac{\partial}{\partial
x}+\frac{y}{2}\sinh2H\left(  x\right)
\frac{\partial}{\partial y};\\
X_{7} &  =\frac{\sinh\left(  2H\left(  x\right)  \right)
}{2H^{\prime}\left( x\right)  }\frac{\partial}{\partial x}+y\left(
1+\cosh^{2}H\left(  x\right) \right)  \frac{\partial}{\partial
y};\;X_{8}=\frac{\sinh^{2}H\left(  x\right)
}{H^{\prime}\left(  x\right)  }y\frac{\partial}{\partial x}+\frac{y}{2}%
\sinh\left(  2H\left(  x\right)  \right)  \frac{\partial}{\partial
y}.
\end{align*}

\bigskip

\noindent We will see that for suitable choices of $H(x)$, the
hyperbolic functions of (\ref{HY1}) define orthonormal systems of
functions in the interval $\left[-1,1\right]$.

\bigskip

\noindent We start for example from the function
\begin{equation}
g\left(  x\right)  =n^{2}\left(  1-x^{2}\right)  \label{orf}%
\end{equation}
From (\ref{GE3}) we have the differential equation
\begin{equation}
y^{\prime\prime}+\frac{x}{1-x^{2}}y^{\prime}+n^{2}\left(
1-x^{2}\right)  y=0.
\label{orf1}%
\end{equation}
It is not difficult to justify that no $n^{th}$-order polynomials
satisfy this equation for $n>0$. The general solution can be
written in terms of hyperbolic functions as
\begin{flalign}
y\left(  x\right)  =C_{1}\sinh\left[  F_{n}\left(  x\right)
\right] +C_{2}\cosh\left[ F_{n}\left(  x\right)  \right],  \label{orf2}%
\end{flalign}
where $F_{n}\left(  x\right) $ is defined as
\begin{equation}
F_{n}\left(  x\right)  = \frac{n}{2}\left(  x\sqrt{x^{2}-1}%
-\ln\left(  x+\sqrt{x^{2}-1}\right)  \right). \label{orf3}%
\end{equation}
Making the substitution $u=$ $\left(  x\sqrt{x^{2}-1}-\ln\left(
x+\sqrt {x^{2}-1}\right)  \right)  $, it is straightforward to
verify that
\begin{equation}
\int_{-1}^{1}\cosh\left( F_{n}\left(  x\right)\right) \cosh\left(  F_{m}\left(  x\right)\right)  \sqrt{1-x^{2}%
}dx=\frac{\mathrm{i}}{2}\int_{-\mathrm{i}\,\pi}^{0}\cosh\left(  \frac{n}%
{2}v\right)  \cosh\left(  \frac{m}{2}v\right)  dv\, \label{orf4}%
\end{equation}
holds. The latter integral can be easily solved, and for $n\neq m$
we obtain that
\begin{align}
\frac{\mathrm{i}}{2}\int_{-\mathrm{i}\,\pi}^{0}\cosh\left(  \frac{n}%
{2}v\right)  \cosh\left(  \frac{m}{2}v\right)  dv\,  &  =\frac{\mathrm{i}}%
{2}\,\left[  \frac{\sinh\left(  \frac{m-n}{2}v\right)
}{m-n}+\frac
{\sinh\left(  \frac{m+n}{2}v\right)  }{m+n}\right]  _{-\mathrm{i\,}\pi}%
^{0}.\,\nonumber\\
&  =\frac{1}{2}\,\left[  \frac{\sin\left(  \frac{m-n}{2}\pi\right)  }%
{m-n}+\frac{\sin\left(  \frac{m+n}{2}\pi\right)  }{m+n}\right]  .
\label{orf5b}%
\end{align}
Now observe that if $n$ and $m$ have different parity, i.e.,
$n=2p$ and $m=2q+1$, then
\begin{equation}
\frac{1}{2}\,\left[  \frac{\sin\left(  \frac{\pi}{2}\right)  }{2q+1-2p}%
+\frac{\sin\left(  \frac{\pi}{2}\right)  }{2q+2p+1}\right]
=-\frac
{2q+1}{\left(  -2q-1+2p\right)  \left(  2q+2p+1\right)  }\neq0, \label{orf7}%
\end{equation}
whereas for the case of $n$, $m$ having the same parity, the
integral (\ref{orf4}) vanishes identically. An analogous result is
obtained if we compute the integrals for the hyperbolic sine
functions.

\bigskip

\noindent As a consequence, two possibilities are given to
construct an orthogonal system of functions with $g(x)$ of type
(\ref{orf}):

\begin{enumerate}
\item  For $g\left(  x\right)  =4n^{2}\left(  1-x^{2}\right)  $,
the fundamental solutions
\begin{align}
P_{n}\left(  x\right)   &  =\cosh\left[  n\left(
x\sqrt{x^{2}-1}-\ln\left(
x+\sqrt{x^{2}-1}\right)  \right)  \right]  ,\label{orf8a}\\
Q_{n}\left(  x\right)   &  =\sinh\left[  n\left(
x\sqrt{x^{2}-1}-\ln\left(
x+\sqrt{x^{2}-1}\right)  \right)  \right]  \label{orf8b}%
\end{align}
of the ODE\
\[
y^{\prime\prime}+\frac{x}{1-x^{2}}y^{\prime}+4n^{2}\left(
1-x^{2}\right) y=0
\]
define orthogonal systems of functions for the weight function
$w\left( x\right)  =\sqrt{1-x^{2}}$ in the interval $\left[
-1,1\right]  $.

\item  For $g\left(  x\right)  =\left(  2n+1\right)  ^{2}\left(
1-x^{2}\right)  $, the fundamental solutions
\begin{align}
P_{n}\left(  x\right)   &  =\cosh\left[  \frac{2n+1}{2}\left(  x\sqrt{x^{2}%
-1}-\ln\left(  x+\sqrt{x^{2}-1}\right)  \right)  \right]  ,\label{orf9a}\\
Q_{n}\left(  x\right)   &  =\sinh\left[  \frac{2n+1}{2}\left(  x\sqrt{x^{2}%
-1}-\ln\left(  x+\sqrt{x^{2}-1}\right)  \right)  \right]  \label{orf9b}%
\end{align}
of the ODE\
\[
y^{\prime\prime}+\frac{x}{1-x^{2}}y^{\prime}+\left(  2n+1\right)
^{2}\left( 1-x^{2}\right)  y=0
\]
define orthogonal systems of functions for the weight function
$w\left( x\right)  =\sqrt{1-x^{2}}$ in the interval $\left[
-1,1\right]  $.
\end{enumerate}

\bigskip

\noindent Considering different choices of the function $g(x)$,
other orthogonal systems of functions can formally be obtained as
solutions of the ODE (\ref{GE3}).

\bigskip
\bigskip

\section{Non-linear deformations }

\bigskip

\noindent We have seen previously that the linear homogeneous ODE
(\ref{GE3}) admits five Noether symmetries if the equation is
derived, via the Helmholtz condition, from the Lagrangian
(\ref{rey12}). The corresponding constants of the motion are
obtained from (\ref{Noe4}), where the two constants linear in the
velocity $y^{\prime}$ generate the remaining invariants. This
fact, to a certain extent, is a direct consequence of the maximal
symmetry of the equation, implying in particular that it is
linearizable \cite{PER}. In this situation, we can ask how to
modify the ODE (\ref{GE3}) by addition of a ``forcing" term such
that the maximal symmetry is broken, but such that the resulting
equation preserves a given subalgebra of Noether symmetries. In
order to avoid those variational symmetries with a constant of the
motion linear in $y^{\prime}$, this subalgebra should be chosen as
the Levi subalgebra $\mathcal{L}_{NS}$. For our specific purposes,
the problem can be formulated in the following terms: Does there
exist a (non-linear) ODE and a Lagrangian $L^{\prime}$ such that
the vector fields $X_{6}$ and $X_{8}$ of (\ref{GE5}) are Noether
symmetries?

\bigskip

\noindent Consider an arbitrary function $G\left(
x,y,y^{\prime}\right) $ and define the extended Lagrangian
\begin{equation}
L_{0}=\frac{1}{2\sqrt{g\left(  x\right)  }}\left(  \left(
y^{\prime}\right) ^{2}-g\left(  x\right)  y^{2}\right)  -G\left(
x,y,y^{\prime}\right)  .
\label{rey31}%
\end{equation}
The equation of motion associated to (\ref{rey31}) has the form
\begin{equation}
y^{\prime\prime}-\frac{g^{\prime}\left(  x\right)  }{2g\left(
x\right) }y^{\prime}+g\left(  x\right)  y-\sqrt{g\left(  x\right)
}\left(  \frac {d}{dx}\left(  \frac{\partial G}{\partial
y^{\prime}}\right)  -\frac{\partial
G}{\partial y}\right)  =0, \label{rey32}%
\end{equation}
which can be interpreted as a ``deformation'' of the ODE
(\ref{GE3}) by the forcing term $G\left( x,y,y^{\prime}\right)  $.
We now require that $X_{6}$ and $X_{8}$ from (\ref{GE5}) are
Noether symmetries of $L_{0}$, imposing additionally that the
symmetry condition (\ref{Noe3}) is satisfied for the same function
$V\left( x,y\right)  $ valid for the Lagrangian (\ref{rey12}).
This will imply in particular that both differential equations
share exactly the same symmetry generators, hence the addition of
the forcing term can actually be seen as a symmetry breaking. On
the other hand, by the properties of commutators, the deformed
equation will have al least a $\frak{sl}(2,\mathbb{R})$-subalgebra
of Noether symmetries.

\bigskip

\noindent We observe again that, as the vector fields $X_{6}$ and
$X_{8}$ are obtained by interchanging the role of $T\left(
x\right) $ and $U\left( x\right)  $, it suffices to compute the
symmetry condition for only one of these symmetries. We make the
computations for $X_{6}$. The Noether symmetry condition
\begin{equation}
\dot{X}_{6}\left(  L_{0}\right)  +A\left(  \xi\right)
L_{0}-A\left( V\right)
\end{equation}
with $V\left(  x,y\right)  $ as given in (\ref{rey14}) leads,
after some simplification using the constraint (\ref{rey8}), to
the following
partial differential equation for $G\left(  x,y,y^{\prime}\right)  :$%
\begin{align}
&  \left(  2T\left(  x\right)  U\left(  x\right)  +\frac{T\left(
x\right) U^{\prime}\left(  x\right)  g\left(  x\right)
}{2}\right)  G\left( x,y,y^{\prime}\right)  +y\,T\left(  x\right)
U\left(  x\right) \frac{\partial G}{\partial y}-\frac{T\left(
x\right)  U^{\prime}\left(
x\right)  }{g\left(  x\right)  }\frac{\partial G}{\partial x}\nonumber\\
&  +\left(  y\left(  T\left(  x\right)  U^{\prime}\left(  x\right)
+T^{\prime}\left(  x\right)  U\left(  x\right)  \right)
-y^{\prime}\,T\left( x\right)  \left(  U\left(  x\right)
+\frac{g^{\prime}\left(  x\right) U^{\prime}\left(  x\right)
}{g\left(  x\right)  ^{2}}\right)  \right) \frac{\partial
G}{\partial y^{\prime}}=0.\label{rey33}
\end{align}
Albeit complicated in form, this PDE can be solved. As we are
interested in those solutions being valid for both the symmetries
$X_{6}$ and $X_{8}$, suppose that the Noether condition
(\ref{Noe3}) is also satisfied for $X_{8}$. The corresponding PDE
for $G\left( x,y,y^{\prime}\right) $ is obtained from
(\ref{rey33}) replacing $T\left(  x\right)  $ by $U\left( x\right)
$. Taking the difference of these two equations leads to the
auxiliary equation\footnote{Recall that $\mathbf{W}$ is the
Wronskian of (\ref{GE3}).}
\begin{equation}
-\frac{\mathbf{W}}{2g\left(  x\right)  ^{2}}\left(
g^{\prime}\left( x\right)  \,y^{\prime}\,\frac{\partial
G}{\partial y^{\prime}}+2g\left( x\right)  \frac{\partial
G}{\partial x}-g^{\prime}\left(  x\right)  G\left(
x,y,y^{\prime}\right)  \right)  =0. \label{rey34}%
\end{equation}
The general solution is easily found to be
\begin{equation}
G\left(  x,y,y^{\prime}\right)  =\varphi\left(  y,\frac{y^{\prime}}%
{\sqrt{g\left(  x\right)  }}\right)  \sqrt{g\left(  x\right)  }, \label{rey35}%
\end{equation}
with $\varphi$ an arbitrary function of its arguments. Inserting
the latter into (\ref{rey33}) and solving for $\varphi$ shows that
the only functions $G\left(  x,y,y^{\prime}\right)  $ for which
$X_{6}$ and $X_{8}$ are Noether symmetries of the Lagrangian
$L_{0}$ in (\ref{rey31}) are
\begin{equation}
G\left(  x,y,y^{\prime}\right)  =\frac{\alpha\,\sqrt{g\left(  x\right)  }%
}{y^{2}},\;\alpha\in\mathbb{R.} \label{rey36}%
\end{equation}
The perturbed equation of motion preserving the
$\frak{sl}(2,\mathbb{R})$-subalgebra of Noether symmetries is
therefore
\begin{equation}
y^{\prime\prime}-\frac{g^{\prime}\left(  x\right)  }{2g\left(
x\right) }y^{\prime}+g\left(  x\right)  y-2\alpha\frac{g\left(
x\right)  }{y^{3}}=0.
\label{rey37}%
\end{equation}

\noindent We observe that if we only require invariance by either
$X_{6}$ or $X_{8}$, the possibilities are wider. However, in these
cases the forcing term will always contain explicitly the function
$T(x)$ and $U(x)$, in addition to $g(x)$. For this reason we leave
this case aside.

\bigskip

\begin{proposition}
For arbitrary functions $g\left(  x\right)  \neq0$, the nonlinear
ODE (\ref{rey37}) possesses a Lie algebra of point symmetries
isomorphic to $\frak{sl}\left(  2,\mathbb{R}\right)  $. Moreover,
any point symmetry is a Noether symmetry for the Lagrangian
\begin{equation}
L_{0}=\frac{1}{2\sqrt{g\left(  x\right)  }}\left(  \left(
y^{\prime}\right) ^{2}-g\left(  x\right)  y^{2}\right)
-\frac{\alpha\,\sqrt{g\left(  x\right)
}}{y^{2}}.\label{rey38}%
\end{equation}
\end{proposition}

\bigskip

\begin{proof}
For the equation of motion associated to the Lagrangian $L_{0}$ we
have the auxiliary function
\begin{equation}
\omega\left(  x,y,y^{\prime}\right)  =\frac{g^{\prime}\left(
x\right) }{2g\left(  x\right)  }y^{\prime}-g\left(  x\right)
y+2\alpha\frac{g\left( x\right)  }{y^{3}}.\label{mel}
\end{equation}
The symmetry condition for point symmetries obtained from
(\ref{SBE}) is given by the system
\begin{align}
& \frac{\partial^{2}\xi}{\partial y^{2}}=0;\quad
\frac{g^{\prime}\left( x\right)  }{g\left( x\right)
}\frac{\partial\xi}{\partial y}+2\frac {\partial^{2}\xi}{\partial
x\partial y}-\frac{\partial^{2}\eta}{\partial
y^{2}}=0;\label{rey38a}\\
& \frac{\partial^{2}\xi}{\partial x^{2}}+\frac{g^{\prime}\left(
x\right) }{2g\left(  x\right)  }\frac{\partial\xi}{\partial
x}+3g\left(  x\right)
\left(  \frac{2\alpha}{y^{3}}-y\right)  \frac{\partial\xi}{\partial y}%
+\frac{g^{\prime}\left(  x\right)  }{2g\left(  x\right)  }\left(
1-\frac{g^{\prime}\left(  x\right)  }{g\left(  x\right)  }\right)
\xi\left(
x,y\right)  -2\frac{\partial^{2}\eta}{\partial x\partial y}=0;\label{rey38b}\\
& g\left(  x\right)  \left(  y-\frac{2\alpha}{y^{3}}\right) \left(
\frac{\partial\eta}{\partial y}-2\frac{\partial\xi}{\partial
x}-g^{\prime }\left(  x\right) \xi\left(  x,y\right)  \right)
-\left(  1+\frac{6\alpha }{y^{4}}\right)  g\left(  x\right)
\eta\left(  x,y\right)-\frac{\partial^{2}\eta}{\partial
x^{2}}\nonumber\\
& +\frac{g^{\prime }\left(  x\right) }{2g\left(  x\right)
}\frac{\partial\eta}{\partial x} =0.\label{rey38c}
\end{align}
From the first two equations we immediately obtain that
\begin{flalign}
\xi\left(  x,y\right)  =F_{11}\left(  x\right)  y+F_{12}\left(
x\right) ;\;\eta\left(  x,y\right)  =\frac{y^{2}}{2}\left(
\frac{g^{\prime}\left( x\right)  F_{11}\left(  x\right)  }{g\left(
x\right)  }+2F_{11}^{\prime }\left(  x\right)
\right)+F_{21}\left(  x\right)  y+F_{22}\left(
x\right). \label{rey39}%
\end{flalign}
Inserting these functions into the equations (\ref{rey38b}) and
(\ref{rey38c}) and separating with respect to the powers of $y$,
the free terms provide us with the two constraints $F_{11}\left(
x\right) =F_{22}\left(  x\right) =0$. This simplification further
leads to the relation
\begin{equation}
F_{12}\left(  x\right)  g^{\prime}\left(  x\right)
+2F_{12}^{\prime}\left( x\right)  g\left(  x\right)  -4g\left(
x\right)  F_{21}\left(  x\right)
=0,\label{rey40}%
\end{equation}
enabling us to write the components $\xi$ and $\eta$ of a point
symmetry $X$ as
\begin{align}
\xi\left(  x,y\right)    & =F_{12}\left(  x\right)  ,\\
\eta\left(  x,y\right)    & =\frac{y}{4}\left(
2F_{12}^{\prime}\left(
x\right)  +\frac{g^{\prime}\left(  x\right)  }{g\left(  x\right)  }%
F_{12}\left(  x\right)  \right)  .
\end{align}
In order to satisfy the equation (\ref{rey38c}), the function
$F_{12}\left( x\right)  $ must be a solution to the differential
equation
\begin{align}
& \frac{d^{3}F_{12}}{dx^{3}}+\frac{4g^{\prime\prime}\left(
x\right)  g\left( x\right)  +16g\left(  x\right)  ^{3}-5\left(
g^{\prime}\left(  x\right) \right)  ^{2}}{4g\left(  x\right)
^{2}}\frac{dF_{12}}{dx}+\frac
{g^{\prime\prime\prime}\left(  x\right)  }{2g\left(  x\right)  }%
-\frac{7g^{\prime}\left(  x\right)  g^{\prime\prime}\left(
x\right)
}{4g\left(  x\right)  ^{3}}\nonumber\\
& \frac{+8g^{\prime}\left(  x\right)  g\left(  x\right)
^{3}+5\left( g^{\prime}\left(  x\right)  \right)  ^{3}}{4g\left(
x\right)  ^{3}}=0.\label{rey41}
\end{align}
This shows that the dimension of the symmetry algebra is at most
$3$. Now, since the Noether symmetries $X_{6},X_{8}$ and $\left[
X_{6},X_{8}\right]$ are also point symmetries of the equation
\cite{Ste}, they automatically satisfy the ODE\ (\ref{rey41}).
From the commutator of the Noether symmetries it follows at once
that $\mathcal{L}\simeq\mathcal{L}_{NS}\simeq\frak{sl}\left(
2,\mathbb{R}\right)  $.
\end{proof}

\bigskip

\noindent This result implies in particular that the perturbed or
deformed differential equation (\ref{rey37}) is not linearizable,
and hence genuinely non-linear. It is worthy to be mentioned that
differential equations invariant under $\frak{sl}(2,\mathbb{R})$
have been analyzed by various authors in connection with
non-linear equations, as well as in the context of systems
admitting constants of the motion of a specific type
\cite{C44,CZ,GOV,HAG,PIL}.

\bigskip

\noindent We now determine the constants of the motion of the
Lagrangian (\ref{rey38}) using formula (\ref{Noe4}): for the
symmetry $X_{6}$, the appropriate function $V\left(  x,y\right)  $
is given by (\ref{rey14}) for $h\left(  x\right)  =0$, resulting
in the expression
\begin{flalign}
\psi_{1}=&-\frac{\left( U^{\prime}\left(  x\right)
T^{\prime}\left( x\right)  +2U\left(  x\right) T\left(  x\right)
g\left(  x\right) \right)  }{g\left(  x\right) \sqrt{g\left(
x\right) }}\,yy^{\prime}+\frac{U\left(  x\right)  T^{\prime
}\left( x\right)  }{2\sqrt{g\left(  x\right)  }}y^{2}+\frac{\alpha
T\left(
x\right)  U^{\prime}\left(  x\right)  }{\sqrt{g\left(  x\right)  }y^{2}%
}\nonumber\\
 & -\frac{T\left(  x\right)  U^{\prime}\left(  x\right)
}{g\left( x\right)  ^{\frac{3}{2}}}\left(  y^{\prime}\right)
^{2}.\label{rey42}%
\end{flalign}
For $X_{8}$, the function $V\left(  x,y\right)  $ is the same, and
the constant of the motion $\psi_{2}$ is obtained by simply
permuting $U\left( x\right)  $ and $T\left(  x\right)  $. In
particular, taking the difference $\psi_{2}-\psi_{1}$ we obtain
the simplified invariant
\begin{equation}
\psi_{2}-\psi_{1}=\frac{\mathbf{W}}{\sqrt{g\left(  x\right)
}}\left(
\frac{\left(  y^{\prime}\right)  ^{2}}{g\left(  x\right)  }+\frac{y^{2}}%
{2}-\frac{\alpha}{y^{2}}\right)  .\label{rey43}%
\end{equation}
Now, for $\left[  X_{6},X_{8}\right]  =-a_{3}\left(
3X_{1}-2X_{7}\right)  $ the function $V\left(  x,y\right)  $
satisfying condition (\ref{Noe3}) is $V\left(  x,y\right)
=\frac{2y^{2}T\left(  x\right)  T^{\prime}\left( x\right)
}{\sqrt{g\left(  x\right)  }}$ and the constant of the motion
\begin{flalign}
\psi_{3}=-\frac{T\left(  x\right)  T^{\prime}\left(  x\right)
\left(  \left(
y^{\prime}\right)  ^{2}+g\left(  x\right)  \left(  y^{2}+\frac{2\alpha}{y^{2}%
}\right)  \right)  +y\,y^{\prime}\left(  4g\left(  x\right)
T^{2}\left( x\right)  -3g\left(  x\right)  +2\left(
T^{\prime}\left(  x\right)
^{2}\right)  \right)  }{g\left(  x\right)  \sqrt{g\left(  x\right)  }}.\label{rey44}%
\end{flalign}
\noindent Clearly the constants $\psi_{0}=\psi_{2}-\psi_{1}$ and
$\psi_{3}$ are independent, and both quadratic in $y^{\prime}$. In
contrast to the non-deformed ODE (\ref{GE3}), for $\alpha\neq0$
these constants of the motion are not obtainable from linear
invariants of the equation. Technically, we could use these
invariants to reduce the order of the equation \cite{Anco,CIC,OV}.
However, for generic functions $g\left( x\right)  $ the reduced
equation has an even more intricate form, and does not greatly
simplify the integration of the non-linear equation. With the
exception of $g\left(  x\right) =1$, that is easily seen to lead
to the classical Pinney equation \cite{PIN}, for arbitrary
non-constant $g\left( x\right) $ it is more practical to use
numerical methods for the resolution.

\bigskip

\noindent As an example to illustrate this fact, we consider
$g\left( x\right)  =x$ and the deformed equation
\begin{equation}
y^{\prime\prime}-\frac{y^{\prime}}{2x}+x\,y-\frac{2\alpha\,x}{y^{3}%
}=0.\label{rey45}%
\end{equation}
For $\alpha=0$ we recover the non-deformed ODE\ with general
solution
\[
y\left(  x\right)  =C_{1}\sin\left(  \frac{2}{3}x\sqrt{x}\right)  +C_{2}%
\cos\left(  \frac{2}{3}x\sqrt{x}\right)
\]
and Wronskian $\mathbf{W}=-\sqrt{x}$. For $\alpha\neq0$ the
constants of the motion of (\ref{rey37}) are
\begin{align*}
\psi_{0}  & =-\frac{1}{\sqrt{x}}\left(  \frac{\left(  y^{\prime}\right)  ^{2}%
}{x}+\frac{y^{2}}{2}-\frac{\alpha}{y^{2}}\right)  ,\\
\psi_{3}  & =\frac{-\sqrt{x}\left(  \left(  y^{\prime}\right)  ^{2}%
y^{2}-xy^{4}-2\alpha x\right)  \sin\left(
\frac{4}{3}x\sqrt{x}\right) +xy^{3}y^{\prime}\cos\left(
\frac{4}{3}x\sqrt{x}\right)  }{2x\sqrt{x}^{2}}.
\end{align*}

\noindent In spite of their apparent simplicity, the fact that
both constants of the motion depend explicitly on the independent
variable $x$ makes their use rather complicated, so that we skip
this step.\newline In the following Figure we compare the
solutions of (\ref{rey45}) for the values $\alpha=1$ and
$\alpha=0$ with the same initial conditions (the dashed line
corresponds to the solution with $\alpha=0$):%
\begin{figure}[h!]
\centering
\includegraphics[scale=0.75]{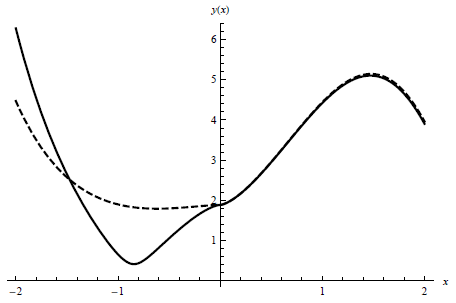}
\caption{$y(x)$ for the non-linear ODE (\ref{rey45}) and the
linearizable ODE.}
\end{figure}

\bigskip
\bigskip

\subsection{Non-linear systems in $N=2$ dimensions}

\bigskip

\noindent Just as the scalar ODE (\ref{GE3}) has been perturbed
using a subalgebra of Noether symmetries, we can consider the
problem of deforming systems of differential equations along the
same lines. We recall that, in different contexts, variations of
this ansatz have already been considered in the literature (see
\cite{HAS} and references therein), although usuallly related to
the time-dependent oscillator equations.

\bigskip

\noindent We start from the decoupled system in $N=2$ dimensions
given by:
\begin{align}
y_{1}^{\prime\prime}-\frac{g^{\prime}\left(  x\right)  }{2g\left(
x\right)
}y_{1}^{\prime}+g\left(  x\right)  y_{1}  & =0,\nonumber\\
y_{2}^{\prime\prime}-\frac{g^{\prime}\left(  x\right)  }{2g\left(
x\right) }y_{2}^{\prime}+g\left(  x\right)  y_{2}  &
=0.\label{ret1}
\end{align}
Clearly this system is linearizable and reducible to the free
particle system $\left\{
z_{1}^{\prime\prime}=0,\,z_{2}^{\prime\prime}=0\right\}  $, hence
the Lie algebra of point symmetries is isomorphic to the rank
three simple Lie algebra $\frak{sl}\left(  4,\mathbb{R}\right)  $
of dimension $15$ \cite{Anco,Ste}. The system (\ref{ret1}) also
arises as the equations of motion associated to the Lagrangian
\begin{equation}
L=\frac{1}{2\sqrt{g\left(  x\right)  }}\left(  \left(
y_{1}^{\prime}\right)
^{2}+\left(  y_{2}^{\prime}\right)  ^{2}-g\left(  x\right)  \left(  y_{1}%
^{2}+y_{2}^{2}\right)  \right)  .\label{ret2}%
\end{equation}
As for the scalar case, a point symmetry
\[
X=\xi\left(  x,y_{1},y_{2}\right)  \frac{\partial}{\partial
x}+\eta_{1}\left( x,y_{1},y_{2}\right)  \frac{\partial}{\partial
y_{1}}+\eta_{2}\left(
x,y_{1},y_{2}\right)  \frac{\partial}{\partial y_{2}}%
\]
is a Noether symmetry of (\ref{ret1}) if the constraint
(\ref{Noe3}) is satisfied for some function $V\left(
x,y_{1},y_{2}\right)  $. The constant of the motion associated to
$X$ is then given by
\begin{equation}
\psi=\xi\left[  y_{1}^{\prime}\frac{\partial L}{\partial y_{1}^{\prime}}%
+y_{2}^{\prime}\frac{\partial L}{\partial y_{2}^{\prime}}-L\right]
-\eta _{1}\frac{\partial L}{\partial
y_{1}^{\prime}}+\eta_{2}\frac{\partial
L}{\partial y_{2}^{\prime}}+V\left(  x,y_{1},y_{2}\right)  .\label{ret3}%
\end{equation}
As (\ref{ret1}) has maximal symmetry, we know that the subalgebra
of Noether symmetries must have dimension $8$ \cite{Arti}. Now,
instead of solving the symmetry condition (\ref{Noe3}), we use the
results obtained for the scalar case (\ref{GE3}) to determine the
Noether symmetries. We first consider the case of Noether
symmetries of the form
\begin{equation}
Y=\eta_{1}\left(  x,y_{1},y_{2}\right)  \frac{\partial}{\partial y_{1}}%
+\eta_{2}\left(  x,y_{1},y_{2}\right)  \frac{\partial}{\partial y_{2}%
}.\label{ret3a}%
\end{equation}
For this choice, the symmetry condition equals
\begin{align}
& \sqrt{g\left(  x\right)  }\left(  \left(  y_{1}^{\prime}\right)  ^{2}%
\frac{\partial\eta_{1}}{\partial y_{1}}+\left(
y_{2}^{\prime}\right)
^{2}\frac{\partial\eta_{2}}{\partial y_{2}}\right)  +\frac{y_{1}^{\prime}%
y_{2}^{\prime}}{\sqrt{g\left(  x\right)  }}\left(  \frac{\partial\eta_{1}%
}{\partial y_{2}}+\frac{\partial\eta_{2}}{\partial y_{1}}\right)
+\frac {1}{\sqrt{g\left(  x\right)  }}\left(
\frac{\partial\eta_{1}}{\partial
x}+\frac{\partial\eta_{2}}{\partial x}\right)  \nonumber\\
& -\left( \frac{\partial V}{\partial y_{1}}+\frac{\partial
V}{\partial y_{2}}\right)-\left( \frac{\partial V}{\partial
x}+\sqrt{g\left(  x\right) }\left(
y_{1}\,\eta_{1}+y_{2}\,\eta_{2}\right)  \right)=0. \label{ret3b}
\end{align}
From the quadratic powers in $y_{1}^{\prime}$ and $y_{2}^{\prime}$
it follows at once that
\begin{equation}
\eta_{1}\left(  x,y_{1},y_{2}\right)  =k\,y_{2}+f_{12}\left(
x\right) ,\;\eta_{2}\left(  x,y_{1},y_{2}\right)
=-k\,y_{1}+f_{13}\left(  x\right)
.\;\label{ret3c}%
\end{equation}
Inserting these expressions into (\ref{ret3b}) we further obtain
$V\left(
x,y_{1},y_{2}\right)  =\left(  y_{1}\,f_{12}\left(  x\right)  +y_{2}%
\,f_{13}\left(  x\right)  \right)  $, and the symmetry condition
reduces to
\begin{flalign}
\frac{y_{1}\left(  f_{12}^{\prime\prime}\left(  x\right)
-\frac{g^{\prime }\left(  x\right)  }{2g\left(  x\right)
}f_{12}^{\prime}\left(  x\right) +g\left(  x\right)  f_{12}\left(
x\right)  \right)  +y_{2}\left( f_{13}^{\prime\prime}\left(
x\right)  -\frac{g^{\prime}\left(  x\right) }{2g\left(  x\right)
}f_{13}^{\prime}\left(  x\right)  +g\left(  x\right)
f_{13}\left(  x\right)  \right)  }{\sqrt{g\left(  x\right)  }}=0.\label{ret3d}%
\end{flalign}
It follows at once that the functions $f_{12}(x)$ and $f_{13}(x)$
must be solutions to the ODE (\ref{GE3}).\newline This proves
that, for the special type (\ref{ret3a}) of vector fields we
obtain five independent Noether symmetries%
\begin{flalign}
Y_{1}=U\left(  x\right)  \frac{\partial}{\partial
y_{1}},\;Y_{2}=U\left( x\right)  \frac{\partial}{\partial
y_{2}},\;Y_{3}=T\left(  x\right) \frac{\partial}{\partial
y_{1}},\;Y_{4}=T\left(  x\right)  \frac{\partial
}{\partial y_{2}},\;Y_{5}=y_{2}\frac{\partial}{\partial y_{1}}-y_{1}%
\frac{\partial}{\partial y_{2}}.\label{ret3e}%
\end{flalign}
In order to obtain the three remaining symmetries, we apply the
results of the preceding sections. A routine computation shows
that the vector fields
\begin{align}
Z_{1}  & =-\frac{T\left(  x\right)  U^{\prime}\left(  x\right)
}{g\left( x\right)  }\frac{\partial}{\partial x}+\frac{\left(
T^{\prime}\left( x\right)  U^{\prime}\left(  x\right)  +2g\left(
x\right)  T\left(  x\right) U\left(  x\right)  \right)  }{g\left(
x\right)  }\left(  y_{1}\frac{\partial
}{\partial y_{1}}+y_{2}\frac{\partial}{\partial y_{2}}\right)  ,\label{ret4a}%
\\
Z_{2}  & =-\frac{U\left(  x\right)  T^{\prime}\left(  x\right)
}{g\left( x\right)  }\frac{\partial}{\partial x}+\frac{\left(
T^{\prime}\left( x\right)  U^{\prime}\left(  x\right)  +2g\left(
x\right)  T\left(  x\right) U\left(  x\right)  \right)  }{g\left(
x\right)  }\left(  y_{1}\frac{\partial }{\partial
y_{1}}+y_{2}\frac{\partial}{\partial y_{2}}\right)  \label{ret4b}
\end{align}
are Noether symmetries of (\ref{ret1}) for the function%
\begin{equation}
V\left(  x,y_{1},y_{2}\right)  =\frac{U\left(  x\right)
T^{\prime}\left( x\right)  +T\left(  x\right)  U^{\prime}\left(
x\right)  }{2\sqrt{g\left(
x\right)  }}\left(  y_{1}^{2}+y_{2}^{2}\right)  .\label{ret5}%
\end{equation}
The vector fields $Z_{1},Z_{2}$ and $Z_{3}=\left[
Z_{1},Z_{2}\right] $ are independent, and thus $\left\{
Z_{1},Z_{2},Z_{3},Y_{1},\cdots,Y_{5}\right\}  $ form a basis of
the Lie algebra $\mathcal{L}_{NS}$ of Noether symmetries of the
system (\ref{ret1}). In particular,  $\left\{
Z_{1},Z_{2},Z_{3}\right\}  $ generate a copy of $\frak{sl}\left(
2,\mathbb{R}\right)  $ isomorphic to the Levi subalgebra of
$\mathcal{L}_{NS}$. We further observe that $Y_{5}$ corresponds to
the infinitesimal generator of a rotation in the plane \cite{GKO}.

\bigskip

\noindent Now we analyze the existence of Lagrangians
\begin{equation}
L_{0}=\frac{1}{2\sqrt{g\left(  x\right)  }}\left(  \left(
y_{1}^{\prime }\right)  ^{2}+\left(  y_{2}^{\prime}\right)
^{2}-g\left(  x\right)  \left( y_{1}^{2}+y_{2}^{2}\right)
-G\left(  x,y_{1},y_{2},y_{1}^{\prime
},y_{1}^{\prime}\right)\right)  \label{ret6}%
\end{equation}
such that the vector fields $Z_{1}$ and $Z_{2}$ are Noether
symmetries. The procedure is completely analogous to that of
scalar ODEs previously considered, for which reason we omit the
detailed computations. Imposing that (\ref{Noe3}) is satisfied for
$Z_{1}$ leads, after some heavy algebraic
manipulation,\footnote{Using that $T\left( x\right)  $ and
$U\left(  x\right)  $ are solutions to the ODE (\ref{GE3}), as
well as the
constraint (\ref{rey8}).} to the PDE:%
\begin{align}
&  2T\left(  x\right)  \left(  U^{\prime}\left(  x\right)
g^{\prime}\left( x\right)  +2U\left(  x\right)  g\left(  x\right)
^{2}\right)  G\left(
x,y_{1},y_{2},y_{1}^{\prime},y_{1}^{\prime}\right)  -2T\left(
x\right) U\left(  x\right)  g\left(  x\right)  U^{\prime}\left(
x\right)
\frac{\partial G}{\partial x}\nonumber\\
&  +\left(  2g\left(  x\right)  ^{2}y_{1}\left(  T\left(  x\right)
U^{\prime }\left(  x\right)  +T^{\prime}\left(  x\right)  U\left(
x\right)  \right) -2T\left(  x\right)  \left(  U^{\prime}\left(
x\right)  g^{\prime}\left( x\right)  +2g\left(  x\right)
^{2}U\left(  x\right)  \right)  y_{1}^{\prime
}\right)  \frac{\partial G}{\partial y_{1}}\nonumber\\
&  +\left(  2g\left(  x\right)  ^{2}y_{2}\left(  T\left(  x\right)
U^{\prime }\left(  x\right)  +T^{\prime}\left(  x\right)  U\left(
x\right)  \right) -2T\left(  x\right)  \left(  U^{\prime}\left(
x\right)  g^{\prime}\left( x\right)  +2g\left(  x\right)
^{2}U\left(  x\right)  \right)  y_{2}^{\prime
}\right)  \frac{\partial G}{\partial y_{1}}\nonumber\\
&  2T\left(  x\right)  U\left(  x\right)  g\left(  x\right)
^{2}\left(
y_{1}\frac{\partial G}{\partial y_{1}}+y_{2}\frac{\partial G}{\partial y_{2}%
}-\frac{U^{\prime}\left(  x\right)  }{g\left(  x\right)
}\frac{\partial G}{\partial x}\right)=0.\label{ret7}
\end{align}

\noindent For the vector field $Z_{2}$, the corresponding PDE
satisfied by $G$ is obtained from (\ref{ret7}) permuting $T\left(
x\right) $ and $U\left( x\right)  $. In this form, however, the
equations are of little use, as all intervening functions are
unknown. We can transform the PDEs using the constraints satisfied
by $T\left(  t\right)  ,U\left(  t\right)  $ and $g\left(
t\right) $ in order to obtain an equivalent pair of differential
equations. Skipping the routine computations, such a set is given
by the equations
\begin{flalign}
2g\left(  x\right)  \frac{\partial G}{\partial
t}+y_{1}^{\prime}g^{\prime }\left(  x\right)  \frac{\partial
G}{\partial y_{1}^{\prime}}+y_{2}^{\prime
}g^{\prime}\left(  x\right)  \frac{\partial G}{\partial y_{2}^{\prime}%
}-2g^{\prime}\left(  x\right)  G\left(  x,y_{1},y_{2},y_{1}^{\prime}%
,y_{1}^{\prime}\right) & =0,\label{ret8a}\\
-2T\left(  x\right)  U\left(  x\right)  \left(
y_{1}\frac{\partial G}{\partial y_{1}}+y_{2}\frac{\partial G}{\partial y_{2}%
}-G\right) +\left(  T\left(  x\right)  U\left(  x\right)  y_{1}^{\prime}-y_{1}%
A_{0}\right)  \frac{\partial G}{\partial y_{1}^{\prime}} &
\nonumber\\+\left( T\left( x\right)  U\left( x\right)
y_{2}^{\prime}-y_{2}A_{0}\right)  \frac{\partial G}{\partial
y_{2}^{\prime}}   & =0,\label{ret8b}%
\end{flalign}
where $A_{0}=$  $T\left(  x\right)  U^{\prime}\left(  x\right)
+T^{\prime }\left(  x\right)  U\left(  x\right)  $. The first of
these equations has the general solution
\begin{equation}
G\left(  x,y_{1},y_{2},y_{1}^{\prime},y_{1}^{\prime}\right)
=g\left( x\right)  \,\Phi\left(
y_{1},y_{2},\frac{y_{1}^{\prime}}{\sqrt{g\left( x\right)
}},\frac{y_{2}^{\prime}}{\sqrt{g\left(  x\right)  }}\right)
.\label{ret9a}%
\end{equation}
Inserting the latter into equation (\ref{ret8b}) and analyzing the
terms depending on $y_{1}^{\prime},y_{2}^{\prime}$, it is not
difficult to verify that the condition
\begin{equation}
\frac{\partial\Phi}{\partial
y_{1}^{\prime}}=\frac{\partial\Phi}{\partial
y_{2}^{\prime}}=0\label{ret9b}%
\end{equation}
must be satisfied. Therefore the integrability condition reduces
to the linear first-order PDE
\begin{equation}
y_{1}\frac{\partial\Phi}{\partial
y_{1}}+y_{2}\frac{\partial\Phi}{\partial y_{2}}-\Phi\left(
y_{1},y_{2}\right)  =0\label{endk}
\end{equation}
with solution $\Phi\left(  y_{1},y_{2}\right)  =\varphi\left(  y_{2}y_{1}%
^{-1}\right)  y_{1}^{-2}$. This shows that the most general
function $G$ satisfying the system (\ref{ret8a})-(\ref{ret8b}) is
given by
\begin{equation}
G\left(  x,y_{1},y_{2},y_{1}^{\prime},y_{1}^{\prime}\right)
=\frac{g\left( x\right)  }{y_{1}^{2}}\,\varphi\left(
\frac{y_{2}}{y_{1}}\right)
.\label{ret10}%
\end{equation}

\bigskip

\begin{proposition}
For any non-zero function $\varphi\left(
\frac{y_{2}}{y_{1}}\right)  $, the
non-linear system%
\begin{align}
y_{1}^{\prime\prime}-\frac{g^{\prime}\left(  x\right)  }{2g\left(
x\right)
}y_{1}^{\prime}+g\left(  x\right)  y_{1}+\frac{g\left(  x\right)  }{y_{1}^{3}%
}\varphi\left(  \frac{y_{2}}{y_{1}}\right)  +\frac{g\left(  x\right)  y_{2}%
}{2y_{1}^{4}}\varphi^{\prime}\left(  \frac{y_{2}}{y_{1}}\right)
&
=0,\label{ret11a}\\
y_{2}^{\prime\prime}-\frac{g^{\prime}\left(  x\right)  }{2g\left(
x\right)
}y_{2}^{\prime}+g\left(  x\right)  y_{2}-\frac{g\left(  x\right)  }{2y_{1}%
^{3}}\varphi^{\prime}\left(  \frac{y_{2}}{y_{1}}\right)    & =0\label{ret11b}%
\end{align}
possesses exactly three Noether symmetries.
\end{proposition}

\bigskip

\noindent The proof is essentially the same as that of Proposition
3 for the scalar case. The system corresponds to the equations of
motion associated to the Lagrangian
\begin{equation}
L_{0}=\frac{1}{2\sqrt{g\left(  x\right)  }}\left(  \left(
y_{1}^{\prime }\right)  ^{2}+\left(  y_{2}^{\prime}\right)
^{2}-g\left(  x\right)  \left(
y_{1}^{2}+y_{2}^{2}\right)  -\frac{g\left(  x\right)  }{y_{1}^{2}}%
\,\varphi\left(  \frac{y_{2}}{y_{1}}\right)  \right)  \label{ret12}%
\end{equation}
Successive reduction of the Noether symmetry condition
(\ref{Noe3}) shows that any such symmetry has the components
\begin{align*}
\xi\left(  x,y_{1},y_{2}\right)    & =\phi\left(  x\right)  ,\\
\eta_{1}\left(  x,y_{1},y_{2}\right)    & =\left( \frac{\phi\left(
x\right) g^{\prime}\left(  x\right)  }{4g\left( x\right)
}+\frac{\phi^{\prime}\left(
x\right)  }{2}\right)  y_{1},\\
\eta_{2}\left(  x,y_{1},y_{2}\right)    & =\left( \frac{\phi\left(
x\right) g^{\prime}\left(  x\right)  }{4g\left( x\right)
}+\frac{\phi^{\prime}\left( x\right)  }{2}\right) y_{2},
\end{align*}
where the function $\phi\left(  x\right)  $ satisfies the
third-order equation (\ref{rey41}). As the solutions to the latter
correspond exactly to the symmetries $Z_{1},Z_{2}$ and $Z_{3}$
preserved by the deformation, we conclude that the dimension of
the Noether symmetry algebra is three, hence it must be isomorphic
to $\frak{sl}(2,\mathbb{R})$. As a consequence, the system
(\ref{ret11a})-(\ref{ret11b}) cannot be linearizable.

\bigskip

\noindent As happened in the scalar case, the integration of the
deformed system is quite cumbersome in spite of the two constants
of the motion, as these contain explicitly the independent
variable $x$ and their expression differs from being an easy one.

\bigskip

\noindent As an example to illustrate one of these deformed
systems, we consider the auxiliary functions
$g(x)=\frac{49}{1-x^{2}}$ and $\varphi(\frac{y_{2}}{y_{1}})=\alpha
\frac{y_{2}^{2}}{y_{1}^{2}}$. We are thus deforming the uncoupled
system consisting of two Chebyshev equations. For the chosen
forcing term, the deformed system equals
\begin{align}
y_{1}^{\prime\prime}-\frac{x}{\left(1-
x^{2}\right)}y_{1}^{\prime}+\frac{49}{\left(1-x^2\right)}y_{1}-\frac{14\alpha
y_{2}^{2}}{\sqrt{x^2-1}y_{1}^{5}}
=0,\label{ret14a}\\
y_{2}^{\prime\prime}-\frac{x}{\left(1-
x^{2}\right)}y_{2}^{\prime}+\frac{49}{\left(1-x^2\right)}y_{2}+\frac{7\alpha
y_{2}}{\sqrt{x^2-1}y_{1}^{4}} =0.\label{ret14b}
\end{align}

\noindent Solving numerically the system for $\alpha=0.21$ and the
initial conditions $y_{1}(0)=-1$, $y_{1}^{\prime}(0)=0$,
$y_{2}(0)=0$, $y_{2}^{\prime}(0)=-7$,\footnote{For $\alpha=0$ the
solutions are clearly the Chebyshev polynomials
$y_{1}(t)=U_{7}(t)$, $y_{2}=T_{7}(t)$.} the solutions $y_{1}(x)$
and $y_{2}(x)$ give rise to the following graphical
representation:

\begin{figure}[h!]
\centering
\includegraphics[scale=0.75]{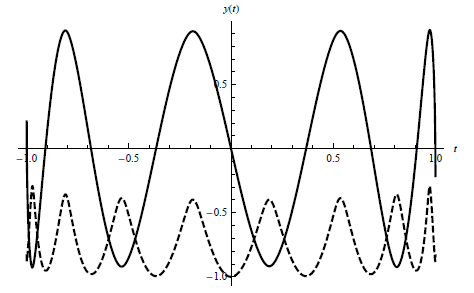}
\caption{Solutions $y_{1}(x)$ (dashed) and $y_{2}(x)$  for the
system (\ref{ret14a})-(\ref{ret14b}).}
\end{figure}

\noindent Both solutions are approximatively oscillations, however
with varying frequency. The resulting trajectory in the plane
$\left\{y_{1},y_{2}\right\}$ has a relatively complicated
structure, as shows the following plot.

\begin{figure}[h!]
\centering
\includegraphics[scale=0.75]{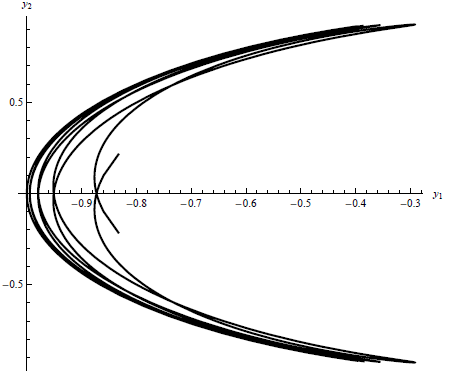}
\caption{Plane trajectory of the solutions $\left(y_{1}(x),
y_{2}(x)\right)$ of system (\ref{ret14a})-(\ref{ret14b}).}
\end{figure}

\newpage


\section*{CV abreviado}

A continuación presentamos una relación de los trabajos publicados, enviados a publicación y en proceso de ejecución relacionados con la temática de esta memoria. Asimismo se indican las conferencias y comunicaciones presentadas con relación a esta temática.

\medskip

\subsection*{Publicaciones}

Campoamor-Stursberg, R. and  Fern\'andez-Saiz, E. (2015). Realizations of sl(3;R) in Terms of Chebyshev Polynomials and Orthogonal Systems of Functions. Symmetry Breaking and Variational Symmetries. {\em In book: Ordinary and Partial Differential Equations}. Chapter: 3. Nova Science Publishers.

\vspace{0.5cm}

Ballesteros, A.; Campoamor-Stursberg, R.: Fernández-Saiz, E.; Herranz, F. J. and de Lucas, J. (2018). Poisson–Hopf algebra deformations of Lie–Hamilton systems. {\em Journal of Physics A: Mathematical and Theoretical}, {\bf 51}(6), 065202.

 \vspace{0.5cm}
 
Ballesteros, A.; Campoamor-Stursberg, R.;  Fern\'andez-Saiz, E.; Herranz, F. J. and de Lucas, J. (2018). 
A unified approach to Poisson-Hopf deformations of Lie-Hamilton systems based on sl(2). {\em Dobrev, V. (ed.). Quantum Theory and Symmetries with Lie Theory and Its Applications in Physics} Volume 1, 2018. pp. 347--366.

\vspace{0.5cm}

Esen, O.; Fernández-Saiz, E.; Sardón, C. and  Zając, M. (2020). Geometry and solutions of an epidemic SIS model permitting fluctuations and quantization, submitted  {\em arXiv preprint 	arXiv:2008.02484 [q-bio.PE]}.

\vspace{0.5cm}

Ballesteros, A.; Campoamor-Stursberg, R.: Fernández-Saiz, E.; Herranz, F. J. and de Lucas, J. A refinement of Poisson--Hopf deformations of Lie--Hamilton systems: The setup of
explicit deformed superposition rules and applications to the oscillator algebra. {\em In progress}.

\vspace{0.5cm}

Campoamor-Stursberg, R.: Fernández-Saiz, E.; Herranz, F. J. and Sardon, C.  An alternative epidemic SIS model based on the Poincaré algebra $\mathfrak{h}_4$. {\em In progress}.

 \medskip

\subsection*{Conferencias y comunicaciones}

Fernández--Saiz, E. (2016). Realizations of $\mathfrak{sl}(3;\mathbb{R})$ in terms of Chebyshev polynomials. Poster session in {\em 10th International Summer School on Geometry, Mechanics, and Control}, in Madrid.

 \vspace{0.5cm}
 
 Fernández--Saiz, E. (2016). Lie symmetries and differential equations: Harrison-Estabrook method. Seminar in Burgos

\vspace{0.5cm}

Ballesteros, A.; Campoamor-Stursberg, R.: Fernández-Saiz, E.; Herranz, F. J. and de Lucas, J. (2017). Quantum algebras in Lie–Hamilton systems. Poster session in {\em 11th International Young Researcher Workshop on Geometry, Mechanics, and Control}, in La Laguna.

\vspace{0.5cm}

Ballesteros, A.; Campoamor-Stursberg, R.: Fernández-Saiz, E.; Herranz, F. J. and de Lucas, J. (2017). Quantum algebras in Lie–Hamilton systems: oscilator system. Poster session in {\em V Iberoamerican Meeting on Geometry, Mechanics and Control}, in La Laguna (Tenerife).

\vspace{0.5cm}

 Campoamor-Stursberg, R. and Fernández--Saiz, E. (2017). Realizations of sl(3,R): Chebyshev polynomials. Poster session in {\em A Workshop to honour Prof. Orlando Ragnisco in his 70th anniversary}, in Burgos.

 \vspace{0.5cm}
 
 Ballesteros, A.; Campoamor-Stursberg, R.: Fernández-Saiz, E.; Herranz, F. J. and de Lucas, J. (2017). Poisson--Hopf algebra deformation of Lie systems. Talk in {\em X. International Symposium Quantum Theory and Symmetries \& 12-th edition of the International Workshop "Lie Theory and Its Applications in Physics"}, in Varna.
 
 \vspace{0.5cm}

Fernández--Saiz, E. (2018). Poisson-Hopf algebra deformations of Lie systems. Scientific Divulgation Seminar in Madrid

\vspace{0.5cm}
 
Ballesteros, A.; Campoamor-Stursberg, R.: Fernández-Saiz, E.; Herranz, F. J. and de Lucas, J. (2018). Poisson--Hopf algebra deformation of Lie systems. Talk in {\em 13th Young Researchers Workshop}, in Madrid.

\vspace{0.5cm}

Fernández--Saiz, E. (2018). Sistemas de Lie y sus aplicaciones. Seminar (Red de Doctorandos en Matemáticas UCM), in Madrid.

\vspace{0.5cm}

Fernández--Saiz, E. (2018). Estructuras de Poisson: un billete de ida y vuelta a la mecánica cuántica. Scientific Divulgation Seminar ({\em 3ª Jornada PhDay Complutense - Ciencias Matemáticas}), in Madrid.

  \medskip

\subsection*{Participaci\'on en organizaci\'on de congresos}

Fernández--Saiz, E. (2016) in the Organizing Committee of {\em X Workshop of Young Researchers in Mathematics}, in Madrid.

\vspace{0.5cm}

Fernández--Saiz, E. (2017) in the Organizing Committee of {\em  Nonlinear Integrable Systems}, in Burgos.



\end{document}